\documentclass[twocolumn]{aastex62}
\usepackage{color}
\usepackage{soul} 
\usepackage{url}
\usepackage{rotating}

\newcommand{\red}{\textcolor{red}}

\newcommand{\nmodelsnospec}{{\red{29}}}
\newcommand{\nclustersnospec}{{\red{19}}}

\newcommand{\lenstool}{{\tt{Lenstool}}}

\newcommand{\Lenstool}{{\tt{Lenstool}}}

\newcommand{\hst}{{\it HST}}
\newcommand{\HST}{{\it HST}}
\newcommand{\JWST}{{\it JWST}}
\newcommand{\Nmodels}{$110$}
\newcommand{\Nclusters}{$74$} 
\newcommand{\NRELICSlenstool}{$22$}
\newcommand{\NRELICSLTM}{$10$}
\newcommand{\NRELICSGLAFIC}{$21$}
\newcommand{\NSGASLENSTOOL}{$35$}

\newcommand{\Lstrength}{$\mathcal{A}_{|\mu|\geq 3}^{lens}$}
\newcommand{\LstrengthN}{$\mathcal{A}_{|\mu|\geq 3}^{0.5}$}
\newcommand{\Lstren}{$\mathcal{A}_{|\mu|\geq 3}$} 
\newcommand{\Earea}{$\pi\theta_E^2$}
\newcommand{\MSL}{$M_{\rm SL}(200 {\rm kpc})$}
\newcommand{\Slope}{$S_{50-200}$}
\usepackage{comment}

\DeclareUnicodeCharacter{2212}{-}
\usepackage{graphicx}

\graphicspath{{./}{figures/}}

\received{--}
\revised{--}
\accepted{--}
\submitjournal{ApJ}

\shorttitle{Lensing Strength}
\shortauthors{Fox et al.}

\begin{document}

\title{The Strongest Cluster Lenses: An Analysis of the Relation Between Strong Gravitational Lensing Strength and the Physical Properties of Galaxy Clusters}

\correspondingauthor{Carter Fox}
\email{carfox@umich.edu}

\author[0000-0001-8316-9482]{Carter Fox}
\author[0000-0003-3266-2001]{Guillaume Mahler}
\author[0000-0002-7559-0864]{Keren Sharon}
\author[0000-0002-7868-9827]{Juan D. Remolina Gonz\'alez}
\affiliation{Department of Astronomy, University of Michigan, 1085 South University Ave, Ann Arbor, MI 48109, USA}



\begin{abstract}

Strong gravitational lensing provides unique opportunities to investigate the mass distribution at the cores of galaxy clusters and to study high redshift galaxies. Using \Nmodels\ strong lensing models of \Nclusters\ cluster fields from the \textit{Hubble} Frontier Fields (HFF), Reionization Lensing Cluster Survey (RELICS), and Sloan Giant Arcs Survey (SGAS), 
we evaluate the lensing strength of each cluster (area with $|\mu|\geq3$ for $z_s=9$, normalized to a lens redshift of $z=0.5$). We assess how large scale mass, projected inner core mass, and the inner slope of the projected mass density profile relate to lensing strength. While we do identify a possible trend between lensing strength and large scale mass (Kendall $\tau=0.26$ and Spearman $r=0.36$), we find that the inner slope ($50$ kpc~$\leq r \leq~200$ kpc) of the projected mass density profile has a higher probability of correlation with lensing strength and can set an upper bound on the possible lensing strength of a cluster (Kendall $\tau=0.53$ and Spearman $r=0.71$). As anticipated, we find that the lensing strength correlates with the effective Einstein area and a large ($\gtrsim 30\farcs0$) radial extent of lensing evidence is a strong indicator of a powerful lens. We attribute the spread in the {relation} to the complexity of individual lensing clusters, which is well captured by the lensing strength estimator. These results can help to more efficiently design future observations to use clusters as cosmic telescopes.

\end{abstract}



\section{Introduction} \label{sec:intro}
Clusters of galaxies are located at the nodes of the cosmic filaments and represent the densest structures of dark matter; their merging history and evolution shape the properties of their mass distribution. The density profiles of massive galaxy clusters show strong self-similarity up to z $\sim$ 2 outside of cluster cores \citep{2017ApJ...843...28M}, but depart from this self-similarity within cluster cores. These high-density core regions are also where evidence of strong lensing is observed in clusters above a characteristic density, in the form of multiple instances of magnified and distorted images of lensed galaxies -- dubbed arcs, or giant arcs. 

Strong lensing galaxy clusters can play a key role in our understanding of the Universe, from mapping the distribution of matter in its densest regions, to the study of the earliest galaxies that they magnify.

An apparent discrepancy between the observed and expected number of giant arcs, known as the ``arc statistics problem'', has motivated studies of the correlation between strong lensing efficiency, cosmology, and cluster properties (see \citealt{2013SSRv..177...31M} for a review). Cosmological analyses reveal that the number of giant arcs is sensitive to $\Omega_m$ and $\sigma_8$ (\citealt{2008ApJ...676..753W}; \citealt{2006MNRAS.372L..73L}; \citealt{2008A&A...486...35F}; \citealt{2009MNRAS.392..930O}; \citealt{2016MNRAS.457.2738B}), self-interacting dark matter cross section (\citealt{2001MNRAS.325..435M}; \citealt{2001ApJ...555..504W}), primordial non-Gaussianity (\citealt{2009MNRAS.392..930O}; \citealt{2011MNRAS.415.1913D}), and dark energy (\citealt{2003A&A...409..449B}; \citealt{2005A&A...442..413M}). 
\cite{2012MNRAS.421.3343G} found that halo concentration is the most important factor for the lensing cross sections for production of large arcs, with more concentrated halos having larger strong lensing cross sections.
\cite{2016ApJ...817...85X} claim that the cluster mass-concentration relation could be the main driver of the lensing efficiency. Observations of strong lensing clusters \citep{2003ApJ...593...48G} suggest that they might belong to a subpopulation of clusters with especially high lensing cross sections, which could explain the excess of observed arcs. \citet{2020ApJ...889..189S} found little to no correlation between total cluster mass and the number of lensed high-$z$ ($z\sim6-8$) candidates, across a large survey of massive clusters. 
Evidence is therefore emerging that the properties that contribute to strong lensing cross section are both numerous and complex.

By leveraging the ability of lensing clusters to magnify the background Universe,
the search for high redshift galaxies has become a particular interest in recent years (for a broad review see \citealt{2011A&ARv..19...47K}). It is 
the main science goal of projects like \textit{Hubble} Frontier Fields (HFF; \citealt{2017ApJ...837...97L}) and the Reionization Lensing Cluster Survey (RELICS; \citealt{2019ApJ...884...85C}), and of particular interest in preparation for the \textit{James Webb Space Telescope} (\JWST). It is therefore imperative to identify predictors of lensing strength, to increase our efficiency in selecting the best cosmic telescopes for followup. 

Surveys to identify, catalog, and study the lines of sight of strong lensing clusters -- either for cluster science or for their use as cosmic telescopes -- employ two major discovery approaches: lensing-selected and non-lensing selected. Lensing-selected surveys rely on inspection of relatively shallow, large data sets, to identify instances of strong lensing evidence (e.g., SDSS, \citealt{2020ApJS..247...12S}; Dark Energy Survey (DES), \citealt{2017ApJS..232...15D}), usually in the form of highly magnified bright giant arcs. The second approach is to examine deep or high-resolution optical imaging data of galaxy clusters that were selected for follow-up observation based on some criteria, usually high total mass as indicated from optical, X-ray, or sub-mm mass proxies. 
Examples of such surveys include the follow-up of the MAssive Cluster Survey (MACS; \citealt{2001ApJ...553..668E}), the South Pole Telescope clusters (\HST\ SNAP 16017; PI: Gladders), the Local Cluster Substructure Survey (LoCuSS; \citealt{2005MNRAS.359..417S}), and the Red-Sequence Cluster Survey (RCS; \citealt{2005ApJS..157....1G}). 
Both approaches can use either manual or machine-assisted identification of lensing evidence.

In this paper we investigate the relation between physical cluster properties and lensing strength, in order to assess some of the commonly used indicators, and identify new indicators of strong lensing efficiency. We base our investigation on a large sample of observed strong lensing clusters at a wide range of redshifts and masses. Our study combines a total of \Nmodels\ strong lensing models of \Nclusters\ cluster fields. Discerning the most important cluster properties for lensing efficiency enhances our understanding of galaxy clusters, and aids in the design of future surveys for cluster lenses. 

This paper is organized as follows. In the remainder of the introduction, we review previous work related to proxies of lensing efficiency.  In Section \ref{sec:sample}, we describe the sample of strong lensing models that we use for our analysis. In Section \ref{sec:methods}, we discuss our methods, define the lensing strength indicator and the cluster properties used in our analysis (large scale mass, core mass, inner slope, effective Einstein area, and the distance to the farthest bright arc).
In Section \ref{results_discussion}, we present and discuss our results. We conclude in Section \ref{sec:summary}.
Throughout this paper, we adopt a standard $\Lambda$-CDM cosmology with $\Omega_m =0.3$, $\Omega_{\Lambda}=0.7$, and $h=0.7$ for models based on \Lenstool, Light-Traces-Mass, GRALE, and GRAVLENS methods. GLAFIC-based models of clusters included in the RELICS program use the parameters found by the Wilkinson Microwave Anisotropy Probe
\citep{2011ApJS..192...18K}, with $\Omega_m =0.272$, $\Omega_{\Lambda}=0.728$, and $h=0.704$. We refer the interested reader to \citet{2015ApJ...802L...9B} for an assessment of the impact of cosmological parameter uncertainties on mass and magnification errors. $M_{\Delta}$ refers to the 3D mass within the radius at which the average density is $\Delta$ times the critical density of the Universe at the redshift of the cluster. 

\subsection{Contributors to Strong Lensing Efficiency} \label{sec:compare_prev_study}

Numerous studies endeavored to identify key lens characteristics that influence strong lensing efficiency, usually coded as the abundance of lenses, or the appearance, distributions, or properties of giant arcs. 

There is a general agreement in the literature that for a given total cluster mass, strong lenses are expected to be more highly concentrated \citep[e.g.,][]{2013SSRv..177...31M,1995ApJ...438..514M,2001ApJ...559..544M,2001ApJ...559..572O,2007A&A...473..715F,2011ApJ...737...74G,2012MNRAS.421.3343G,2016MNRAS.460.4453S,2001ApJ...559..572O}. 
The slope of the mass distribution affects not only the abundance of strong lensing features, but also the lensing configuration; flat inner profiles are found to produce images that are less distorted \citep{1998ApJ...508L..47F,1998MNRAS.294..299W,2015ApJ...805..184M}.

The geometry of the lensing halo also plays a role in increasing the strong lensing power. 
\cite{2003MNRAS.340..105M,2007A&A...461...25M} show that the ellipticty of the lens affects its lensing efficiency, and might have higher impact than asymmetries and substructures to boost the lensing signal.
\citet{2007ApJ...654..714H} report that triaxiality increases the number of giant arcs, compared to spherically symmetric halos that fail to reproduce the lensing signal \citep{2003astro.ph.12072D,2004A&A...418..413B}. 
Studies comparing simulations and observations \citep{2004ApJ...609...50D, 2005ApJ...633..768H} show that realistic triaxiality and substructure can reproduce the observed lensing signal at low redshift clusters, claiming that the discrepancy at higher redshift is due to the redshift dependence of the cluster mass function.
Moreover, an alignment of the major axis of a triaxial halo along the line-of-sight would increase the projected mass slope of the inner profile \citep{2003MNRAS.340..105M,2007A&A...461...25M} and thus increase the strong lensing efficiency.

Although the dark matter component dominates the cluster mass, its galaxies and baryonic components have a non-negligible effect, directly or indirectly, on its lensing cross section. 
Studying the cluster stellar component, \citealt{2008MNRAS.386.1845H} show that a higher level of substructure increases lensing efficiency. 
\citet{2005APh....24..257H} show in simulated clusters that the lensing cross section increases with the mass of the central cluster galaxy.
However, these cluster properties are not independent of each other, and may work in opposite directions: for example, \citet{2010MNRAS.404..325R} present observational evidence that as the central galaxy grows, the mass fraction of substructures and their stellar mass content diminishes. 

Physical processes, such as baryonic cooling, influence the mass profile  \citep{2008ApJ...687...22R,2008ApJ...676..753W,2018MNRAS.473.1736K} and have a potential to indirectly affect the lensing potential. 
\citet{2008MNRAS.386.1845H} found that adding stellar component to dark matter simulation can boost the lensing signal, up to a factor of 2 for $10^{12}-10^{13}$M$_{\odot}$ dark matter halos. Furthermore,
\cite{2012MNRAS.427..533K} and \cite{2015MNRAS.452..343S,2015MNRAS.454.2277S} found using simulation that active galactic nuclei feedback and baryonic physics at the core of the cluster could flatten the profile, effectively reducing strong lensing efficiency, predominantly for the lowest mass clusters.
Nevertheless, \cite{2013ApJ...772...24B} found no significant difference in the cooling signatures of the intracluster medium between lensing-selected and non-lensing-selected clusters. 

Large scale dynamical processes, such as cluster mergers, also influence lensing efficiency. The Einstein radii are larger in mergers, especially when the two merging cores are separated by $d<0.3 \mathrm{Mpc}$ \citep{2012A&A...547A..66R}.
This claim is confirmed by \citet{2004MNRAS.349..476T}, who demonstrate that the cross section for giant arcs can increase by an order of magnitude during the merging process, and that radial arcs are more common when substructure crosses the main cluster center. 

Several studies highlighted aspects that may artificially affect the measured abundance of lensing incidents, and care should be taken when comparing observed lensing abundance and lens properties to theoretical predictions. 
Shape-based identification of arcs is sensitive to the distribution of intrinsic source shapes (e.g., \citealt{2003ApJ...584..691Z}), and to  seeing \citep{2001AJ....121...10C}, which circularizes the lensed images. 
The intrinsic morphology of the lensed galaxies can affect their detection efficiency in a way that may be redshift dependent. 
For example, \citet{2003ApJ...593...48G} suggests that the clumpiness nature of sources at higher redshift might help to better identify lensing features in cluster observations and explain some of the discrepancy observed at higher redshifts between observations and theoretical predictions. 
Structure along the line-of-sight can boost the lensing cross section, decoupled from the properties of the main lensing plane.
For example, \cite{2014ApJ...783...41B} found excess structure along the line-of-sight to strong lensing clusters.  

This paper builds upon these previous investigations. We focus our analysis on clusters that are already known to be strong lenses, with a wide range of total masses and redshifts. By selection, the projected mass density at the cores of these clusters exceeds the critical density for strong lensing; and they are likely drawn from the higher end of the concentration distribution for their mass. 
We define a lensing strength indicator that may be less relevant to studies of arcs statistics, but instead, is directly applicable to surveys that utilize strong lensing clusters as cosmic telescopes. 

\section{Strong Lensing Models} \label{sec:sample}

This paper investigates how lensing strength relates to physical properties of galaxy clusters. We gather a sample of strong lensing clusters with publicly available models
from the HFF \citep{2017ApJ...837...97L}, RELICS \citep{2019ApJ...884...85C}, and Sloan Giant Arcs Survey (SGAS; \citealt{2020ApJS..247...12S}). The final sample, which consists of \Nclusters\ cluster fields with a total of \Nmodels\ lens models, spans a wide range in mass and redshift. Other repositories of strong lensing clusters with public lens models are available in the literature. For example, the Cluster Lensing And Supernova survey with Hubble (CLASH) \citep{2012ApJS..199...25P}, Local Cluster Substructure Survey (LoCuSS; \citealt{2005MNRAS.359..417S}, \citealt{2010MNRAS.404..325R}), and can be used in future work to complement this analysis.
The four HFF clusters that are part of the CLASH sample are included in this work.

Table~\ref{tab:all_clusts_table} lists the cluster fields used in this analysis and their relevant information. Figure \ref{fig:mass_vs_z} shows their mass-redshift distribution and in the Appendix we provide a gallery of color images for each cluster. 
We discuss the surveys from which the clusters we include in our sample were drawn and the publicly available lens models below. 
 
Strong lens modeling algorithms solve for the mass distribution in the lens plane, based on lensing evidence and modeling assumptions. The parametric approach models a cluster lens with a few large scale components that represent one or more cluster-scale dark matter halos, combined with small scale components that use mass-luminosity relations to model galaxy-scale halos. These halos are usually physically-motivated mass profiles, such as Navarro-Frenk-White (NFW; \citealt{1997ApJ...490..493NFW}) or isothermal. ``Free-form'' models (also known as non-parametric) solve for the mass distribution with no underlying assumption on its radial dependence. Modeling algorithms also vary in their assumption of correlation between mass and light. Several modeling methods exist in the literature; we refer the interested reader to a recent comparison by \citet{2017MNRAS.472.3177M}. 
This paper uses models computed with several algorithms, for which public models are available: \Lenstool\ (\citealt{2007NJPh....9..447J}; \citealt{1996ApJ...471..643K}; \citealt{2009MNRAS.395.1319J}), Light-Traces-Mass (LTM; \citealt{2009MNRAS.396.1985Z}), GLAFIC \citep{2010PASJ...62.1017O}, GRALE (\citealt{2006MNRAS.367.1209L,2014MNRAS.439.2651M}), and GRAVLENS \citep{2001astro.ph..2340K}.

In some cases, we include in our analysis multiple models of the same cluster that were computed with different modeling algorithms. This approach may highlight systematic uncertainties that are not captured in the statistical modeling uncertainty of each model. We find that despite the differences between models, the results do not contradict the overall findings. 
None of the lens models used in this work include multi-plane lensing to account for possible structure along the line of sight. We note that while structures along the line-of-sight as well as difference in the choice of cosmology can influence the lensing potential, these perturbations only marginally affect our results.

\begin{figure}
    \centering
    \includegraphics[width=\linewidth]{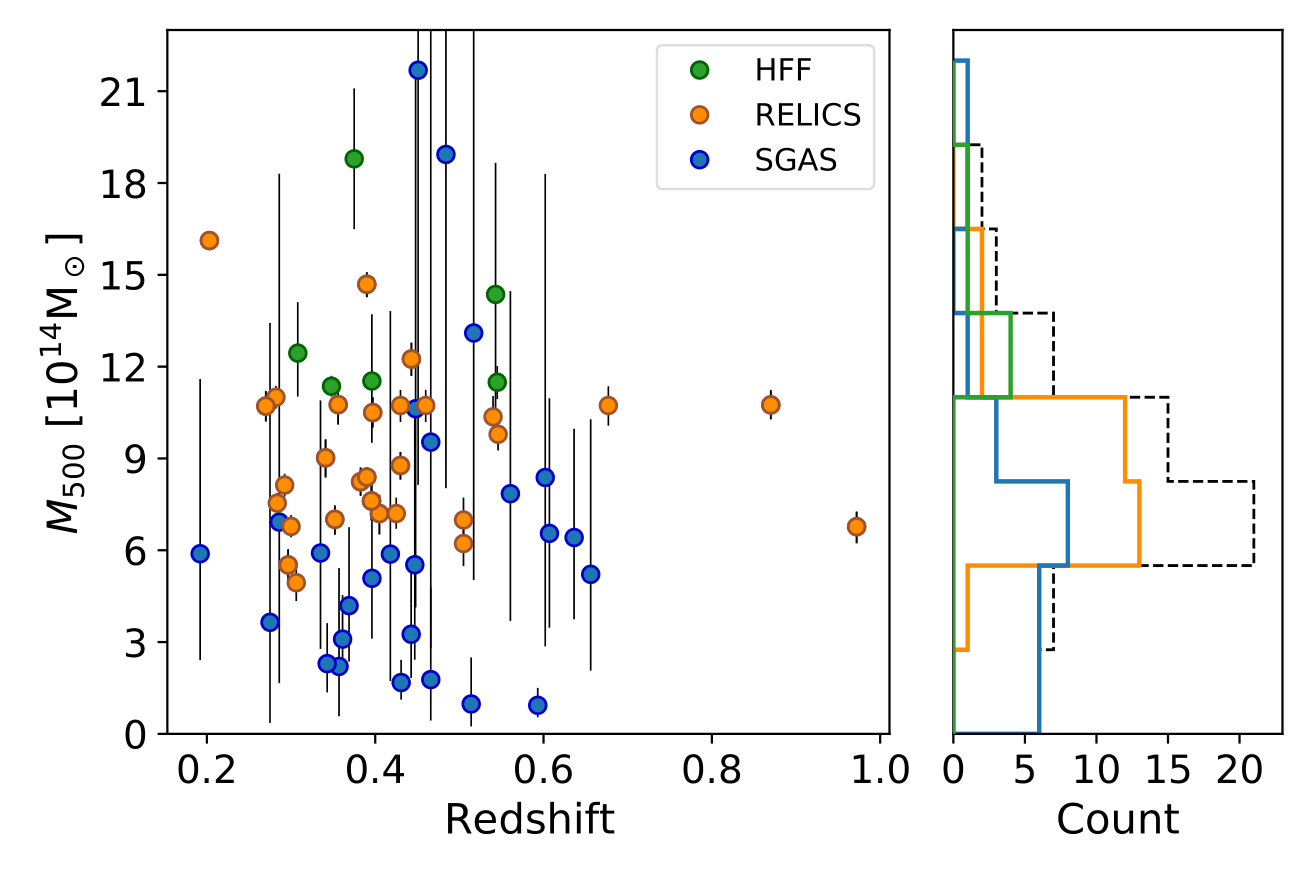}
    \caption{The mass vs. redshift distribution of the clusters included in this work, separated by the three surveys from which the sample was assembled. See Section \ref{sec:totalmass} for description of the $M_{500}$ estimates, which are measured independent of the lensing analyses.}
    \label{fig:mass_vs_z}
\end{figure}

\subsection{Frontier Fields} \label{sec:sample_hff}
The \textit{Hubble} Frontier Fields (HFF; \citealt{2017ApJ...837...97L}) is a Director's Discretionary time campaign, using the \textit{Hubble Space Telescope (HST)} and the \textit{Spitzer Space Telescope}, to obtain deep imaging data of six strong-lensing galaxy clusters. The science goal of this project is to study the high-$z$ Universe, by combining the resolution power of \hst\ and strong lensing magnification in lines of sight of six massive clusters of galaxies. The strategy, of extremely deep multiband observations of each field, enables a high-completeness detection of cluster member galaxies and faint background sources. It resulted in the best constrained strong lensing lines of sight to date, with hundreds of lensing constraints per field. Through extensive community effort, the strong lensing potentials of these fields were modeled by several teams using different modeling approaches. The lens models are publicly available to the community on the Mikulski Archive for Space Telescopes (MAST)\footnote{https://doi.org/10.17909/T9KK5N}.
We use in our analysis public HFF lens models that satisfy the following criteria: the model is available on MAST as v4 or above (but see a comment below); provides a field of view large enough to encompass the region of interest for our work, where the magnification is $|\mu|\geq3$ (see Section \ref{sec:lensingstrength}); and the maps provided have a resolution of $0\farcs4$ pixel$^{-1}$ or better.
The models used in this work (see Table~\ref{tab:all_clusts_table}) are described in \citet{2014ApJ...797...48J,2016MNRAS.457.2029Jcats1149,cam,2019MNRAS.485.3738Lcatsabell370,2016ApJ...819..114K,2018ApJ...855....4K,2020MNRAS.492..503RKeetonmodels}. We use the v4.1 (v4 when not available) lens models in all cases except for Abell~2744, where we use Sharon v3 and CATS v3.1, which are confined to the main cluster core and do not extend to the Northwest structure (See \citealt{2018MNRAS.473..663M}). The largest differences usually appear in the lensing strength values, which is expected. \citet{2017MNRAS.472.3177M} show that for highly constrained clusters, such as the HFF clusters, the surface mass densities and magnifications show strong similarity across the different modeling techniques. Similarly, \citet{2018ApJ...863...60R} finds excellent agreement between the v4 models of one of the HFF clusters. \citet{2020MNRAS.494.4771R} highlights a 6\% bias and 40\% scatter at $|\mu|=3$ that points to underestimation of statistical uncertainties of the HFF models. 

\subsection{RELICS} \label{sec:sample_relics}
The Reionization Lensing Cluster Survey (RELICS; \citealt{2019ApJ...884...85C}) observed galaxy cluster fields, taking advantage of lensing magnification in order to detect galaxies at high redshifts ($z\gtrsim6$). While its main science objective was similar to that of the HFF project, RELICS used a complementary strategy of a shallow survey with a large number of lines of sight to achieve its goals.  
RELICS observed 41 clusters with \hst, over half of which were selected from the \textit{Planck} catalog for their high mass, based on an $M_{500}$ estimate derived from the Sunyaev-Zel'dovich effect (SZ; \citealt{1970Ap&SS...7....3S}). Known lensing features assisted the selection of the remaining clusters, as well as other mass estimates, such as X-ray and weak lensing \citep{2019ApJ...884...85C}. 

Clusters observed in the RELICS program have been modeled by the RELICS team with three parametric modeling techniques: \Lenstool\ \citep{2007NJPh....9..447J}, LTM \citep{2009MNRAS.396.1985Z} and GLAFIC \citep{2010PASJ...62.1017O}. The lens models are made publicly available to the community on MAST \footnote{https://doi.org/10.17909/T9SP45}.  
Not all the lines of sight in the RELICS program resulted in successful lens models. This is primarily due to the lack of identified lensing constraints in some of the clusters. 
RELICS models are typically based on a handful of lensed galaxies -- compared to hundreds for HFF. This may result in under-constrained models (e.g., \citealt{2018ApJ...859..159C}). 
We include in our analysis all publicly available models that were either published, or satisfy the following criteria:
the model-predicted lensed images are consistent with the observed lensing evidence to within $1\farcs5$; and the model does not produce major mass components or critical curves that protrude into regions with no constraints or a physical motivation. 
This resulted in all the available \lenstool\ models (\NRELICSlenstool), all the available LTM models, (\NRELICSLTM), and most of the GLAFIC models (\NRELICSGLAFIC). Several clusters were modeled by more than one method, which helps in indicating systematic uncertainties that are due to modeling approaches. However, it is beyond the scope of this paper to conduct a thorough investigation of the differences that may exist between different models of the same cluster. Some examples of these comparisons between the different RELICS lens models can be found in \citet{2019ApJ...874..132A,2020ApJ...898....6A}.

Three of the GLAFIC lens models did not have a large enough field-of-view to capture all the area of moderate magnification, $|\mu| \geq 3$, which is required for our analysis (see section \ref{sec:lensingstrength}). For these models, we report the measured value, but add an additional 5\% to the upper error bar, based on a power law extrapolation of the best model magnification maps. We mark these models in Table~\ref{tab:all_clusts_table} and indicate them with a dashed black border in plots including lensing strength. Lastly, Abell~1758a has two major components (Northwest and Southeast). Although contained in one model, the amplification areas of the lensing strength we compute (see Section \ref{sec:lensingstrength}) for each component do not intersect. We therefore split the model into Northwest and Southeast components.

\subsection{SGAS}
The science goal of the Sloan Giant Arcs Survey (SGAS, \citealt{2020ApJS..247...12S}) is to study the star formation history and physical properties of highly-magnified galaxies at cosmic noon ($1\gtrsim z \gtrsim 3$), using the magnification of strong lensing clusters to both amplify the light from these galaxies, and to resolve individual star forming regions. The SGAS clusters were selected from the Sloan Digital Sky Survey (SDSS; \citealt{2017AJ....154...28B}) based on the identification of strong lensing evidence  (\citealt{2011ApJS..193....8B}; \citealt{2020ApJS..247...12S}). This sample extends to lower cluster masses than RELICS and HFF. We use \NSGASLENSTOOL\ public \Lenstool\ models of SGAS clusters from \citet{2020ApJS..247...12S}, excluding SDSS~J1002$+$2031 and SDSS~J1527$+$0652, which were categorized by \citet{2020ApJS..247...12S} as being too poorly constrained (see \citealt{2020ApJS..247...12S} for more details). 

\section{Methods}

In this section, we define the lensing strength indicator (\LstrengthN) and the three cluster properties investigated in this work as predictors of lensing strength: large scale mass ($M_{500}$), strong lensing projected core mass ($M_{\rm SL, core}$), and the inner slope ($S_{50-200}$) of the projected mass density profile. We also determine the effective Einstein radius ($e\theta_E$), often used to judge lensing strength, for comparison with these other cluster properties and our lensing strength indicator. Lastly, for SGAS clusters and suitable RELICS and HFF clusters, we measure the distance between the BCG and the farthest bright arc, a common indicator of lensing strength from imaging data without a lens model, to compare with our lensing strength measurement. 

Each publicly available lens model was provided with a range of mass maps that sample the posterior probability distribution of the lens model, usually $\sim$100 maps, for computing uncertainties. For all quantities we compute from the lens models, statistical uncertainties are given as the interval that contains $68\%$ of the values computed from these ``range'' models. 

\label{sec:methods}
\subsection{Lensing Strength} \label{sec:lensingstrength}
The magnification ($\mu$) of a point in the image plane is the factor by which a source is magnified as a result of its light passing through the lens plane. The magnification is defined as follows,

\begin{equation}
    \mu \equiv ((1-\kappa)^2 - \gamma^2)^{-1},
\end{equation}
where $\kappa$ and $\gamma$ are the convergence and shear of the lensing potential, respectively. 

Several authors have used a definition of ``lensing strength'' as the image-plane area in which a source at redshift $z_s$ is magnified by a factor $\mu$ or above \citep[e.g.,][and references therein]{2014MNRAS.444..268R,2015ApJ...800...38G,cam,2018ApJ...859..159C,2019ApJ...884...85C}. In this paper, we focus our analysis on $z_s=9$ and $|\mu|\geq3$. 
We define the lensing strength of a cluster as the area in the image plane where a source at redshift 9 is magnified by at least a factor of 3, denoted as \Lstrength.
A threshold of $|\mu|\geq3$ allows for significant magnification, which is important for studies of the background Universe, and remains close enough to the strong lensing core, where the constraints are. Additionally, the uncertainty in the overall magnification of a lens is dominated by the inaccuracy in regions of high magnification ($|\mu|>10$), or high length-to-width\footnote{the length-to-width ratio for large arcs is proportional to the magnification $\mu$ for similar convergence $\kappa$ and can be approximated as $L/W \simeq {\lambda_2}/{\lambda_1} \simeq 4\mu(1-\kappa)^2$ \citep{1994A&A...287....1B}.} ratios (l/w~$>$~10), which was demonstrated in several studies 
\citep[e.g.,][]{2012MNRAS.421.3343G,2018ApJ...859..159C,2020MNRAS.494.4771R}.

A source redshift of $z_s = 9$ is chosen in order to hone our investigation onto the applicability for discovery of high redshift galaxies, as this is the typical redshift sought by lensing-assisted high-$z$ searches (e.g., RELICS, HFF). Moreover, for sources beyond $z_s\sim6$, the magnification map does not change considerably. 

Finally, we normalize \Lstrength\ to a lens redshift of $z=0.5$, by multiplying it by the ratio of the angular diameter distances squared, to account for the geometric dependence of the area on the distance to the lens. 
\begin{equation}
    \mathcal{A}_{|\mu|\geq 3}^{0.5} \equiv \mathcal{A}_{|\mu|\geq 3}^{lens} \biggl ( \frac{D_{L}}{D_{0.5}} \biggr )^2
\end{equation}
where \LstrengthN\ is the lensing strength normalized to $z=0.5$,
$D_{L}$ is the angular diameter distance from the observer to the lens, and $D_{0.5}$ is the angular diameter distance from the observer to $z=0.5$.

\begin{figure*}
    \begin{center}
    \includegraphics[width=\textwidth]{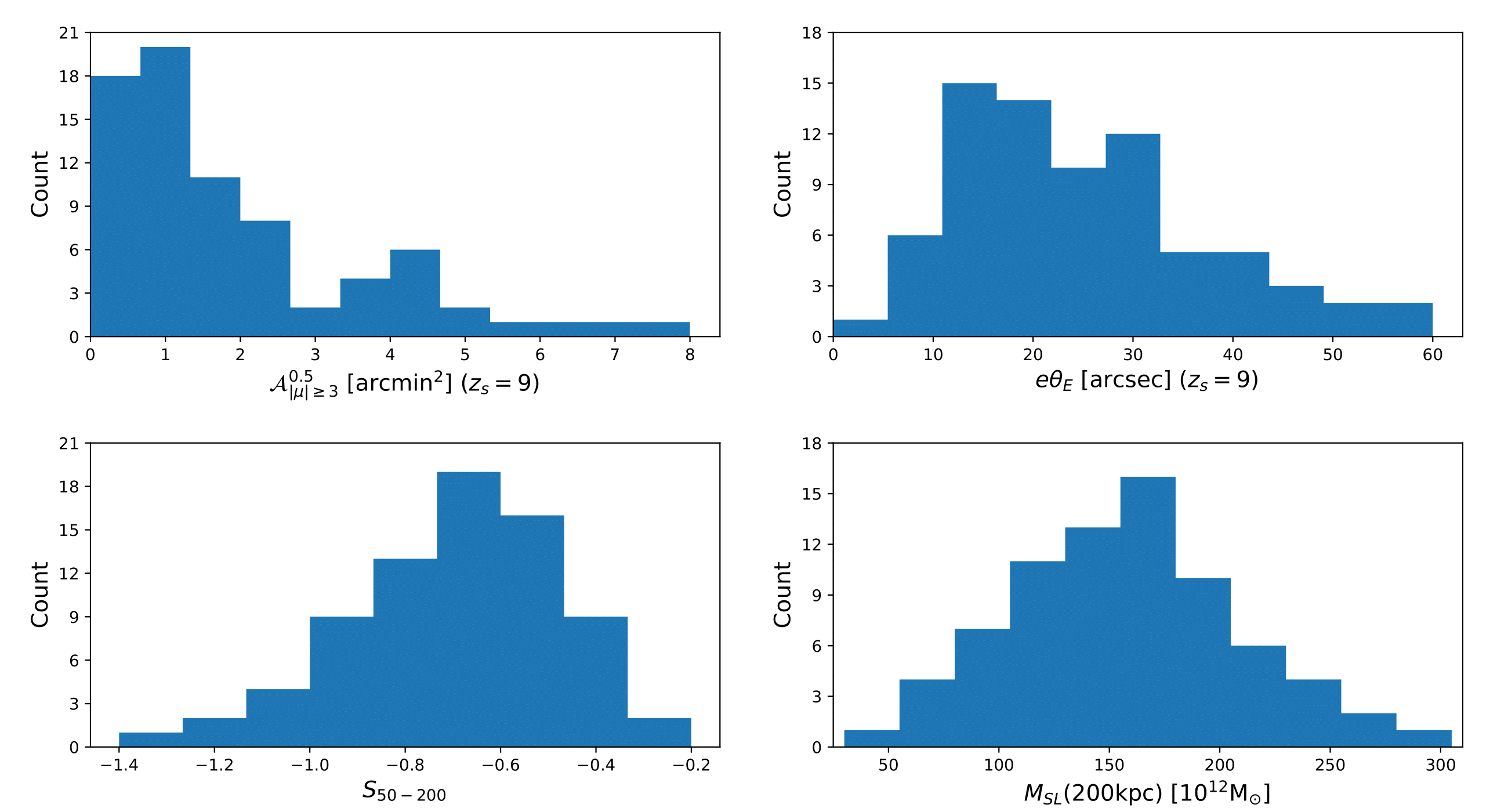}
    \caption{Distribution of cluster properties computed from the strong lensing models. \textit{Top left}: Normalized lensing strength, defined as the image plane area with $|\mu| \geq 3$ for $z_s=9$ and corrected for the redshift dependence of area projection onto the sky. \textit{Top right}: Effective Einstein radius for $z_s=9$. \textit{Bottom left}: Slope of the projected mass density profile, measured between $50-200$ kpc from the BCG. \textit{Bottom right}: Core mass, defined as the projected mass within $200$ kpc of the BCG. \Lenstool\ models are used where available (using the Sharon models for the HFF clusters), LTM if there is no \Lenstool\ model, and GLAFIC if there is no \Lenstool\ or LTM model. }
    \label{fig:all_hists}
    \end{center}
\end{figure*}

In Table~\ref{tab:all_clusts_table}, we list the non-normalized lensing strength (\Lstrength) as well as the normalized lensing strength (\LstrengthN). Figure \ref{fig:all_hists} shows the distribution of normalized lensing strengths across the sample of clusters. In Figure \ref{fig:all_hists}, \Lenstool\ models are used where available (using the Sharon models for the HFF clusters), LTM if there is no \Lenstool\ model, and GLAFIC if there is no \Lenstool\ or LTM model.

The availability of spectroscopic constraints has a significant impact on the magnification uncertainty of a lens model. For example, by analyzing lens models of simulated clusters with and without spectroscopic constraints, \citet{2016ApJ...832...82J} found that models with no spectroscopic redshift constraints have the highest magnification uncertainties, up to $\sim10\%$. This can clearly impact the lensing strength quantity we measure in a non-trivial way. While all the HFF and SGAS models rely on spectroscopic redshift constraints, \nmodelsnospec\ of the RELICS models of \nclustersnospec\ clusters do not (see Table~\ref{tab:all_clusts_table}).  We evaluate a conservative estimate of the extra uncertainty introduced to the computed lensing strength due to the lack of spectroscopic information, as follows. First, we increase by 10\% the magnification maps of the “range” models that have a lensing strength above the best model value, and reduce by 10\% those with a lensing strength below the best model value. We then recompute the 68\% confidence interval from the modified maps, keeping the best model unmodified. Treating the models with a lensing strength above or below the best model separately prevents any from crossing over, such as a model with a lensing strength originally below the best model that would surpass it with the 10\% boost. This keeps the error bar more conservative, as models crossing over could reduce the 68\% confidence interval. Overall, this boosts both sides of the error bar to conservatively account for the non-trivial dependence of the lensing strength on a 10\% uncertainty in the magnification map.
In Figures \ref{fig:merged_lens_strength_vs_stuff_plots}, \ref{fig:m500_vs_slope}, and \ref{fig:e_rad_area_vs_slope}, 
red error bars indicate models without spectroscopic redshift constraints or if it is unknown whether a spectroscopic constraint is used. 

We focus our analysis on the normalized lensing strength, but show comparisons with the non-normalized lensing strength in the Appendix. We note that the conclusions of this work are independent of this choice.

\subsection{Large Scale Mass Estimates} \label{sec:totalmass}

Large scale mass proxies have been used to predict lensing strength in survey design (e.g., LoCuSS, \citealt{2005MNRAS.359..417S}; RELICS, \citealt{2019ApJ...884...85C}). To test how well lensing strength correlates with total mass, we have obtained a large scale mass estimate, reaching beyond the strong lensing regime, for as many clusters in our sample as possible. For clusters included in the RELICS survey, we use the \textit{Planck} SZ $M_{500}$ mass (Table~\ref{tab:all_clusts_table}). 
MACS~J0025.4$-$12 and CL~J0152.7$-$1357 do not have a \textit{Planck} SZ mass and thus are not included in this comparison. Furthermore, Abell~1758a, RXS~J060313.4$+$4212-N, and RXS~J060313.4$+$4212-S are not included in this comparison because the Northern and Southern components of RXS~J060313.4$+$4212 were modeled separately and similarly we split the Abell~1758a model into Northwest and Southwest components, as described in Section \ref{sec:sample_relics}.

HFF clusters Abell S1063 and MACS~J0717.5$+$3745 are included in the \textit{Planck} SZ catalog \citep{2016A&A...594A..27P}, providing a $M_{500}$ estimate. For Abell~370, we use the $M_{500}$ estimate reported in \cite{2011ApJ...729..127U}. X-ray temperatures for MACS~J1149.5$+$2223 and MACS~J0416.1$-$2403 were reported by \citet{2012ApJS..199...25P} and for Abell~2744 by \citet{1998MNRAS.296..392A}. We convert the X-ray temperature to $M_{500}$ using the empirical relation from \citet{2007A&A...474L..37A}.

For the SGAS clusters, we apply well-established methods in the literature to derive $M_{500}$ estimates from the velocity dispersion measurements reported by \cite{2020ApJS..247...12S}, as described below. The velocity dispersions reported by \cite{2020ApJS..247...12S} were derived for each cluster from cluster members with spectroscopic redshifts (N$_{\mathrm{spec}}\geq4$), using the Gapper and bi-weight methods \citep{1990AJ....100...32B}. In order to derive $M_{500}$ estimates from the velocity dispersions in \cite{2020ApJS..247...12S}, we follow four steps. 1)~We first convert the velocity dispersion to $M_{200}$ using the $\sigma_{DM} - M_{200}$ relation from \citet{2008ApJ...672..122E}, assuming a velocity bias $b_v = \sigma_g / \sigma_{DM}=1$, following \citet[and references therein]{2011ApJS..193....8B}. The resulting $M_{200}$ values are consistent with those calculated by \cite{2020ApJS..247...12S}, who used the same method. 2)~We derive the concentration $c_{200}$ and its uncertainty using the concentration-mass relation reported by \citet{2018ApJ...859...55C} for individual halos. 3)~From a normal distribution of $c_{200}$ concentrations, we randomly select 10000 $c_{200}$ values and compute the corresponding $c_{500}$ for each using Appendix C of \citet{2003ApJ...584..702H}, taking $r_{200}$ as the virial radius. 4)~Finally, we use $c_{200}$ and $c_{500}$ to convert $M_{200}$ to $M_{500}$, following \citet{2003ApJ...584..702H}. From the final distribution of $M_{500}$ values of each cluster, we take the median as the $M_{500}$ estimate. The statistical uncertainty is estimated by determining where 68\% of the $M_{500}$ values on either side of the median are included.

We note that the velocity dispersions computed for some SGAS clusters rely on a small number of galaxy redshifts. 7 SGAS clusters have 10 or fewer galaxy redshift measurements included in their velocity dispersion calculations. For velocity dispersions computed by \citet{2020ApJS..247...12S}, the uncertainty is given by the empirical relation reported by \citet{2014ApJ...792...45R}, which depends on the number of galaxy redshifts. In our method for estimating $M_{500}$ from the velocity dispersion, we propagate these uncertainties.
For velocity dispersions listed but not computed by \citet{2020ApJS..247...12S}, (5 of the clusters with 10 or less galaxy redshifts), we acknowledge that the empirical formula reported by \citet{2014ApJ...792...45R} would suggest slightly larger uncertainties. The uncertainty of the $M_{500}$ estimate we compute for these clusters may therefore be underestimated. We refer the reader to \citet{2020ApJS..247...12S} for specifics about these clusters.
We emphasize that our study relies on the entire sample. We do not require very accurate $M_{500}$ estimates for each individual cluster, only a good estimate to help identify trends across a large sample. Studies including \citet{2014ApJ...792...45R} and \citet{1990AJ....100...32B} report that these velocity dispersion estimators are unbiased and we find no indication of a particular bias that would lead to our $M_{500}$ estimates of clusters with only a few galaxy redshifts being systematically biased low or high. We therefore contend that our reported uncertainties encompass the uncertainty introduced by having few galaxy redshifts, with the possible exception of the few clusters with potentially underestimated uncertainties. Furthermore, inaccuracies in $M_{500}$ estimates of clusters based on few galaxy redshifts simply boost the scatter in the trends we investigate, rather than adding a particular bias.

Figure \ref{fig:mass_vs_z} displays the distribution of $M_{500}$ estimates across the sample. Table~\ref{tab:all_clusts_table} 
contains the corresponding values. We focus our analysis on $M_{500}$, but show comparisons with $M_{200}$ in the Appendix. 

\subsection{Projected Core Mass Estimates from Strong Lensing} \label{sec:SLmass}
Strong lensing models provide a detailed description of the mass distribution in the cores of galaxy clusters. We use the projected mass density maps that were derived from the strong lensing models of each cluster; these maps include cluster scale and galaxy scale halos. We obtain a cylindrical core mass measurement, \MSL, by summing the projected mass density within a $200$ kpc aperture centered on the BCG, which is selected by visual inspection. In a few cases, e.g., where there are two or more galaxies with comparable brightnesses, we select the central bright galaxy that is more consistent with the center of the primary lensing potential as determined from the distribution of the lensing evidence. 

The availability of spectroscopic redshift constraints (Section~\ref{sec:lensingstrength}) also impacts the accuracy of mass estimates derived from strong lens models. 
In a detailed study of the CLASH and HFF cluster, RXC~J2248.7$-$4431 (Abell~S1063), \citet{2016A&A...587A..80C} found that lack of information on the source redshift of lensed constraints leads to model-predicted source redshifts biased towards higher values. The derived cluster mass is thus biased towards lower values, due to the degeneracy between total cluster mass and source redshift. They conclude that the impact on the total mass is not very significant, but quantities dependent on cosmological distances would be greatly impacted. In a study of simulated clusters, \citet{2016ApJ...832...82J} found that models with no spectroscopic constraints are biased towards lower masses by 5\%-10\%. Furthermore, \citet{2021ApJ...910..146R} found the mass within the effective Einstein radius to be biased low by 7.22\% without a known background source redshift, using Single-Halo models of clusters in the `Outer Rim' simulation \citep{2019ApJS..245...16H}. The 4\% bias found by \citet{2016A&A...587A..80C} for Abell S1063 is within with these findings. To conservatively account for the bias found in these studies, for the models in our sample with no spectroscopic constraints we add a systematic uncertainty of $0.1\times$\MSL\ to the upper error bar of \MSL.

A correlation between the lensing strength and the mass enclosed in the core is expected, from the basic equations of gravitational lensing. For example, 
The Einstein radius for a spherically symmetric lens is proportional to the square root of the mass enclosed within it (see also Section~\ref{sec:eff_e_rad_results_discussion}).
Figure \ref{fig:all_hists} displays the distribution of core masses across the sample of clusters. Table~\ref{tab:all_clusts_table} 
contains the corresponding core masses for each cluster, as obtained from each modeling method. 

\subsection{Inner Slope of Projected Mass Density Profile} \label{sec:innerslope}
The third cluster property we investigate is the steepness of the projected mass density profile at the core, extending the analysis beyond lensing strength indicators that are based purely on total mass. The inner slope is measured from the strong lensing mass maps. 

As was done in Section~\ref{sec:SLmass}, we use the total projected mass density maps, which include both the cluster scale and galaxy scale components. This allows the profile to reflect physical cluster properties such as substructure, multiple major mass components, and stellar mass. Using the total mass map also helps minimize the degeneracies between the different components of the strong lensing model. 

\begin{figure}[ht]
    \centering
    \includegraphics[width=\linewidth]{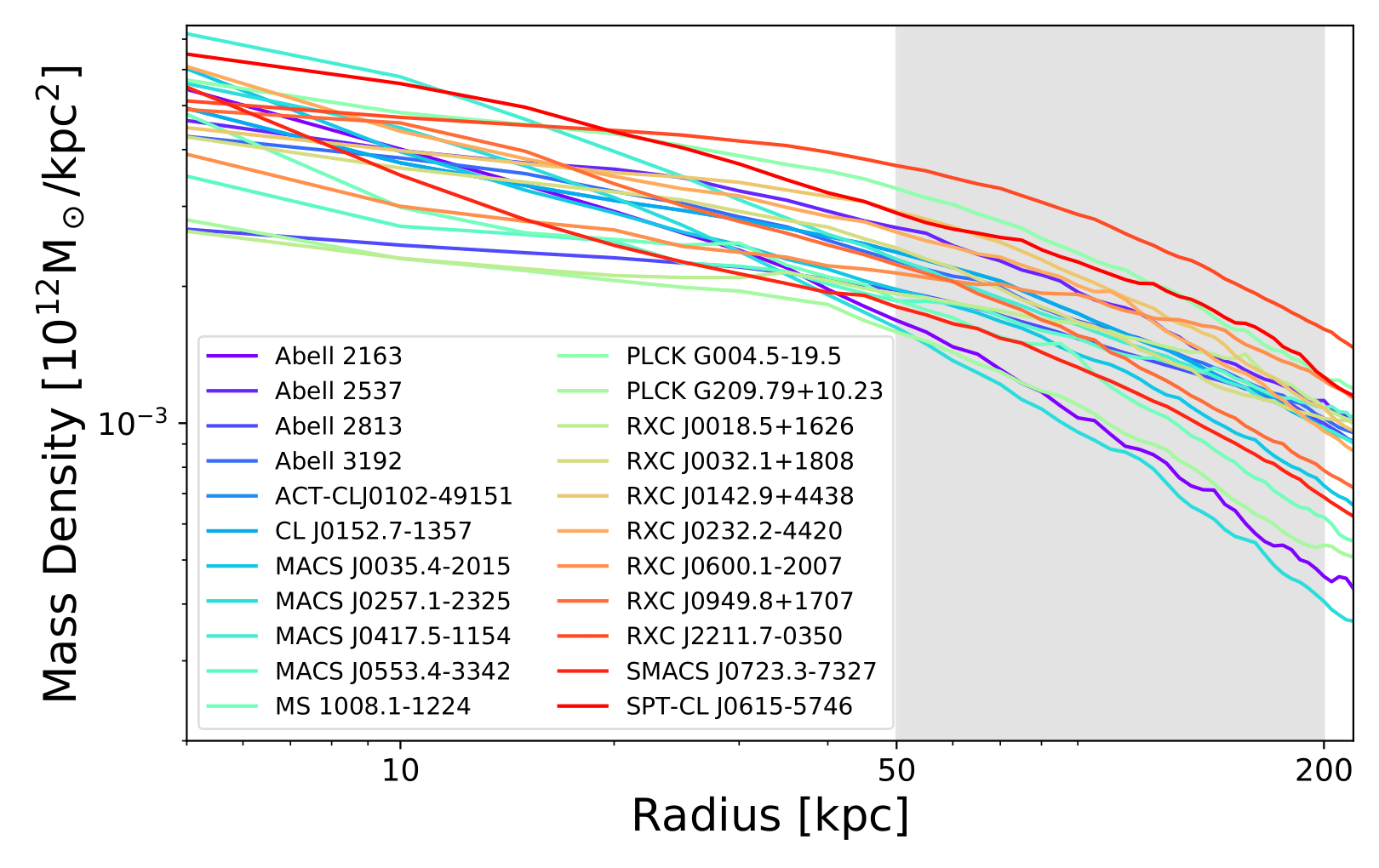}
    \caption{Projected mass density profiles of \Lenstool\ models for clusters of the sample included in the RELICS program. The grey shaded region indicates where the slope is measured ($S_{50-200}$). Clusters show clear differences in their projected density distributions in this region.}
    \label{fig:dens_profs}
\end{figure}

The projected density profile is computed from the average densities inside concentric annuli, centered on the chosen BCG, in steps of $5$ kpc. As an illustration of the range of projected mass profiles, Figure \ref{fig:dens_profs} shows the projected mass density profiles computed for clusters included in the RELICS sample and modeled with \lenstool.  We measure the logarithmic slope between $50$ kpc and $200$ kpc, by fitting the average density profile with a power law using curve$\_$fit in the python module scipy.optimize \footnote{Python scipy.optimize \url{https://docs.scipy.org/doc/scipy/reference/tutorial/optimize.html}}. We denote the slope as $S_{50-200}$. 

The motivation for measuring the slope at projected radii larger than $50$ kpc is to avoid the region typically dominated by the BCG, and to be less sensitive to a possible offset between the BCG and the DM halo. During the modeling procedure, the center of the main dark matter halo is usually allowed to vary. Unless constrained by lensing features at the very center of the cluster (typically radial arcs) the offset between the center of the main dark matter halo and the BCG is not well constrained. \cite{2017MNRAS.472.1972H} explored the BCG offset of 10 dynamically relaxed clusters, and found that the BCG wobbles with an amplitude of $11.82^{+7.3}_{-3.0}$ kpc with respect to the centroid of the DM halo, assuming it can be described as a harmonic oscillator. \cite{2019MNRAS.488.1572H} investigated the median BCG offset for different dark matter cross sections using the BAryons and HAloes of MAssive Systems simulations (BAHAMAS; \citealt{2017MNRAS.465.2936M}) and found the median offset to be below $10$ kpc, lowest for dark matter with no cross section. Taking $50$ kpc as the inner boundary of the slope measurement range allows us to mitigate this uncertainty. Furthermore, most of the light of BCGs lies within $50$ kpc, indicating that most of their stellar mass would be enclosed in this range too.

We note that some studies have used ‘normalized’ profiles (e.g., \citealt{2018ApJ...864...98B} and \citealt{2019A&A...632A..36C} scale the projected mass density by $\pi r_{200c}^2/M_{200c}$ and the radius by $1/r_{200c}$), which may simplify comparisons between clusters. 
We refrain from using a similar normalization here, in order to avoid introducing the large uncertainties on $M_{200c}$ and $r_{200c}$ into measurements at small projected radii. 
The estimated statistical uncertainty of the slope measurement takes into account both the range of slopes across the ``range'' mass maps that were provided with each of the public models (i.e., lens modeling uncertainty), and the uncertainty associated with fitting the slope to the mass map. 
The former samples the posterior probability distribution in the mass distribution related to the modeling process to assess the degree of uncertainty in the slope estimate; the latter assesses the uncertainty of the slope fitted to the best model in the region $50-200$ kpc from the BCG. We add these statistical uncertainties in quadrature.

For the GLAFIC lens models of clusters included in the RELICS program, a linear interpolation is applied to increase the map resolution of the ``range'' models from $0.4$ to $0.05$ arcsec/pixel. This is necessary to accommodate the $5$ kpc step size, enabling a smooth profile that is not significantly affected by pixel size. The interpolated maps are used throughout our analysis. Table~\ref{tab:all_clusts_table} contains the slope ($S_{50-200}$) for each cluster and modeling method. The bottom-left panel of Figure~\ref{fig:all_hists} displays the distribution of slopes across the sample.

\subsection{Effective Einstein Radius $\&$ Area} \label{sec:effective_e_rad}
The fourth property we explore is the effective Einstein radius ($e\theta_E$), defined as the radius of a circle with an area the same as the area enclosed by the main tangential critical curve. The tangential critical curve is the theoretical line of infinite magnification; its naming indicates the primary direction in which the images (arcs) are magnified. To determine the tangential critical curve from the strong lensing models, we compute the inverse of the magnification in the tangential direction, $\mu_t^{-1} = 1 - \kappa - \gamma$, for a background source redshift at $z_s = 9$. Table~\ref{tab:all_clusts_table} 
contains the effective Einstein radius for each cluster and modeling method for $z_s = 9$. The top-right panel of Figure~\ref{fig:all_hists} shows the distribution of effective Einstein radii across the sample. In order to compare with the lensing strength area, we use the area within the main tangential critical curve (effective Einstein area, \Earea).

The effective Einstein area is expected to have a strong correlation with our definition of lensing strength. In the bottom left panel of Figure \ref{fig:merged_lens_strength_vs_stuff_plots}, we compare \LstrengthN\ with \Earea\ for each field. We also plot the expected relationship for three different potentials, at $z_{lens}=0.5$ and no ellipticity. Maroon shows a singular isothermal sphere (SIS). Blue shows a pseudo-isothermal elliptical mass distribution (PIEMD, \citealt{1993ApJ...417..450K}) with a core radius of 40~kpc and a cut radius of 1500~kpc. Purple shows an NFW \citep{1997ApJ...490..493NFW} distribution with a scale radius of 100~kpc. The curves are created by varying the mass of the halo, keeping the core and cut radii constant in the PIEMD case and keeping the scale radius constant in the NFW case. In the SIS model, the two are related by a constant factor. In the PIEMD and NFW models, the relationship is similar, but also dependent on the characteristic radii chosen. For these two cases, 
the radius at which the slope of the curve flattens increases with an increasing core or scale radius.
For many of the clusters in our sample, the mass distribution is much more complicated than a singular halo with no ellipticity, greatly influencing the extent of significantly magnified area. Although  \Lstren\ and \Earea\ are clearly interconnected, we choose to focus on \Lstren\ because it captures a diversity of lens properties effect on the magnified area that the commonly-used Einstein radius may miss.

\subsection{Distance to Farthest Bright Arc} \label{thetaarc}
The observed radial extent of lensing evidence is often used as a proxy of the Einstein radius, and thus the lensing power of a cluster. We test the efficacy of this indicator by comparing the separation between the BCG and the farthest bright arc, which we denote as $\theta_{arc}$, to the lensing strength. By construction, the SGAS clusters are a prime sample for testing this assessment, as they were identified based on the observation of a giant bright arc in their field. We use the \HST\ images to find and measure $\theta_{arc}$ for each SGAS cluster. We measure $\theta_{arc}$ in 17 RELICS and HFF clusters, in which a bright arc can be easily identified; if more than one bright arc is observed, we select the one farthest from the chosen BCG. Because only a fraction of the sample is included in this test, we do not include $\theta_{arc}$ in Table~\ref{tab:all_clusts_table} or Figure \ref{fig:all_hists}.

\subsection{Statistical indicators}\label{sec:stats}
We use the Kendall $\tau$ and Spearman $r$ rank correlation coefficients to assess the existence of correlations between parameters that were evaluated in this work. There are several RELICS and HFF clusters for which we use results from multiple modeling algorithms. When computing Kendall $\tau$ and Spearman $r$ coefficients, we avoid double-counting these clusters by computing the coefficients 10000 times, each time randomly drawing the model used for each cluster with more than one model. We report the mean value of the computed coefficients across the 10000 trials and the mean p-value. There are a few other ways to consider these clusters with multiple models. We tested three methods: 1) Using all models, each having equal weight. 2) Using one model per cluster, e.g., prioritizing \lenstool\ and using other algorithms if the highest-priority one is unavailable. 3) including all models for each cluster, but giving less weight to multiple models of the same cluster, so that each cluster as a whole has equal weight. We find that the computed Kendall $\tau$ and Spearman $r$ coefficients are very similar across these methods and therefore the conclusions are not affected by these choices.

\section{Results and Discussion} \label{results_discussion}
In this section we investigate the dependence of our normalized lensing strength indicator (\LstrengthN) on the large scale mass ($M_{500}$), the strong lensing core mass (\MSL), the inner slope of the projected mass density profile (\Slope), the effective Einstein area (\Earea), and the separation between the BCG and farthest bright arc ($\theta_{arc}$). We discuss our results and compare our findings with previous studies. 

\subsection{Large Scale Mass ($M_{500}$)} \label{sec:total_mass_results_discussion}

In the top-left panel of Figure \ref{fig:merged_lens_strength_vs_stuff_plots} we plot the normalized lensing strength against $M_{500}$, a large scale mass estimate that is measured independently of the lensing analyses. Light grey vertical lines connect models of the same cluster that were obtained with different techniques. We find clusters with low lensing strength across the entire mass range. Furthermore, clusters with a given mass span a broad range of lensing strengths, particularly for $M_{500}\approx11\times10^{14}$ M$_{\odot}$, implying that a high mass does not guarantee a high lensing strength. We compute Kendall $\tau$ and Spearman $r$ coefficients of 0.26 and 0.36, respectively, with p-values $<0.01.$

\citet{2020ApJ...889..189S} reported the high-$z$ candidates found by the RELICS project. They found little to no correlation between $M_{500}$ and the number of $z\sim6-8$ candidates discovered. High mass clusters, such as the massive lens MACS~J0417.5$−$1154, often produced hardly any $z\sim6-8$ candidates. Our result is consistent with this observation, i.e., a high total mass is not necessarily an indicator of a high lensing strength. This result is also in agreement with \citet{2018ApJ...859..159C} who computed lens models for five clusters in the RELICS survey and found that the magnified source-plane area (proxy for lensing efficiency), 
is not a simple function of the total mass of a cluster. 
For instance, Abell~2163 provides a relatively low magnification area, despite having the highest \textit{Planck} SZ mass of the sample.

\subsection{Strong Lensing Core Mass (\MSL)} \label{sec:core_mass_results_discussion}
We observe a 
high probability of correlation ($\tau=0.66$ and $r=0.84$ with p-values $<0.01$) between the inner core mass and normalized lensing strength, shown in the top-right panel of Figure \ref{fig:merged_lens_strength_vs_stuff_plots}. As noted in section \ref{sec:SLmass}, this is expected, as the inner core mass is directly responsible for the strong lensing effect; however, we note a large spread in lensing strengths. Furthermore, we note that the lower bound of the observed lensing strength increases with inner core mass, as can be inferred from the absence of clusters in the bottom right region of the plot. 

\subsection{Inner Slope of Projected Mass Density Profile (\Slope)} \label{slope_results_discussion}

\begin{figure*}
    \centering
    \includegraphics[width=\textwidth]{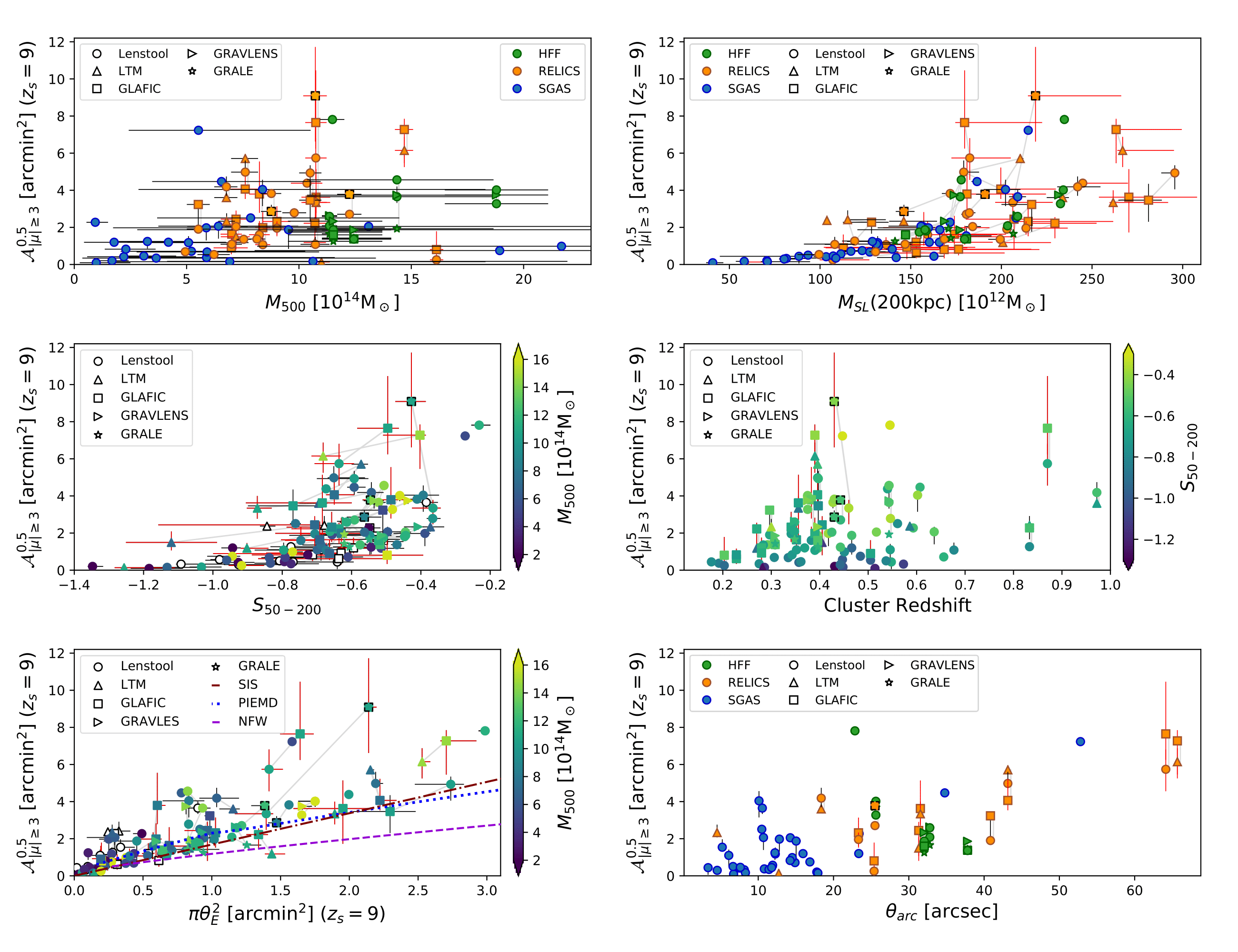}
    \caption{Normalized lensing strength (\LstrengthN) compared with cluster properties. \textit{Top left:} Large scale mass estimate ($M_{500}$), which is measured independently of the lensing analyses.  \textit{Top Right:} Projected mass within $200$ kpc of the BCG (\MSL). \textit{Middle Left:} Inner slope of the projected mass density profile ($S_{50-200}$), color coded by $M_{500}$. Clusters without an $M_{500}$ estimate are plotted in white. \textit{Middle Right:} Cluster redshift, color coded by $S_{50-200}$. 
    \textit{Bottom Left:} Effective Einstein area (\Earea), color coded by $M_{500}$. Clusters without an $M_{500}$ estimate are plotted in white. The curves show the expected relationship for different potentials at a redshift of $z_{lens}=0.5$ and no ellipticity. Maroon shows an SIS distribution; Blue shows a PIEMD, with a core radius of 40~kpc, and a cut radius of 1500~kpc; Purple shows an NFW profile with a scale radius of 100~kpc. \textit{Bottom Right:} Distance between the BCG and farthest bright arc ($\theta_{arc}$). 
    The shape of the data points indicates different lens modeling algorithms, where light grey lines connect models of the same cluster modeled by different algorithms. The three GLAFIC models where 5\% was added to the upper error bar due to the limited field of view are given a dashed black border. In all panels, red error bars indicate models without spectroscopic redshift constraints or if it is not known whether a spectroscopic constraint is used.}
    \label{fig:merged_lens_strength_vs_stuff_plots}
\end{figure*}

We show in the middle left panel of Figure \ref{fig:merged_lens_strength_vs_stuff_plots} the normalized lensing strength against the slope of the projected mass density profile ($S_{50-200}$). Light grey lines connect models of the same cluster produced with different algorithms. Flat profiles (high $S_{50-200}$) show the largest range of lensing strengths, including the highest values. We note that the upper left corner of the diagram, indicating the steep slope and high lensing strength region, is not populated. At steeper slopes (low $S_{50-200}$), the highest lensing strength measured diminishes. This suggests that the slope of the projected mass density profile can be an indicator of the maximum possible lensing strength of a cluster, with the highest lensing strengths only found in clusters with flatter inner profiles. We compute Kendall $\tau$ and Spearman $r$ coefficients of 0.53 and 0.71, respectively, with p-values $<0.01$.
Notably, the measured correlation coefficients are greater than those for the large scale mass, suggesting a higher probability of correlation.

We note that the slope measured in models of the same cluster computed with different techniques often differ (e.g., PLCK~G171.9-40.7, PLCK~G287.0+32.9); however, the differences tend to follow the trend of a flatter slope having a higher lensing strength.

The importance of the shape of the projected mass density profile can help explain why there is no significant correlation between $M_{500}$ and lensing strength. Strong lensing requires surpassing a critical threshold of high surface density. Flatter profiles that already reach the critical threshold could therefore offer a larger magnification area than steeper profiles because the density drops off slower with radius, boosting their lensing strengths.
Comparing two strong-lensing clusters with the same total mass, the cluster with a flatter profile would enable a larger lensing strength region, as more area is above the critical density, despite the peak density being less than that of the cluster with a steeper profile. Cluster physical properties such as substructure, ellipticity, and merging events could flatten the profile, enabling a more elongated area of critical density.
Furthermore, if two strong-lensing clusters have the same inner slope, the cluster with the higher mass could provide a larger lensing strength area by enabling more area to be of critical density.

\begin{figure}[ht]
    \centering
    \includegraphics[width=\linewidth]{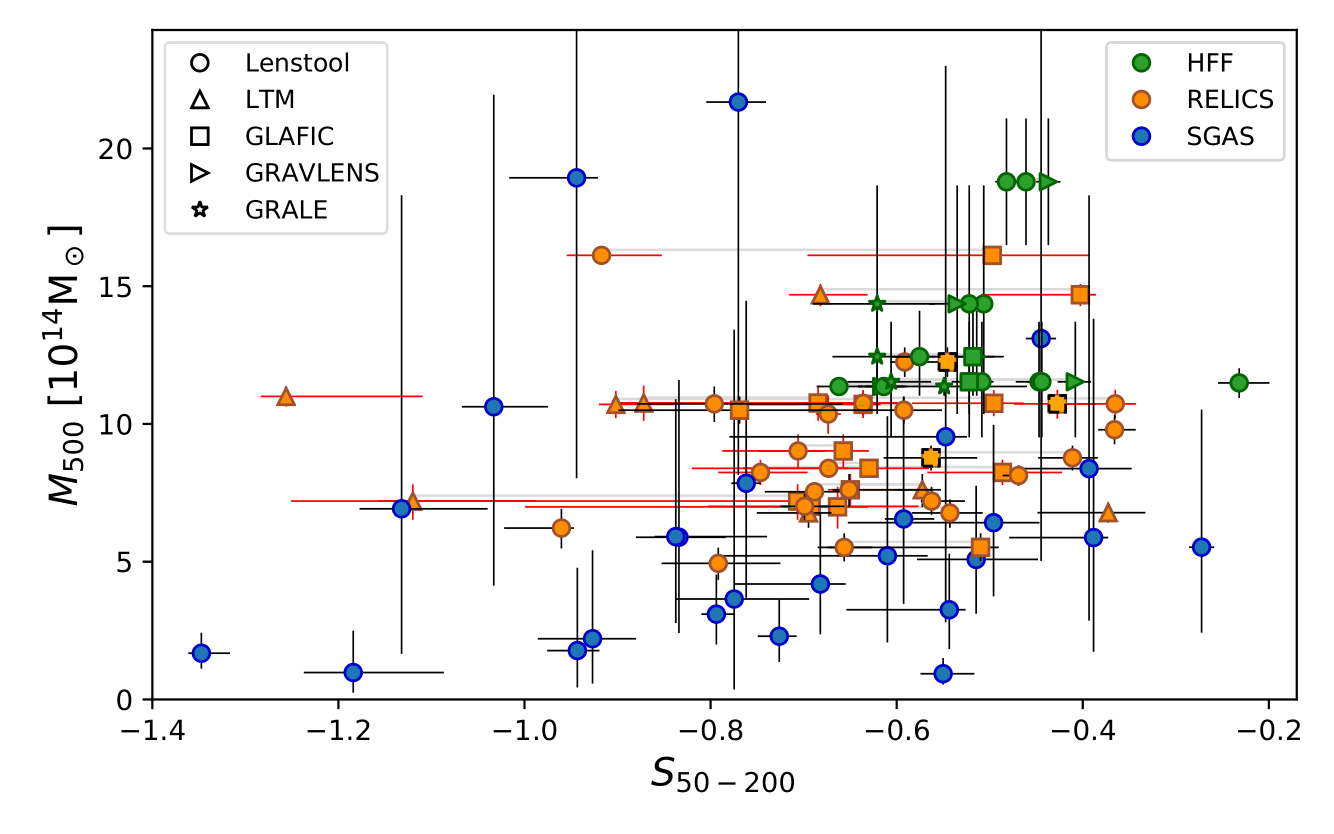}
    \caption{Large scale mass estimate ($M_{500}$) which is measured independently of the lensing analyses, compared to the inner slope of the projected mass density profile, measured in between $50-200$ kpc from the BCG. Color coding and shape refer to different samples of clusters and different lens modeling algorithms, respectively. Light grey horizontal lines connect models of the same cluster, computed with different algorithms. Red error bars indicate models without spectroscopic redshift constraints or if it is not known whether a spectroscopic constraint is used. No obvious correlation is found between the two indicators.}
    \label{fig:m500_vs_slope}
\end{figure}

We see little to no stratification by large scale mass in the normalized lensing strength vs.\,\,inner slope plot (middle-left panel of Figure \ref{fig:merged_lens_strength_vs_stuff_plots}). This further suggests that several physical properties affect the lensing strength of a cluster. $S_{50-200}$ is an attempt to capture several properties by characterizing the shape of the projected mass density profile at the core of the galaxy cluster. The overall trend we find hints to an interesting combination of physical cluster properties leading to the overall lensing strength of the cluster. Future studies can explore more rigorous assessments of the mass density profile and other physical cluster properties to further investigate what dominates lensing strength. Additionally, we note that we find a low probability of correlation between the inner slope and large scale mass (Kendall $\tau$=0.17, Spearman $r$=0.23, with p-values $< 0.1$), as seen in Figure \ref{fig:m500_vs_slope}. $M_{500}$ and $S_{50-200}$ represent very different cluster-centric regimes, suggesting properties of different scales do not need to be well correlated. We find that a high large-scale mass does not imply a flat inner slope or a high lensing strength. 

The age of the structure may also play a role in the relation between lensing strength and the inner slope we measure. As a cluster ages, we expect the mass density profile to become more cuspy as the cluster relaxes, becoming more centralized and drawing more mass into the center of the potential well. We therefore expect a cluster to move diagonally towards the bottom left of the normalized lensing strength vs. inner slope plot (middle left panel of Figure \ref{fig:merged_lens_strength_vs_stuff_plots}) over cosmic time as the slope steepens and lensing strength lessens. In a more disturbed cluster, such as a major merger with a flat profile, over time the magnification area will radically evolve due to the evolution of the 2D projection of the mass distribution \citep{2012A&A...547A..66R}. The cluster may greatly increase in lensing strength through the merging process, but decrease as the profile steepens in the more centralized merged cluster.

Although we correct for the redshift dependence on how the lensing strength area projects onto our sky, cluster redshift can still play a role in lensing strength. We note that the normalized lensing strength - slope relation we observe persists across the range of cluster redshifts, as shown in the middle right panel of Figure \ref{fig:merged_lens_strength_vs_stuff_plots}, which compares the normalized lensing strength with cluster redshift, color coded by $S_{50-200}$. 

The finding that a high lensing strength is only possible in clusters with flatter inner profiles does not conflict with the consensus that lensing clusters tend to be more concentrated. The samples of clusters that are investigated here are lensing-selected, and as such are likely to already be highly concentrated.
The concentration is defined as $c=r_{\Delta}/r_s$; where $r_\Delta$ is the radius of interest and $r_s$ is the scale radius. Higher-concentration means that a higher fraction of the total mass is contained in the cluster core. 
The inner slope, $S_{50-200}$, is measured at much smaller projected radii, and represents the slope of the projected mass density inner to the $r_s$ range.
For reference, $r_{500}$ in our sample ranges from $556$ kpc to $1676$ kpc. \citet{2009MNRAS.392..930O} illustrate these characteristic scales compared to the Einstein radius (see their Fig. 1). 

\subsection{Effective Einstein Radius ($e\theta_E$) $\&$ Area (\Earea)} \label{sec:eff_e_rad_results_discussion}
We plot the normalized lensing strength against the effective Einstein area for $z_s=9$ in the bottom left panel of Figure \ref{fig:merged_lens_strength_vs_stuff_plots}, color coded by $M_{500}$. Although we use the normalized lensing strength here, in the Appendix we show the corresponding plot with both measurements shown in their native cluster redshift, i.e., not normalized for the redshift dependence of area projection onto the sky. We note that the conclusions are independent of this choice. 

We observe a relation between the normalized lensing strength and effective Einstein area, with a scatter indicative of how much the region of $|\mu|\geq3$ extends from the main tangential critical curve, which depends on the physical properties of a cluster that impact the gradient of the magnification. We compute Kendall $\tau$ and Spearman $r$ coefficients of 0.68 and 0.86, respectively, with p-values $<0.01,$ suggesting a strong probability of correlation. The normalized lensing strength and effective Einstein area generally increase with mass, but a high mass does not guarantee a high lensing strength. 

\begin{figure}
    \centering
    \includegraphics[width=\linewidth,trim={.5cm 0 1.3cm 0}]{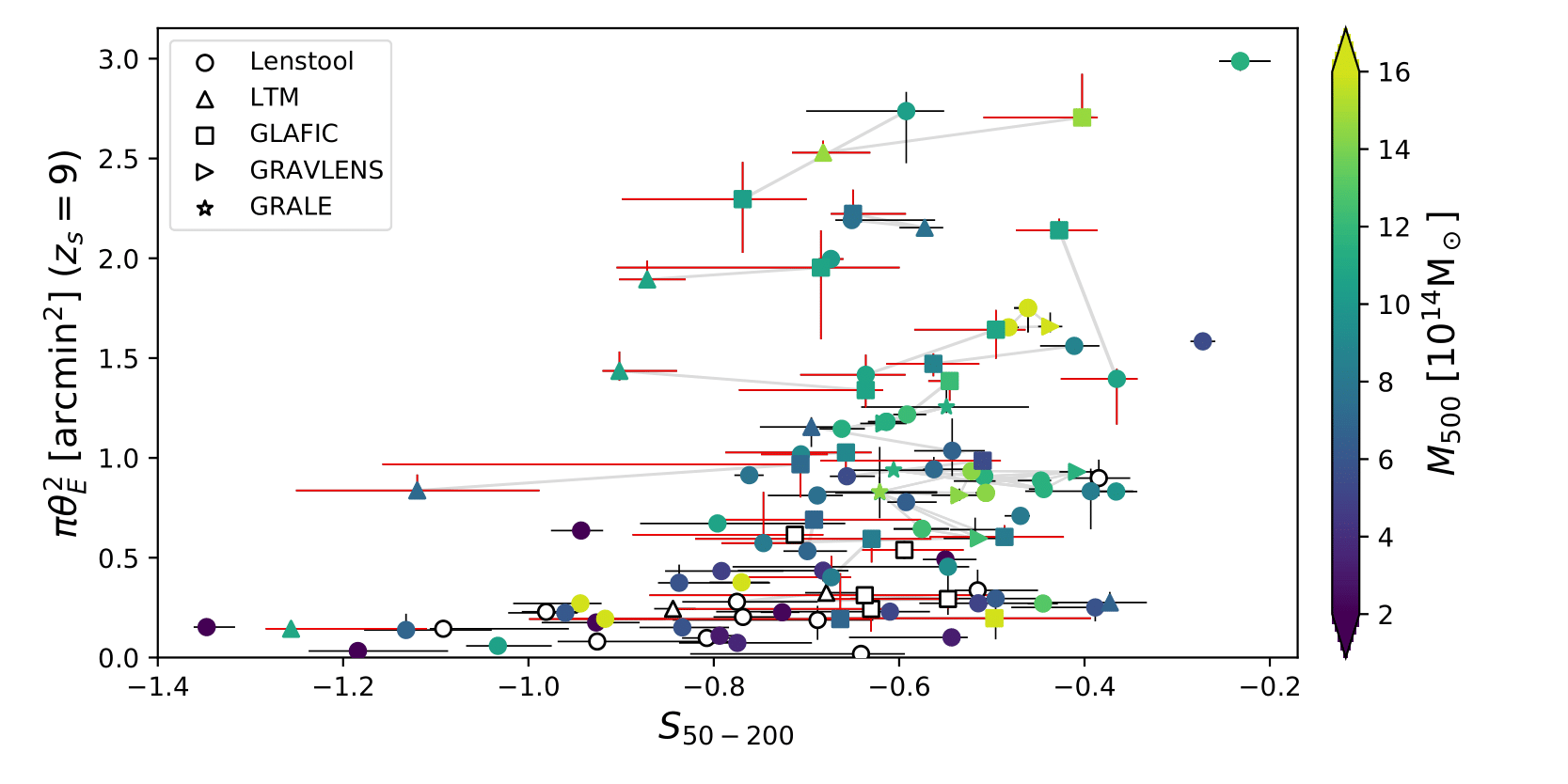}
    \caption{Effective Einstein area for a source at $z_s=9$ compared to the inner slope of the projected mass density profile, measured in between $50-200$ kpc from the BCG. Color coding shows the large scale mass estimate ($M_{500}$) which is measured independent of the lensing analyses. Clusters without an $M_{500}$ estimate are plotted in white. Shape refers to different lens modeling algorithms and light grey lines connect models of the same cluster modeled by different algorithms.
    Red error bars indicate models without spectroscopic redshift constraints or if it is not known whether a spectroscopic constraint is used. This plot shows how effective Einstein area relates less strongly to the slope because it does not capture high magnification regions far from the critical curve.}
    \label{fig:e_rad_area_vs_slope}
\end{figure}

Due to the correlation between the effective Einstein area and the normalized lensing strength, they show similar trends with large scale mass, inner core mass, and the slope of the projected mass density profile. In Figure \ref{fig:e_rad_area_vs_slope} we compare the effective Einstein area for $z_s=9$ with \Slope, color coded by the large scale mass estimate ($M_{500}$) which was measured independently of the lensing analysis.

We find that $e\theta_E$ has a lower probability of correlation with the inner slope than our lensing strength indicator does. This suggests that the slope has a stronger impact on regions farther from the very inner core, a flat profile enabling more area far from the tangential critical curve to be moderately magnified. Our lensing strength indicators, \Lstrength\ and \LstrengthN, better capture the magnification potential of a cluster, accounting for the lower magnification regions that extend farther from the critical curve, and provide an important boost in high-$z$ survey volume.

\subsection{Distance to Farthest Bright Arc ($\theta_{arc}$)} \label{thetaarc2}
In the bottom right panel of Figure \ref{fig:merged_lens_strength_vs_stuff_plots} we plot the normalized lensing strength against the distance between the BCG and the farthest identified bright arc ($\theta_{arc}$). 
As in Section \ref{sec:eff_e_rad_results_discussion}, we use the normalized lensing strength here, but show in the Appendix the corresponding plot with both measurements in their native cluster redshift, i.e., not normalized for the redshift dependence of area projection onto the sky. The conclusions are again independent of this choice. 

We find that clusters with low $\theta_{arc}$ span a broad range of lensing strengths, and could not define powerful lenses. However, all clusters with $\theta_{arc} \gtrsim 30\farcs0$ have moderate to high lensing strengths, implying that an observation of large $\theta_{arc}$ may be used as a direct observational indicator to identify and select powerful lenses. Overall, we compute Kendall $\tau$ and Spearman $r$ coefficients of 0.39 and 0.55, respectively, with p-values $<0.01$.

\section{Summary and Conclusions} \label{sec:summary}

We present an analysis of lensing strength across a sample of \Nmodels\ strong lensing models of \Nclusters\ cluster fields from the HFF, RELICS, and SGAS programs. We use as an indicator of lensing strength the area on the sky with moderate to high magnification that a cluster provides, which facilitates the search for magnified high redshift galaxies. The lensing strength, \LstrengthN, is defined as the image plane area with absolute magnification of at least three for a source at $z_s=9$, corrected for the cluster redshift to account for the redshift dependence of area projection onto the sky. We compare the normalized lensing strength to the large scale mass, inner core projected mass, inner slope of the projected mass density profile, effective Einstein area, and the distance between the BCG and the farthest bright arc detected in the field. The results of this analysis are summarized as follows: 

\begin{itemize}
    \item We find that a high large scale mass ($M_{500}$) does not guarantee a high lensing strength (\LstrengthN). We observe more clusters with a high lensing strength as large scale mass increases, but find low lensing strength clusters across the entire mass range, including high-mass lensing clusters with a low lensing strength.
    \item  As expected, the lensing strength has a high probability of correlation with the projected aperture mass within $200$ kpc of the BCG, computed from the strong lensing model.
    \item We find that the lensing strength indicator, \LstrengthN, has a high probability of correlation with the inner slope of the projected mass density profile, measured in an annulus between $50-200$ kpc from the BCG ($S_{50-200}$). Notably, the probability of correlation with lensing strength is higher than for large scale mass ($M_{500}$).
    Flatter profiles offer a wide range of lensing strengths, whereas the maximum lensing strength observed diminishes for steeper profiles (middle-left panel of Figure \ref{fig:merged_lens_strength_vs_stuff_plots}). No cluster is found to have a steep profile and a high lensing strength, leaving the top-left region of the lensing strength vs. inner slope plot empty. 
    \item The lensing strength indicator, \LstrengthN, has a high probability of correlation with effective Einstein area (\Earea). The scatter between these indicators stems from the intrinsic variance of cluster properties that determine the extent of significant magnification beyond the tangential critical curve. The lensing strength indicator captures the significantly magnified regions that extend farther from the tangential critical curve. 
    \item We find clusters with a short distance between the BCG and farthest bright arc ($30\farcs0 \gtrsim \theta_{arc}$) to have a wide range of lensing strength; however, all clusters with $\theta_{arc} \gtrsim 30\farcs0$ are powerful lenses with a high lensing strength.
\end{itemize}

Studies of the dependence of lensing strength on physical cluster properties can aid the selection of lensing clusters for future surveys, including criteria beyond mass proxies. Maximizing lensing strength provides a greater opportunity to find high redshift candidates that give insight into the early Universe. Furthermore, the absence of clusters with a steep profile and high lensing strength could help constrain cosmology. Our results would support cosmologies that do not produce clusters with steep inner profiles and high lensing strengths at the current age of the Universe. A future study could assess how the lensing strength -- inner slope relation appears in simulations for different cosmologies and how the relation changes as the Universe ages.

This paper demonstrates, from a sample of \Nclusters\ clusters with strong lens models, that the efficacy of galaxy clusters as efficient cosmic telescopes cannot be solely determined from global properties like their total mass or redshift. 
Lensing-assisted high-$z$ surveys that select lenses by mass alone will not be optimized for lensing strength, as some fraction of the high-mass clusters may have low lensing strengths: high lensing efficiency requires strong lensing clusters whose projected mass distribution has a shallow inner slope.

\section*{Acknowledgements}
We thank Dan Coe for useful discussion and helpful comments.
We thank the HFF, RELICS, and SGAS projects for making their lens models publicly available. Some of the High Level Science Products (HLSP) presented in this paper were obtained from the Mikulski Archive for Space Telescopes (MAST). STScI is operated by the Association of Universities for Research in Astronomy, Inc., under NASA contract NAS5-26555.
JDRG acknowledges support by the National Science Foundation Graduate Research Fellowship Program under Grant No. DGE 1256260. GM received funding from the European Union’s Horizon 2020 research and innovation programme under the Marie Skłodowska-Curie grant agreement No MARACAS - DLV-896778.

\begin{longrotatetable}
\begin{deluxetable*}{llllccccccccc}
\tablecolumns{11} 
\tabletypesize{\scriptsize}
\tablewidth{\textwidth}
\tablecaption{\label{tab:all_clusts_table}}
\tablehead{\colhead{} &
            \colhead{$\alpha$}     & 
            \colhead{$\delta$}    & 
            \colhead{}       & 
            \colhead{$M_{500}$}       &
            \colhead{$M_{200}$}       &
            \colhead{}       &
            \colhead{}       &
            \colhead{$M_{\rm SL}(200 {\rm kpc})$}       &
            \colhead{}  &
            \colhead{$e\theta _E$} &
            \colhead{$\mathcal{A}_{|\mu|\geq 3}^{lens}$} & 
            \colhead{$\mathcal{A}_{|\mu|\geq 3}^{0.5}$}    \\[-8pt]
            \colhead{Cluster}       &
            \colhead{J2000}       &
            \colhead{J2000}       &
            \colhead{$z_{lens}$}  &
            \colhead{($10^{14}$M$_{\odot}$)}       &
            \colhead{($10^{14}$M$_{\odot}$)}       &
            \colhead{Alg.}       &
            \colhead{Spec-$z$}       & 
            \colhead{($10^{12}$M$_{\odot}$)}       &
            \colhead{$S_{50-200}$}       &
            \colhead{(arcsec)}    &
            \colhead{(arcmin$^2$)}       &
            \colhead{(arcmin$^2$)}       
}
\startdata
HFF \\
\hline
Abell~2744 & 3.586259 & -30.400173 & 0.308 & 12.44$_{-1.66}^{+1.42}$ & 19.40$_{-2.26}^{+2.66}$ & L-S$^{\rm a}$ & yes & 179.69$_{-2.04}^{+2.00}$ & -0.58$_{-0.03}^{+0.03}$ & 27.18$_{-0.09}^{+0.07}$ & 2.47$_{-0.10}^{+0.10}$ & 1.36$_{-0.05}^{+0.06}$\\ 
 &  &  &  &  &  & G$^{\rm b}$ & yes & 180.79$_{-2.04}^{+1.33}$ & -0.52$_{-0.03}^{+0.02}$ & 27.09$_{-0.34}^{+1.24}$ & 2.50$_{-0.10}^{+0.08}$ & 1.38$_{-0.05}^{+0.04}$\\ 
 &  &  &  &  &  & GL$^{\rm c}$ & yes & 177.17$_{-3.41}^{+2.50}$ & -0.51$_{-0.04}^{+0.03}$ & 26.15$_{-0.21}^{+0.25}$ & 3.37$_{-0.27}^{+0.25}$ & 1.86$_{-0.15}^{+0.14}$\\ 
 &  &  &  &  &  & GR & yes & 172.11$_{-4.07}^{+3.85}$ & -0.62$_{-0.05}^{+0.06}$ & 30.74$_{-2.47}^{+4.03}$ & 2.58$_{-0.21}^{+0.15}$ & 1.43$_{-0.12}^{+0.08}$\\ 
Abell~370 & 39.97133 & -1.58224 & 0.375 & 18.79$_{-2.30}^{+2.30}$ & 31.60$_{-3.86}^{+3.86}$ & L-S$^{\rm a}$ & yes & 234.13$_{-1.82}^{+0.70}$ & -0.46$_{-0.01}^{+0.01}$ & 44.80$_{-1.58}^{+0.00}$ & 5.62$_{-0.42}^{+0.08}$ & 4.02$_{-0.30}^{+0.05}$\\ 
 &  &  &  &  &  & L-CATS$^{\rm d}$ & yes & 232.58$_{-0.53}^{+0.31}$ & -0.48$_{-0.01}^{+0.01}$ & 43.54$_{-0.20}^{+0.02}$ & 4.58$_{-0.06}^{+0.02}$ & 3.27$_{-0.05}^{+0.02}$\\ 
 &  &  &  &  &  & GL$^{\rm c}$ & yes & 231.60$_{-1.35}^{+1.16}$ & -0.44$_{-0.01}^{+0.01}$ & 43.59$_{-0.39}^{+0.91}$ & 5.24$_{-0.25}^{+0.18}$ & 3.75$_{-0.18}^{+0.13}$\\
Abell~S1063 & 342.18321 & -44.530894 & 0.348 & 11.36$_{-0.34}^{+0.34}$ & 17.74$_{-0.54}^{+0.54}$ & L-S$^{\rm a}$ & yes & 208.95$_{-0.83}^{+0.87}$ & -0.61$_{-0.02}^{+0.02}$ & 36.79$_{-0.07}^{+0.23}$ & 3.99$_{-0.04}^{+0.06}$ & 2.59$_{-0.02}^{+0.04}$\\
 &  &  &  &  &  & L-CATS & yes & 203.40$_{-1.59}^{+1.63}$ & -0.66$_{-0.02}^{+0.03}$ & 36.23$_{-0.16}^{+0.17}$ & 3.21$_{-0.07}^{+0.09}$ & 2.08$_{-0.05}^{+0.06}$\\ 
 &  &  &  &  &  & GL$^{\rm c}$ & yes & 208.65$_{-2.36}^{+1.87}$ & -0.62$_{-0.03}^{+0.02}$ & 36.65$_{-0.27}^{+0.23}$ & 4.05$_{-0.19}^{+0.17}$ & 2.63$_{-0.12}^{+0.11}$\\  
 &  &  &  &  &  & GR & yes & 206.59$_{-2.82}^{+3.04}$ & -0.55$_{-0.09}^{+0.09}$ & 37.92$_{-0.46}^{+1.56}$ & 2.55$_{-0.13}^{+0.05}$ & 1.66$_{-0.08}^{+0.03}$\\ 
MACS~J0416.1$-$2403 & 64.038078 & -24.067497 & 0.396 & 11.53$_{-2.18}^{+2.02}$ & 18.08$_{-3.24}^{+3.51}$ & L-S$^{\rm a}$ & yes & 154.70$_{-0.82}^{+1.27}$ & -0.51$_{-0.01}^{+0.01}$ & 32.26$_{-0.09}^{+0.04}$ & 2.28$_{-0.02}^{+0.07}$ & 1.74$_{-0.02}^{+0.05}$\\  
 &  &  &  &  &  & L-Cam$^{\rm e}$ & yes & 157.83$_{-2.21}^{+0.51}$ & -0.44$_{-0.02}^{+0.01}$ & 31.09$_{-0.00}^{+0.59}$ & 2.45$_{-0.10}^{+0.04}$ & 1.87$_{-0.08}^{+0.03}$\\ 
 &  &  &  &  &  & L-CATS & yes & 158.50$_{-0.37}^{+1.50}$ & -0.45$_{-0.03}^{+0.01}$ & 31.87$_{-0.23}^{+0.02}$ & 2.36$_{-0.03}^{+0.05}$ & 1.80$_{-0.02}^{+0.04}$\\ 
 &  &  &  &  &  & G$^{\rm b}$ & yes & 147.14$_{-1.91}^{+3.60}$ & -0.52$_{-0.02}^{+0.03}$ & 31.83$_{-0.02}^{+0.36}$ & 2.09$_{-0.10}^{+0.17}$ & 1.60$_{-0.08}^{+0.13}$\\
 &  &  &  &  &  & GL$^{\rm c}$ & yes & 168.69$_{-5.22}^{+1.55}$ & -0.41$_{-0.02}^{+0.02}$ & 32.65$_{-0.38}^{+0.20}$ & 3.05$_{-0.21}^{+0.05}$ & 2.33$_{-0.16}^{+0.04}$\\ 
 &  &  &  &  &  & GR & yes & 141.33$_{-1.05}^{+1.68}$ & -0.61$_{-0.05}^{+0.04}$ & 32.79$_{-0.29}^{+0.61}$ & 1.64$_{-0.06}^{+0.03}$ & 1.25$_{-0.04}^{+0.03}$\\ 
MACS~J0717.5$+$3745 & 109.398239 & 37.745731 & 0.545 & 11.49$_{-0.55}^{+0.53}$ & 18.24$_{-0.89}^{+0.86}$ & L-S$^{\rm a}$ & yes & 234.73$_{-1.30}^{+1.53}$ & -0.23$_{-0.02}^{+0.03}$ & 58.51$_{-0.47}^{+0.15}$ & 7.15$_{-0.10}^{+0.19}$ & 7.81$_{-0.11}^{+0.21}$\\ 
MACS~J1149.5$+$2223 & 177.39875 & 22.398533 & 0.543 & 14.36$_{-4.30}^{+4.00}$ & 22.92$_{-15.19}^{+12.93}$ & L-S$^{\rm a}$ & yes & 177.85$_{-0.36}^{+2.35}$ & -0.51$_{-0.01}^{+0.01}$ & 30.74$_{-0.35}^{+0.33}$ & 4.19$_{-0.08}^{+0.18}$ & 4.56$_{-0.08}^{+0.20}$\\
 &  &  &  &  &  & L-CATS$^{\rm f}$ & yes & 177.28$_{-1.00}^{+2.40}$ & -0.52$_{-0.01}^{+0.02}$ & 32.71$_{-0.26}^{+0.63}$ & 3.35$_{-0.01}^{+0.23}$ & 3.65$_{-0.01}^{+0.24}$\\ 
 &  &  &  &  &  & GL$^{\rm c}$ & yes & 173.96$_{-6.08}^{+6.55}$ & -0.54$_{-0.03}^{+0.03}$ & 30.52$_{-0.53}^{+0.62}$ & 3.43$_{-0.39}^{+0.43}$ & 3.74$_{-0.42}^{+0.47}$\\  
 &  &  &  &  &  & GR & yes & 170.78$_{-1.82}^{+2.87}$ & -0.62$_{-0.07}^{+0.06}$ & 30.83$_{-0.04}^{+1.35}$ & 1.77$_{-0.13}^{+0.06}$ & 1.93$_{-0.14}^{+0.06}$\\
\hline
\\
RELICS \\
\hline
Abell~1758a~NW$^{\rm g}$ & 203.159975 & 50.55994 & 0.2799 & 8.22$_{-0.28}^{+0.27}$ & 12.66$_{-0.44}^{+0.43}$ & G & unknown & 152.96$_{-6.67}^{+42.26}$ & -0.59$_{-0.05}^{+0.06}$ & 24.84$_{-1.08}^{+1.02}$ & 2.51$_{-0.61}^{+1.15}$ & 1.20$_{-0.29}^{+0.55}$\\ 
Abell~1758a~SE$^{\rm g}$ & 203.21692 & 50.526126 & 0.2799 &  &  & G & unknown & 147.88$_{-45.85}^{+37.70}$ & -0.63$_{-0.22}^{+0.03}$ & 16.70$_{-3.92}^{+2.69}$ & 1.93$_{-1.33}^{+0.68}$ & 0.93$_{-0.64}^{+0.33}$\\ 
Abell~2163 & 243.953949 & -6.144829 & 0.203 & 16.12$_{-0.29}^{+0.30}$ & 25.05$_{-0.46}^{+0.48}$ & L$^{\rm h}$ & no & 107.04$_{-7.24}^{+20.16}$ & -0.92$_{-0.04}^{+0.06}$ & 14.91$_{-0.11}^{+0.42}$ & 0.84$_{-0.19}^{+0.30}$ & 0.25$_{-0.06}^{+0.09}$\\ 
 &  &  &  &  &  & G & unknown & 168.45$_{-37.53}^{+52.51}$ & -0.50$_{-0.20}^{+0.10}$ & 15.02$_{-0.51}^{+0.20}$ & 2.72$_{-1.62}^{+3.33}$ & 0.81$_{-0.48}^{+0.99}$\\ 
Abell~2537 & 347.092607 & -2.19217 & 0.2966 & 5.52$_{-0.51}^{+0.51}$ & 8.43$_{-0.80}^{+0.80}$ & L$^{\rm h}$ & yes & 202.99$_{-0.92}^{+2.90}$ & -0.66$_{-0.02}^{+0.03}$ & 32.25$_{-0.15}^{+0.32}$ & 3.63$_{-0.15}^{+0.44}$ & 1.90$_{-0.08}^{+0.23}$\\ 
 &  &  &  &  &  & G & yes & 216.70$_{-16.67}^{+1.95}$ & -0.51$_{-0.18}^{+0.02}$ & 33.61$_{-0.66}^{+0.12}$ & 6.22$_{-2.82}^{+0.35}$ & 3.23$_{-1.47}^{+0.18}$\\ 
Abell~2813 & 10.852708 & -20.628215 & 0.2924 & 8.13$_{-0.38}^{+0.37}$ & 12.53$_{-0.60}^{+0.58}$ & L & yes & 168.54$_{-1.94}^{+4.95}$ & -0.47$_{-0.02}^{+0.01}$ & 28.52$_{-0.02}^{+0.66}$ & 2.67$_{-0.10}^{+0.10}$ & 1.37$_{-0.05}^{+0.05}$\\ 
Abell~3192 & 59.725312 & -29.925271 & 0.425 & 7.20$_{-0.50}^{+0.52}$ & 11.19$_{-0.80}^{+0.83}$ & L & yes & 184.14$_{-4.04}^{+4.58}$ & -0.56$_{-0.01}^{+0.04}$ & 32.87$_{-0.14}^{+1.08}$ & 2.45$_{-0.10}^{+0.22}$ & 2.04$_{-0.09}^{+0.19}$\\ 
Abell~697 & 130.739821 & 36.366498 & 0.282 & 11.00$_{-0.37}^{+0.37}$ & 17.06$_{-0.59}^{+0.59}$ & LTM$^{\rm i}$ & no & 58.97$_{-0.47}^{+12.66}$ & -1.26$_{-0.03}^{+0.15}$ & 12.85$_{-0.13}^{+0.23}$ & 0.28$_{-0.04}^{+0.08}$ & 0.13$_{-0.02}^{+0.04}$\\ 
Abell~S295 & 41.353387 & -53.029324 & 0.3 & 6.78$_{-0.36}^{+0.37}$ & 10.41$_{-0.57}^{+0.58}$ & LTM$^{\rm i}$ & yes & 146.14$_{-8.01}^{+18.36}$ & -0.37$_{-0.08}^{+0.04}$ & 17.77$_{-0.46}^{+1.71}$ & 4.38$_{-0.49}^{+0.80}$ & 2.33$_{-0.26}^{+0.43}$\\ 
ACT$-$CL~J0102$-$49151 & 15.740679 & -49.272001 & 0.87 & 10.75$_{-0.47}^{+0.48}$ & 17.44$_{-0.78}^{+0.80}$ & L$^{\rm h}$ & no & 182.56$_{-10.13}^{+22.63}$ & -0.64$_{-0.07}^{+0.04}$ & 40.29$_{-0.77}^{+1.43}$ & 3.59$_{-0.74}^{+0.67}$ & 5.74$_{-1.18}^{+1.07}$\\ 
 &  &  &  &  &  & G & unknown & 179.71$_{-5.14}^{+42.76}$ & -0.50$_{-0.09}^{+0.03}$ & 43.38$_{-1.93}^{+1.30}$ & 4.74$_{-0.87}^{+1.74}$ & 7.65$_{-1.41}^{+2.81}$\\ 
 CL~J0152.7$-$1357 & 28.182425 & -13.95515 & 0.833 & \nodata & \nodata & L & yes & 119.09$_{-6.99}^{+7.96}$ & -0.78$_{-0.06}^{+0.09}$ & 17.88$_{-0.68}^{+0.78}$ & 0.81$_{-0.09}^{+0.11}$ & 1.26$_{-0.13}^{+0.17}$\\ 
 &  &  &  &  &  & LTM$^{\rm j}$ & yes & 115.24$_{-0.00}^{+21.01}$ & -0.68$_{-0.02}^{+0.05}$ & 19.29$_{-0.76}^{+2.47}$ & 1.55$_{-0.00}^{+0.32}$ & 2.41$_{-0.00}^{+0.50}$\\ 
 &  &  &  &  &  & G & yes & 128.40$_{-15.82}^{+9.37}$ & -0.55$_{-0.14}^{+0.01}$ & 18.32$_{-1.53}^{+0.25}$ & 1.45$_{-0.39}^{+0.07}$ & 2.28$_{-0.62}^{+0.11}$\\
MACS~J0025.4$-$1222 & 6.364154 & -12.373026 & 0.586 & \nodata & \nodata & LTM$^{\rm i}$ & yes & 103.79$_{-1.50}^{+0.80}$ & -0.84$_{-0.02}^{+0.02}$ & 16.73$_{-0.30}^{+0.40}$ & 2.02$_{-0.05}^{+0.05}$ & 2.37$_{-0.06}^{+0.06}$\\ 
MACS~J0035.4$-$2015 & 8.858893 & -20.262288 & 0.352 & 7.01$_{-0.50}^{+0.45}$ & 10.82$_{-0.79}^{+0.71}$ & L & yes & 146.91$_{-1.35}^{+3.57}$ & -0.70$_{-0.03}^{+0.04}$ & 24.71$_{-0.17}^{+0.32}$ & 1.66$_{-0.06}^{+0.15}$ & 1.09$_{-0.04}^{+0.10}$\\ 
 &  &  &  &  &  & G & unknown & 174.08$_{-7.33}^{+28.47}$ & -0.69$_{-0.11}^{+0.12}$ & 28.13$_{-0.78}^{+0.47}$ & 2.49$_{-0.84}^{+1.50}$ & 1.64$_{-0.55}^{+0.99}$\\ 
MACS~J0159.8$-$0849 & 29.955451 & -8.832999 & 0.405 & 7.20$_{-0.68}^{+0.61}$ & 11.17$_{-1.08}^{+0.97}$ & LTM$^{\rm i}$ & no & 179.81$_{-5.00}^{+28.38}$ & -1.12$_{-0.13}^{+0.13}$ & 30.97$_{-0.50}^{+1.47}$ & 1.90$_{-0.55}^{+0.75}$ & 1.50$_{-0.43}^{+0.59}$\\ 
 &  &  &  &  &  & G & unknown & 206.97$_{-26.62}^{+22.03}$ & -0.71$_{-0.45}^{+0.01}$ & 33.31$_{-2.86}^{+0.47}$ & 3.12$_{-2.09}^{+0.59}$ & 2.44$_{-1.64}^{+0.46}$\\ 
MACS~J0257.1$-$2325 & 44.286469 & -23.434684 & 0.5049 & 6.22$_{-0.74}^{+0.70}$ & 9.69$_{-1.18}^{+1.12}$ & L & yes & 99.44$_{-11.95}^{+1.03}$ & -0.96$_{-0.06}^{+0.01}$ & 16.01$_{-0.76}^{+0.09}$ & 0.53$_{-0.11}^{+0.01}$ & 0.54$_{-0.11}^{+0.01}$\\ 
MACS~J0308.9$+$2645 & 47.2331734 & 26.7605103 & 0.356 & 10.76$_{-0.65}^{+0.63}$ & 16.79$_{-1.04}^{+1.01}$ & LTM$^{\rm k}$ & no & 261.64$_{-2.03}^{+30.31}$ & -0.87$_{-0.03}^{+0.04}$ & 46.59$_{-0.56}^{+1.15}$ & 4.99$_{-0.86}^{+0.98}$ & 3.34$_{-0.57}^{+0.66}$\\ 
 &  &  &  &  &  & G & unknown & 270.29$_{-16.50}^{+37.63}$ & -0.68$_{-0.22}^{+0.08}$ & 47.31$_{-4.36}^{+2.25}$ & 5.45$_{-2.86}^{+2.27}$ & 3.63$_{-1.90}^{+1.51}$\\ 
MACS~J0417.5$-$1154 & 64.39454 & -11.908851 & 0.443 & 12.25$_{-0.55}^{+0.53}$ & 19.32$_{-0.89}^{+0.86}$ & L$^{\rm l}$ & yes & 180.71$_{-2.81}^{+3.24}$ & -0.59$_{-0.02}^{+0.02}$ & 37.35$_{-0.44}^{+0.36}$ & 3.10$_{-0.14}^{+0.16}$ & 2.71$_{-0.13}^{+0.14}$\\ 
 &  &  &  &  &  & G$^{\rm m}$ & yes & 190.98$_{-4.96}^{+0.65}$ & -0.55$_{-0.02}^{+0.01}$ & 39.84$_{-1.42}^{+0.00}$ & 4.33$_{-0.27}^{+0.13}$ & 3.78$_{-0.23}^{+0.11}$\\ 
MACS~J0553.4$-$3342 & 88.330692 & -33.707542 & 0.43 & 8.77$_{-0.46}^{+0.44}$ & 13.70$_{-0.74}^{+0.70}$ & L & yes & 171.76$_{-4.61}^{+3.38}$ & -0.41$_{-0.04}^{+0.03}$ & 42.30$_{-0.32}^{+0.15}$ & 4.53$_{-0.29}^{+0.26}$ & 3.83$_{-0.24}^{+0.22}$\\ 
 &  &  &  &  &  & G$^{\rm m}$ & yes & 146.26$_{-4.52}^{+5.84}$ & -0.56$_{-0.05}^{+0.05}$ & 41.06$_{-0.87}^{+0.73}$ & 3.39$_{-0.20}^{+0.43}$ & 2.86$_{-0.17}^{+0.36}$\\ 
MS~1008.1$-$1224 & 152.634522 & -12.664695 & 0.3062 & 4.94$_{-0.60}^{+0.57}$ & 7.53$_{-0.93}^{+0.89}$ & L & yes & 130.62$_{-3.44}^{+5.46}$ & -0.79$_{-0.06}^{+0.07}$ & 22.28$_{-0.91}^{+0.32}$ & 1.25$_{-0.13}^{+0.16}$ & 0.69$_{-0.07}^{+0.09}$\\ 
PLCK~G004.5$-$19.5$^{\rm n}$ & 289.270977 & -33.522357 & 0.54 & 10.36$_{-0.72}^{+0.68}$ & 16.40$_{-1.17}^{+1.10}$ & L & no & 244.86$_{-0.20}^{+24.94}$ & -0.67$_{-0.01}^{+0.01}$ & 47.83$_{-0.17}^{+0.07}$ & 4.05$_{-0.02}^{+0.05}$ & 4.38$_{-0.02}^{+0.06}$\\ 
PLCK~G171.9$-$40.7 & 48.239437 & 8.369767 & 0.27 & 10.71$_{-0.50}^{+0.49}$ & 16.58$_{-0.79}^{+0.78}$ & LTM$^{\rm k}$ & no & 200.82$_{-2.64}^{+26.52}$ & -0.90$_{-0.02}^{+0.06}$ & 40.57$_{-0.70}^{+1.36}$ & 2.58$_{-0.00}^{+0.61}$ & 1.18$_{-0.00}^{+0.28}$\\ 
 &  &  &  &  &  & G & unknown & 229.52$_{-12.73}^{+22.95}$ & -0.64$_{-0.14}^{+0.02}$ & 39.18$_{-1.29}^{+1.00}$ & 4.89$_{-1.79}^{+0.73}$ & 2.22$_{-0.81}^{+0.33}$\\ 
PLCK~G287.0$+$32.9 & 177.709036 & -28.082135 & 0.39 & 14.69$_{-0.42}^{+0.39}$ & 23.17$_{-0.68}^{+0.63}$ & LTM$^{\rm o}$ & no & 266.87$_{-2.93}^{+28.31}$ & -0.68$_{-0.03}^{+0.05}$ & 53.84$_{-0.11}^{+0.64}$ & 8.19$_{-1.20}^{+0.97}$ & 6.15$_{-0.90}^{+0.73}$\\ 
 &  &  &  &  &  & G & unknown & 263.27$_{-0.08}^{+36.22}$ & -0.40$_{-0.11}^{+0.02}$ & 55.68$_{-0.45}^{+2.25}$ & 9.72$_{-2.43}^{+0.77}$ & 7.27$_{-1.82}^{+0.58}$\\ 
PSZ2~G209.79$+$10.23 & 110.598933 & 7.408627 & 0.677 & 10.73$_{-0.66}^{+0.63}$ & 17.17$_{-1.08}^{+1.04}$ & L & yes & 108.11$_{-5.56}^{+10.91}$ & -0.80$_{-0.08}^{+0.14}$ & 27.74$_{-0.88}^{+0.67}$ & 0.81$_{-0.15}^{+0.30}$ & 1.08$_{-0.21}^{+0.40}$\\ 
RXC~J0018.5$+$1626 & 4.639919 & 16.437871 & 0.546 & 9.79$_{-0.53}^{+0.53}$ & 15.48$_{-0.86}^{+0.86}$ & L & yes & 182.29$_{-1.10}^{+3.38}$ & -0.37$_{-0.02}^{+0.02}$ & 30.87$_{-0.48}^{+0.38}$ & 2.54$_{-0.02}^{+0.22}$ & 2.79$_{-0.02}^{+0.24}$\\ 
RXC~J0032.1$+$1808 & 8.049474 & 18.14366 & 0.3956 & 7.61$_{-0.63}^{+0.57}$ & 11.81$_{-1.00}^{+0.91}$ & L & yes & 179.23$_{-0.82}^{+8.45}$ & -0.65$_{-0.02}^{+0.09}$ & 50.11$_{-0.31}^{+0.62}$ & 6.51$_{-0.09}^{+0.82}$ & 4.98$_{-0.07}^{+0.63}$\\ 
 &  &  &  &  &  & LTM$^{\rm p}$ & yes & 210.40$_{-1.07}^{+1.75}$ & -0.57$_{-0.03}^{+0.02}$ & 49.67$_{-0.17}^{+0.31}$ & 7.47$_{-0.18}^{+0.04}$ & 5.71$_{-0.13}^{+0.03}$\\ 
 &  &  &  &  &  & G & unknown & 199.67$_{-2.61}^{+30.38}$ & -0.65$_{-0.02}^{+0.06}$ & 50.47$_{-0.29}^{+1.37}$ & 5.34$_{-0.70}^{+1.53}$ & 4.06$_{-0.53}^{+1.16}$\\ 
RXC~J0142.9$+$4438 & 25.730087 & 44.634676 & 0.341 & 9.02$_{-0.64}^{+0.60}$ & 14.00$_{-1.02}^{+0.95}$ & L$^{\rm h}$ & no & 213.58$_{-2.46}^{+23.00}$ & -0.71$_{-0.04}^{+0.03}$ & 34.15$_{-0.23}^{+0.17}$ & 3.10$_{-0.55}^{+0.58}$ & 1.96$_{-0.35}^{+0.37}$\\ 
 &  &  &  &  &  & G & unknown & 214.77$_{-11.71}^{+46.90}$ & -0.66$_{-0.13}^{+0.03}$ & 34.31$_{-1.57}^{+0.36}$ & 3.68$_{-1.27}^{+1.31}$ & 2.32$_{-0.80}^{+0.82}$\\ 
RXC~J0232.2$-$4420 & 38.077317 & -44.346666 & 0.2836 & 7.54$_{-0.32}^{+0.33}$ & 11.59$_{-0.50}^{+0.52}$ & L & yes & 199.37$_{-6.63}^{+1.09}$ & -0.69$_{-0.05}^{+0.03}$ & 30.50$_{-0.66}^{+0.11}$ & 2.76$_{-0.26}^{+0.08}$ & 1.36$_{-0.13}^{+0.04}$\\ 
RXC~J0600.1$-$2007 & 90.034008 & -20.135751 & 0.46 & 10.73$_{-0.54}^{+0.51}$ & 16.89$_{-0.87}^{+0.82}$ & L & no & 206.32$_{-9.74}^{+21.98}$ & -0.37$_{-0.06}^{+0.02}$ & 39.99$_{-3.28}^{+0.72}$ & 3.67$_{-0.62}^{+0.48}$ & 3.35$_{-0.87}^{+0.44}$\\ 
 &  &  &  &  &  & G$^{\rm m,q}$ & unknown & 218.82$_{-4.18}^{+47.28}$ & -0.43$_{-0.05}^{+0.04}$ & 49.53$_{-0.00}^{+0.67}$ & 10.79$_{-2.93}^{+3.13}$ & 9.09$_{-2.47}^{+2.63}$\\ 
RXC~J0911.1$+$1746 & 137.797997 & 17.774794 & 0.5049 & 6.99$_{-0.79}^{+0.73}$ & 10.92$_{-1.26}^{+1.17}$ & G & unknown & 112.81$_{-9.76}^{+68.38}$ & -0.66$_{-0.34}^{+0.03}$ & 14.85$_{-0.57}^{+8.86}$ & 0.89$_{-0.35}^{+0.71}$ & 0.90$_{-0.35}^{+0.72}$\\ 
RXC~J0949.8$+$1707 & 147.4659 & 17.119601 & 0.3826 & 8.24$_{-0.46}^{+0.46}$ & 12.80$_{-0.73}^{+0.73}$ & L & no & 159.34$_{-0.00}^{+38.42}$ & -0.75$_{-0.05}^{+0.05}$ & 25.63$_{-0.00}^{+5.73}$ & 2.19$_{-0.00}^{+1.00}$ & 1.60$_{-0.00}^{+0.74}$\\ 
 &  &  &  &  &  & G & unknown & 181.02$_{-6.57}^{+43.57}$ & -0.49$_{-0.08}^{+0.06}$ & 26.31$_{-0.51}^{+1.27}$ & 5.20$_{-1.39}^{+2.40}$ & 3.80$_{-1.01}^{+1.75}$\\ 
RXC~J2211.7$-$0350 & 332.941368 & -3.828952 & 0.397 & 10.50$_{-0.49}^{+0.50}$ & 16.43$_{-0.79}^{+0.80}$ & L$^{\rm h}$ & yes & 295.64$_{-7.49}^{+1.89}$ & -0.59$_{-0.11}^{+0.04}$ & 56.01$_{-2.67}^{+0.98}$ & 6.43$_{-1.15}^{+0.54}$ & 4.93$_{-0.88}^{+0.41}$\\ 
 &  &  &  &  &  & G & yes & 281.15$_{-16.63}^{+7.33}$ & -0.77$_{-0.13}^{+0.07}$ & 51.30$_{-3.01}^{+2.08}$ & 4.53$_{-1.52}^{+1.15}$ & 3.47$_{-1.17}^{+0.88}$\\ 
RXS~J060313.4$+$4212$-$N & 90.819495 & 42.244882 & 0.228 & 10.76$_{-0.43}^{+0.45}$ & 16.60$_{-0.68}^{+0.71}$ & G & unknown & 153.09$_{-25.12}^{+48.68}$ & -0.64$_{-0.23}^{+0.13}$ & 18.91$_{-1.57}^{+1.03}$ & 1.76$_{-0.81}^{+1.52}$ & 0.62$_{-0.29}^{+0.54}$\\ 
RXS~J060313.4$+$4212$-$S & 90.851319 & 42.158696 & 0.228 &  &  & G & unknown & 176.38$_{-26.50}^{+40.05}$ & -0.71$_{-0.18}^{+0.03}$ & 26.55$_{-0.75}^{+0.58}$ & 2.33$_{-0.94}^{+0.45}$ & 0.83$_{-0.33}^{+0.16}$\\ 
SMACS~J0723.3$-$7327 & 110.826755 & -73.454628 & 0.39 & 8.39$_{-0.34}^{+0.33}$ & 13.05$_{-0.54}^{+0.53}$ & L & no & 136.36$_{-0.00}^{+25.68}$ & -0.67$_{-0.09}^{+0.02}$ & 21.46$_{-0.00}^{+2.84}$ & 1.42$_{-0.21}^{+0.53}$ & 1.07$_{-0.16}^{+0.40}$\\ 
 &  &  &  &  &  & G & unknown & 157.22$_{-22.45}^{+36.19}$ & -0.63$_{-0.19}^{+0.06}$ & 26.10$_{-2.60}^{+0.89}$ & 2.65$_{-1.29}^{+1.12}$ & 1.98$_{-0.96}^{+0.84}$\\ 
SPT$-$CL~J0615$-$5746 & 93.965434 & -57.780113 & 0.972 & 6.77$_{-0.54}^{+0.49}$ & 10.92$_{-0.89}^{+0.81}$ & L$^{\rm r}$ & yes & 242.03$_{-11.26}^{+12.61}$ & -0.54$_{-0.04}^{+0.04}$ & 34.45$_{-2.10}^{+2.69}$ & 2.47$_{-0.29}^{+0.33}$ & 4.19$_{-0.49}^{+0.56}$\\ 
 &  &  &  &  &  & LTM & yes & 233.74$_{-8.75}^{+3.49}$ & -0.69$_{-0.06}^{+0.02}$ & 36.38$_{-1.58}^{+0.77}$ & 2.12$_{-0.09}^{+0.00}$ & 3.61$_{-0.15}^{+0.00}$\\ 
\hline 
\\
SGAS$^{\rm s}$ \\
\hline
SDSS~J0004$-$0103 & 1.216424 & -1.0544069 & 0.514 & 0.98$_{-0.73}^{+1.52}$ & 1.50$_{-1.13}^{+2.38}$ & L & yes & 40.82$_{-3.64}^{+5.85}$ & -1.18$_{-0.05}^{+0.10}$ & 6.10$_{-0.88}^{+1.07}$ & 0.09$_{-0.02}^{+0.02}$ & 0.09$_{-0.02}^{+0.02}$\\ 
SDSS~J0108$+$0624 & 17.1751119 & 6.4121008 & 0.54778 & \nodata & \nodata & L & yes & 88.36$_{-43.64}^{+0.07}$ & -0.64$_{-0.18}^{+0.05}$ & 4.56$_{-1.64}^{+1.99}$ & 0.40$_{-0.35}^{+0.00}$ & 0.44$_{-0.39}^{+0.00}$\\ 
SDSS~J0146$-$0929 & 26.733363 & -9.497921 & 0.4469 & \nodata & \nodata & L & yes & 111.47$_{-4.24}^{+4.85}$ & -0.98$_{-0.03}^{+0.04}$ & 16.20$_{-0.26}^{+0.31}$ & 0.65$_{-0.05}^{+0.07}$ & 0.57$_{-0.05}^{+0.06}$\\ 
SDSS~J0150$+$2725 & 27.503546 & 27.426761 & 0.30619 & \nodata & \nodata & L & yes & 180.65$_{-3.64}^{+13.90}$ & -0.52$_{-0.02}^{+0.07}$ & 19.64$_{-0.53}^{+3.01}$ & 2.82$_{-0.18}^{+0.78}$ & 1.54$_{-0.10}^{+0.43}$\\ 
SDSS~J0333$-$0651 & 53.269401 & -6.856349 & 0.5729 & \nodata & \nodata & L & yes & 81.42$_{-3.76}^{+11.68}$ & -1.09$_{-0.01}^{+0.14}$ & 12.85$_{-0.46}^{+0.60}$ & 0.28$_{-0.03}^{+0.13}$ & 0.33$_{-0.04}^{+0.15}$\\ 
SDSS~J0851$+$3331 & 132.911942 & 33.518369 & 0.3689 & 4.19$_{-1.83}^{+2.56}$ & 6.58$_{-2.88}^{+3.96}$ & L & yes & 160.28$_{-11.80}^{+3.44}$ & -0.68$_{-0.09}^{+0.03}$ & 22.33$_{-1.38}^{+0.64}$ & 1.71$_{-0.33}^{+0.11}$ & 1.20$_{-0.23}^{+0.08}$\\ 
SDSS~J0915$+$3826 & 138.912799 & 38.449517 & 0.3961 & 5.08$_{-1.97}^{+2.67}$ & 8.04$_{-3.09}^{+4.11}$ & L & yes & 131.77$_{-5.45}^{+10.21}$ & -0.51$_{-0.06}^{+0.07}$ & 17.64$_{-0.31}^{+0.68}$ & 1.54$_{-0.25}^{+0.45}$ & 1.18$_{-0.19}^{+0.35}$\\ 
SDSS~J0928$+$2031 & 142.018891 & 20.52919 & 0.192 & 5.89$_{-3.48}^{+5.71}$ & 9.23$_{-5.49}^{+8.91}$ & L & yes & 141.97$_{-10.55}^{+10.72}$ & -0.83$_{-0.05}^{+0.05}$ & 13.14$_{-0.56}^{+0.44}$ & 1.37$_{-0.23}^{+0.23}$ & 0.37$_{-0.06}^{+0.06}$\\ 
SDSS~J0952$+$3434 & 148.167609 & 34.579473 & 0.357 & 2.20$_{-1.63}^{+3.21}$ & 3.42$_{-2.56}^{+5.03}$ & L & yes & 103.35$_{-6.97}^{+7.84}$ & -0.93$_{-0.06}^{+0.05}$ & 14.17$_{-0.19}^{+0.46}$ & 0.62$_{-0.09}^{+0.11}$ & 0.42$_{-0.06}^{+0.07}$\\ 
SDSS~J0957$+$0509 & 149.4132969 & 5.158845 & 0.448 & 10.62$_{-6.49}^{+11.32}$ & 17.22$_{-10.61}^{+18.53}$ & L & yes & 58.18$_{-4.64}^{+2.54}$ & -1.03$_{-0.03}^{+0.06}$ & 8.18$_{-0.48}^{+0.24}$ & 0.19$_{-0.02}^{+0.03}$ & 0.17$_{-0.02}^{+0.02}$\\ 
SDSS~J1038$+$4849 & 159.681586 & 48.821594 & 0.4308 & 1.68$_{-0.56}^{+0.74}$ & 2.59$_{-0.84}^{+1.12}$ & L & yes & 70.65$_{-1.55}^{+2.20}$ & -1.35$_{-0.01}^{+0.03}$ & 13.22$_{-0.12}^{+0.11}$ & 0.24$_{-0.01}^{+0.02}$ & 0.20$_{-0.01}^{+0.02}$\\ 
SDSS~J1050$+$0017 & 162.6663711 & 0.285224 & 0.5931 & 0.93$_{-0.39}^{+0.57}$ & 1.44$_{-0.60}^{+0.87}$ & L & yes & 171.67$_{-1.59}^{+1.97}$ & -0.55$_{-0.02}^{+0.03}$ & 23.73$_{-0.17}^{+0.17}$ & 1.92$_{-0.11}^{+0.15}$ & 2.27$_{-0.13}^{+0.17}$\\ 
SDSS~J1055$+$5547 & 163.769173 & 55.806467 & 0.466 & 1.77$_{-1.34}^{+3.01}$ & 2.76$_{-2.10}^{+4.79}$ & L & yes & 165.78$_{-3.08}^{+3.00}$ & -0.94$_{-0.03}^{+0.02}$ & 27.00$_{-0.53}^{+0.52}$ & 1.29$_{-0.07}^{+0.06}$ & 1.20$_{-0.07}^{+0.06}$\\ 
SDSS~J1110$+$6459 & 167.573863 & 64.996641 & 0.656 & 5.21$_{-3.15}^{+5.07}$ & 8.45$_{-5.14}^{+8.22}$ & L & yes & 117.01$_{-7.38}^{+4.40}$ & -0.61$_{-0.19}^{+0.04}$ & 16.22$_{-0.81}^{+0.80}$ & 0.55$_{-0.06}^{+0.03}$ & 0.71$_{-0.08}^{+0.03}$\\ 
SDSS~J1115$+$1645 & 168.7684547 & 16.7605789 & 0.537 & \nodata & \nodata & L & yes & 93.38$_{-3.07}^{+6.93}$ & -0.81$_{-0.03}^{+0.03}$ & 10.59$_{-0.40}^{+1.10}$ & 0.47$_{-0.04}^{+0.09}$ & 0.50$_{-0.05}^{+0.09}$\\ 
SDSS~J1138$+$2754 & 174.537311 & 27.908537 & 0.451 & 21.69$_{-13.55}^{+23.51}$ & 36.00$_{-22.78}^{+38.98}$ & L & yes & 132.24$_{-3.42}^{+2.57}$ & -0.77$_{-0.03}^{+0.03}$ & 20.78$_{-0.31}^{+0.20}$ & 1.10$_{-0.06}^{+0.04}$ & 0.98$_{-0.06}^{+0.04}$\\ 
SDSS~J1152$+$0930 & 178.19748 & 9.504093 & 0.517 & 13.10$_{-8.07}^{+14.03}$ & 21.50$_{-13.35}^{+23.32}$ & L & yes & 156.56$_{-2.41}^{+0.55}$ & -0.44$_{-0.02}^{+0.02}$ & 17.60$_{-0.33}^{+0.09}$ & 1.98$_{-0.08}^{+0.02}$ & 2.05$_{-0.08}^{+0.02}$\\ 
SDSS~J1152$+$3313 & 178.000771 & 33.228271 & 0.3612 & 3.10$_{-1.10}^{+1.44}$ & 4.81$_{-1.68}^{+2.20}$ & L & yes & 107.26$_{-0.06}^{+1.75}$ & -0.79$_{-0.02}^{+0.02}$ & 11.18$_{-0.03}^{+0.08}$ & 0.67$_{-0.00}^{+0.03}$ & 0.46$_{-0.00}^{+0.02}$\\ 
SDSS~J1156$+$1911 & 179.02267 & 19.186919 & 0.54547 & \nodata & \nodata & L & yes & 131.14$_{-40.05}^{+13.96}$ & -0.69$_{-0.16}^{+0.06}$ & 14.68$_{-3.90}^{+2.76}$ & 1.01$_{-0.54}^{+0.29}$ & 1.10$_{-0.59}^{+0.31}$\\ 
SDSS~J1207$+$5254 & 181.899649 & 52.916438 & 0.275 & 3.65$_{-3.29}^{+9.78}$ & 5.70$_{-5.16}^{+15.72}$ & L & yes & 108.72$_{-17.54}^{+15.07}$ & -0.77$_{-0.06}^{+0.08}$ & 9.16$_{-0.39}^{+0.80}$ & 0.73$_{-0.22}^{+0.31}$ & 0.34$_{-0.10}^{+0.15}$\\ 
SDSS~J1209$+$2640 & 182.348771 & 26.679503 & 0.5606 & 7.85$_{-4.16}^{+6.62}$ & 12.73$_{-6.79}^{+10.80}$ & L & yes & 208.00$_{-0.36}^{+1.13}$ & -0.76$_{-0.02}^{+0.02}$ & 32.35$_{-0.10}^{+0.22}$ & 2.23$_{-0.03}^{+0.02}$ & 2.51$_{-0.03}^{+0.02}$\\ 
SDSS~J1329$+$2243 & 202.39391 & 22.721064 & 0.4427 & 3.26$_{-1.43}^{+2.03}$ & 5.12$_{-2.24}^{+3.15}$ & L & yes & 128.87$_{-13.96}^{+1.47}$ & -0.54$_{-0.11}^{+0.02}$ & 10.75$_{-0.27}^{+0.14}$ & 1.42$_{-0.44}^{+0.04}$ & 1.24$_{-0.38}^{+0.04}$\\ 
SDSS~J1336$-$0331 & 204.00035 & -3.524963 & 0.17637 & \nodata & \nodata & L & yes & 162.81$_{-12.07}^{+5.45}$ & -0.77$_{-0.03}^{+0.04}$ & 15.26$_{-1.05}^{+0.39}$ & 1.86$_{-0.31}^{+0.16}$ & 0.45$_{-0.08}^{+0.04}$\\ 
SDSS~J1343$+$4155 & 205.886851 & 41.917627 & 0.418 & 5.88$_{-4.15}^{+7.94}$ & 9.34$_{-6.65}^{+12.78}$ & L & yes & 159.73$_{-27.96}^{+3.76}$ & -0.39$_{-0.09}^{+0.02}$ & 16.96$_{-2.37}^{+0.45}$ & 2.42$_{-0.84}^{+0.12}$ & 1.98$_{-0.68}^{+0.10}$\\ 
SDSS~J1420$+$3955 & 215.166797 & 39.918593 & 0.607 & 6.55$_{-3.09}^{+4.42}$ & 10.61$_{-4.96}^{+7.11}$ & L & yes & 186.56$_{-3.04}^{+1.52}$ & -0.59$_{-0.02}^{+0.03}$ & 29.87$_{-0.62}^{+0.30}$ & 3.69$_{-0.19}^{+0.19}$ & 4.47$_{-0.23}^{+0.23}$\\ 
SDSS~J1439$+$1208 & 219.790757 & 12.140434 & 0.42734 & \nodata & \nodata & L & yes & 209.01$_{-4.78}^{+4.56}$ & -0.38$_{-0.03}^{+0.03}$ & 32.09$_{-1.61}^{+1.65}$ & 4.35$_{-0.38}^{+0.43}$ & 3.65$_{-0.32}^{+0.36}$\\ 
SDSS~J1456$+$5702 & 224.003682 & 57.03898 & 0.484 & 18.93$_{-10.91}^{+17.94}$ & 31.14$_{-18.04}^{+29.87}$ & L & yes & 123.23$_{-10.24}^{+2.10}$ & -0.94$_{-0.07}^{+0.02}$ & 17.62$_{-0.68}^{+0.13}$ & 0.78$_{-0.16}^{+0.04}$ & 0.75$_{-0.15}^{+0.04}$\\ 
SDSS~J1522$+$2535 & 230.719852 & 25.590965 & 0.602 & 8.38$_{-5.52}^{+9.92}$ & 13.70$_{-9.12}^{+16.52}$ & L & yes & 202.27$_{-12.03}^{+8.28}$ & -0.39$_{-0.07}^{+0.05}$ & 30.90$_{-3.54}^{+2.09}$ & 3.36$_{-0.75}^{+0.44}$ & 4.04$_{-0.90}^{+0.52}$\\ 
SDSS~J1531$+$3414 & 232.79429 & 34.240312 & 0.335 & 5.91$_{-3.14}^{+4.99}$ & 9.36$_{-4.99}^{+7.91}$ & L & yes & 127.39$_{-0.47}^{+15.83}$ & -0.84$_{-0.02}^{+0.10}$ & 20.71$_{-0.24}^{+2.52}$ & 1.09$_{-0.03}^{+0.42}$ & 0.68$_{-0.02}^{+0.26}$\\ 
SDSS~J1604$+$2244 & 241.042271 & 22.738579 & 0.286 & 6.92$_{-5.26}^{+11.38}$ & 10.95$_{-8.41}^{+18.28}$ & L & yes & 70.92$_{-3.64}^{+23.60}$ & -1.13$_{-0.05}^{+0.09}$ & 12.57$_{-0.00}^{+3.71}$ & 0.31$_{-0.02}^{+0.22}$ & 0.15$_{-0.01}^{+0.11}$\\ 
SDSS~J1621$+$0607 & 245.384938 & 6.121973 & 0.3429 & 2.30$_{-0.94}^{+1.32}$ & 3.55$_{-1.44}^{+2.01}$ & L & yes & 139.79$_{-3.55}^{+1.38}$ & -0.73$_{-0.02}^{+0.02}$ & 16.12$_{-0.49}^{+0.13}$ & 1.30$_{-0.08}^{+0.03}$ & 0.83$_{-0.05}^{+0.02}$\\ 
SDSS~J1632$+$3500 & 248.042757 & 35.008267 & 0.466 & 9.53$_{-6.73}^{+13.46}$ & 15.46$_{-11.04}^{+22.31}$ & L & yes & 166.03$_{-45.46}^{+3.78}$ & -0.55$_{-0.23}^{+0.02}$ & 22.82$_{-6.05}^{+0.69}$ & 2.02$_{-1.12}^{+0.13}$ & 1.87$_{-1.04}^{+0.12}$\\ 
SDSS~J1723$+$3411 & 260.90068 & 34.199481 & 0.44227 & \nodata & \nodata & L & yes & 80.21$_{-2.88}^{+10.24}$ & -0.93$_{-0.04}^{+0.14}$ & 9.59$_{-0.22}^{+0.41}$ & 0.34$_{-0.03}^{+0.11}$ & 0.30$_{-0.01}^{+0.14}$\\ 
SDSS~J2111$-$0114 & 317.830623 & -1.239844 & 0.6363 & 6.42$_{-2.68}^{+3.55}$ & 10.40$_{-4.30}^{+5.69}$ & L & yes & 155.66$_{-17.94}^{+5.98}$ & -0.50$_{-0.16}^{+0.05}$ & 18.39$_{-6.37}^{+0.36}$ & 1.63$_{-0.53}^{+0.21}$ & 2.07$_{-0.67}^{+0.27}$\\ 
SDSS~J2243$-$0935 & 340.836359 & -9.588595 & 0.447 & 5.53$_{-3.10}^{+4.99}$ & 8.81$_{-4.97}^{+8.04}$ & L & yes & 214.79$_{-1.06}^{+0.62}$ & -0.27$_{-0.01}^{+0.01}$ & 42.60$_{-0.39}^{+0.38}$ & 8.19$_{-0.23}^{+0.19}$ & 7.23$_{-0.20}^{+0.17}$\\ 
\enddata 
\vspace{+.3cm}
All clusters included in this paper, separated by strong lensing programs (HFF, RELICS, and SGAS). $\alpha$ and $\delta$ refer to the right ascension and declination of the BCG selected in the analysis. $z_{lens}$ refers to the redshift of the cluster. $M_{500}$ and $M_{200}$ are the large scale mass estimates, measured independent of the lensing analyses. Alg. refers to the modeling algorithm used: L={\tt{Lenstool}}, LTM=Light-Traces-Mass, G=GLAFIC, GL=GRAVLENS, and GR=GRALE. For the HFF {\tt{Lenstool}} models, letters after the dash indicate the lens modeling team, where 
L-S=Sharon, L-CATS=CATS, and L-Cam=Caminha. 
Spec-$z$ indicates whether or not the lens model employs a spectroscopic redshift constraint or if it is unknown whether a spectroscopic constraint is used. $M_{\rm SL}(200 {\rm kpc})$ refers to the mass within a $200$ kpc aperture from the BCG, computed from the strong lens model. $S_{50-200}$ refers to the logarithmic slope of the projected mass density profile in the range of $50-200$ kpc from the BCG. $e\theta _E$ is the effective Einstein radius for $z_s=9$. \Lstrength\ is the non-corrected lensing strength and \LstrengthN\ is the distance-corrected lensing strength, for $z_s=9$.

\tablecomments{
References for HFF models: $^{\rm a}$ \cite{2014ApJ...797...48J};
$^{\rm b}$ \cite{2018ApJ...855....4K};
$^{\rm c}$ \cite{2020MNRAS.492..503RKeetonmodels};
$^{\rm d}$ \cite{2019MNRAS.485.3738Lcatsabell370};
$^{\rm e}$ \cite{cam};
$^{\rm f}$ \cite{2019MNRAS.485.3738Lcatsabell370} \\[1pt]
$^{\rm g}$ Abell~1758a has two major components (Northwest and Southeast). Although contained in one model, the amplification areas of the lensing strength we compute (see section \ref{sec:lensingstrength}) for each component do not intersect. We therefore split the model into Northwest and Southeast components. RXS~J060313.4$+$4212 has a Northern and Southern component, modeled separately. Because they have been separated into components, we do not include the \textit{Planck} SZ M$_{500}$ mass of these two clusters in total mass comparisons. \\[1pt] 
References for RELICS models: $^{\rm h}$ \cite{2018ApJ...859..159C};
$^{\rm i}$ \cite{2018ApJ...863..145C};
$^{\rm j}$ \cite{2019ApJ...874..132A};
$^{\rm k}$ \cite{2018ApJ...858...42A};
$^{\rm l}$ \cite{2019ApJ...873...96M};
$^{\rm o}$ \cite{2017ApJ...839L..11Z};
$^{\rm p}$ \cite{2020ApJ...898....6A};
$^{\rm r}$ \cite{2018ApJ...863..154P}\\[1pt] 
$^{\rm m}$ Denotes the three GLAFIC models in which an additional 5\% was added to the upper error bar of the lensing strength to account for the magnification map not including the entire $|\mu| \geq 3$ region. \\[1pt] 
$^{\rm n}$PLCK~G004.5$-$19.5 lens redshift was updated to $z=0.519$ in \cite{2017AA607}. The cluster lens was modeled at the redshift reported in the table. \\[1pt]
$^{\rm q}$ The GLAFIC model of RXC~J0600.1$-$2007 employs an updated redshift of 0.43 from more recent spectroscopy with the Multi Unit Spectroscopic Explorer on the Very Large Telescope.\\[1pt] 
Reference for SGAS models: $^{\rm s}$ \cite{2020ApJS..247...12S} \\[1pt]
The availability of lensing constraints in the GLAFIC models of RELICS clusters is based on Table A1 of \citet{2020MNRAS.496.2591O}.\\[1pt]
This table is available at the Strasbourg astronomical Data Center (CDS) via anonymous ftp to cdsarc.u-strasbg.fr (130.79.128.5) or via \url{https://cdsarc.unistra.fr/viz-bin/cat/J/ApJ/928/87}}
\medskip
\end{deluxetable*}
\end{longrotatetable}



\bibliographystyle{yahapj}
\bibliography{biblio_lens_strength}

\appendix

\section{Gallery of Clusters} 
\begin{figure}
\center
    \includegraphics[width=0.23\textwidth]{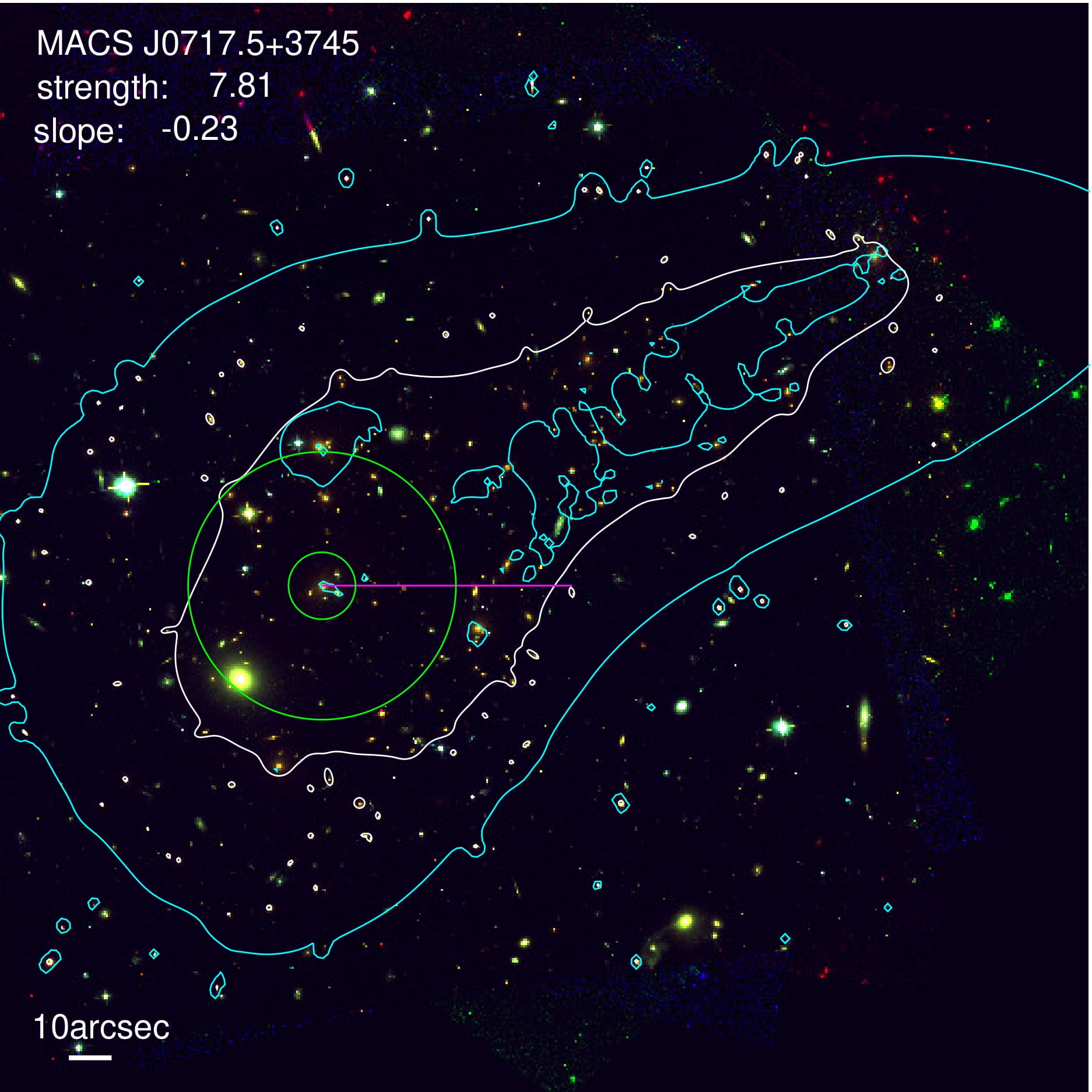}
    \includegraphics[width=0.23\textwidth]{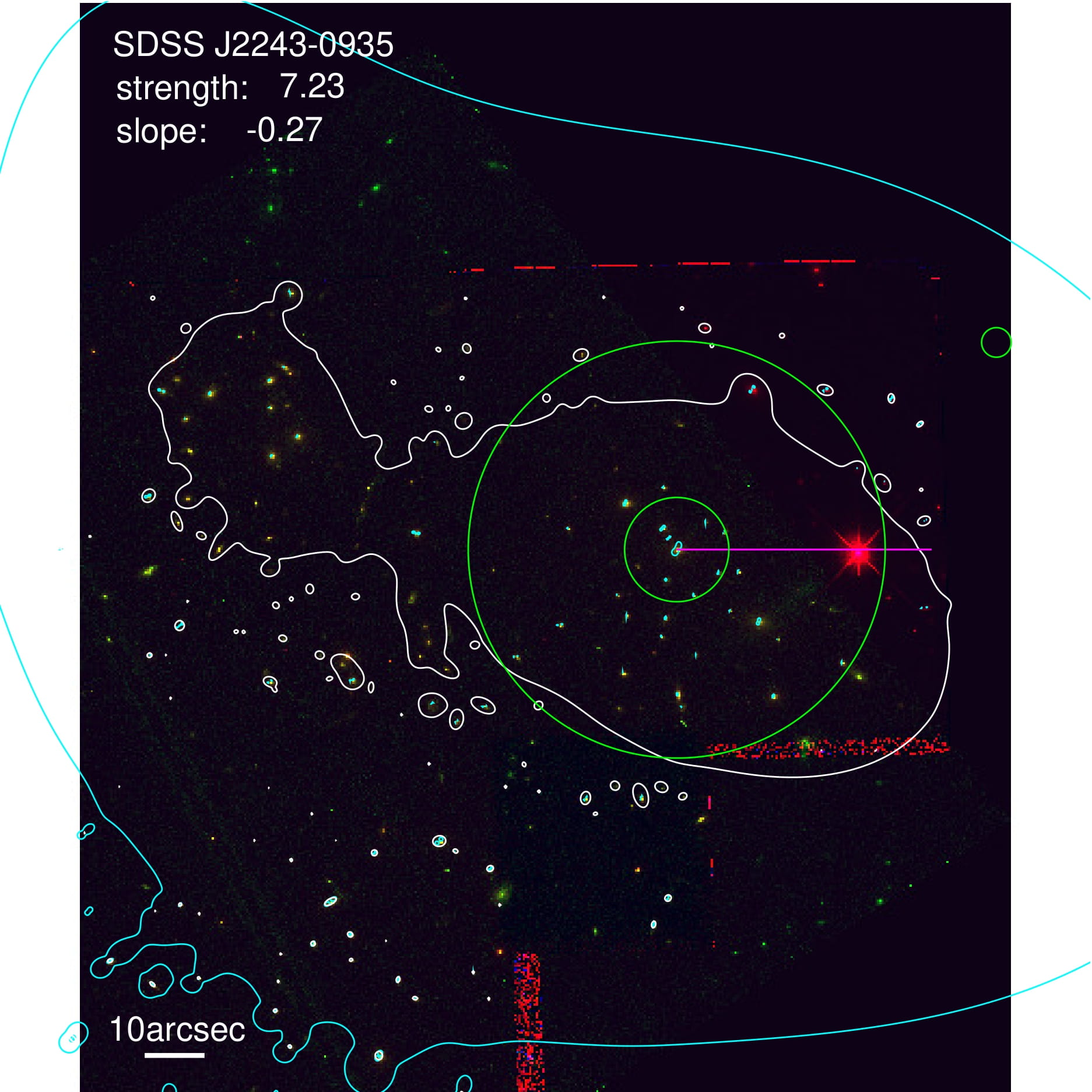}
    \includegraphics[width=0.23\textwidth]{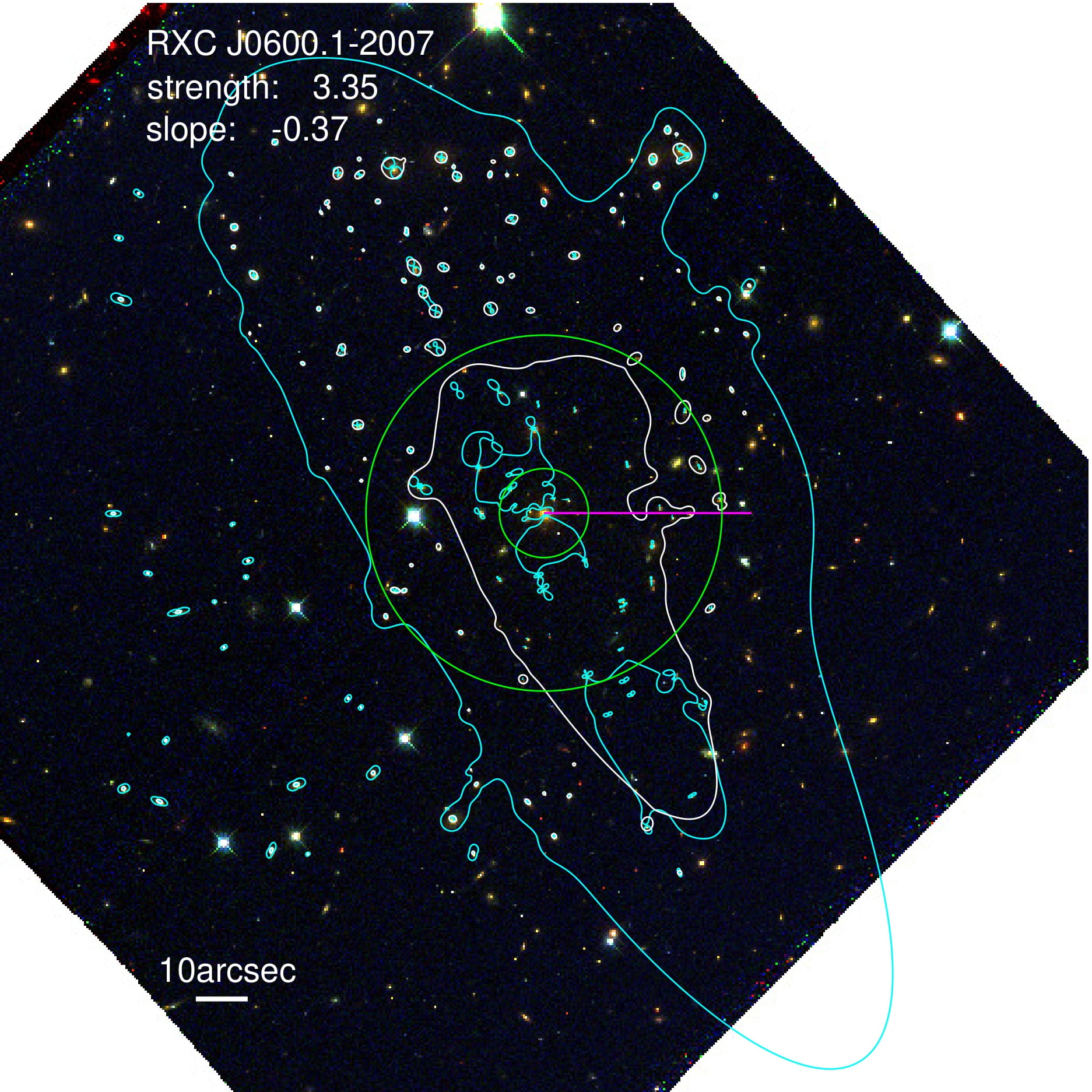}
    \includegraphics[width=0.23\textwidth]{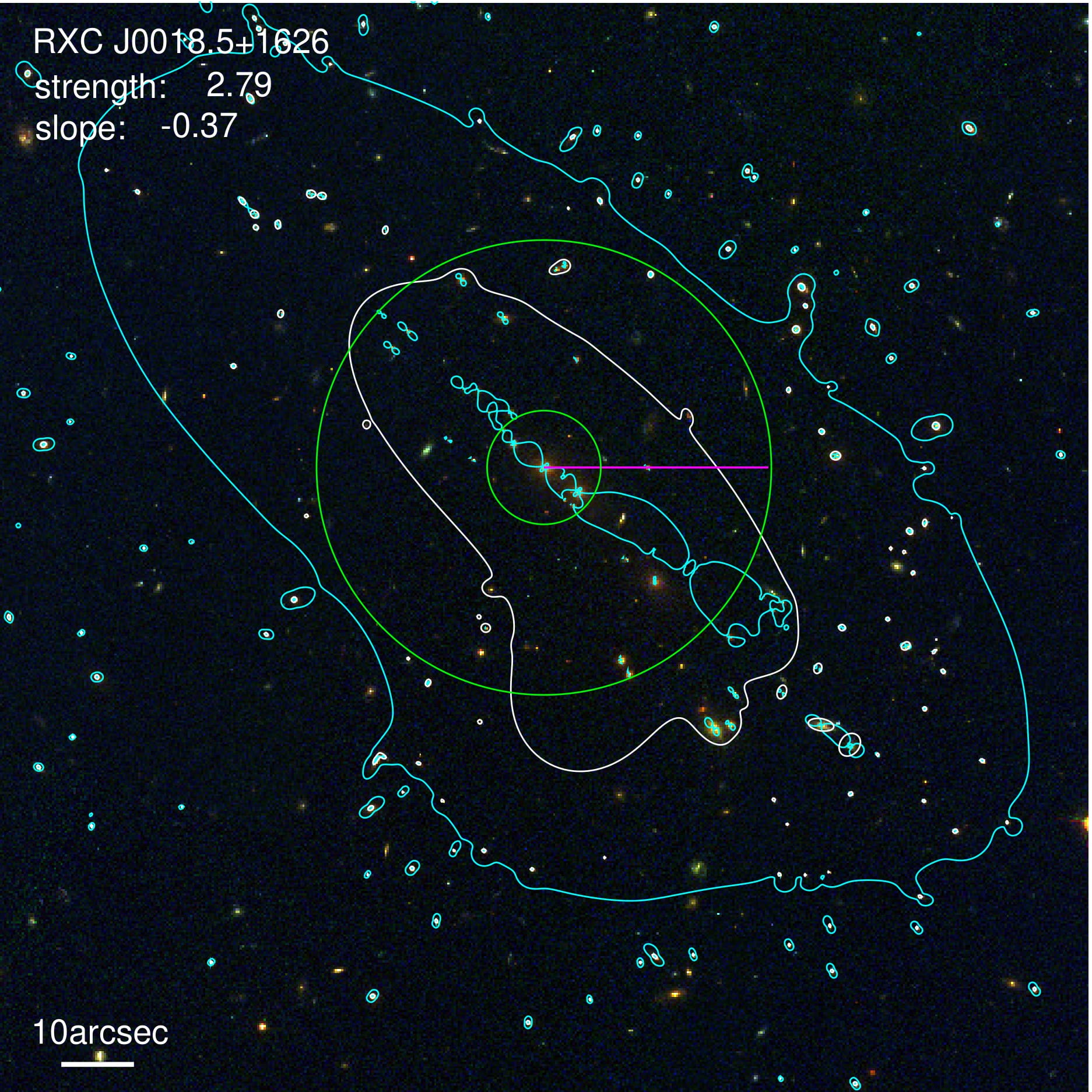}\\
    \includegraphics[width=0.23\textwidth]{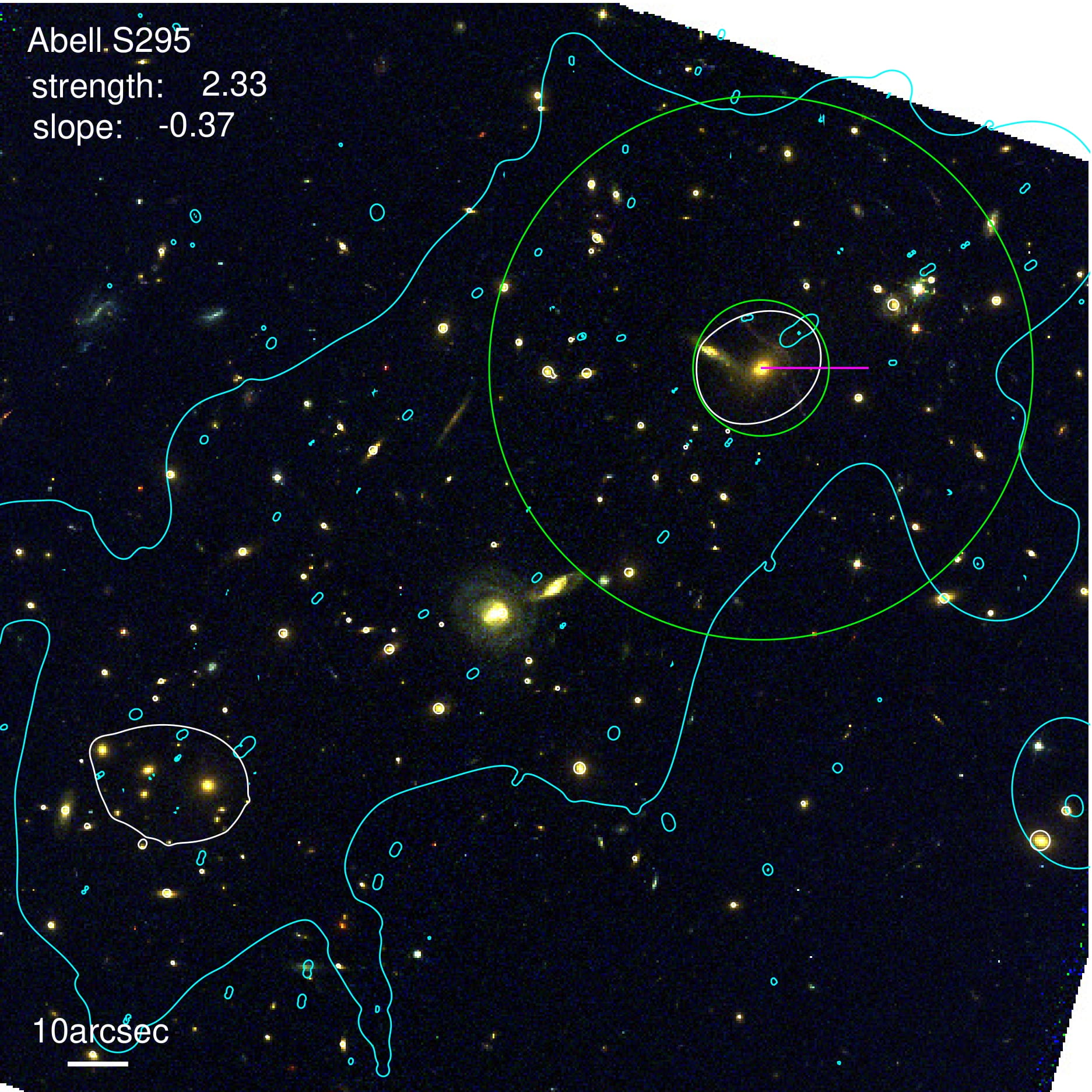}
    \includegraphics[width=0.23\textwidth]{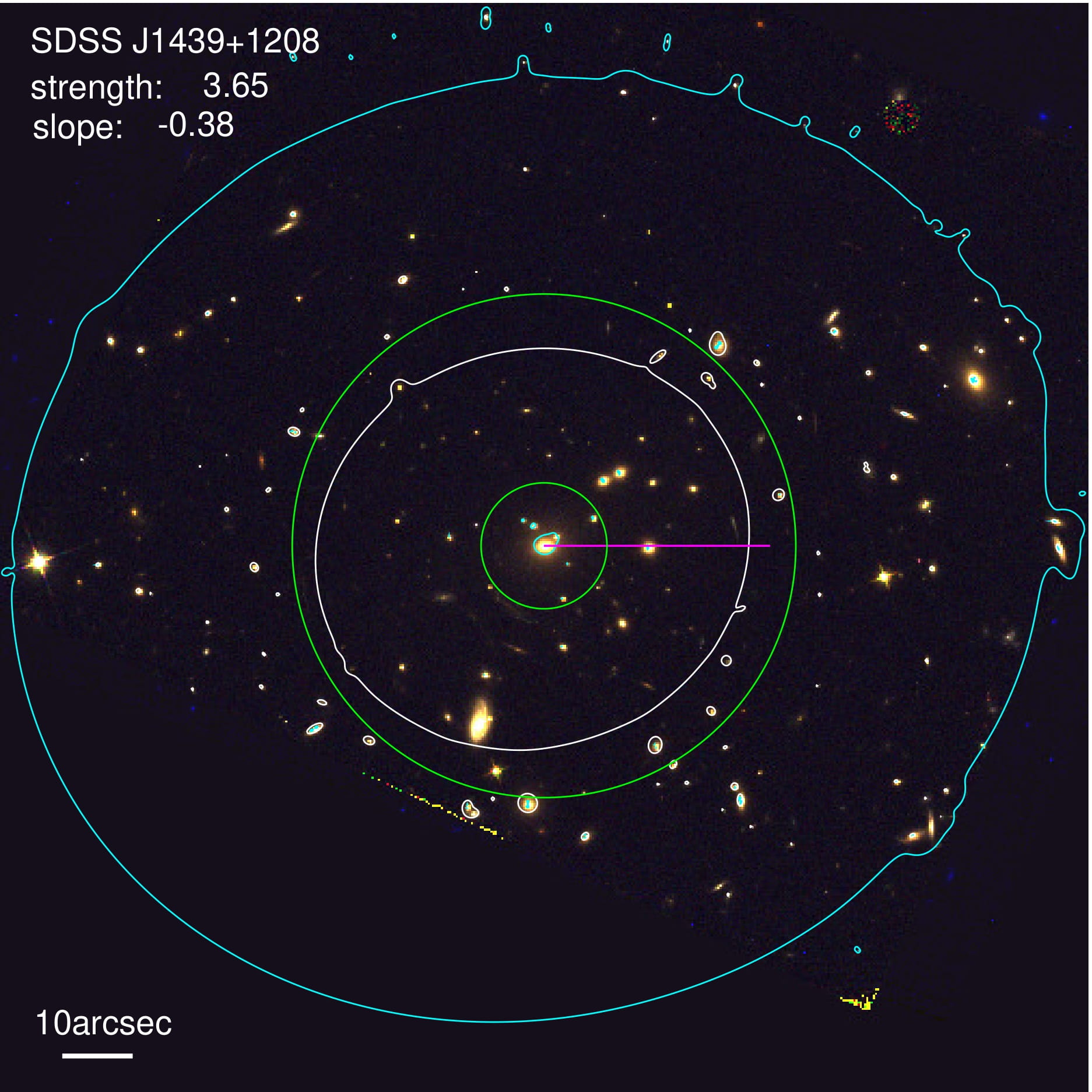}
    \includegraphics[width=0.23\textwidth]{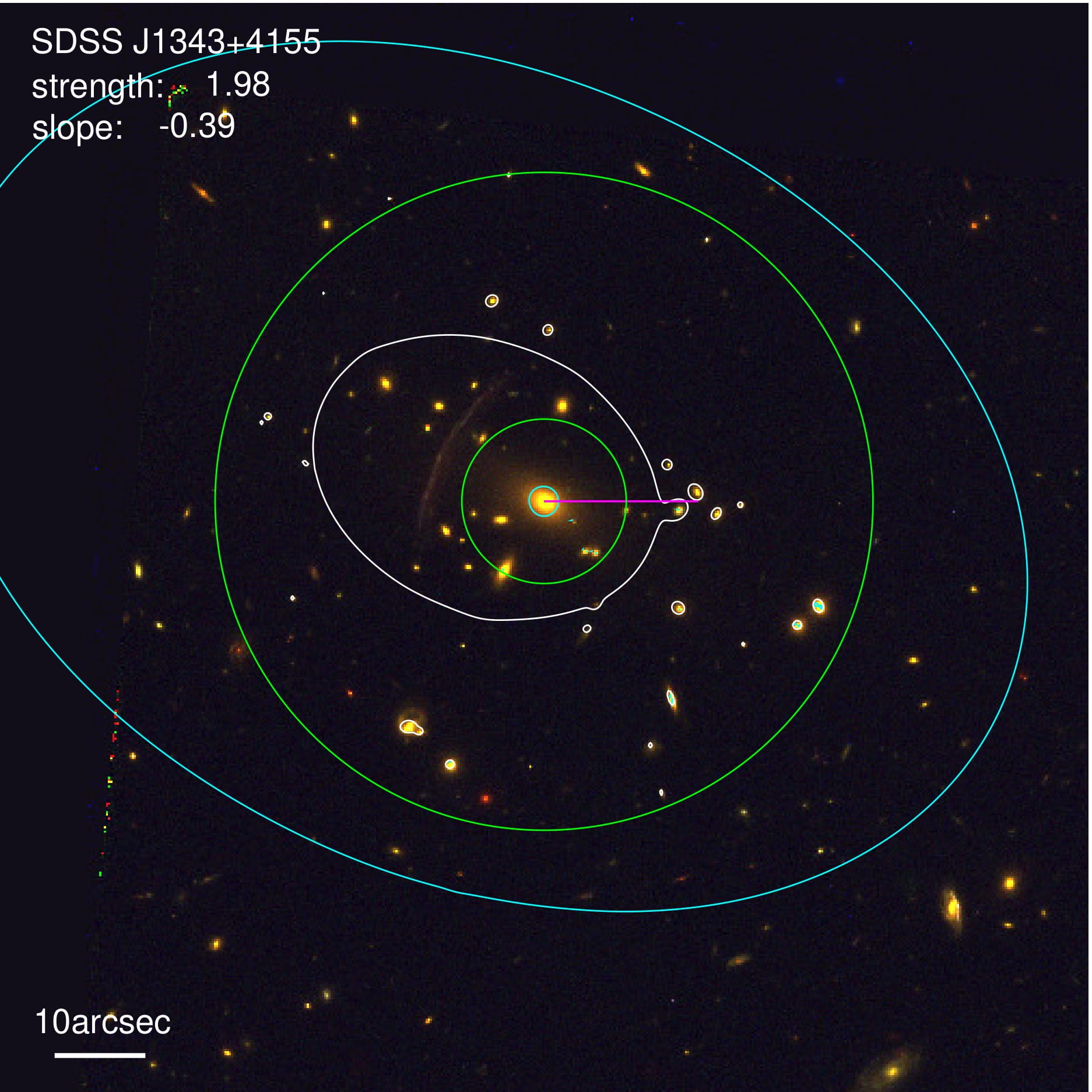}
    \includegraphics[width=0.23\textwidth]{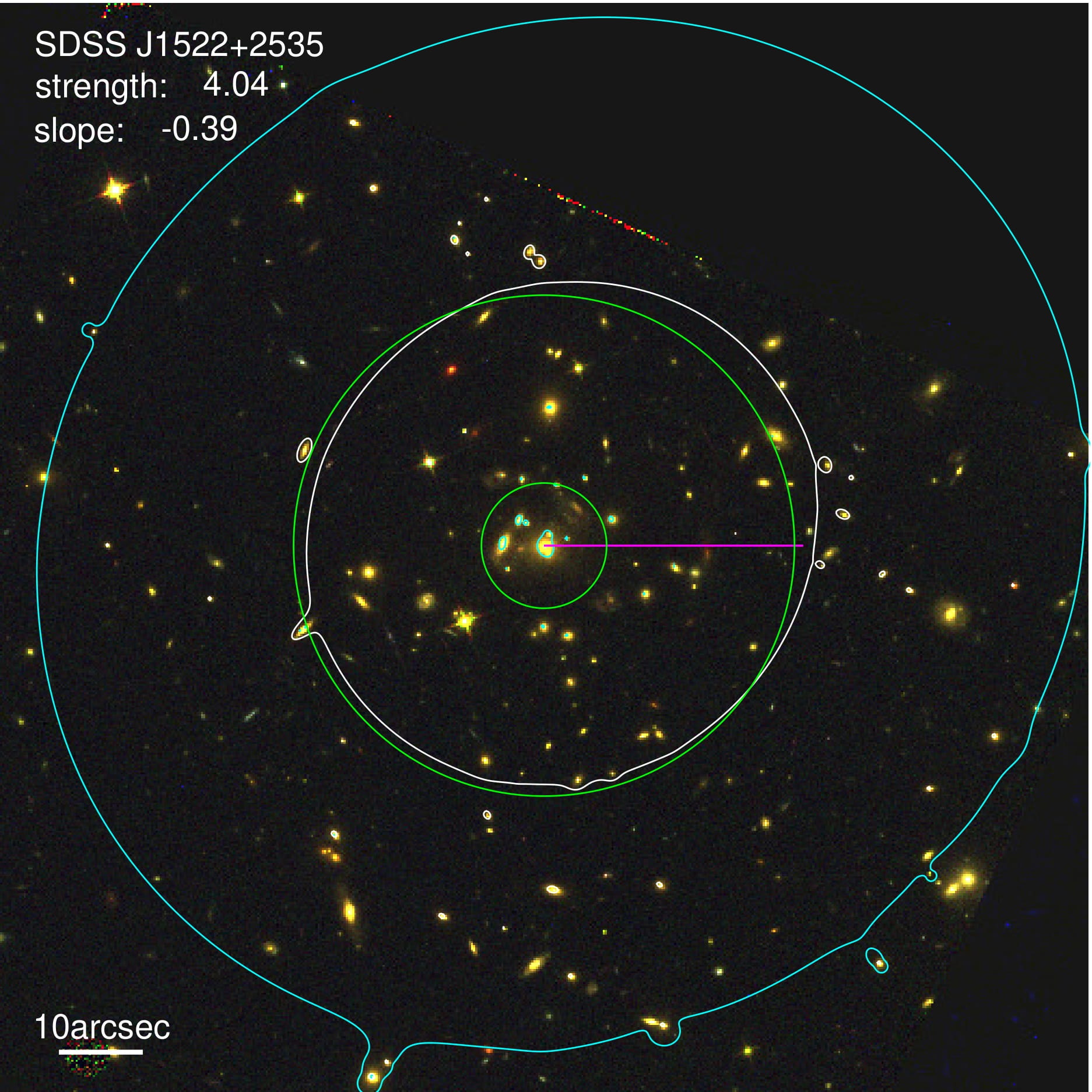}\\
    \includegraphics[width=0.23\textwidth]{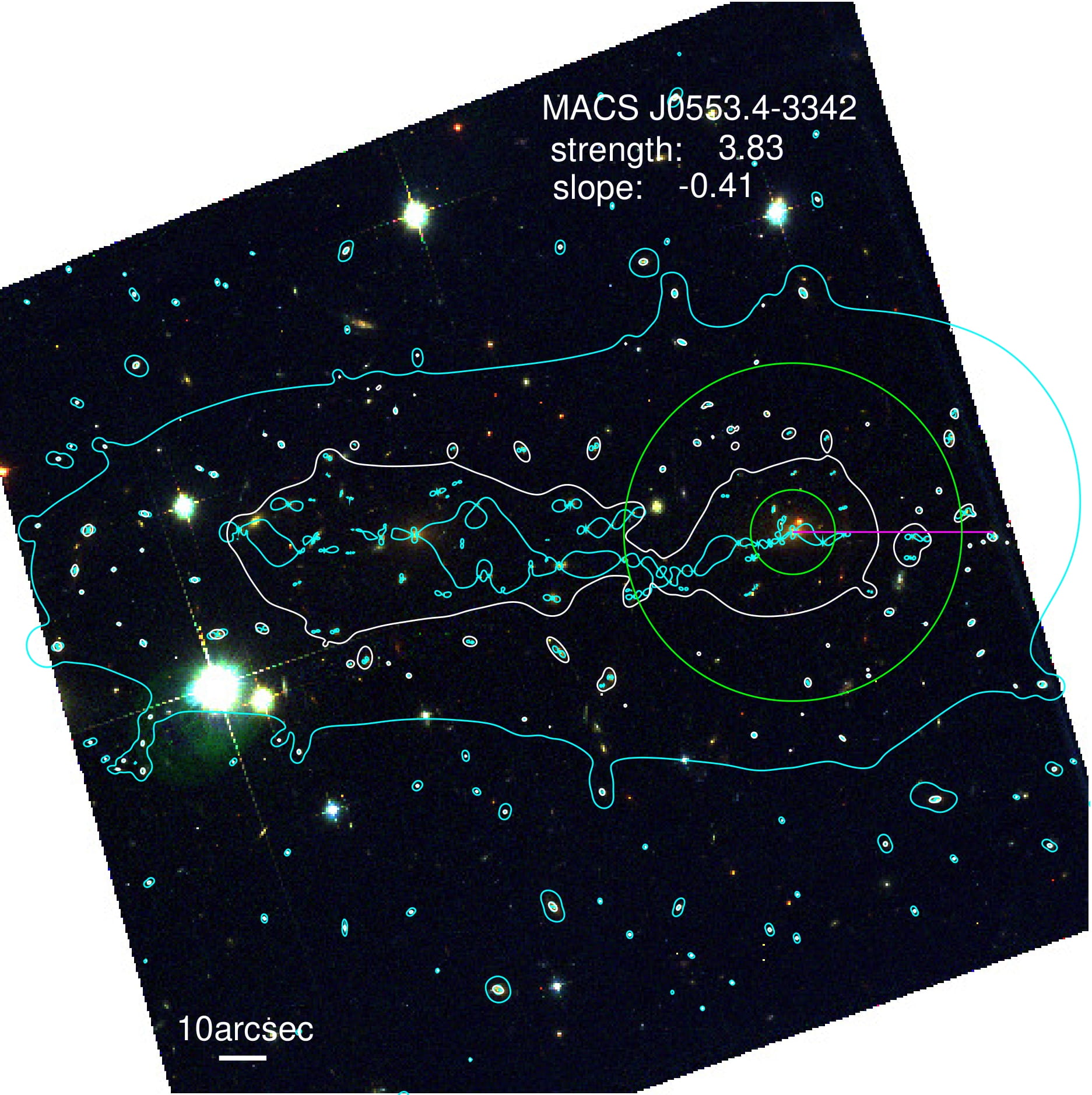}
    \includegraphics[width=0.23\textwidth]{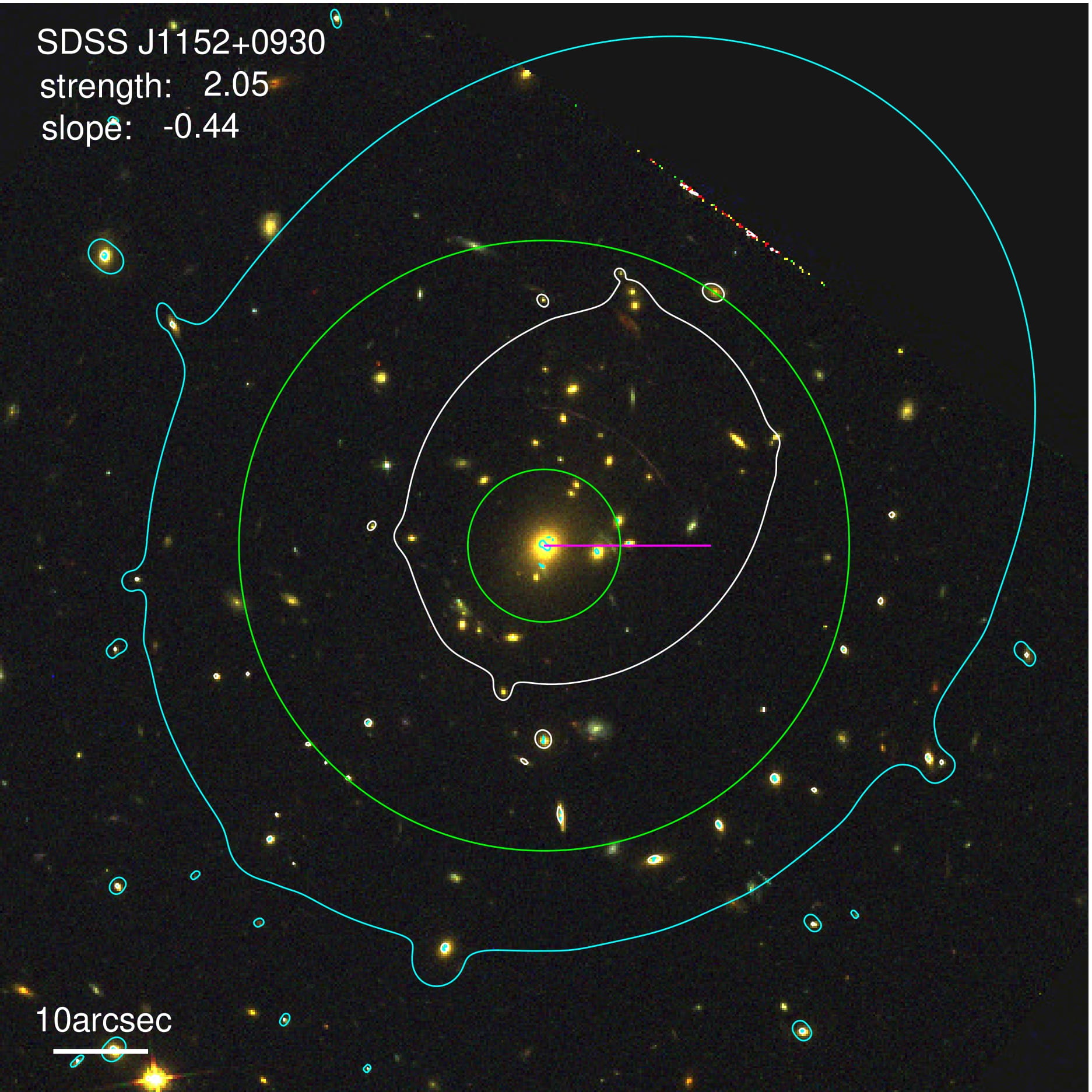}
    \includegraphics[width=0.23\textwidth]{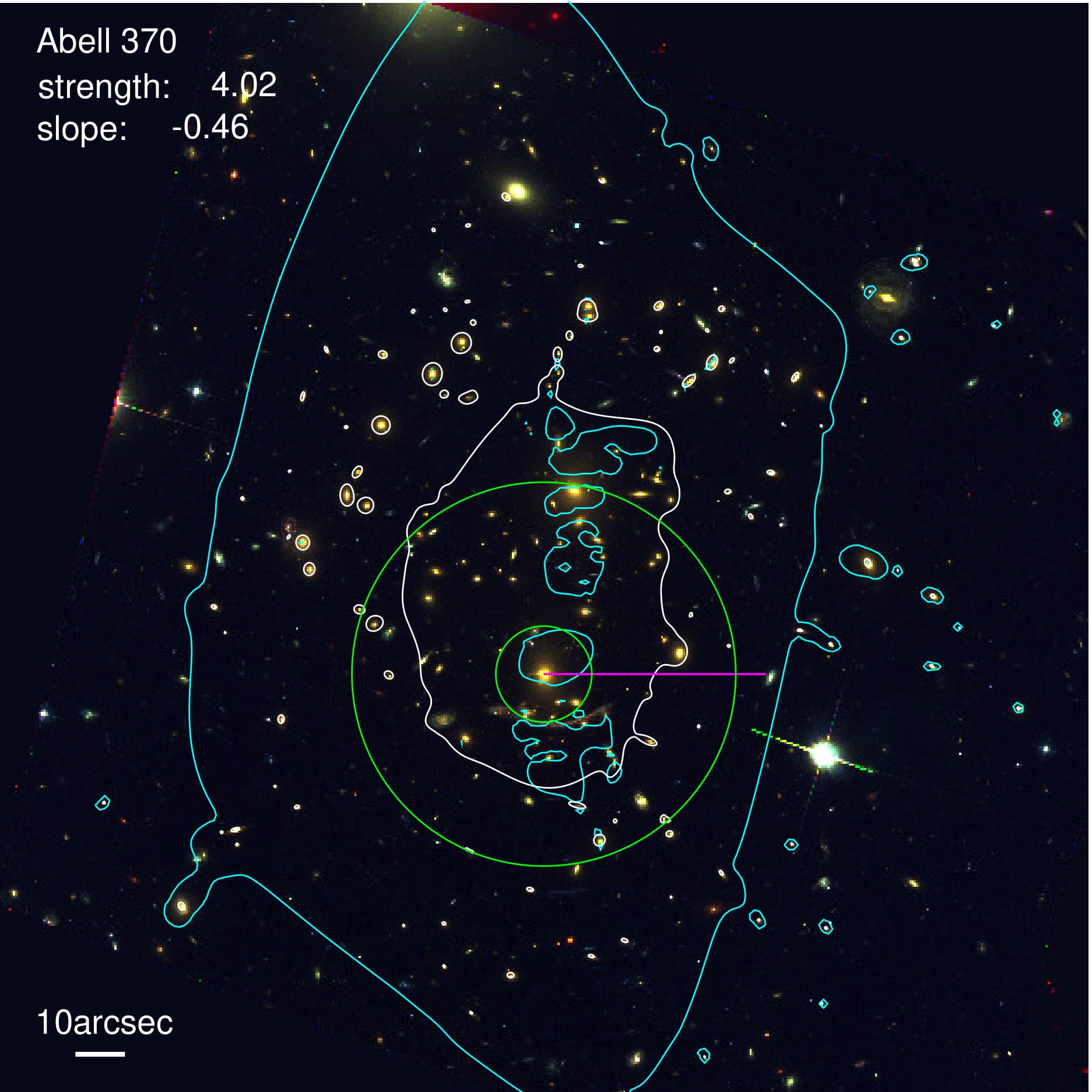}
    \includegraphics[width=0.23\textwidth]{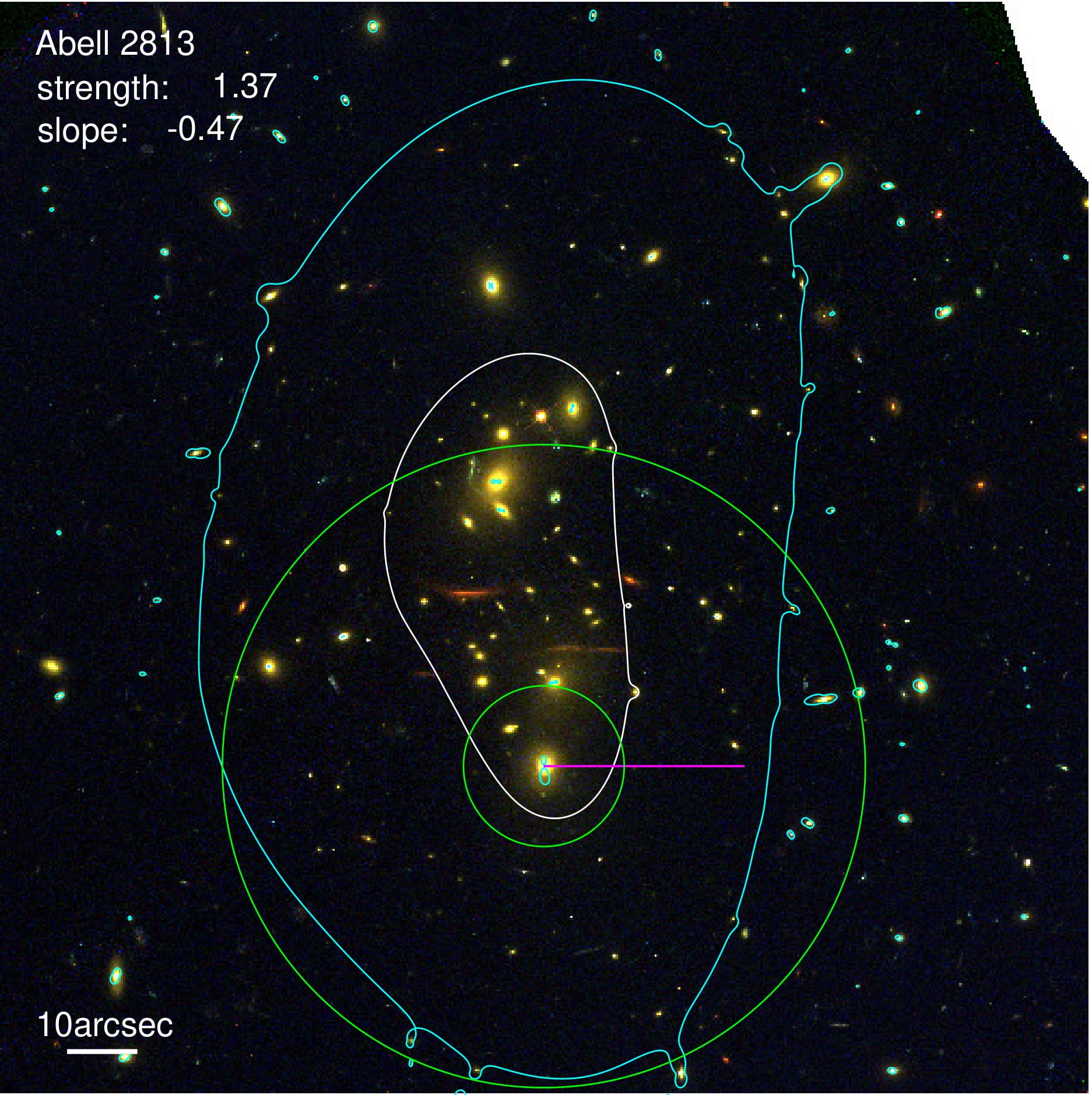}\\
    \includegraphics[width=0.23\textwidth]{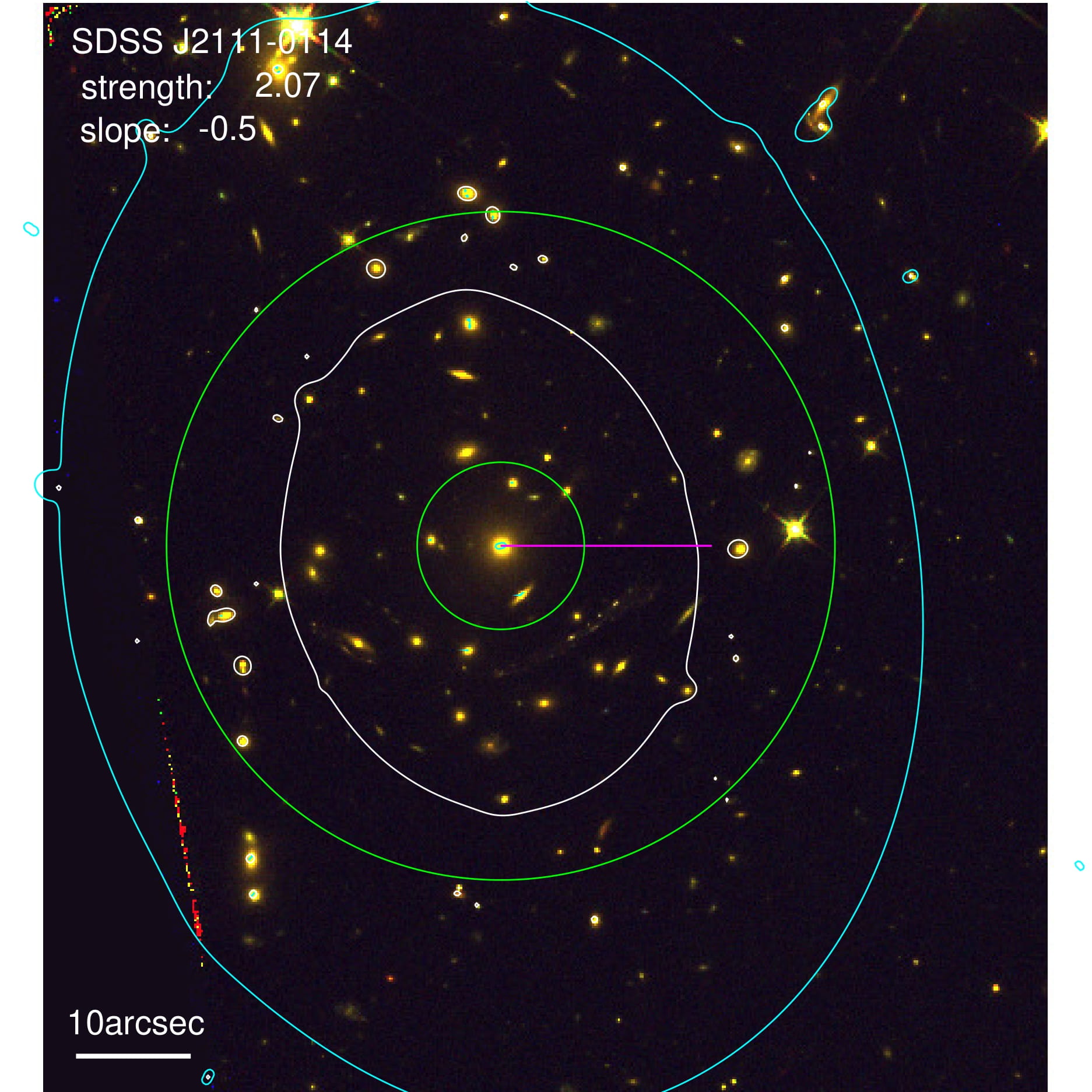}
    \includegraphics[width=0.23\textwidth]{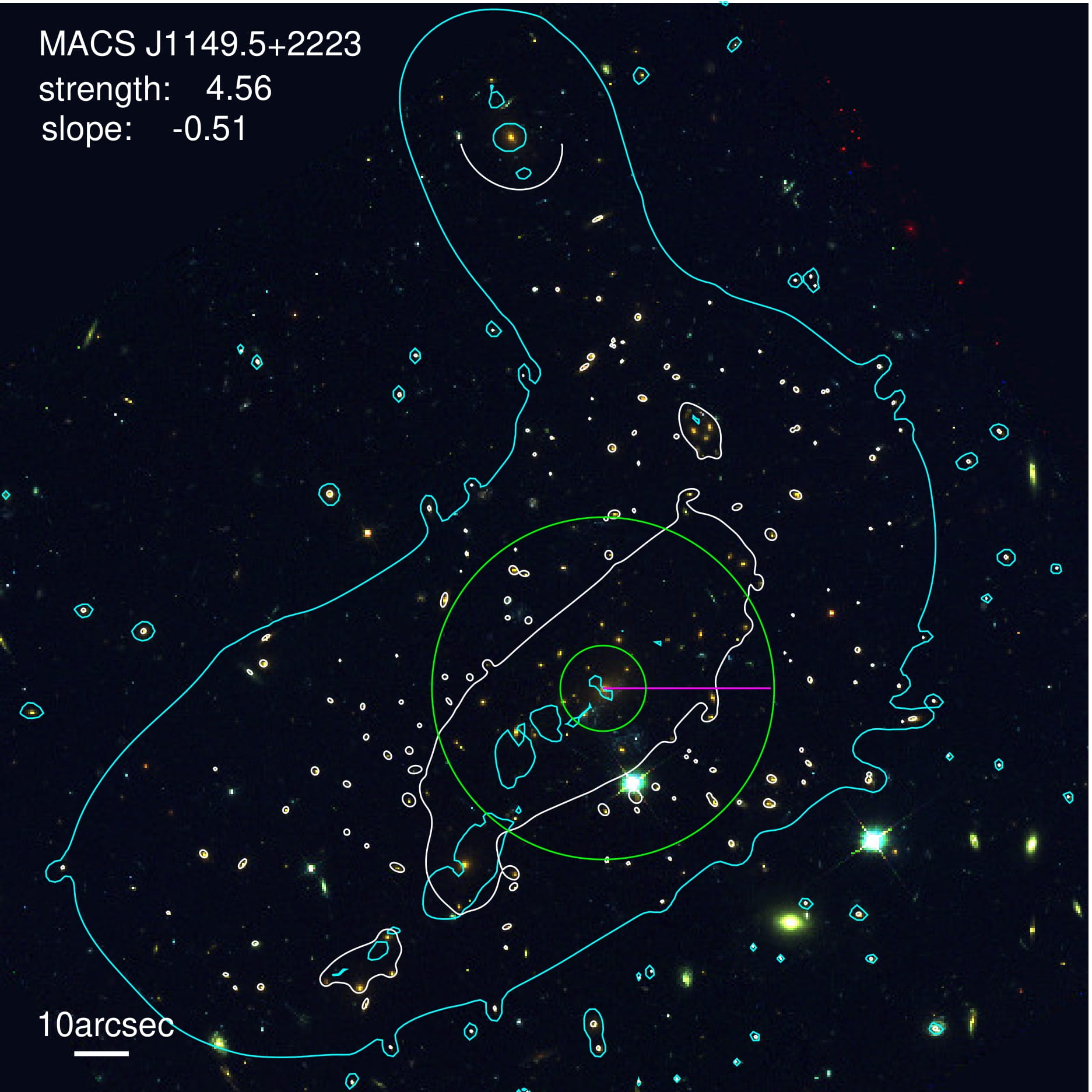}
    \includegraphics[width=0.23\textwidth]{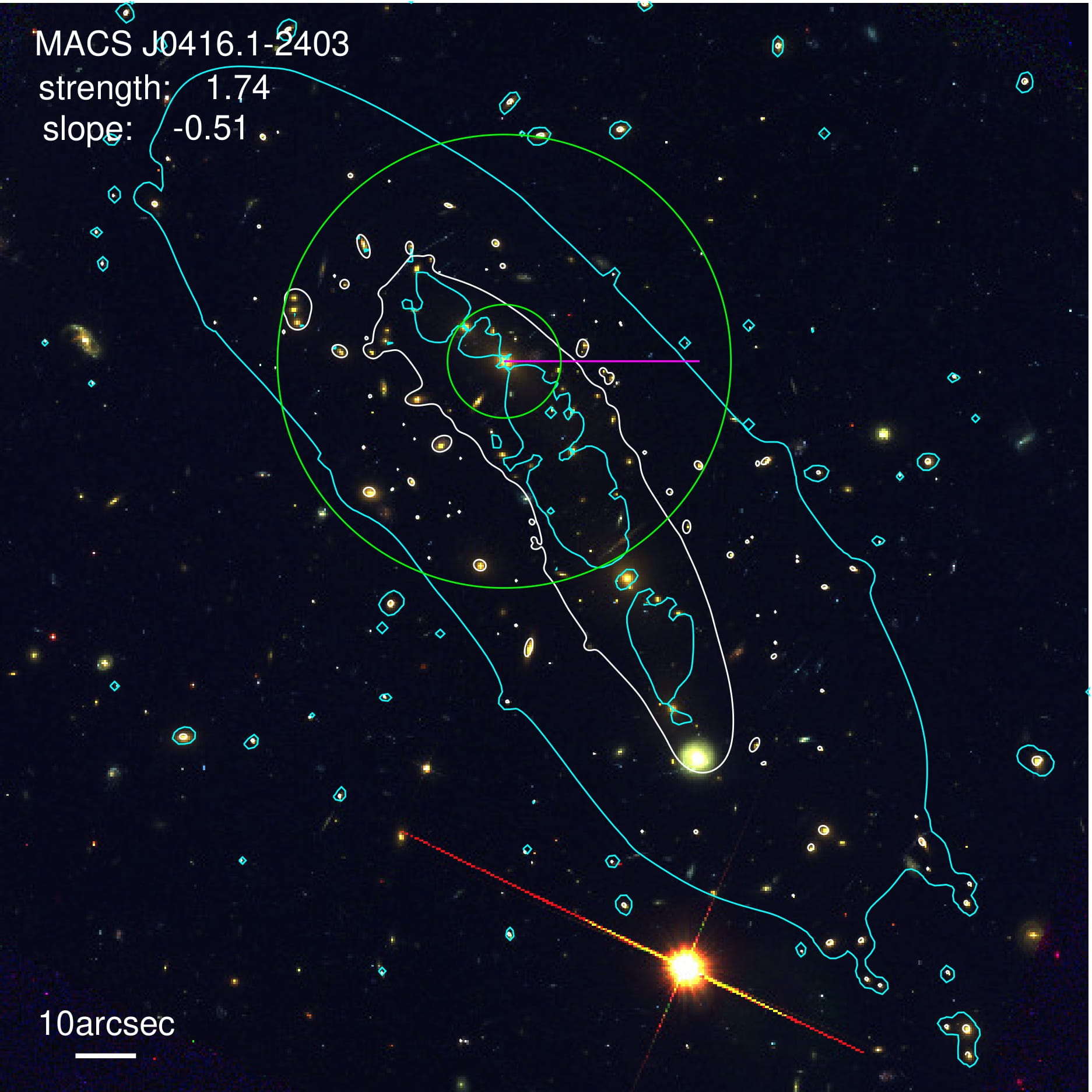}
    \includegraphics[width=0.23\textwidth]{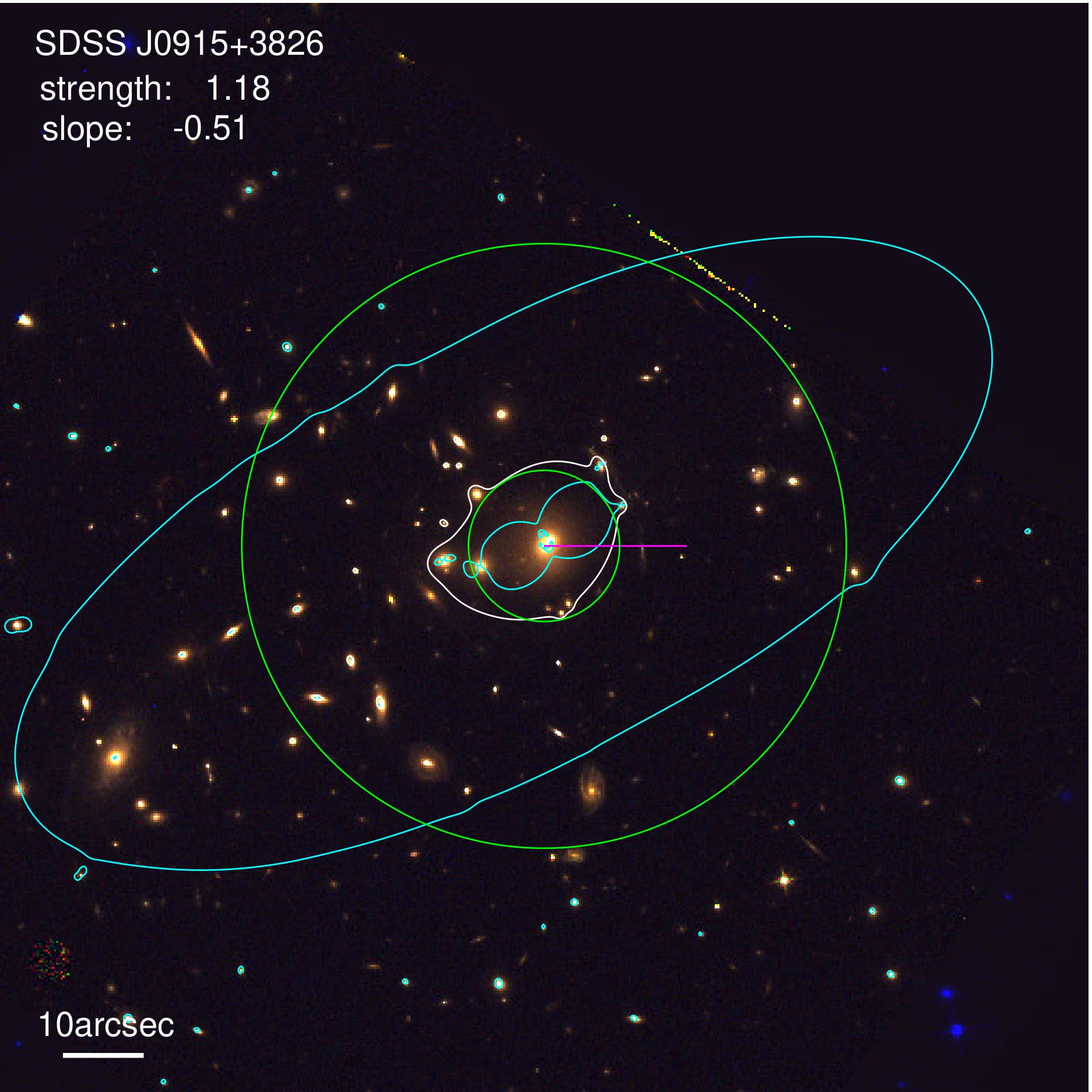}\\
    \includegraphics[width=0.23\textwidth]{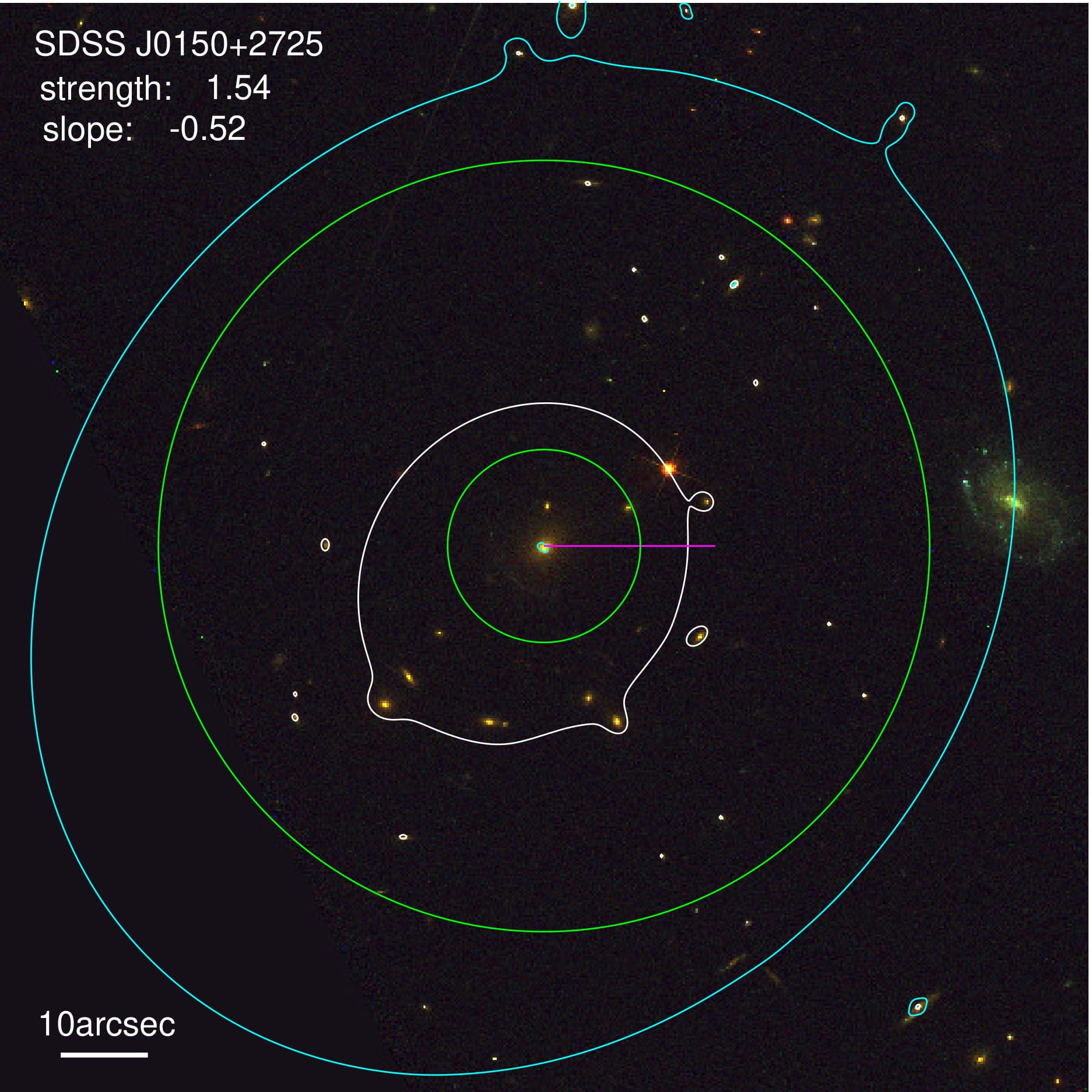}
    \includegraphics[width=0.23\textwidth]{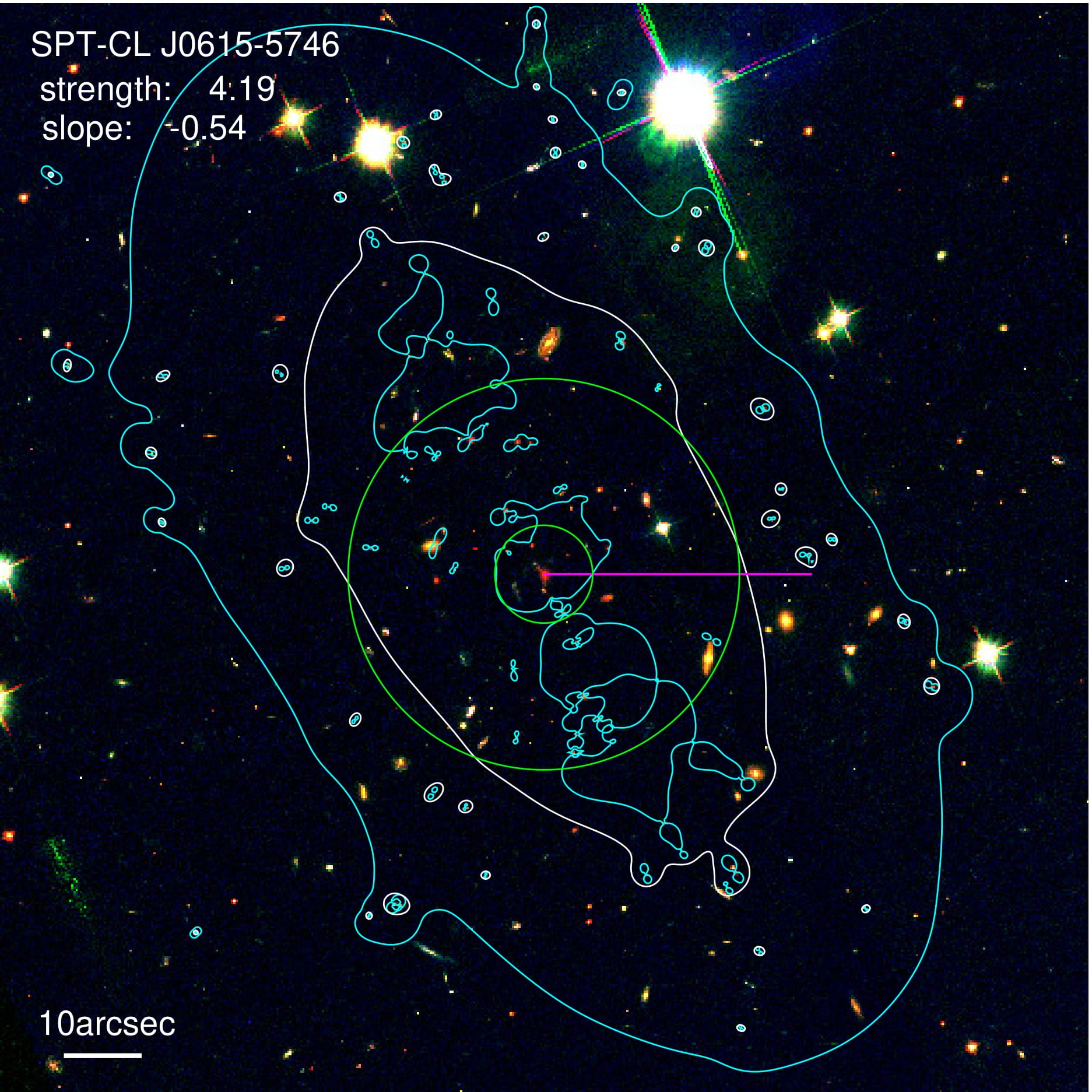}
    \includegraphics[width=0.23\textwidth]{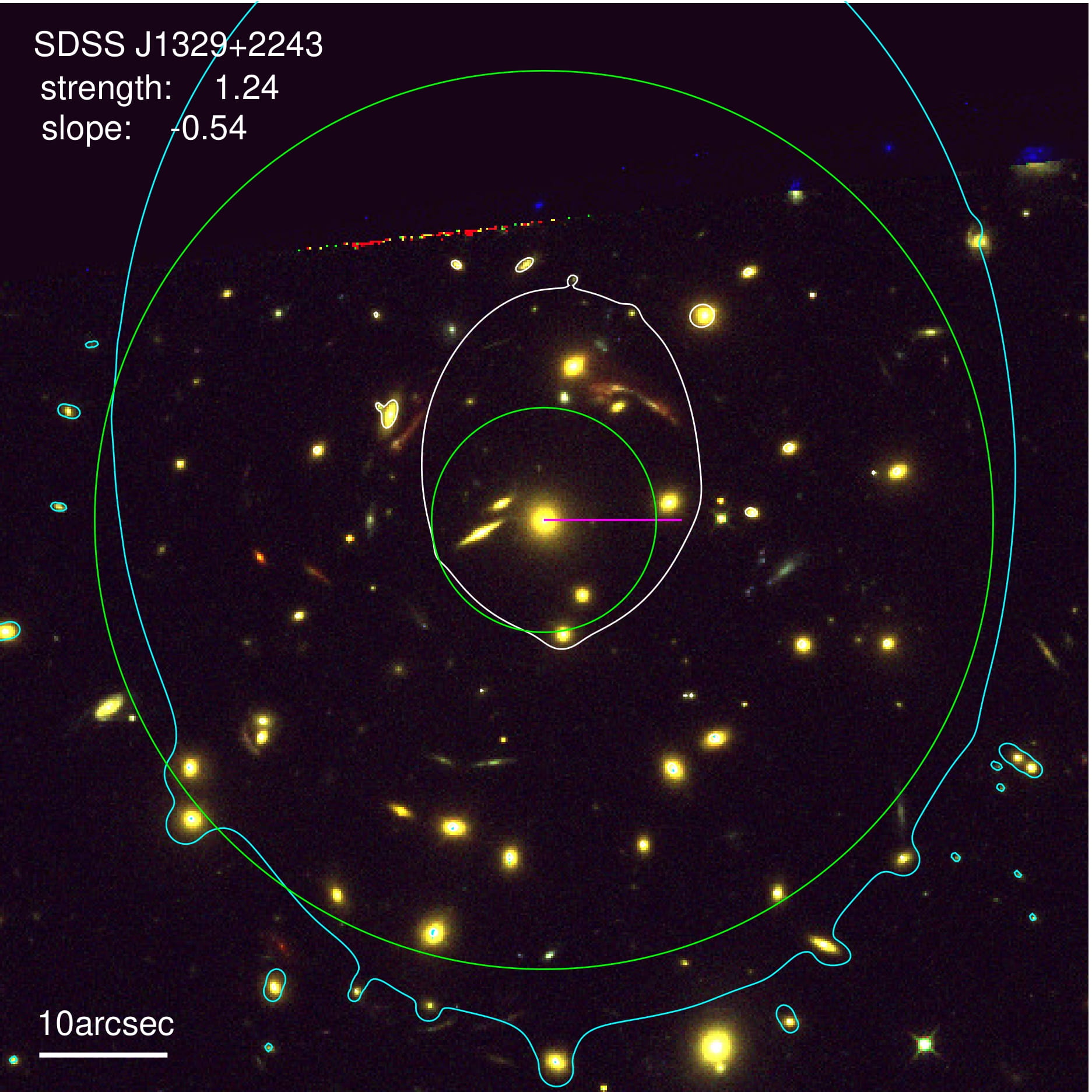}
    \includegraphics[width=0.23\textwidth]{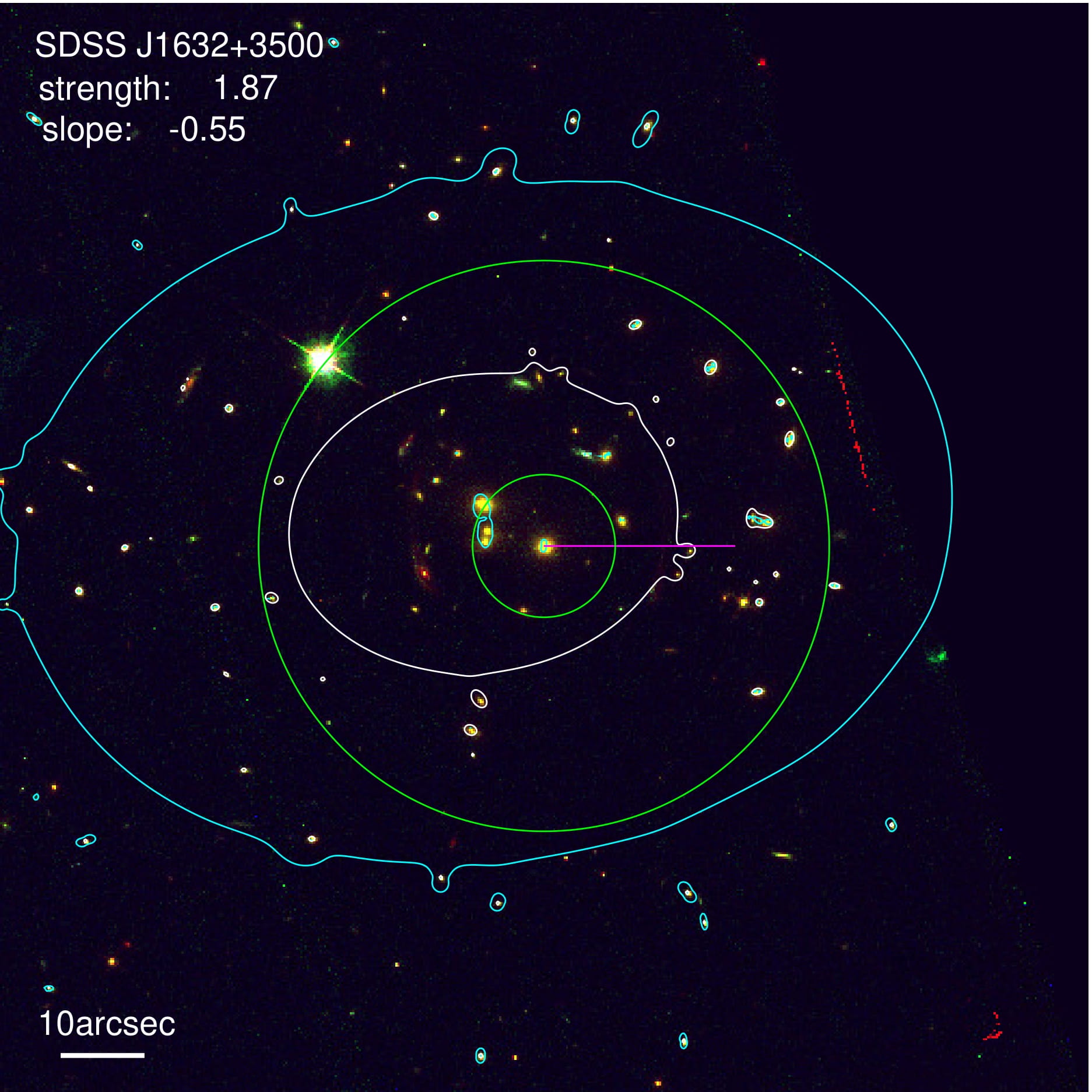}\\
    \caption{Composite color images of the clusters 
    MACS~J0717.5$+$3745, SDSS~J2243$−$0935, RXC~J0600.1$-$2007, RXC~J0018.5$+$1626, Abell~S295, SDSS~J1439$+$1208, SDSS~J1343$+$4155, SDSS~J1522$+$2535, MACS~J0553.4$-$3342, SDSS~J1152$+$0930, Abell~370, and Abell~2813, SDSS~J2111$−$0114, MACS~J1149.5$+$2223, MACS~J0416.1$-$2403, SDSS~J0915$+$3826, SDSS~J0150$+$2725, SPT$-$CL~J0615$-$5746, SDSS~J1329$+$2243, and SDSS~J1632$+$3500.
    Green circles show radii of $50$ kpc and $200$ kpc from the BCG, the range for the inner slope measurement. For each cluster, the normalized lensing strength (\LstrengthN) and inner slope ($S_{50-200}$) are listed. White contours mark the mass density at $\kappa=1$, where $\kappa$ is scaled to $D_{LS}/D_S = 1$. Cyan contours mark the magnification at $|\mu|\geq3$. The effective Einstein radius is shown in magenta. \Lenstool\ models are used where available (using the Sharon models for the HFF clusters), LTM if there is no \Lenstool\ model, and GLAFIC if no \Lenstool\ or LTM model is available. Clusters are ordered left to right and top to bottom by slope, from flattest to steepest. North is up and East is left in all panels. A $10\farcs0$ scale is found in each figure for size reference.}
    \label{fig:gallery1}
\end{figure}

\begin{figure*}
\center
    \includegraphics[width=0.23\textwidth]{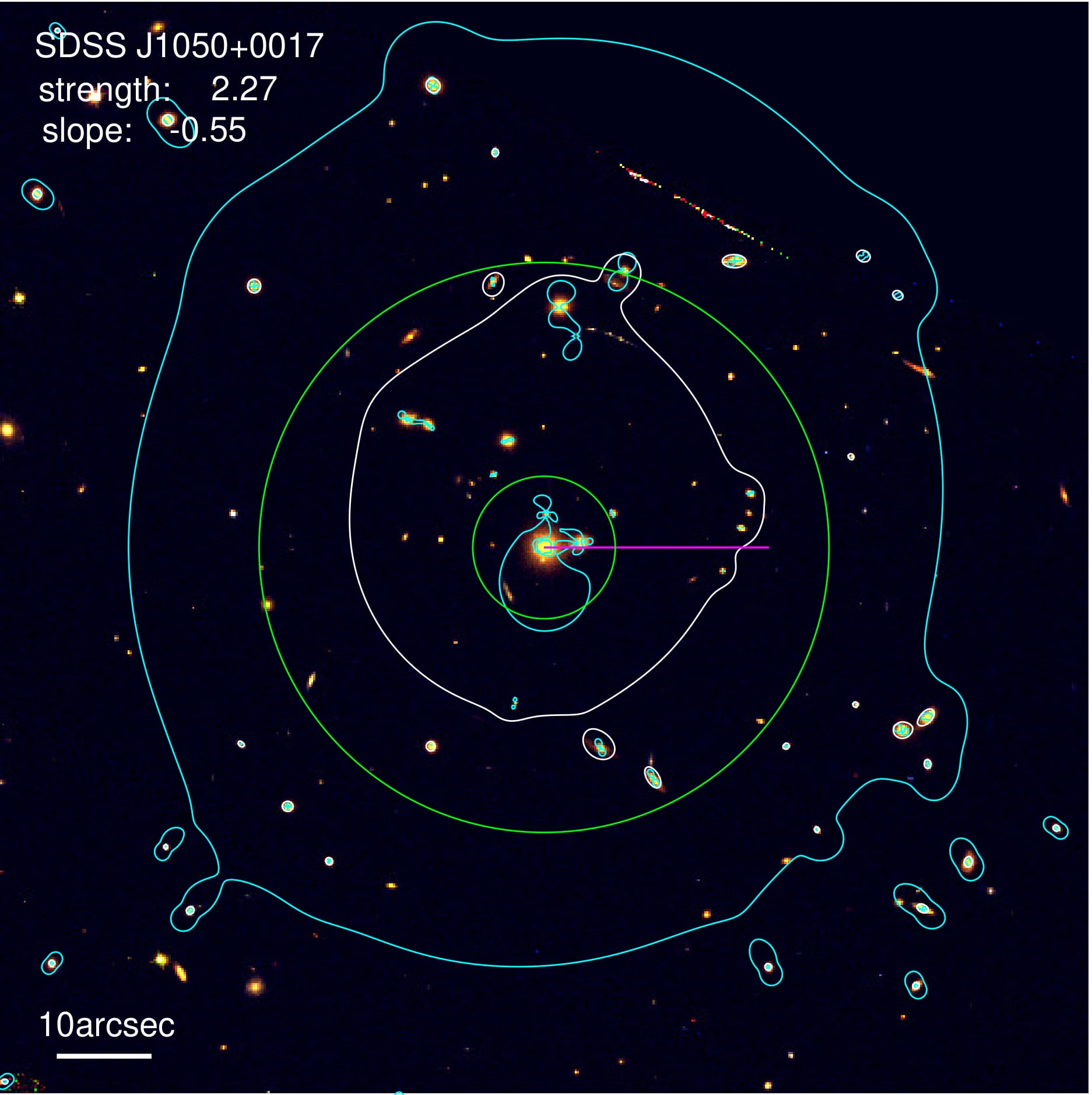}
    \includegraphics[width=0.23\textwidth]{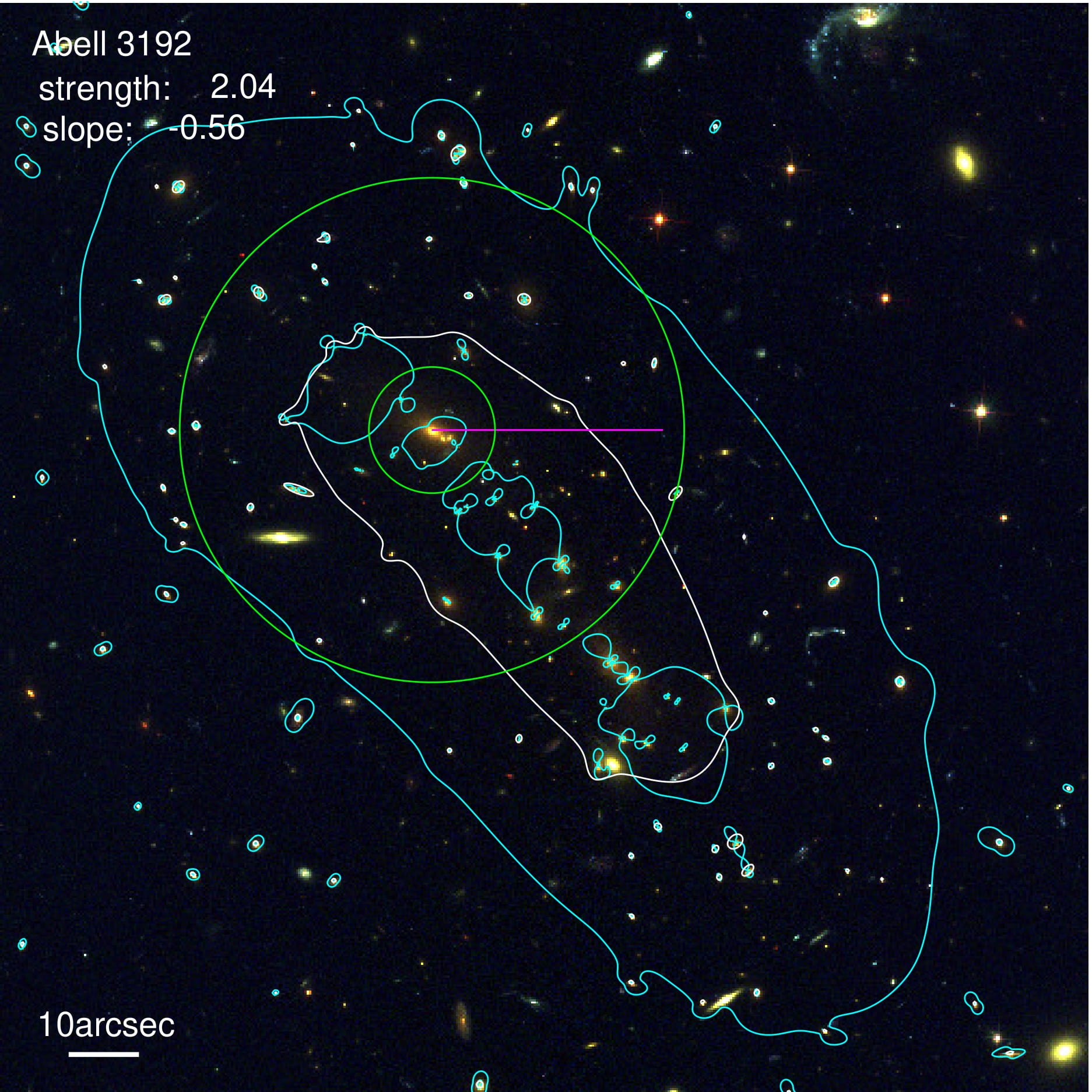}
    \includegraphics[width=0.23\textwidth]{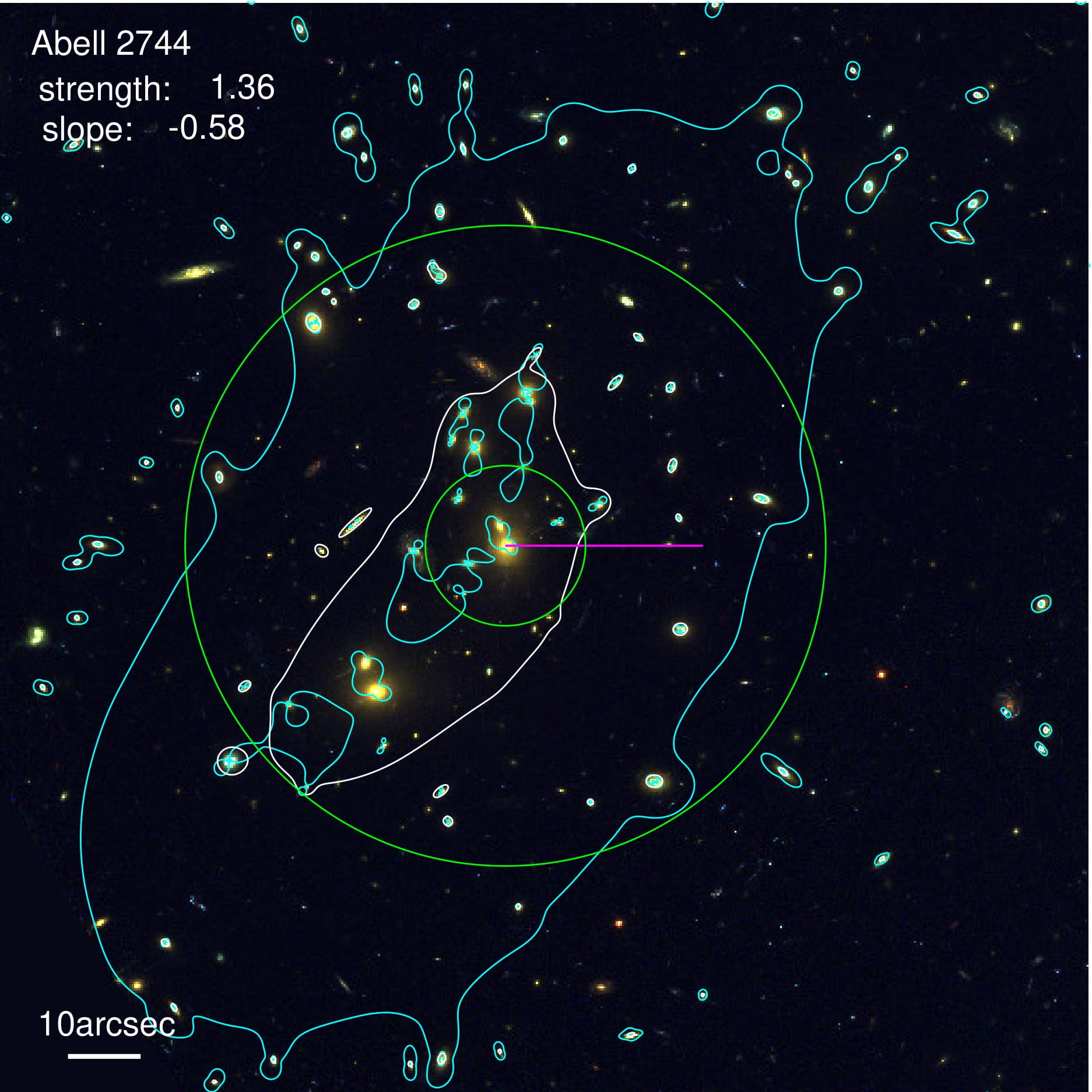}
    \includegraphics[width=0.23\textwidth]{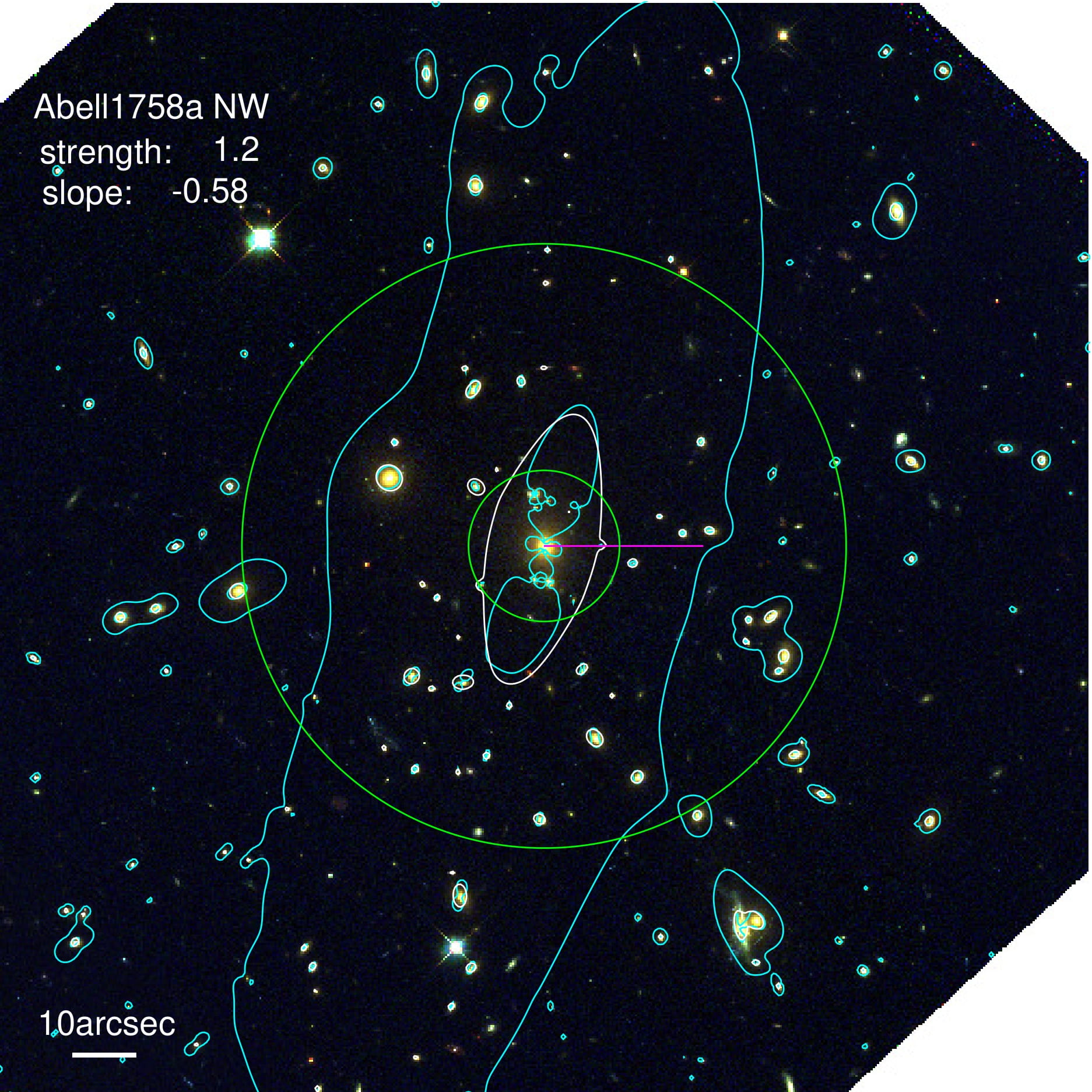}\\
    \includegraphics[width=0.23\textwidth]{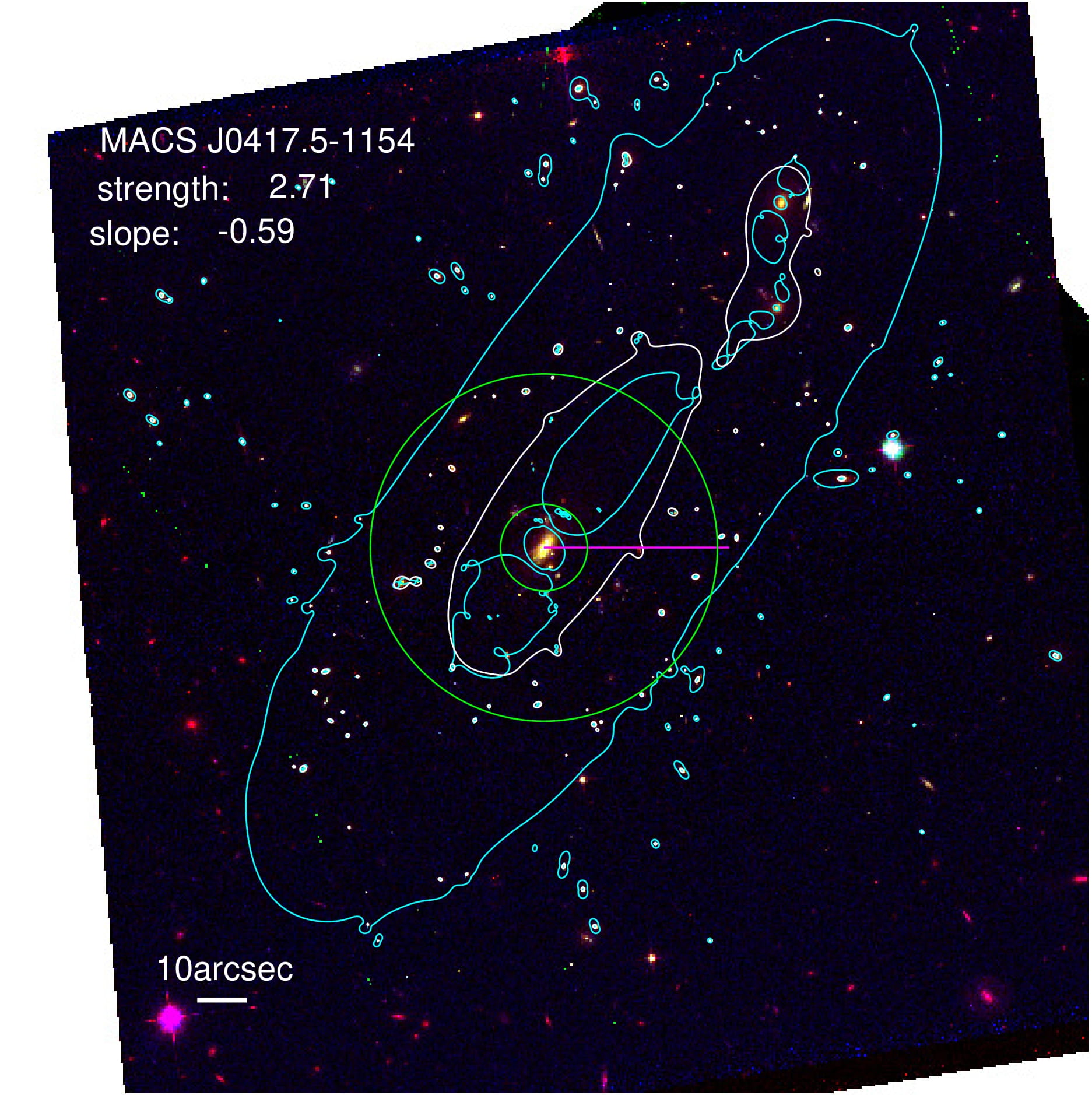}
    \includegraphics[width=0.23\textwidth]{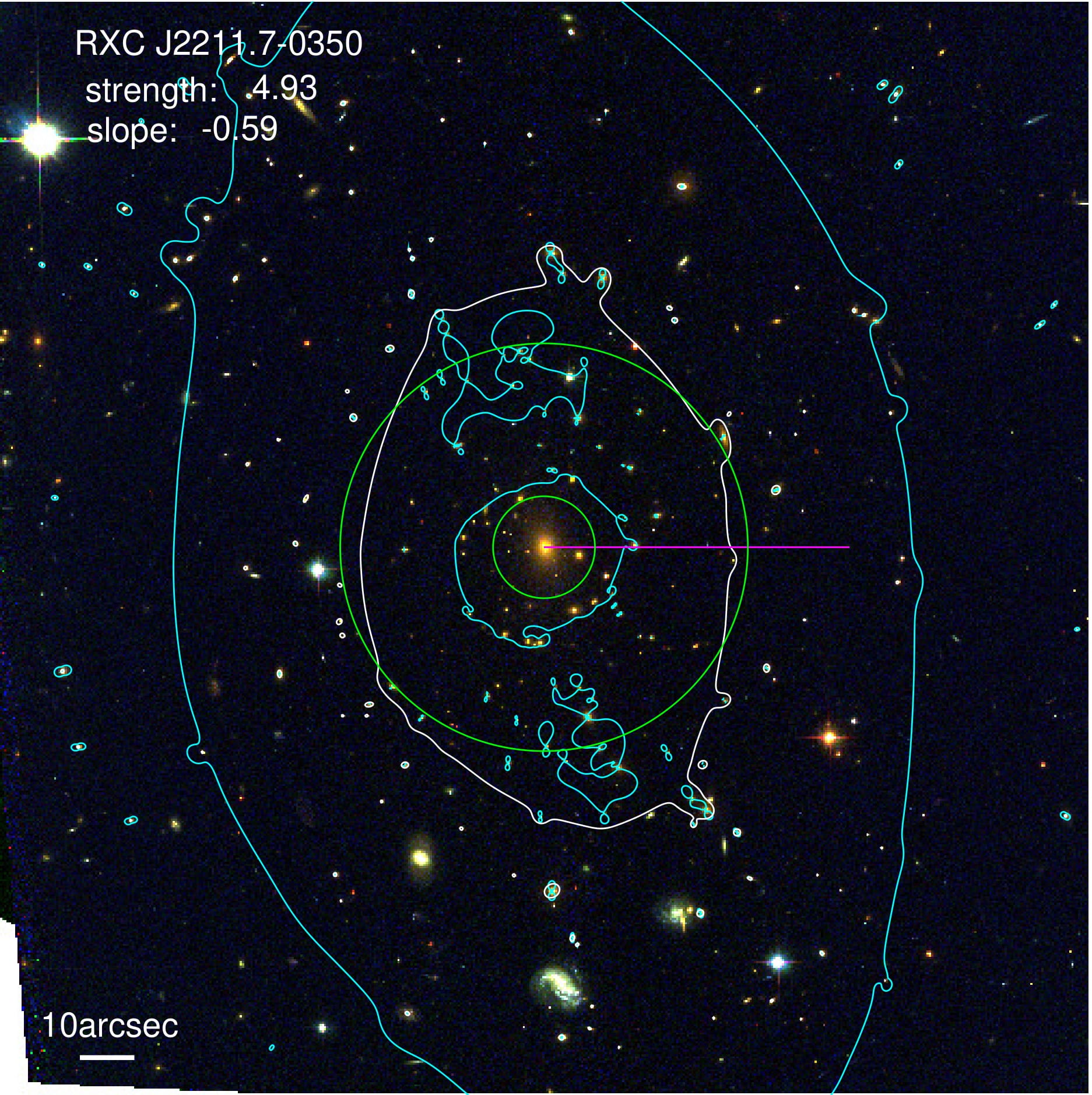}
    \includegraphics[width=0.23\textwidth]{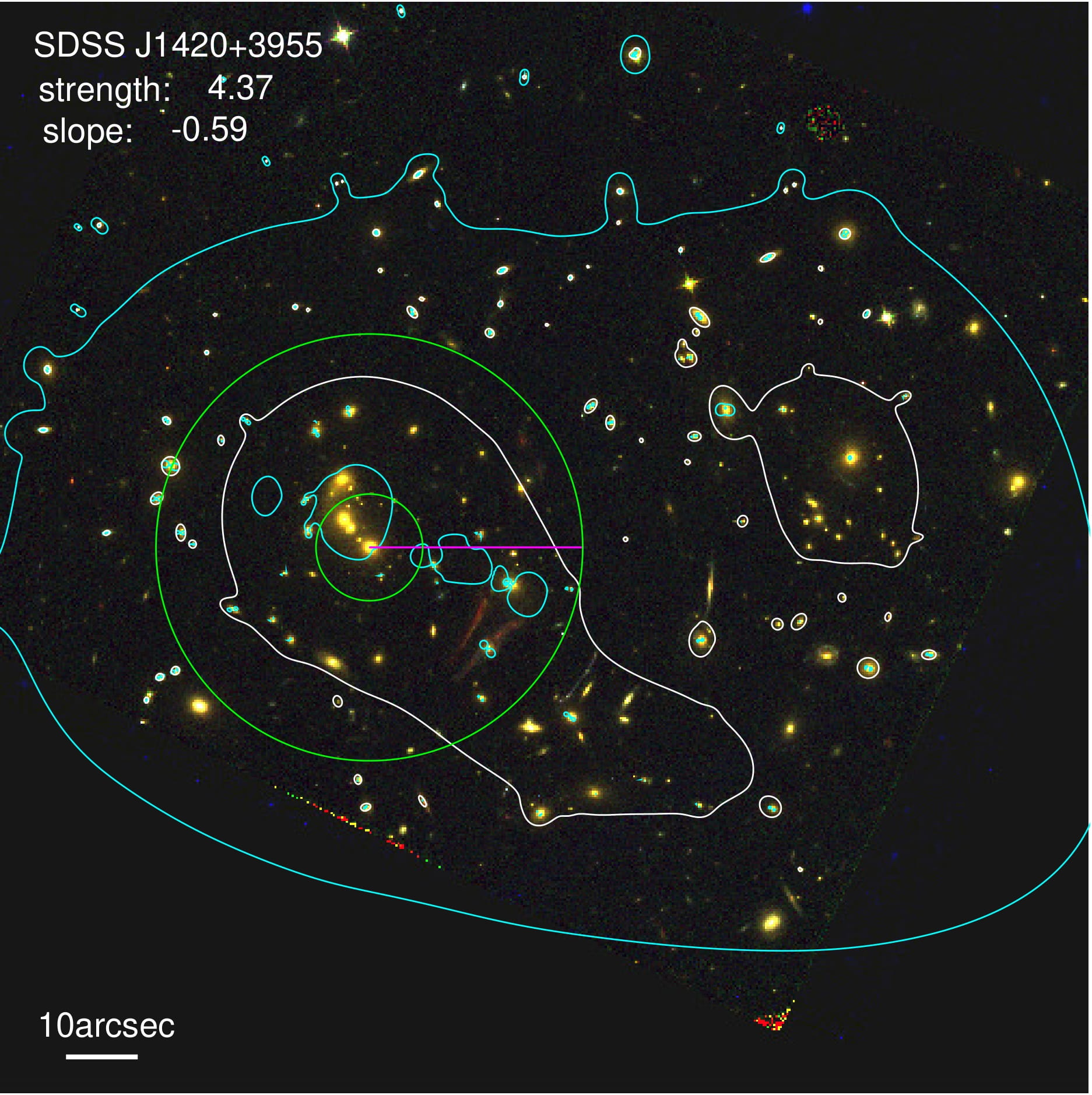}
    \includegraphics[width=0.23\textwidth]{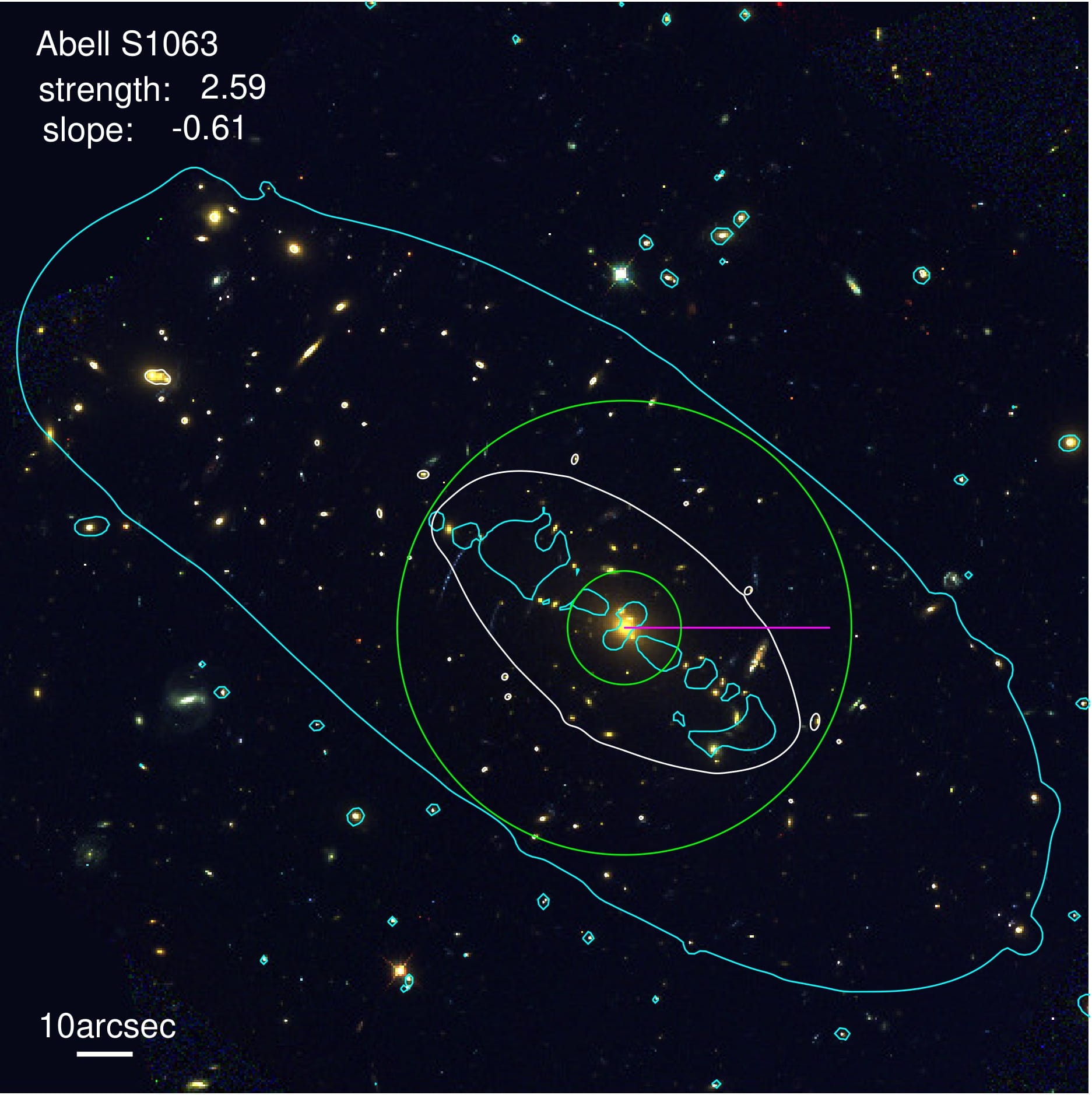}\\
    \includegraphics[width=0.23\textwidth]{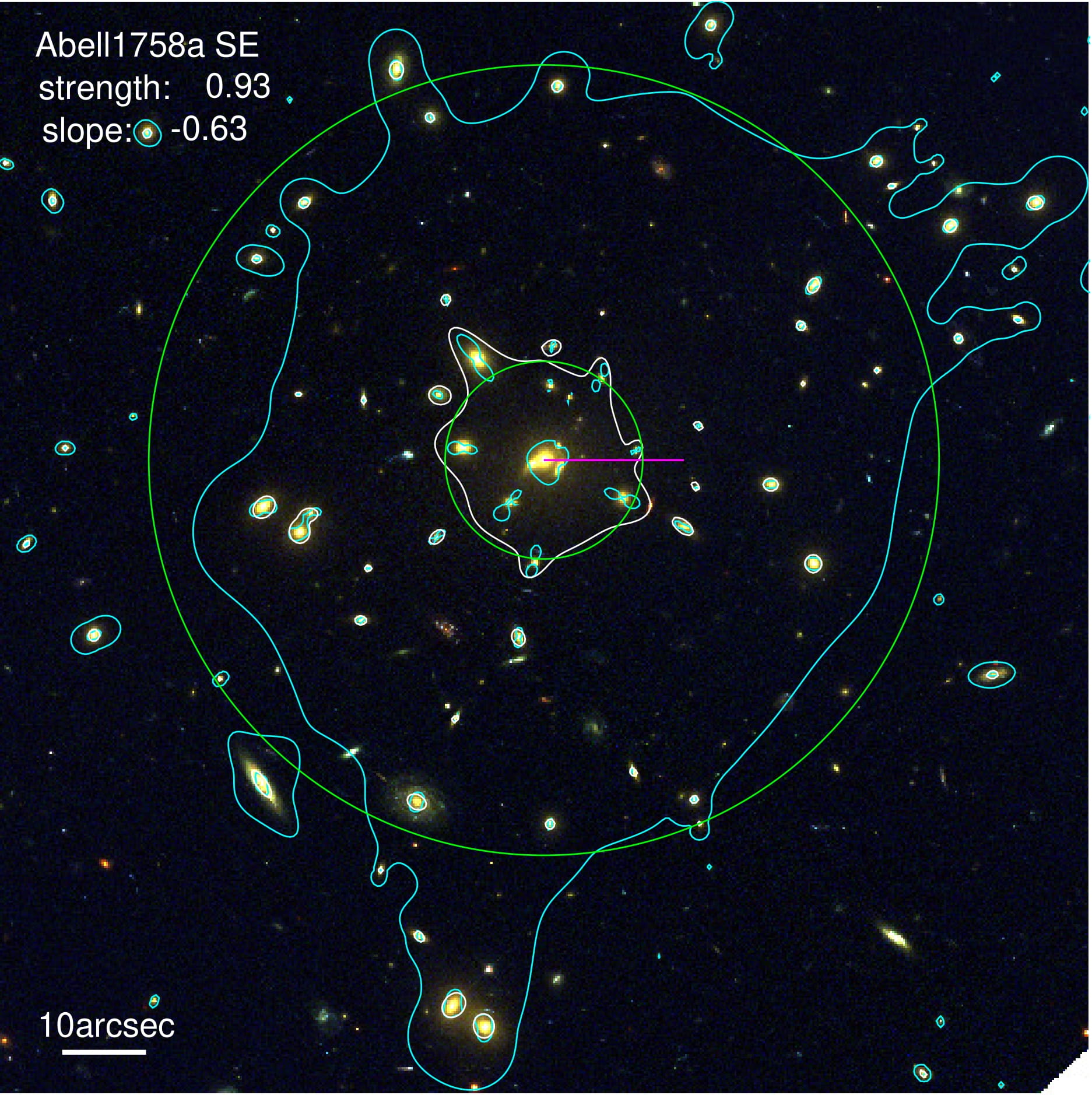}
    \includegraphics[width=0.23\textwidth]{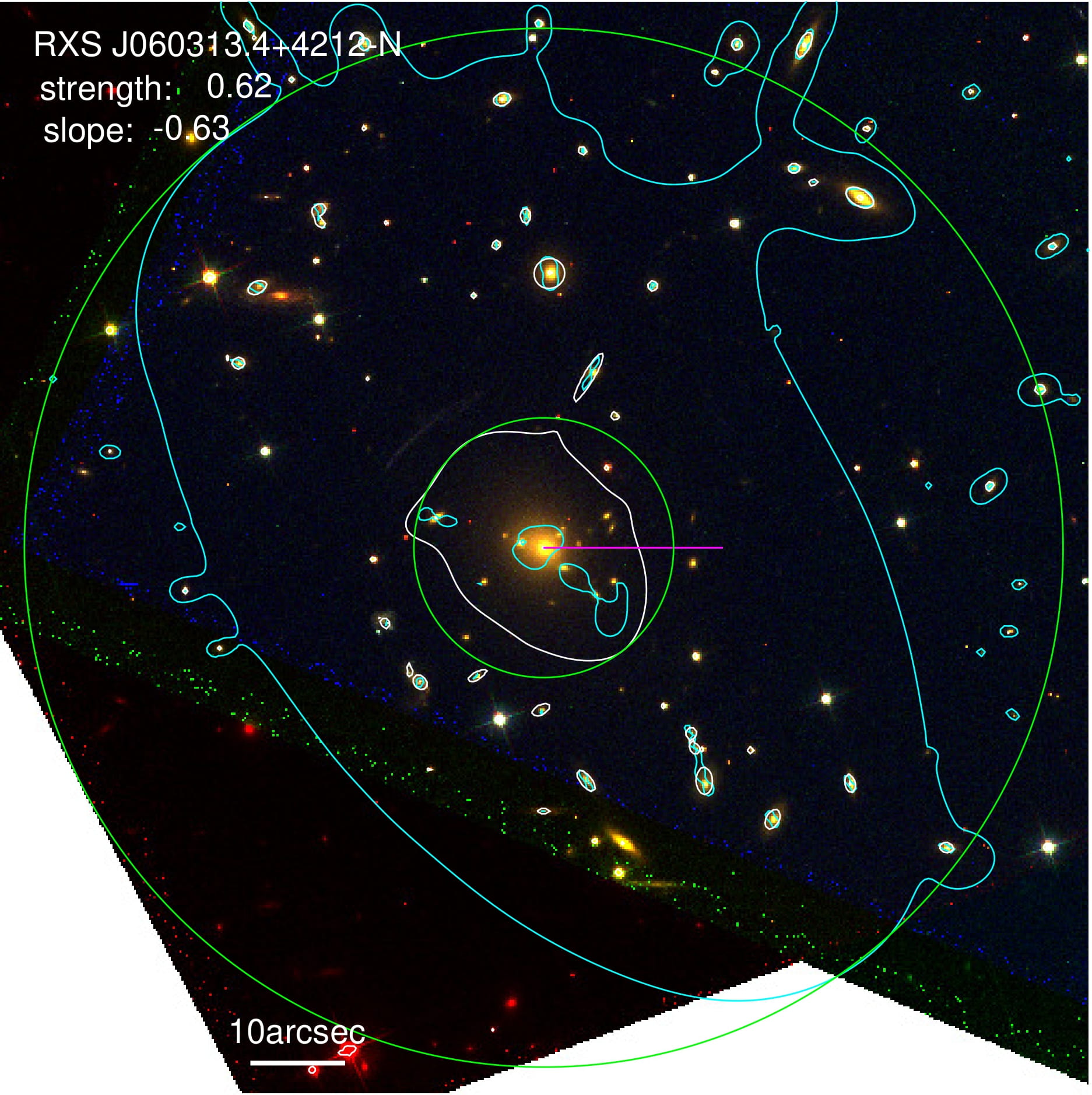}
    \includegraphics[width=0.23\textwidth]{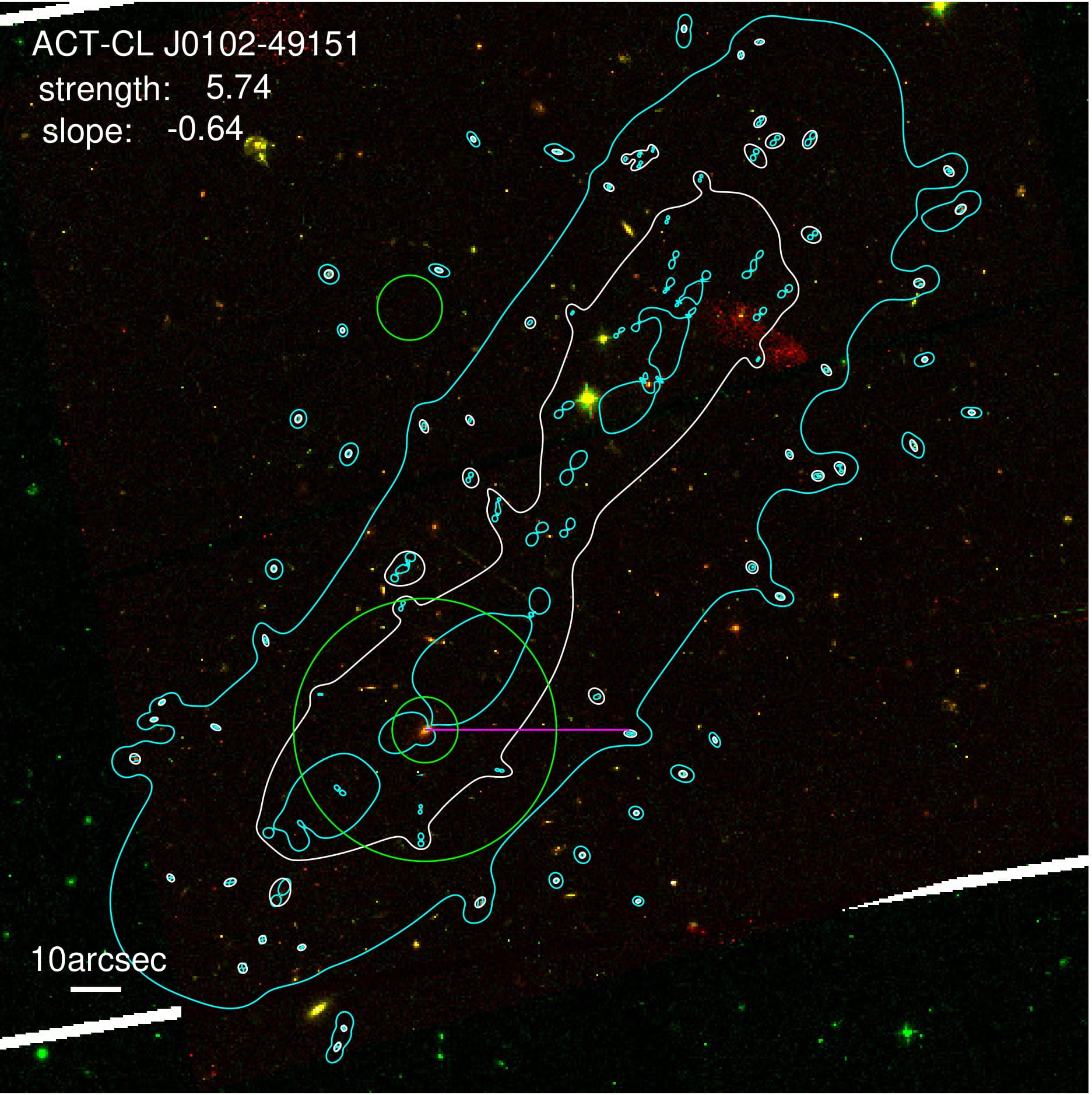}
    \includegraphics[width=0.23\textwidth]{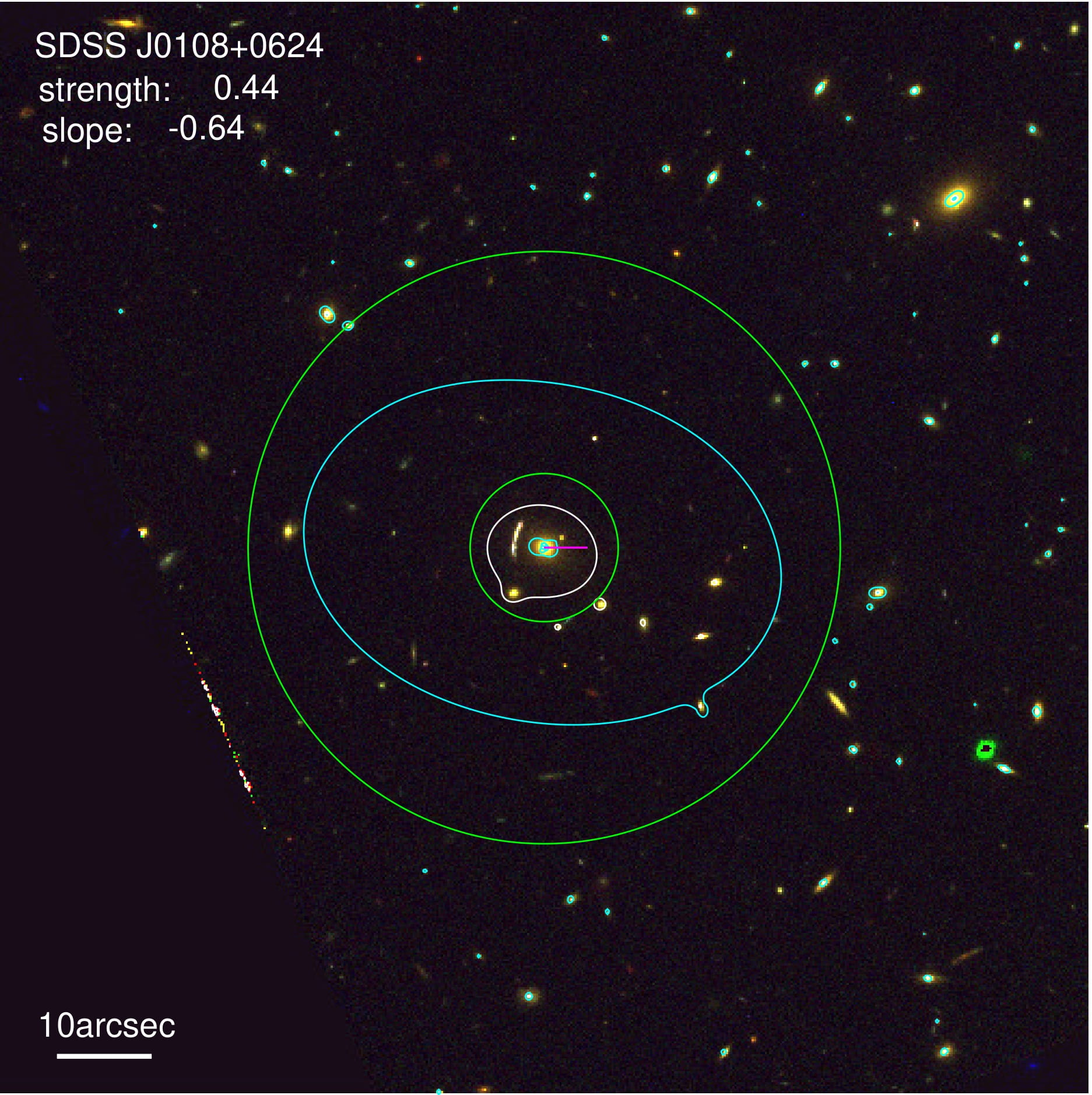}\\
    \includegraphics[width=0.23\textwidth]{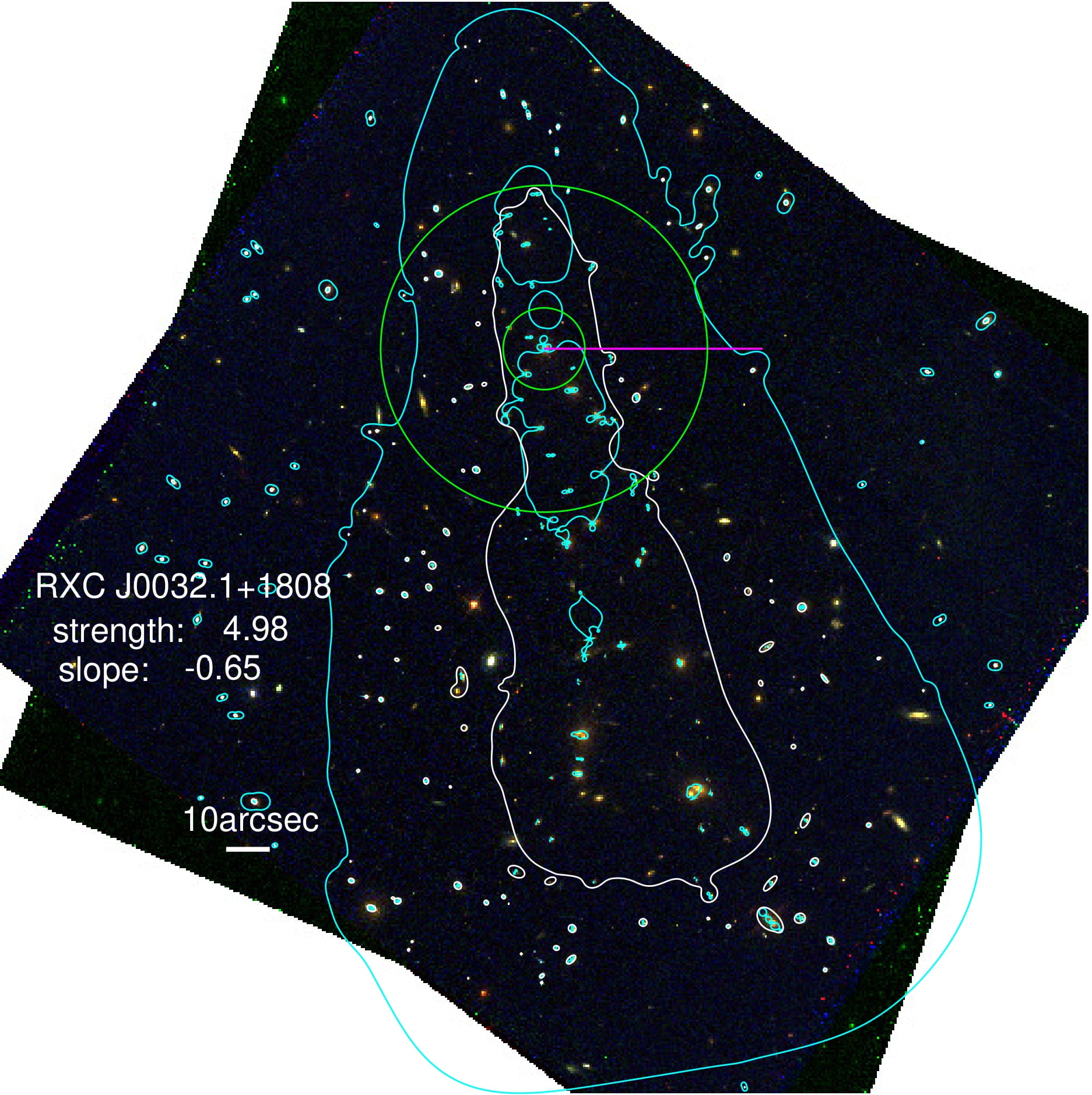}
    \includegraphics[width=0.23\textwidth]{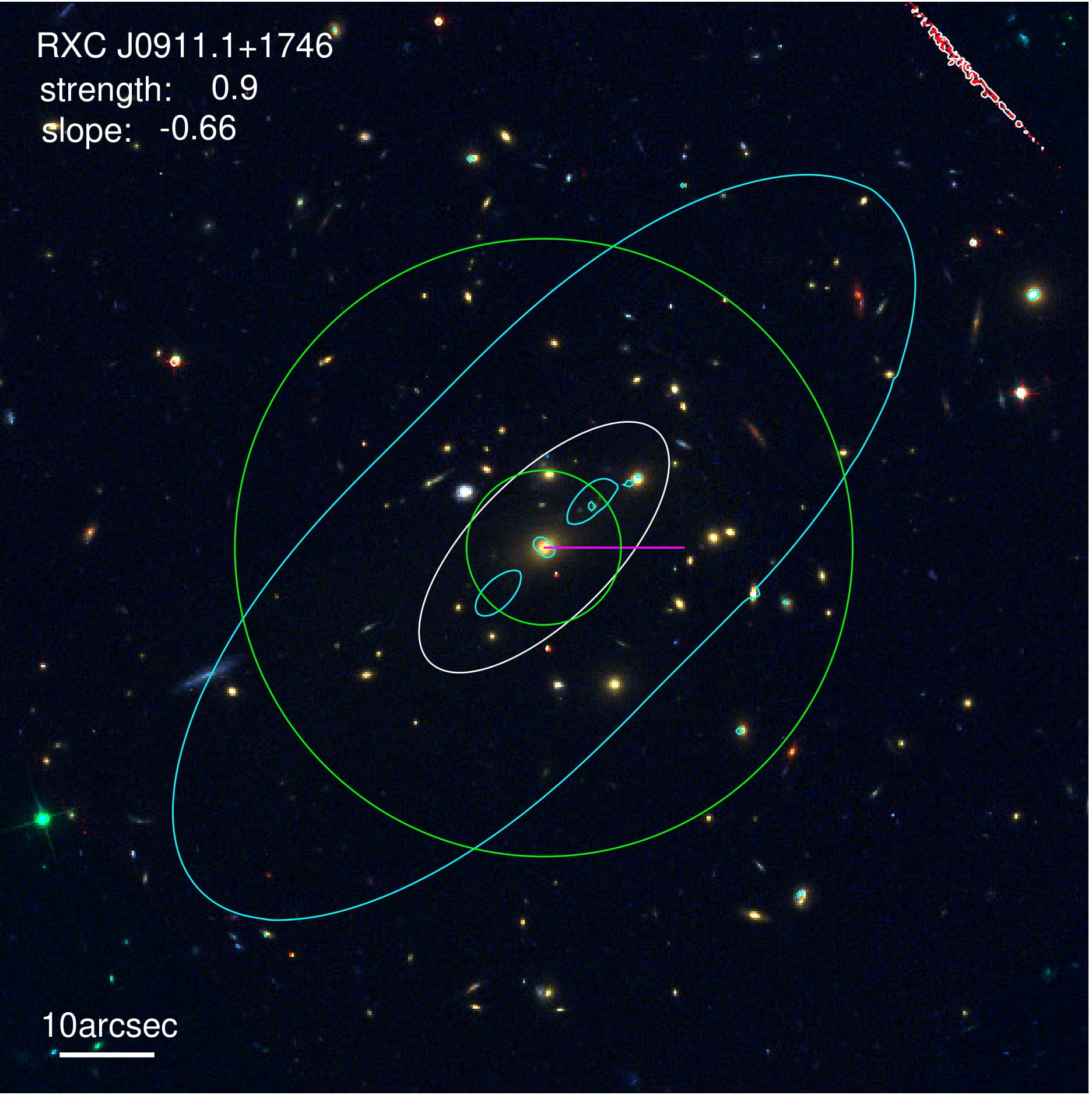}
    \includegraphics[width=0.23\textwidth]{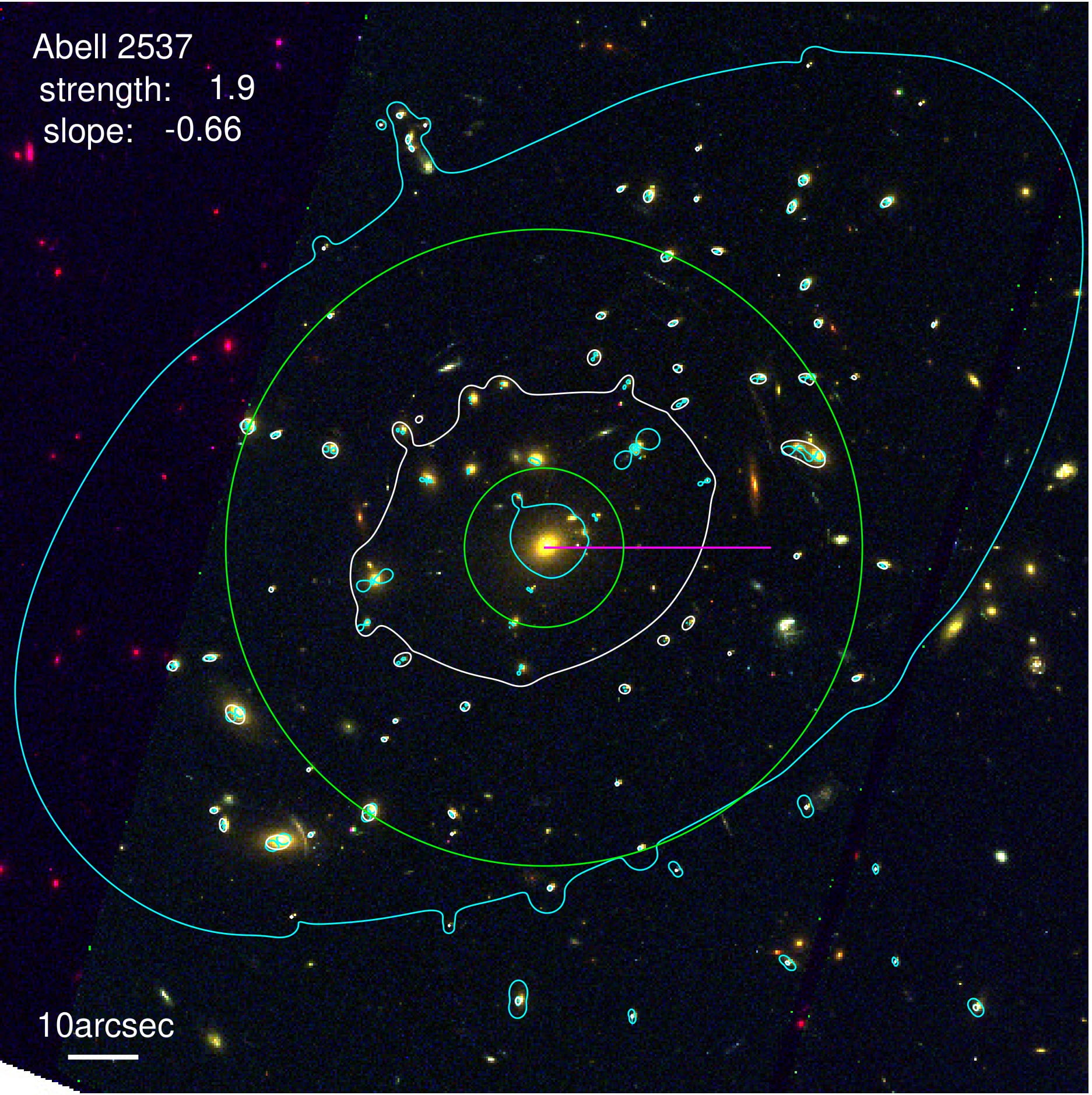}
    \includegraphics[width=0.23\textwidth]{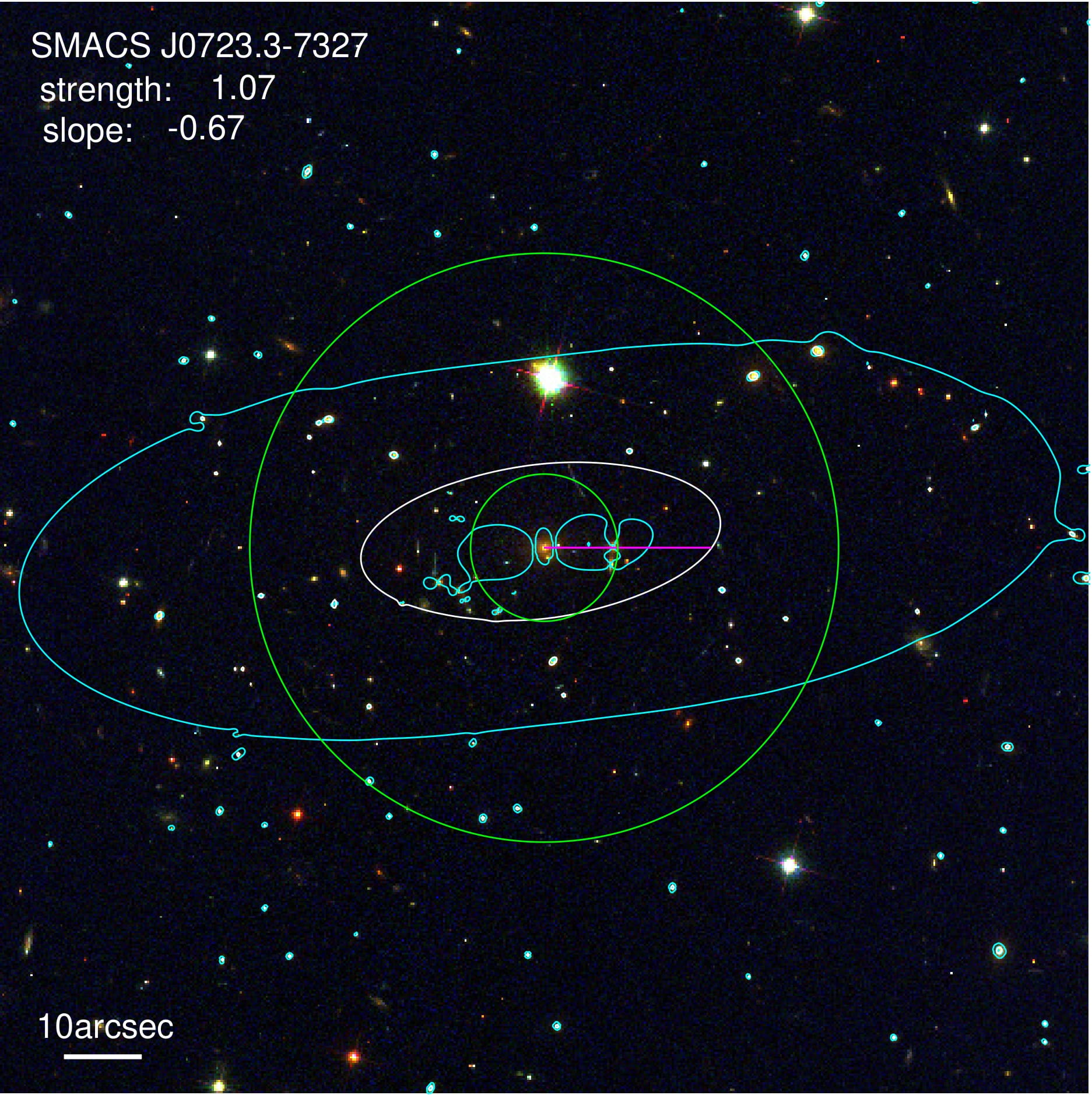}\\  
    \includegraphics[width=0.23\textwidth]{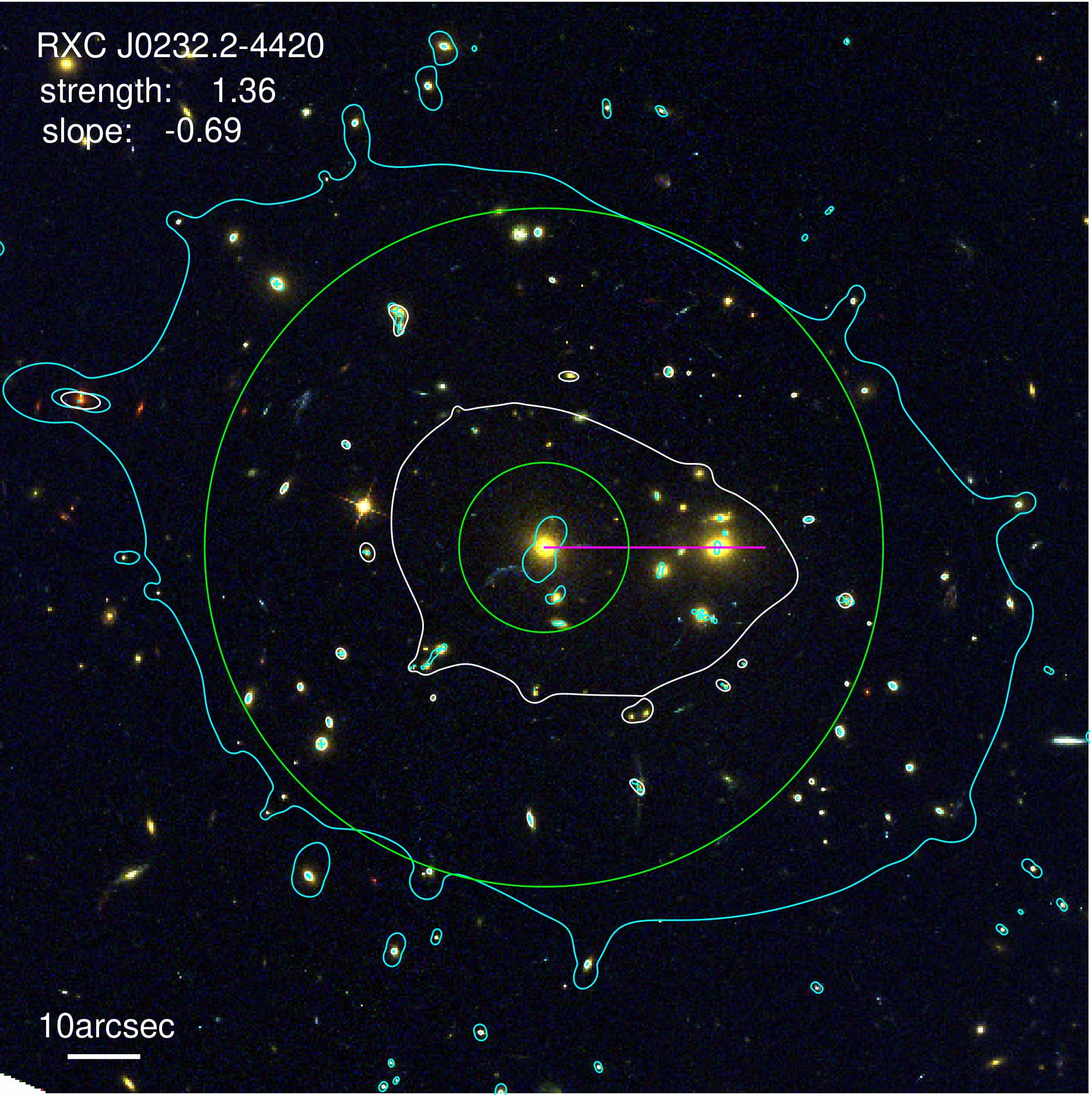}
    \includegraphics[width=0.23\textwidth]{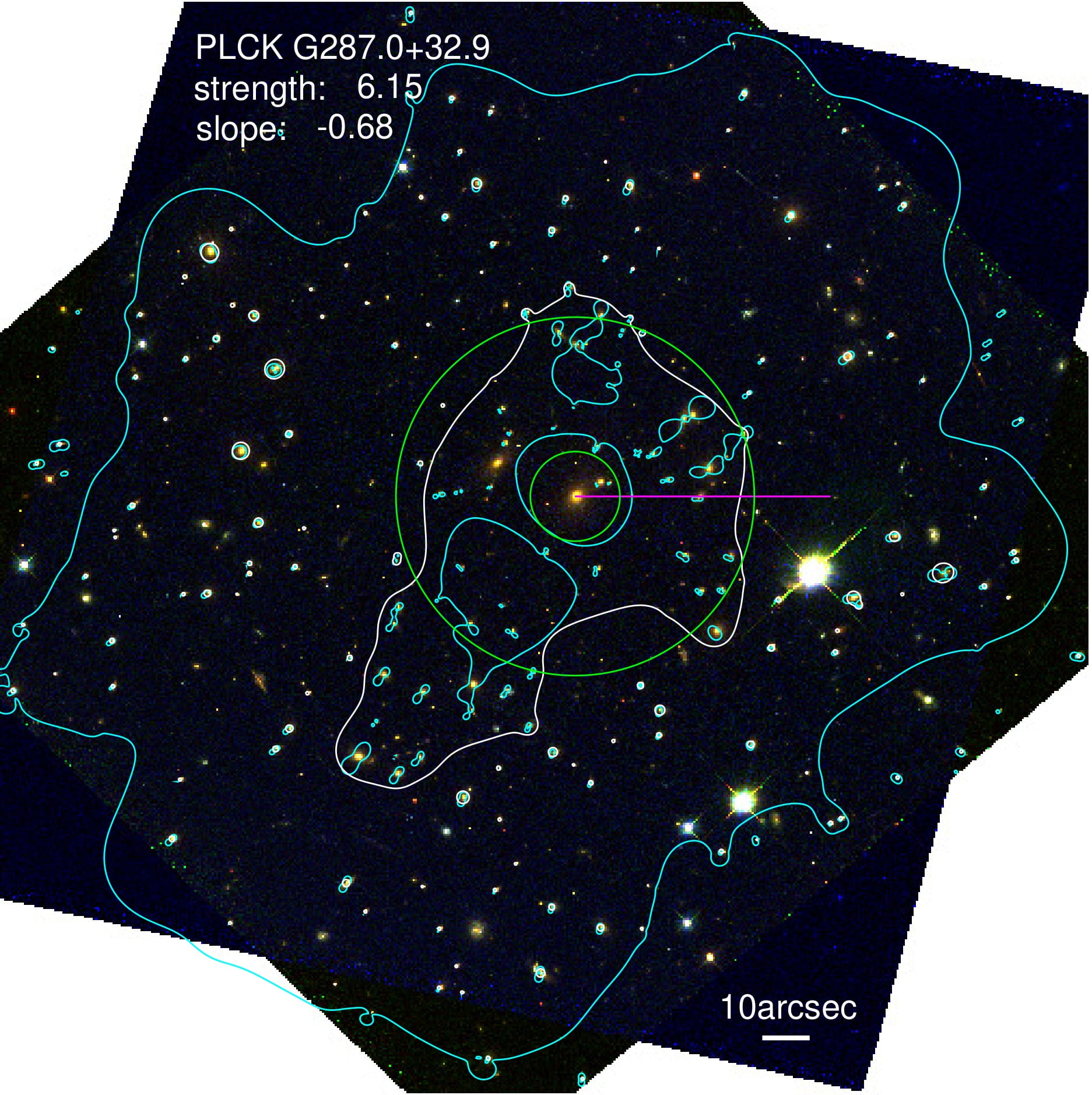}
    \includegraphics[width=0.23\textwidth]{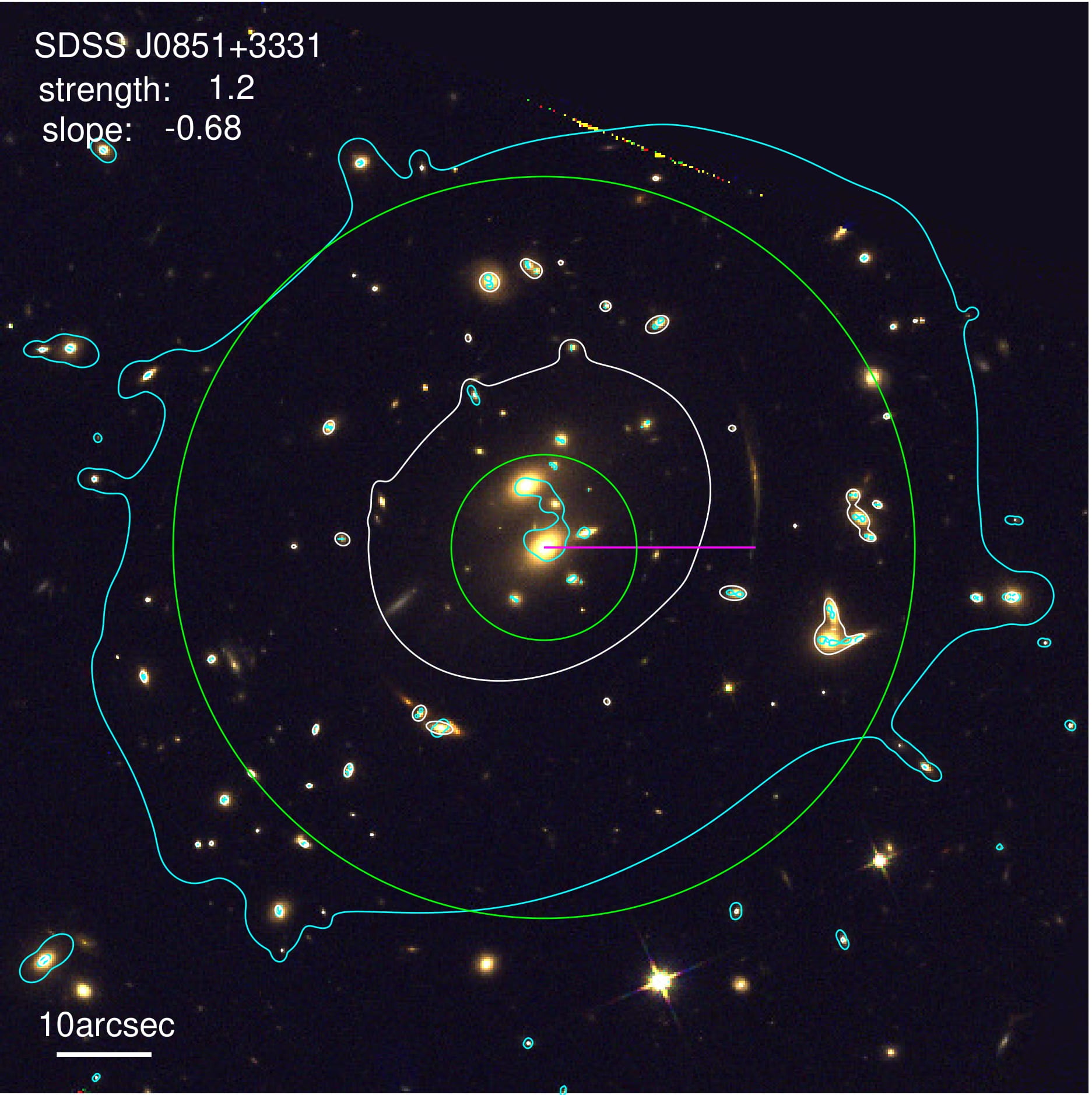}
    \includegraphics[width=0.23\textwidth]{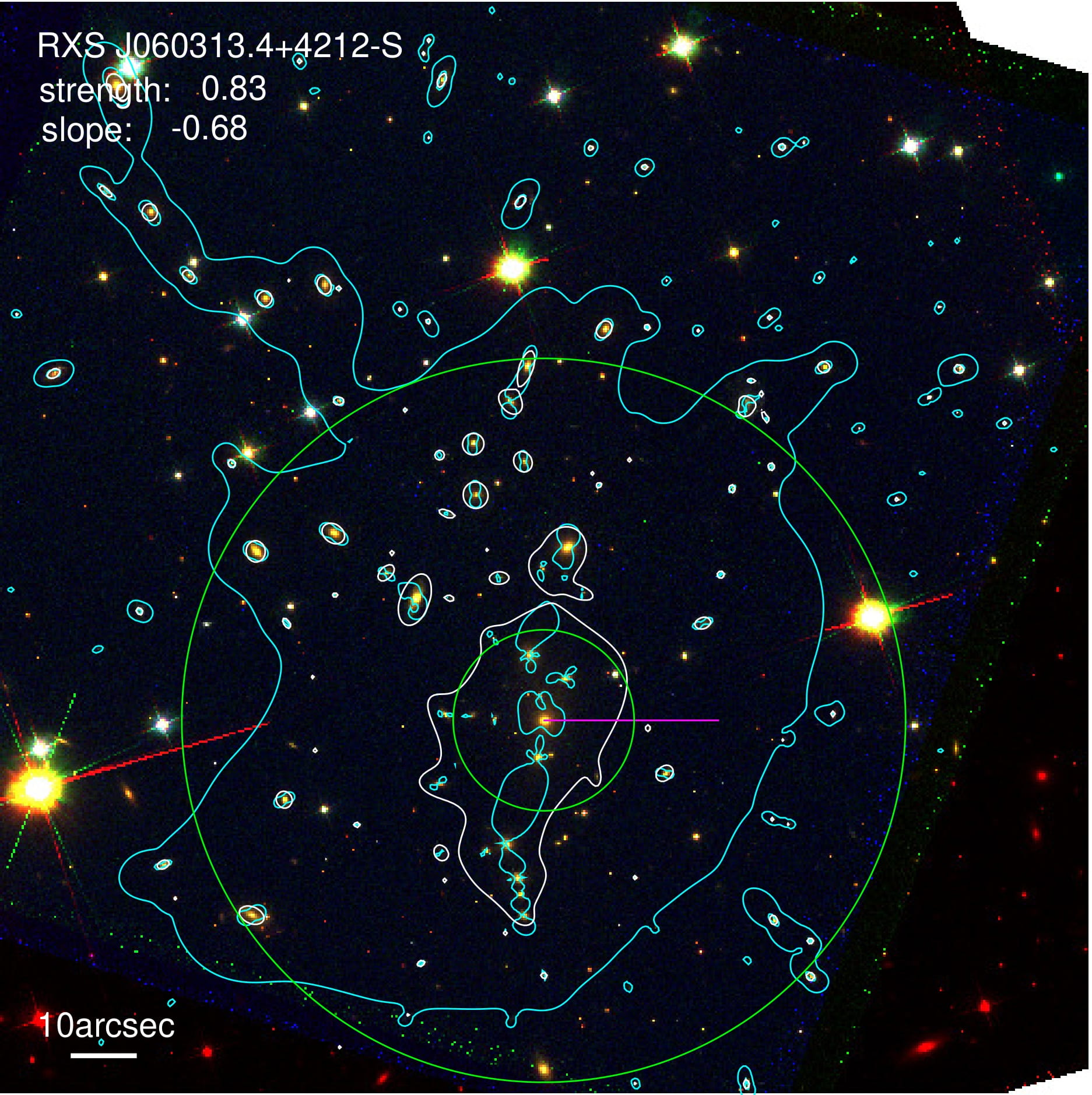}\\
    \caption{Same as Figure \ref{fig:gallery1}, for the clusters SDSS~J1050$+$0017, Abell~3192, Abell~2744, Abell~1758a~NW, MACS~J0417.5$-$1154, RXC~J2211.7$-$0350, SDSS~J1420$+$3955, Abell~S1063, Abell~1758a~SE, RXS~J060313.4$+$4212$-$N, ACT$-$CLJ0102$-$49151, SDSS~J0108$+$0624, RXC~J0032.1$+$1808, RXC~J0911.1$+$1746, Abell~2537, SMACS~J0723.3$-$7327, RXC~J0232.2$-$4420, PLCK~G287.0$+$32.9, SDSS~J0851$+$3331, and RXS~J060313.4$+$4212$-$S.
    }
    \label{fig:gallery2}
\end{figure*}

\begin{figure*}
\center
    \includegraphics[width=0.23\textwidth]{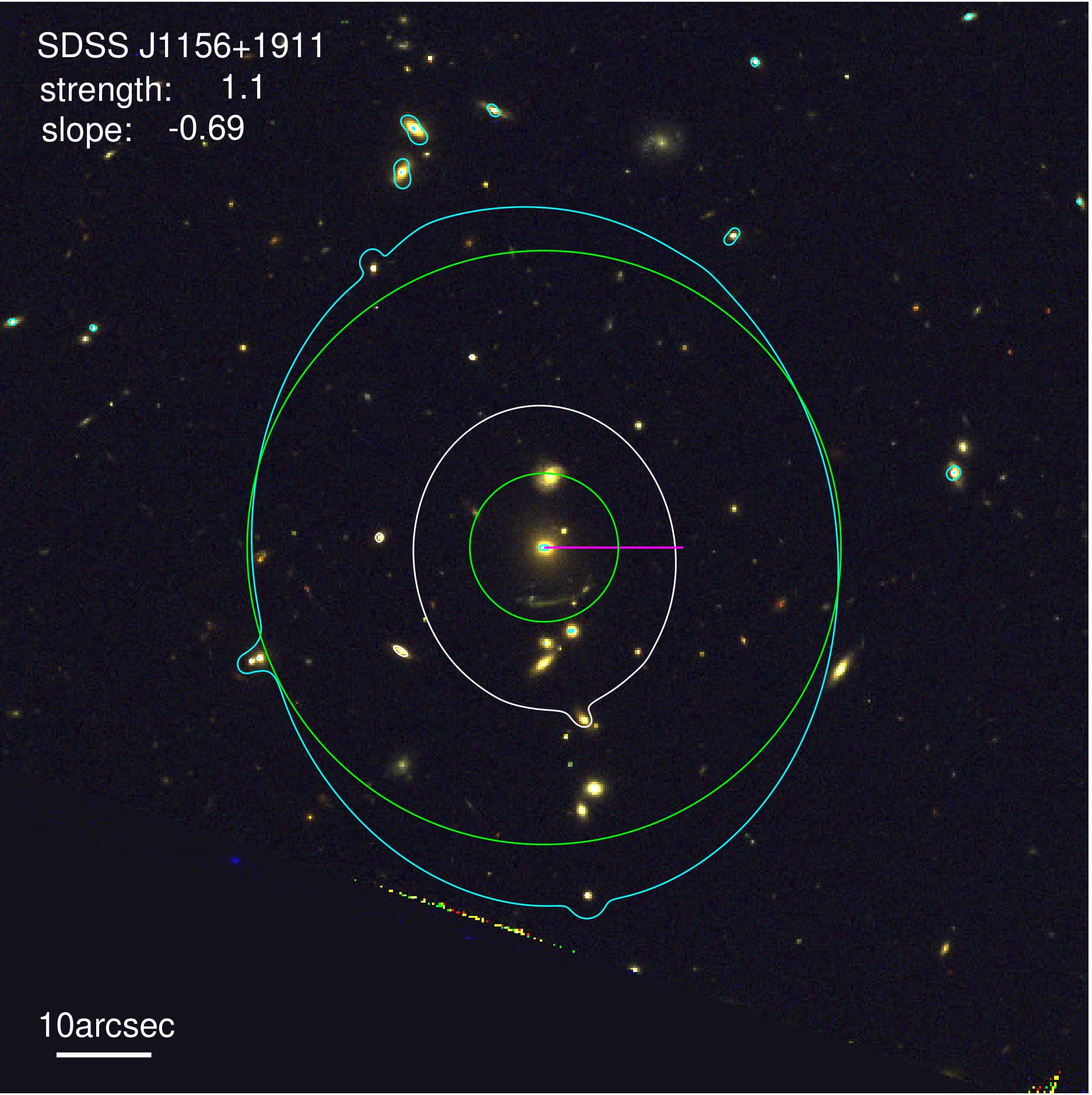}
    \includegraphics[width=0.23\textwidth]{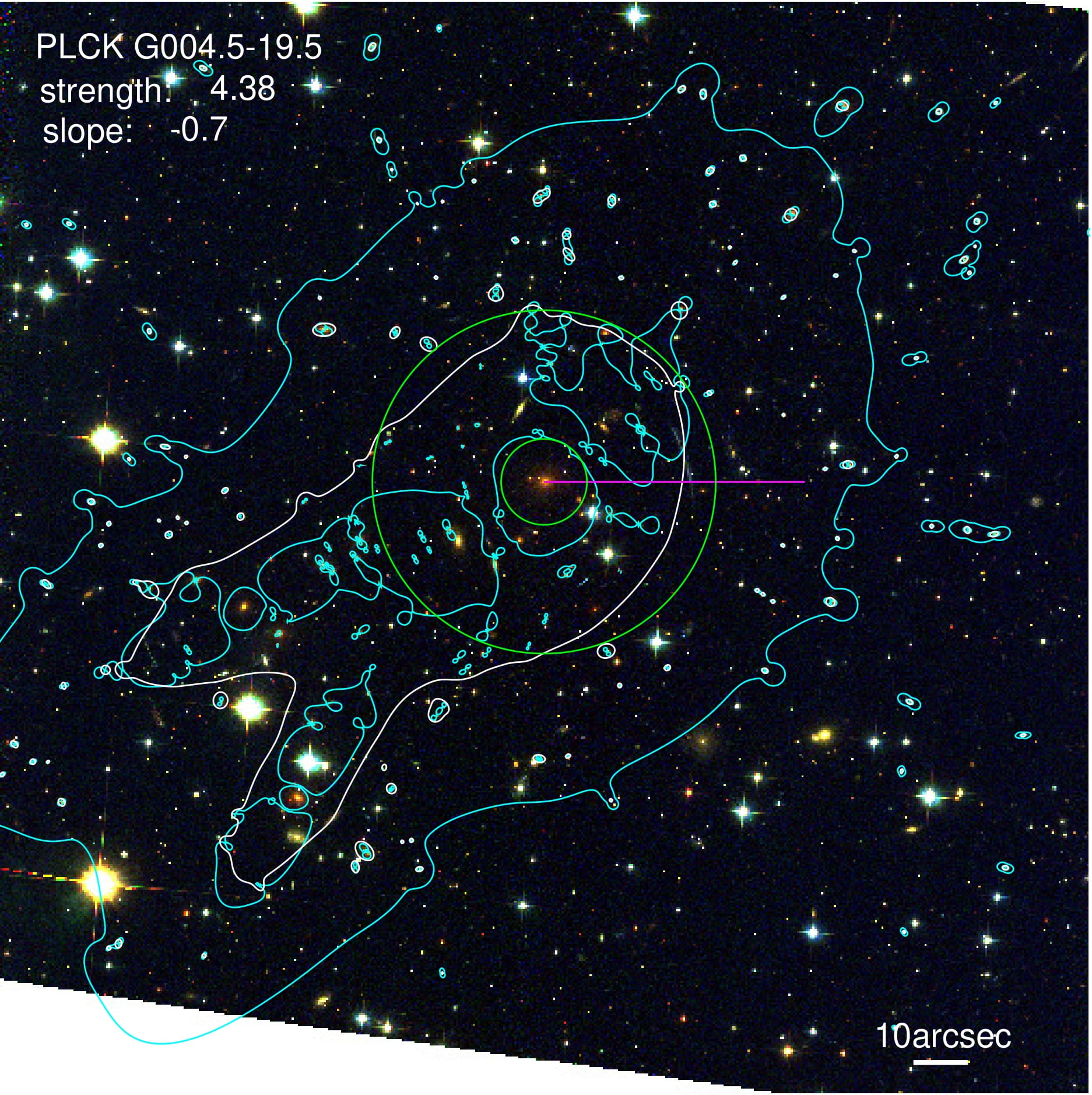}
    \includegraphics[width=0.23\textwidth]{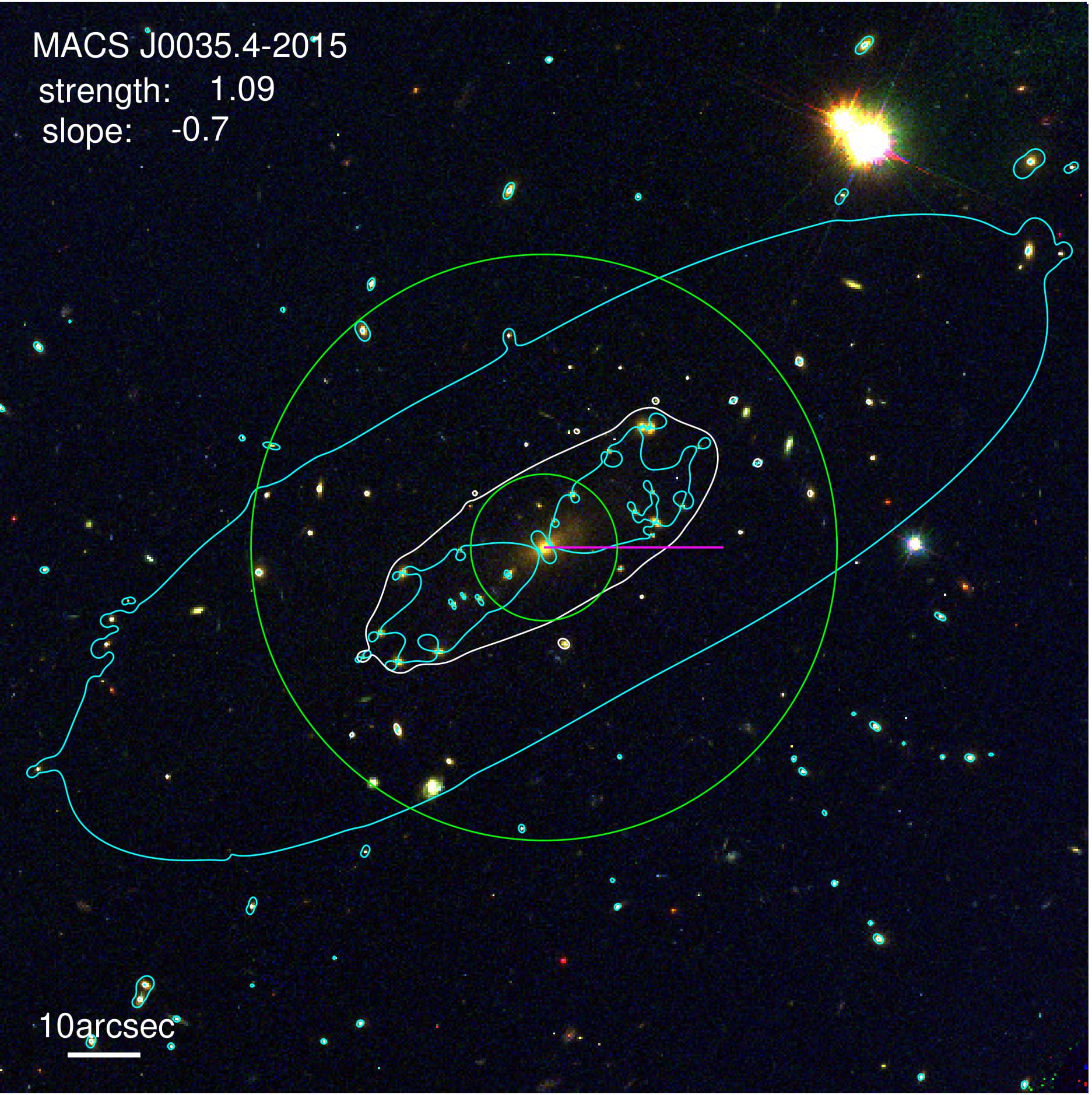}
    \includegraphics[width=0.23\textwidth]{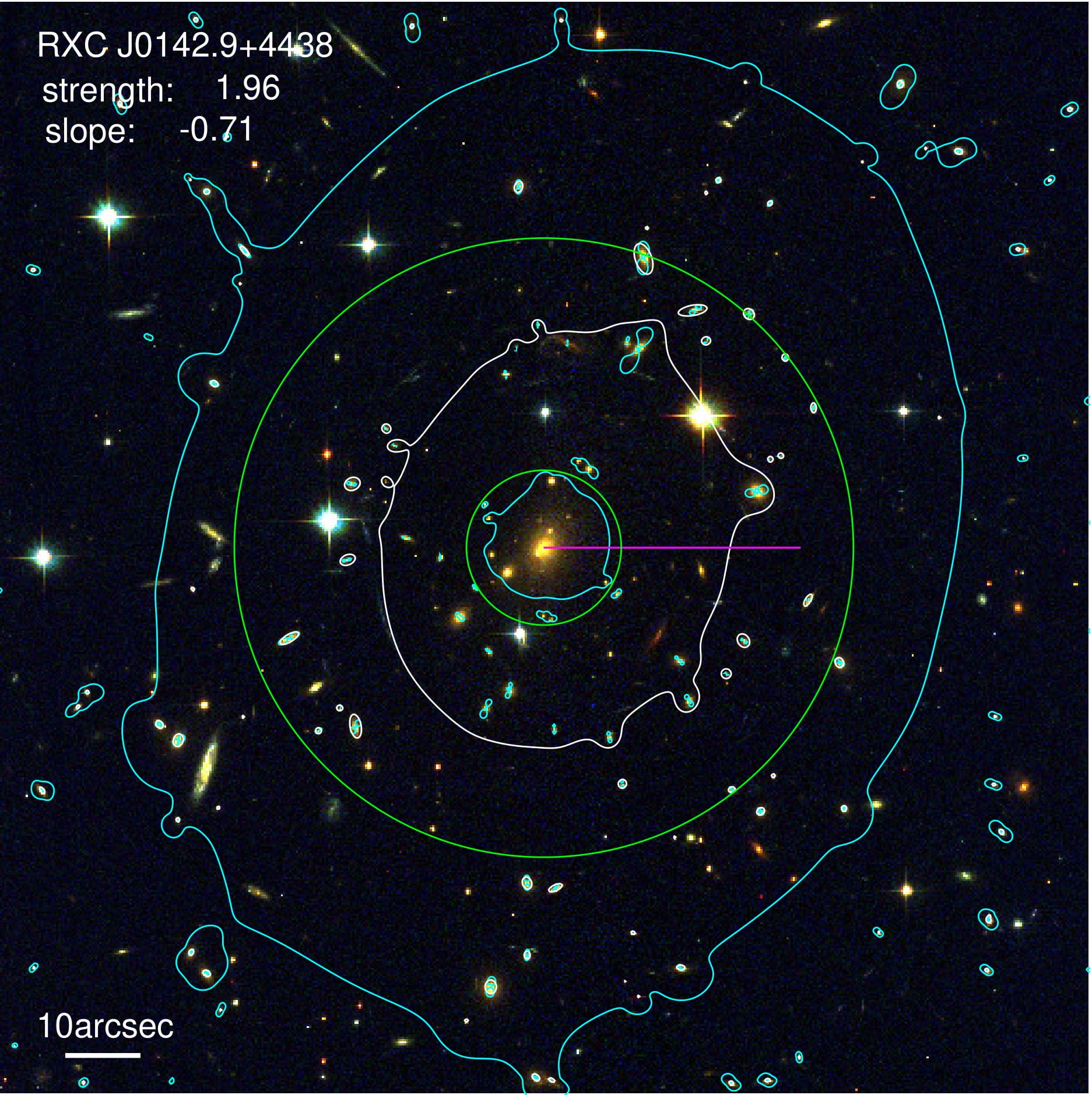}\\
    \includegraphics[width=0.23\textwidth]{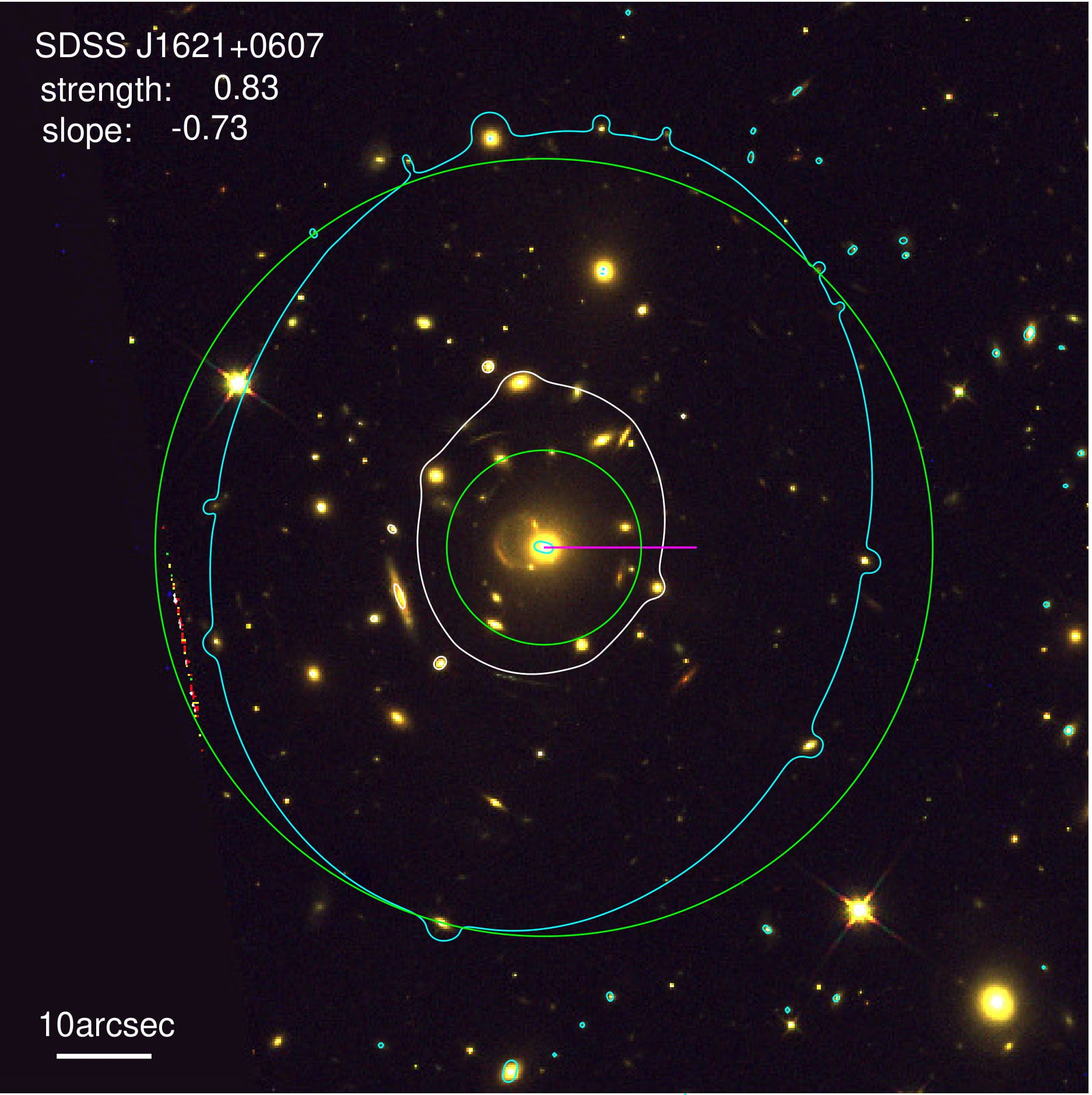}
    \includegraphics[width=0.23\textwidth]{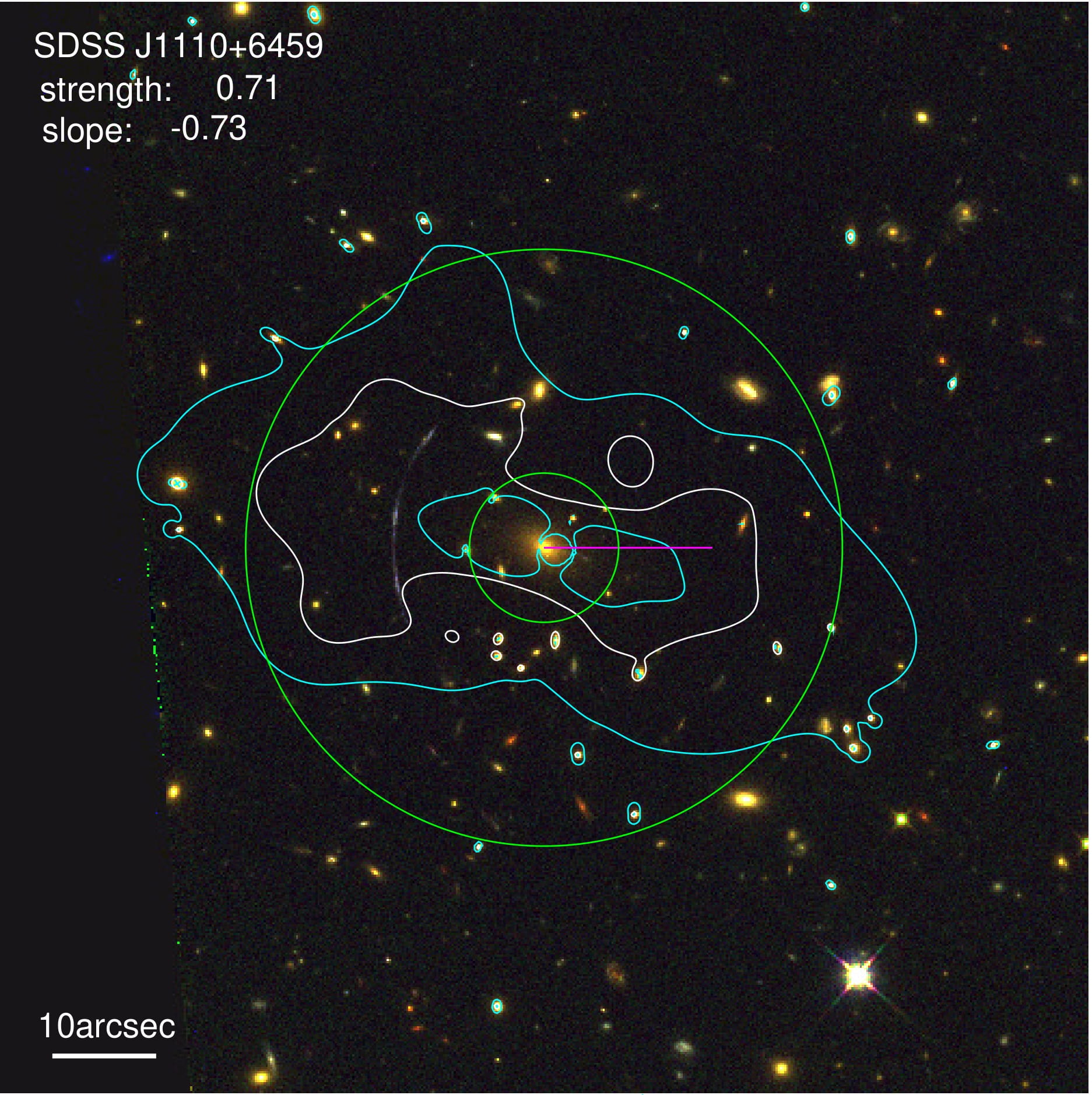}
    \includegraphics[width=0.23\textwidth]{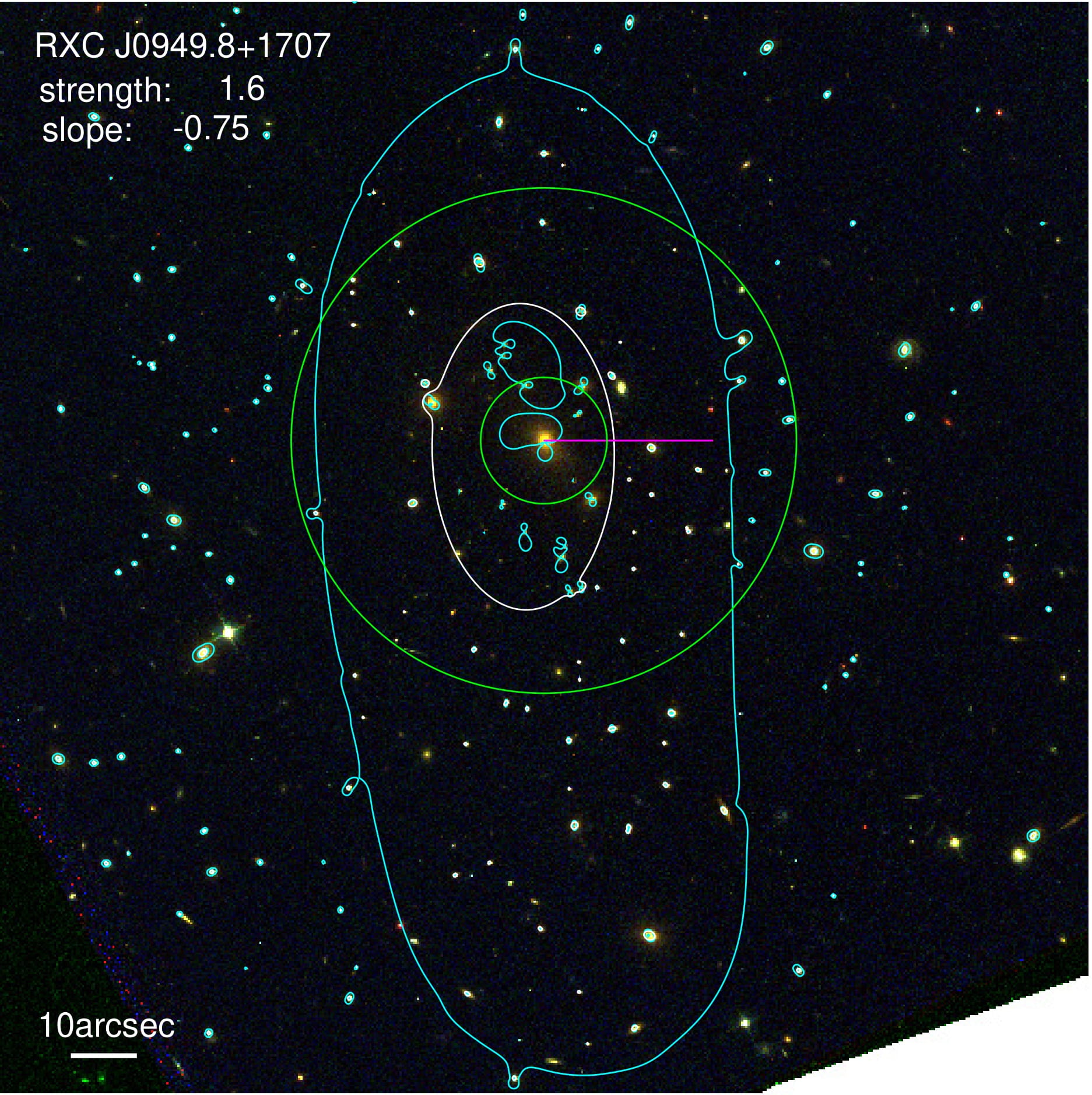}
    \includegraphics[width=0.23\textwidth]{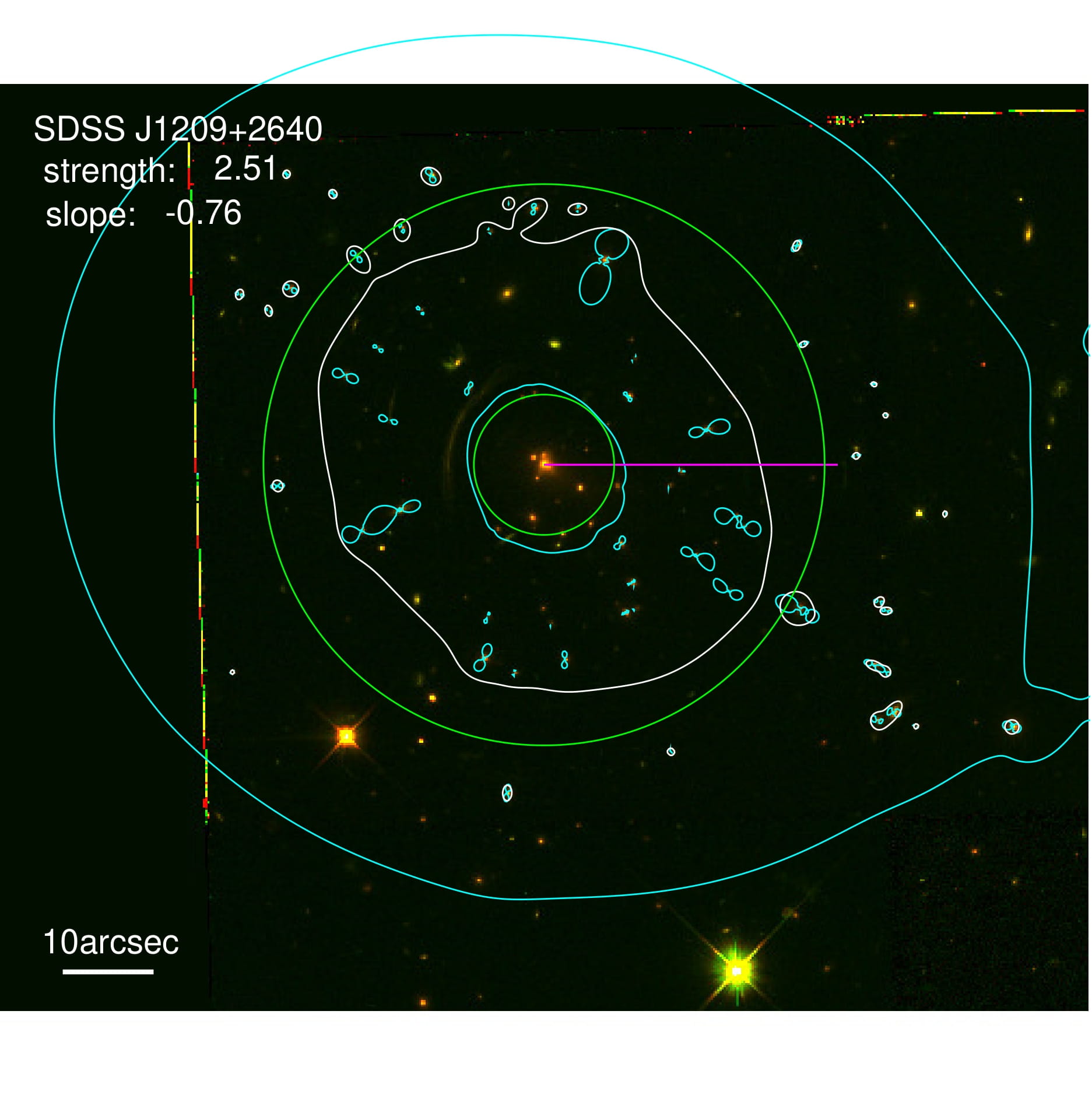}\\
    \includegraphics[width=0.23\textwidth]{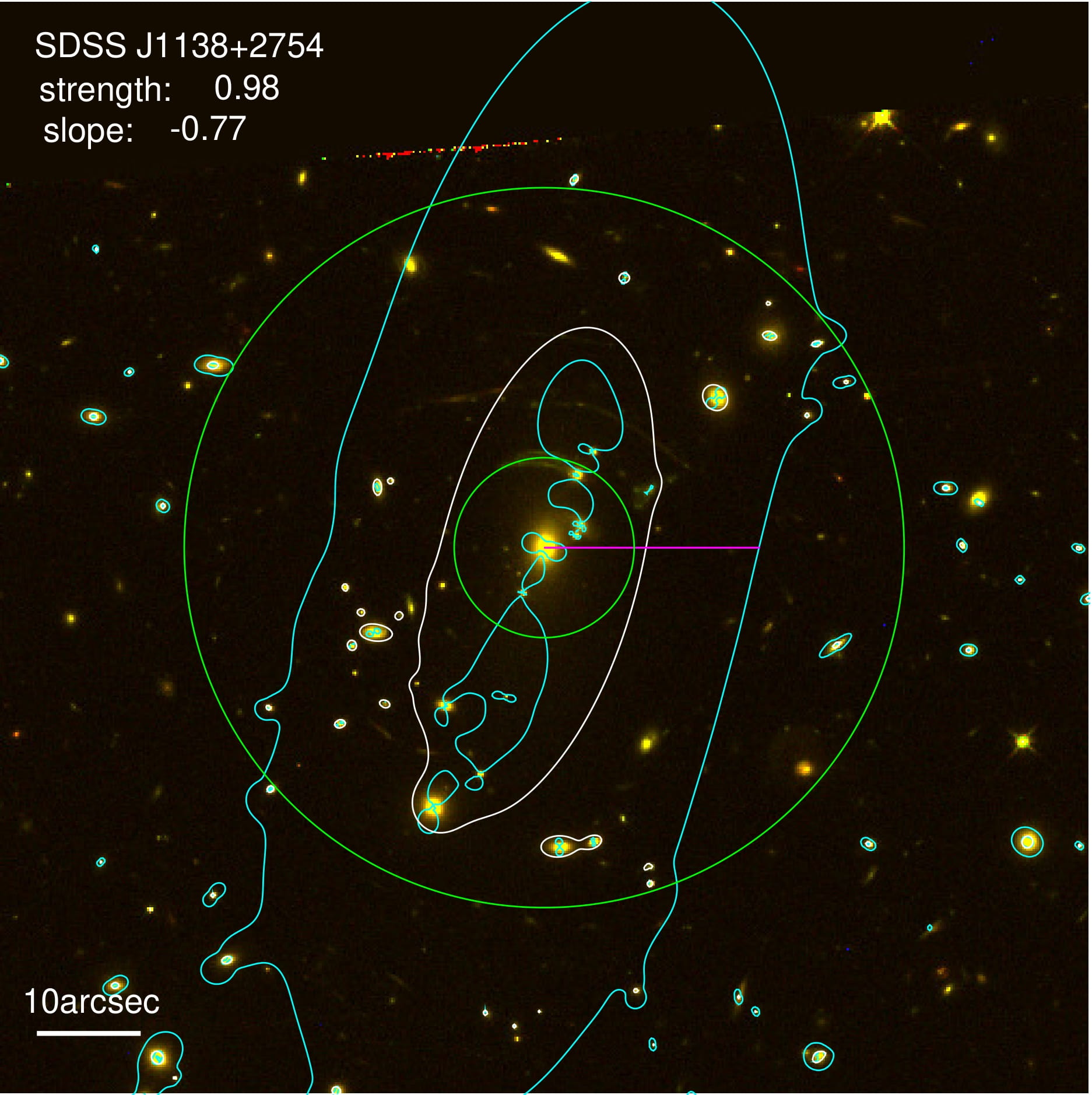}
    \includegraphics[width=0.23\textwidth]{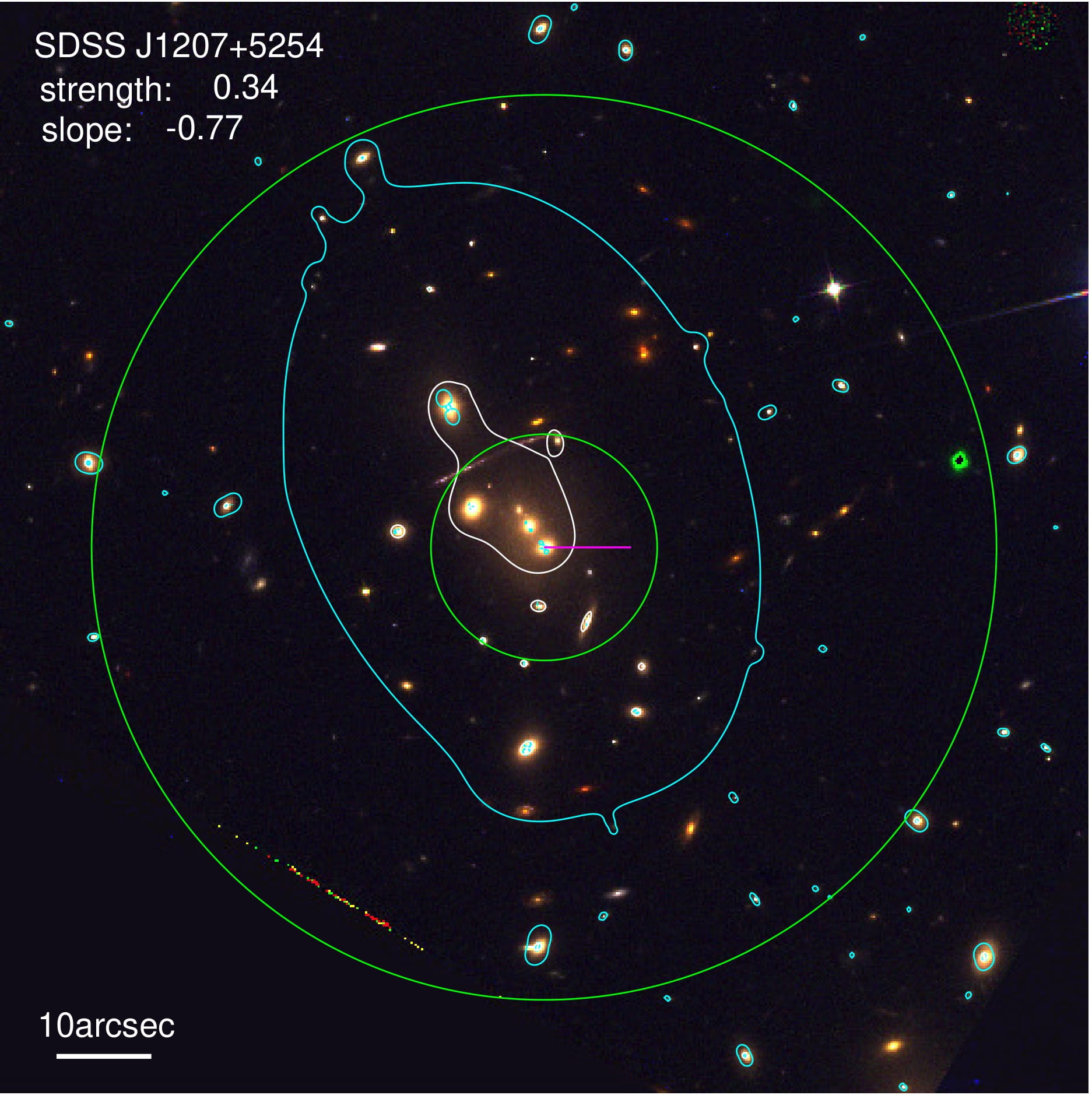}
    \includegraphics[width=0.23\textwidth]{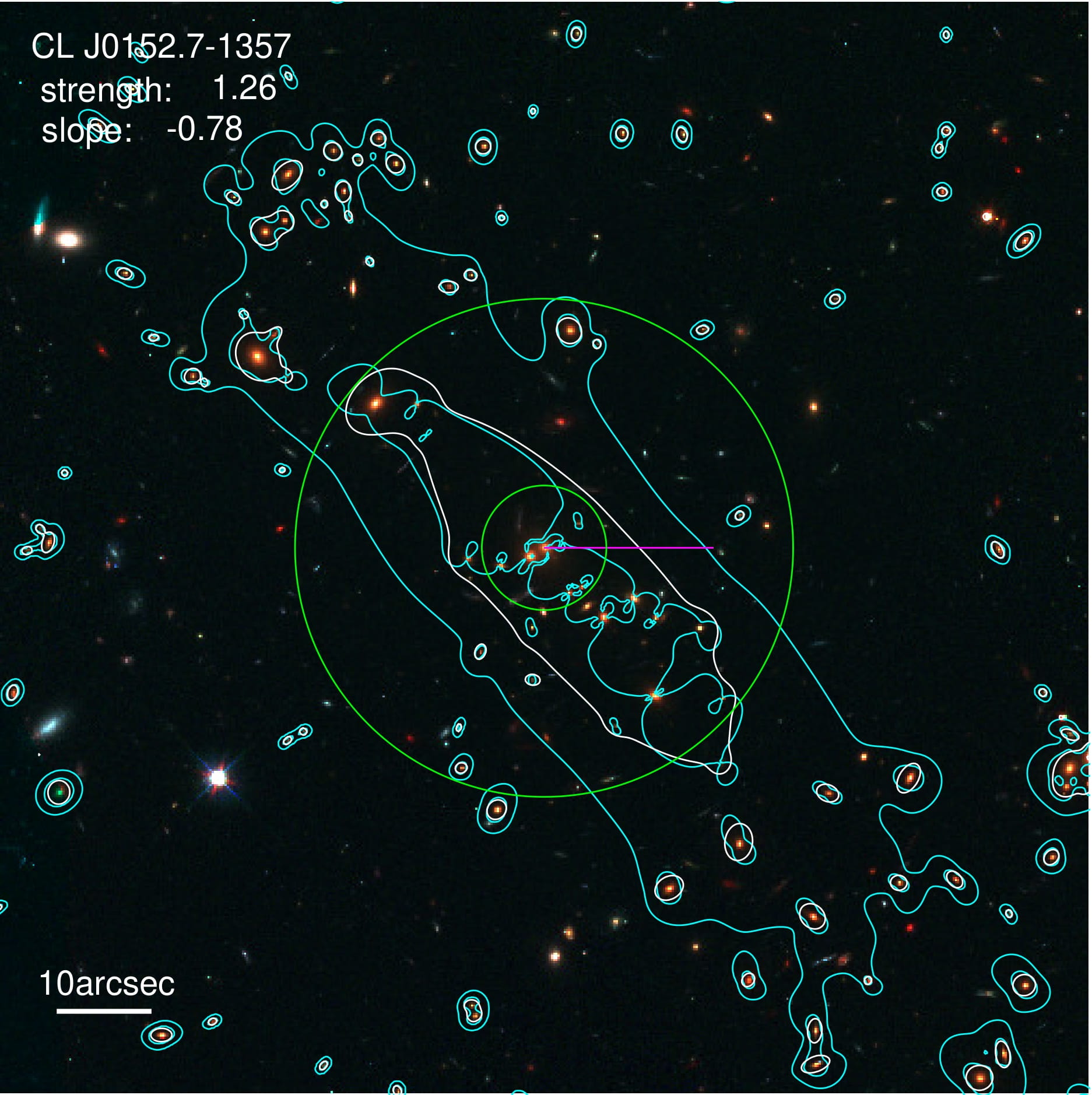}
    \includegraphics[width=0.23\textwidth]{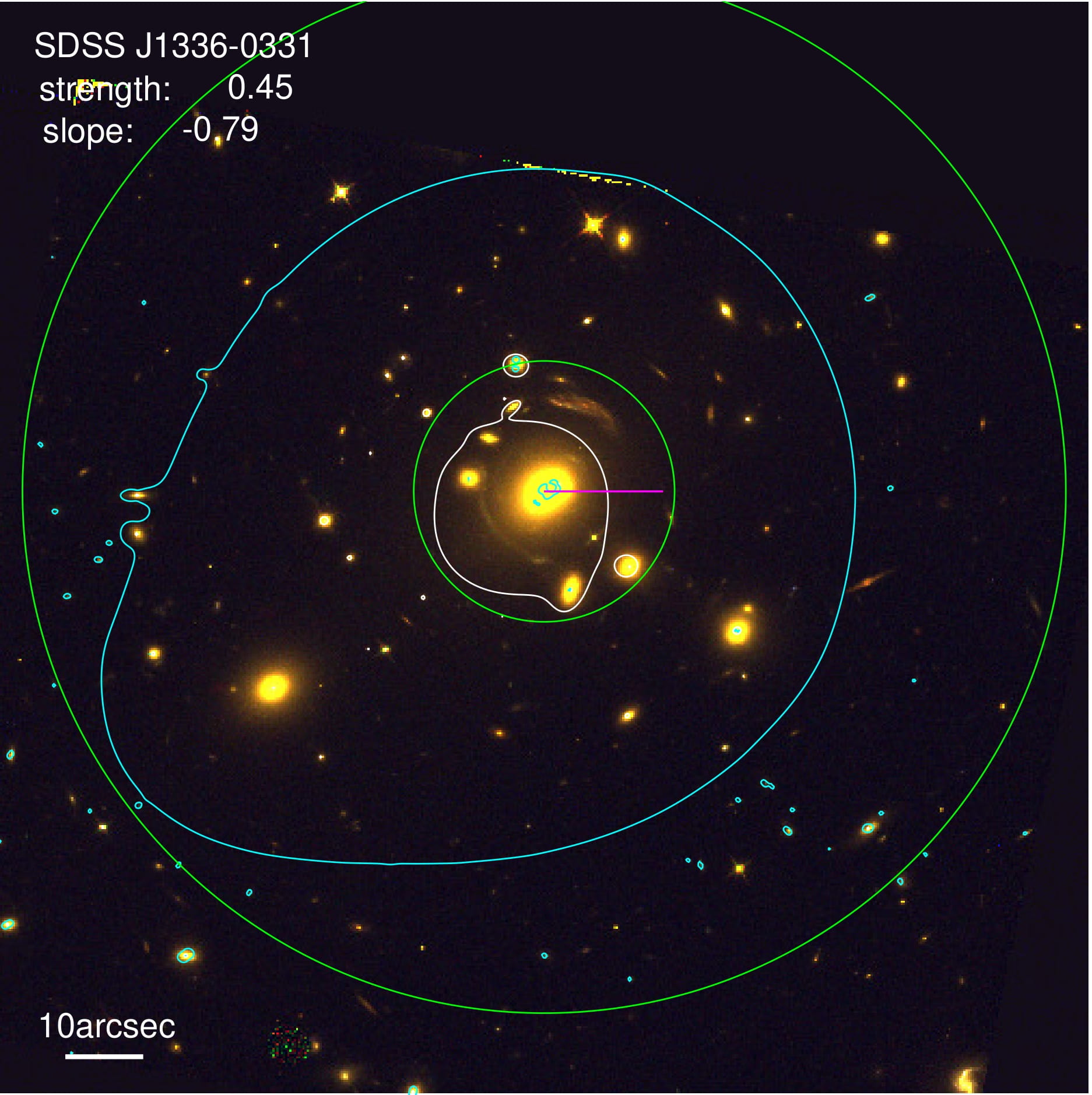}\\
    \includegraphics[width=0.23\textwidth]{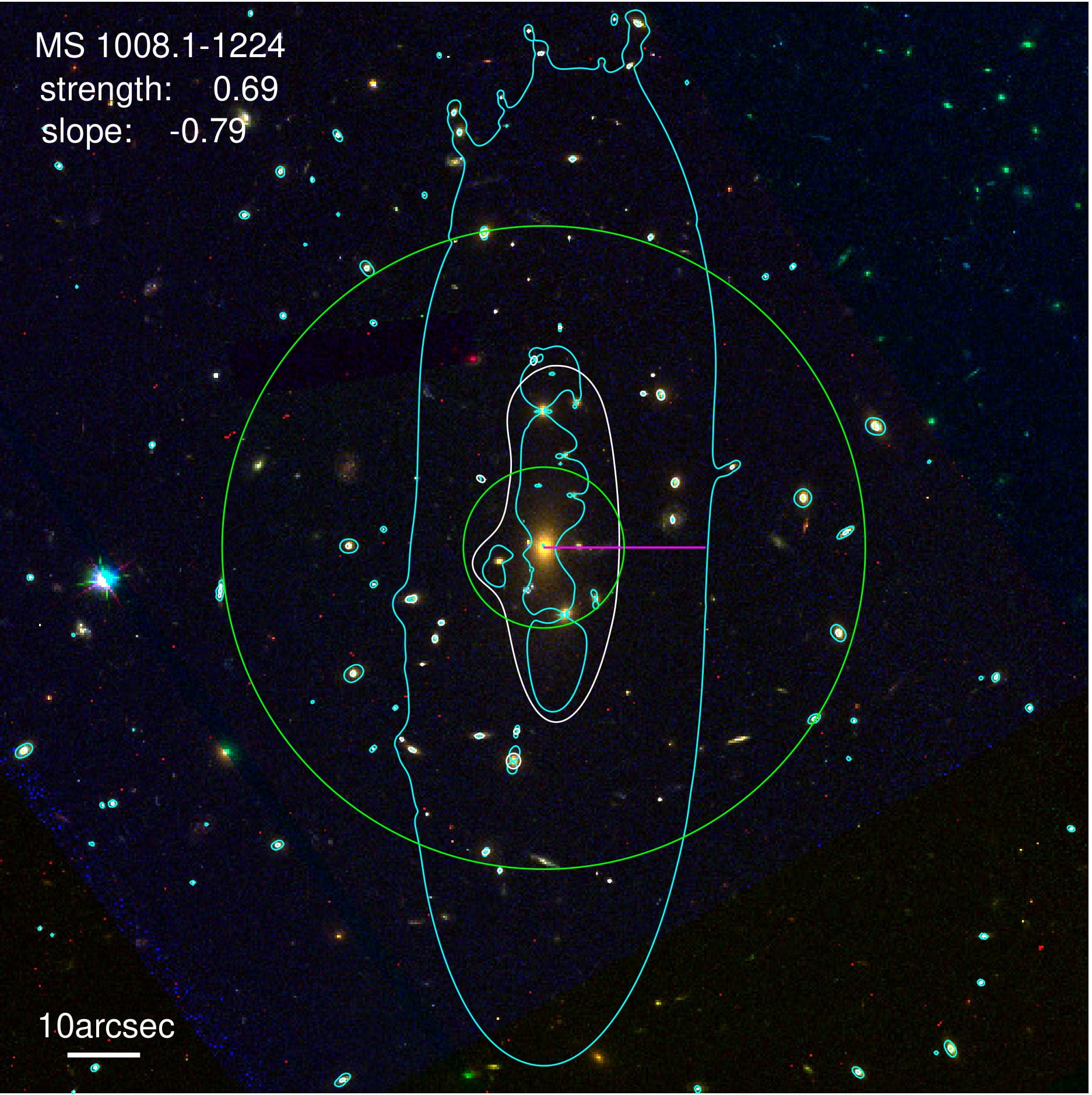}
    \includegraphics[width=0.23\textwidth]{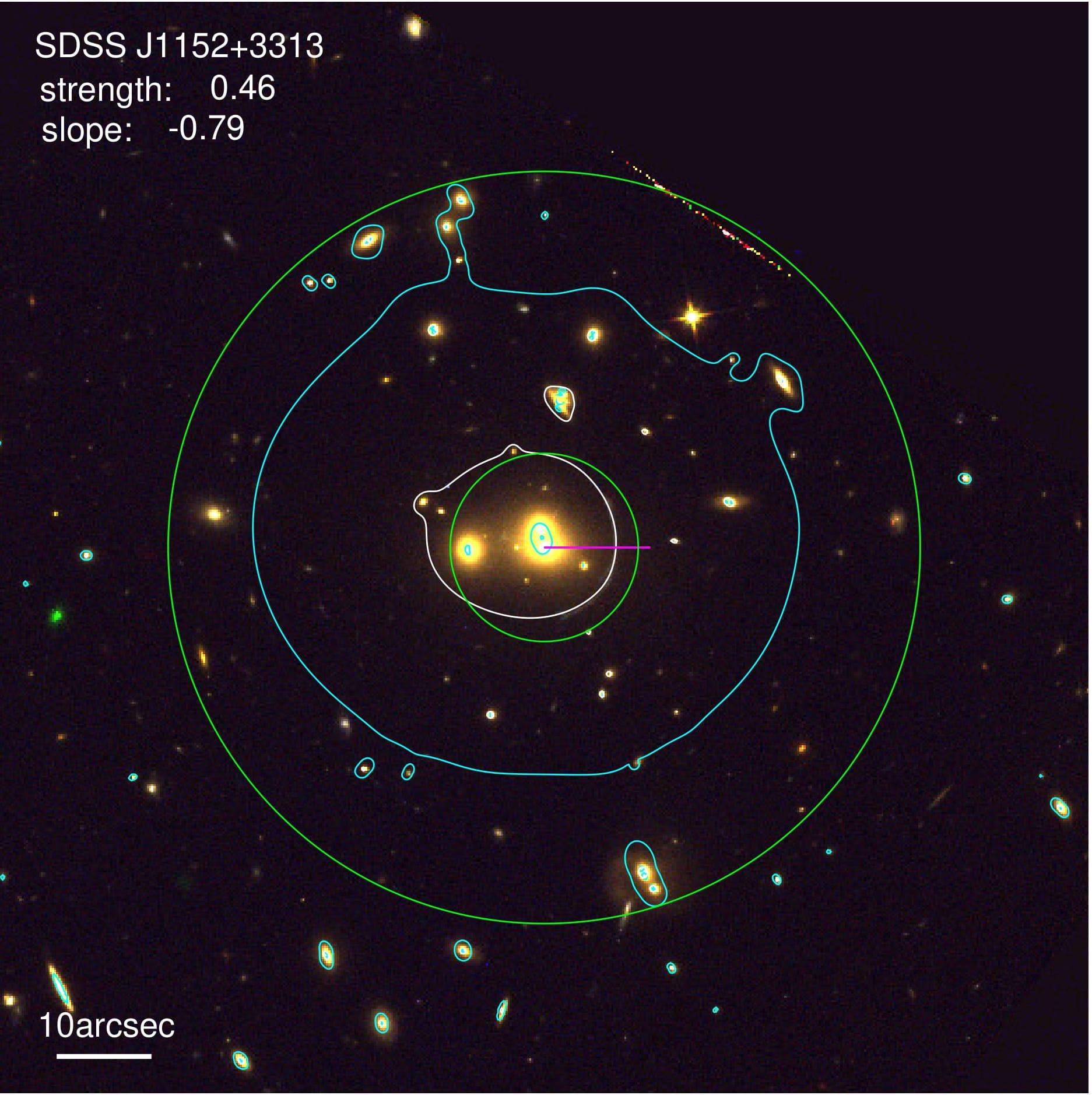}
    \includegraphics[width=0.23\textwidth]{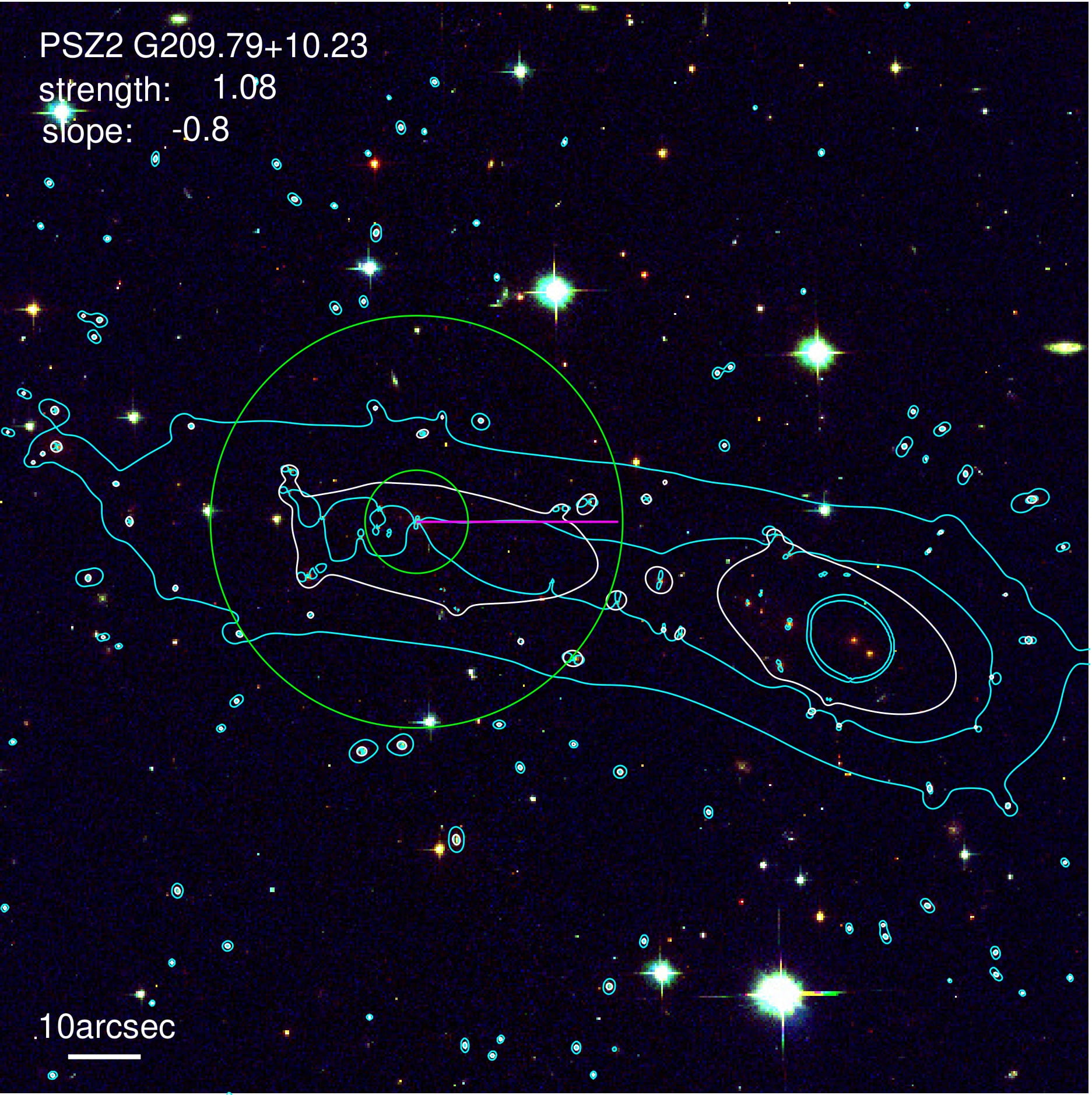}
    \includegraphics[width=0.23\textwidth]{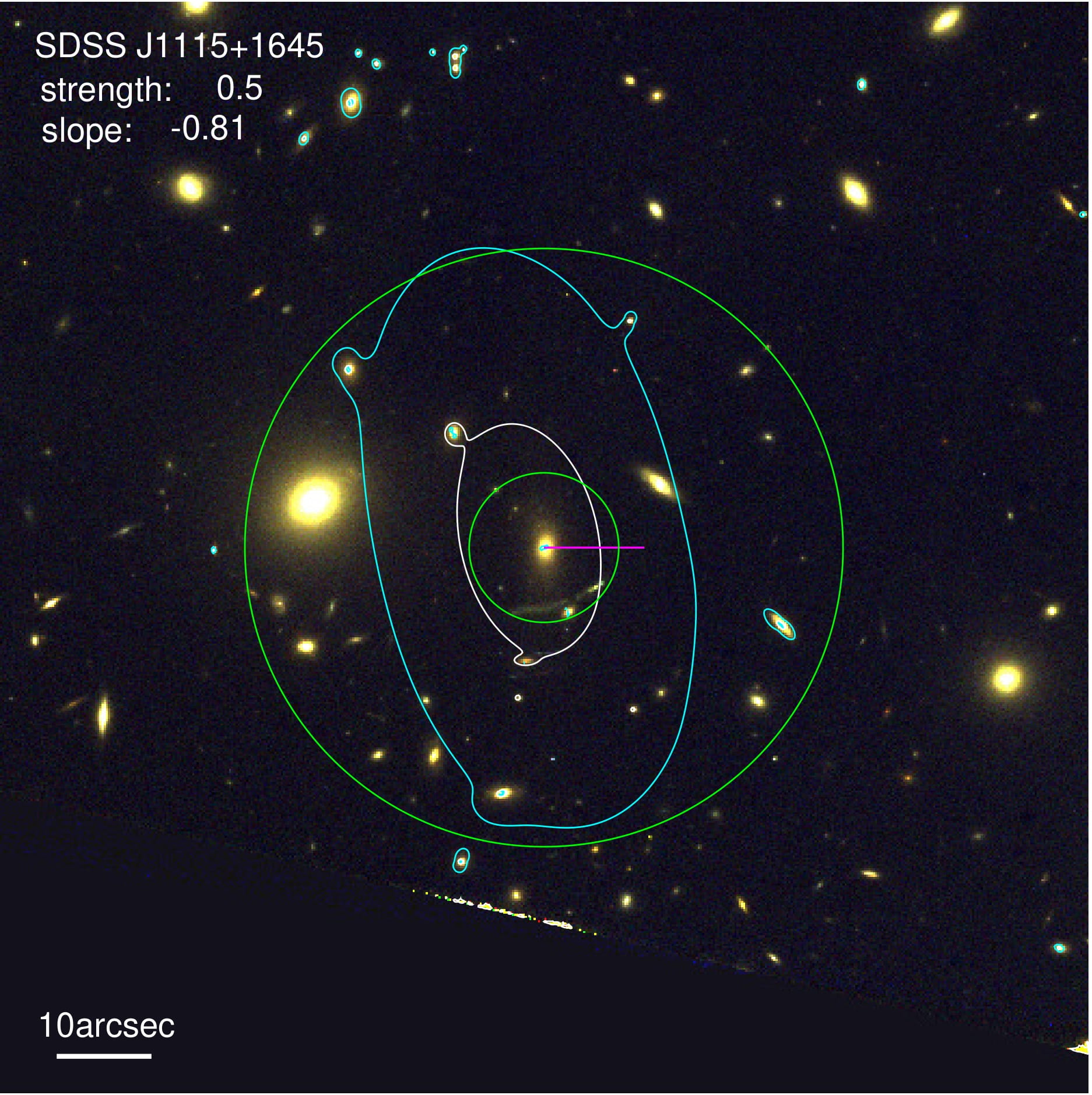}\\
    \includegraphics[width=0.23\textwidth]{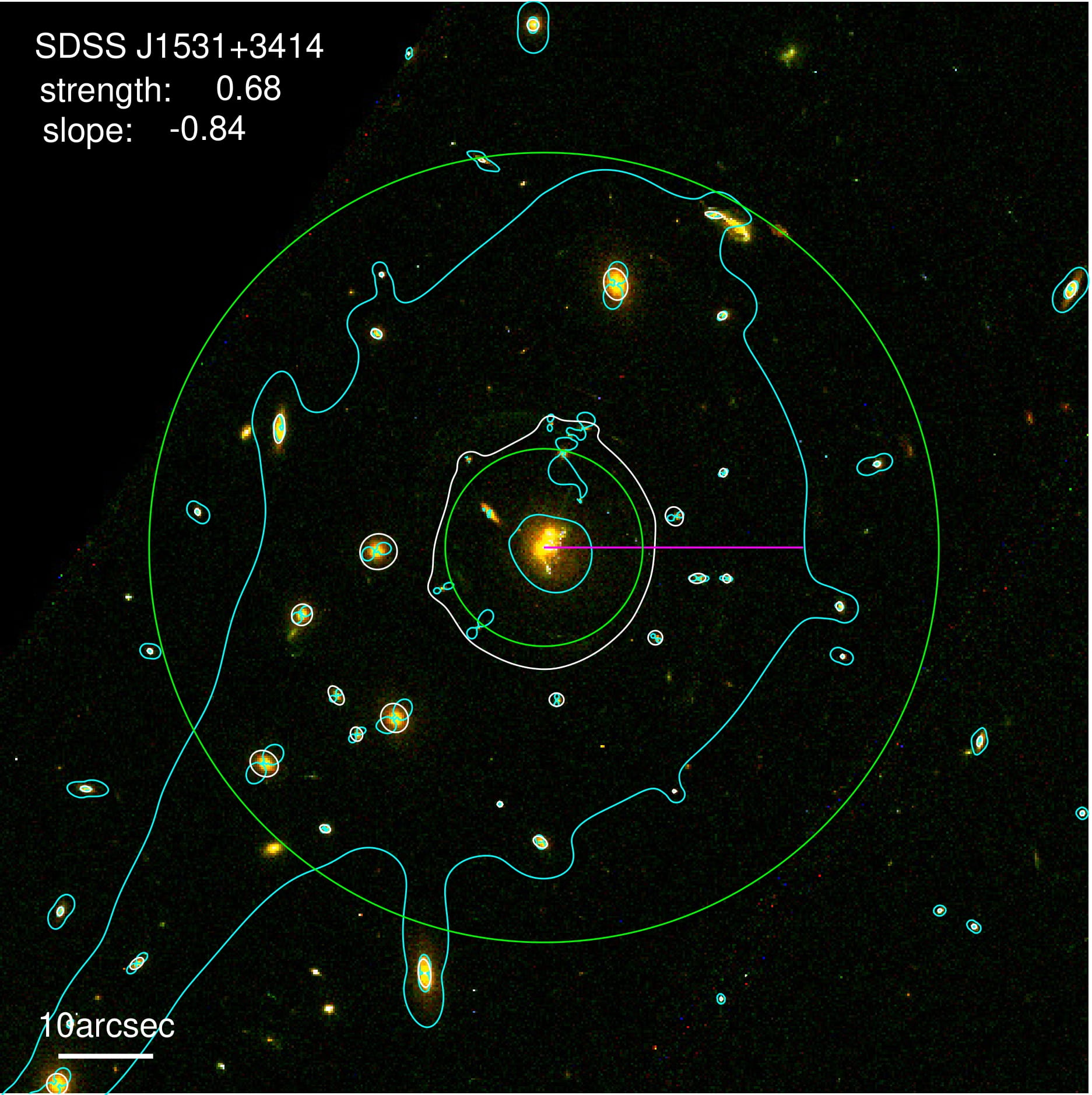}
    \includegraphics[width=0.23\textwidth]{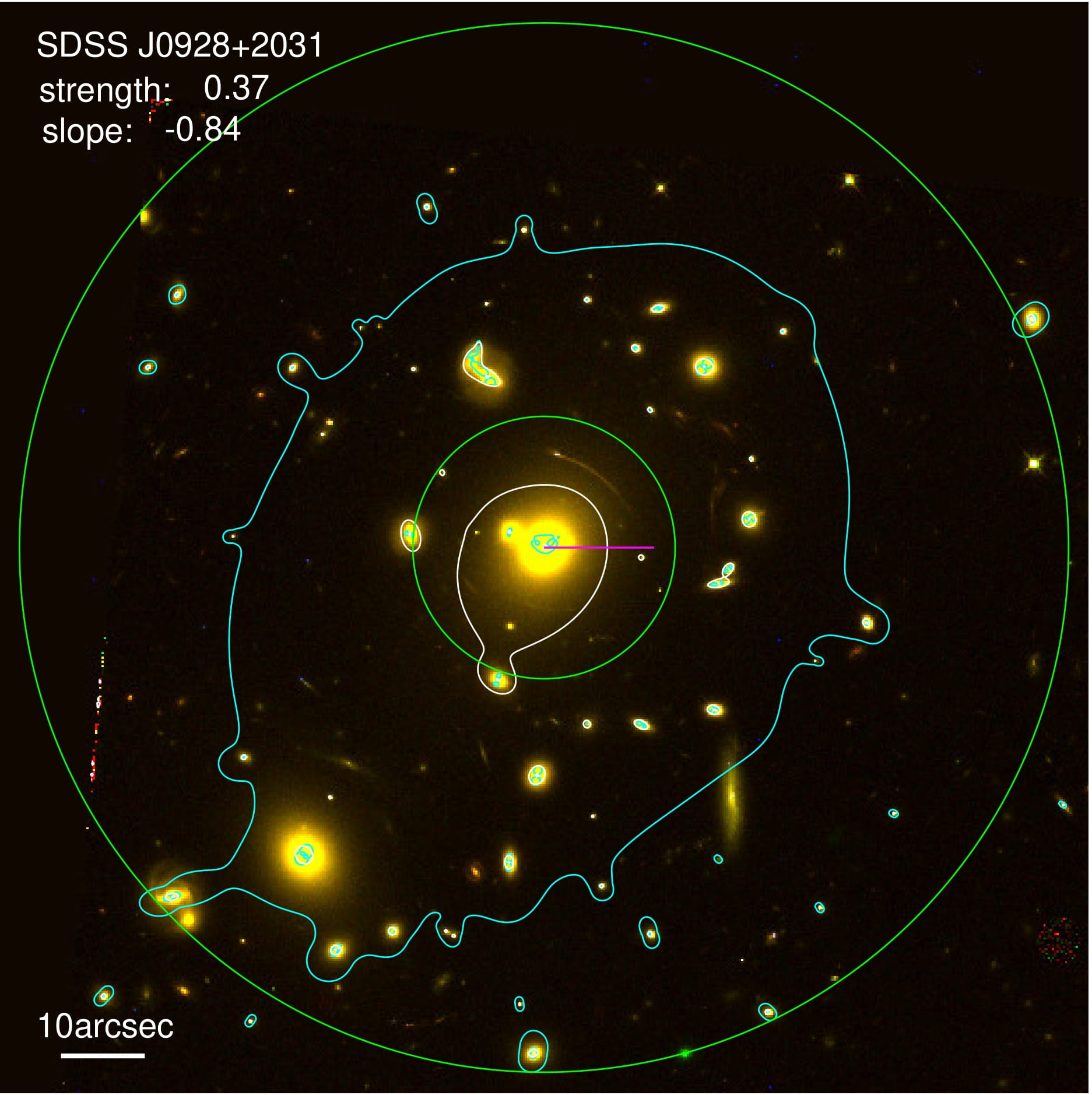}    \includegraphics[width=0.23\textwidth]{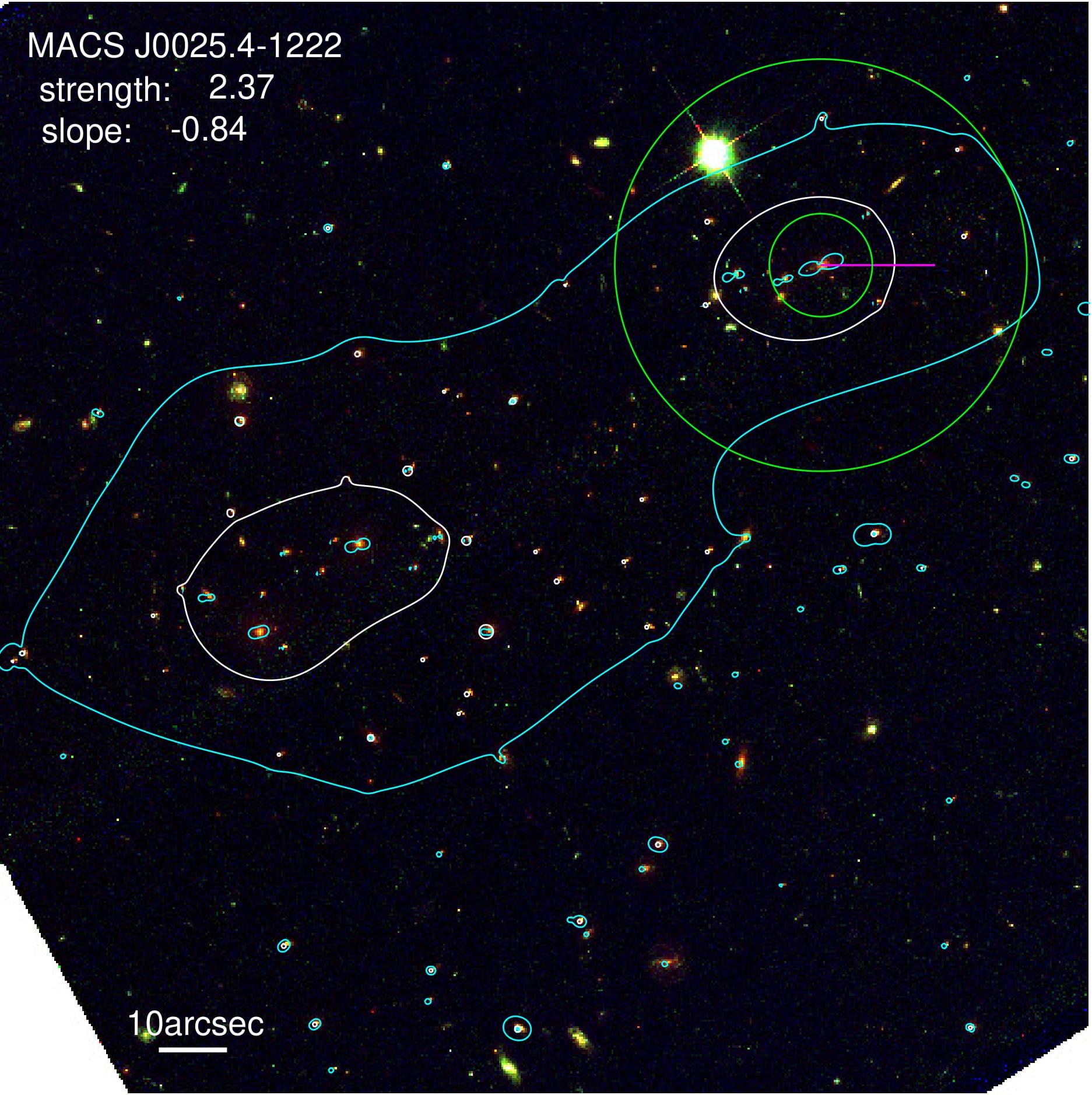}
    \includegraphics[width=0.23\textwidth]{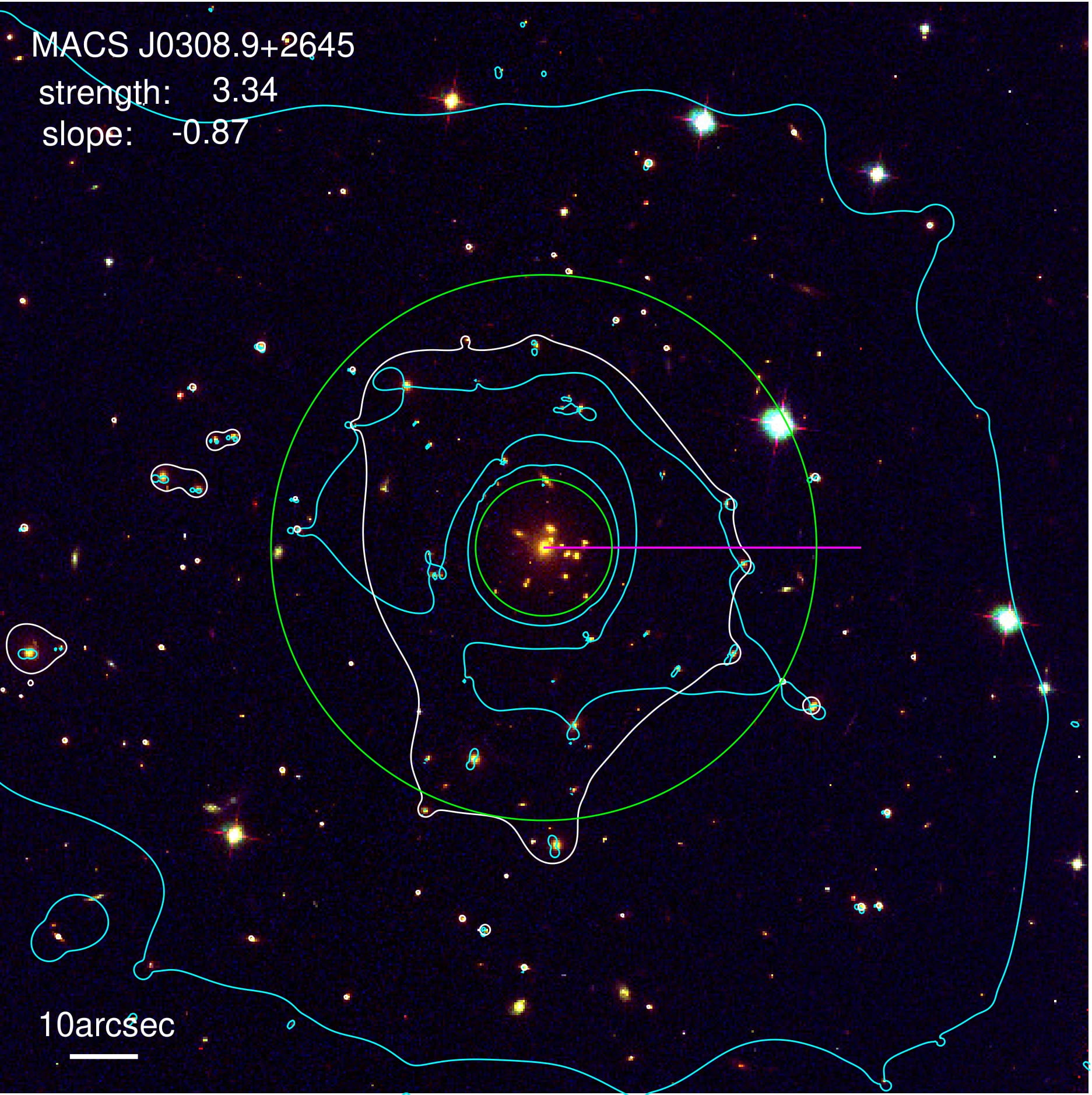}\\
    \caption{Same as Figure \ref{fig:gallery1}, for the clusters SDSS~J1156$+$1911, PLCK~G004.5$-$19.5, MACS~J0035.4$-$2015, RXC~J0142.9$+$4438, SDSS~J1621$+$0607, SDSS~J1110$+$6459, RXC~J0949.8$+$1707, SDSS~J1209$+$2640, CL~J0152.7$-$1357, SDSS~J1336$−$0331, MS~1008.1$-$1224, SDSS~J1152$+$3313, PLCK~G209.79$+$10.23, SDSS~J1115$+$1645, SDSS~J1531$+$3414, SDSS~J0928$+$2031, MACS~J0025.4$-$1222, and MACS~J0308.9$+$2645.
    }
    \label{fig:gallery3}
\end{figure*}

\begin{figure*}
\center
    \includegraphics[width=0.23\textwidth]{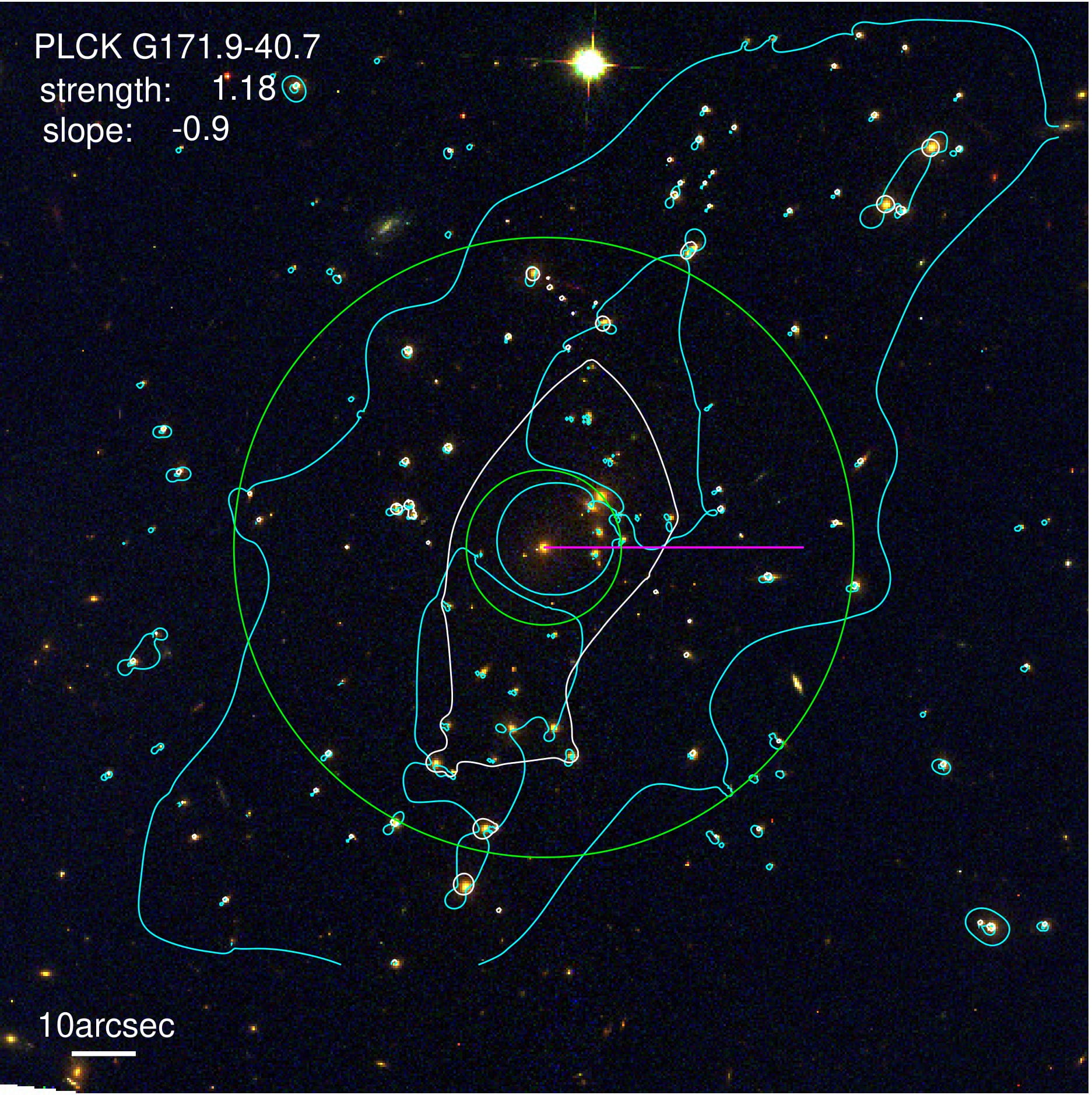} 
    \includegraphics[width=0.23\textwidth]{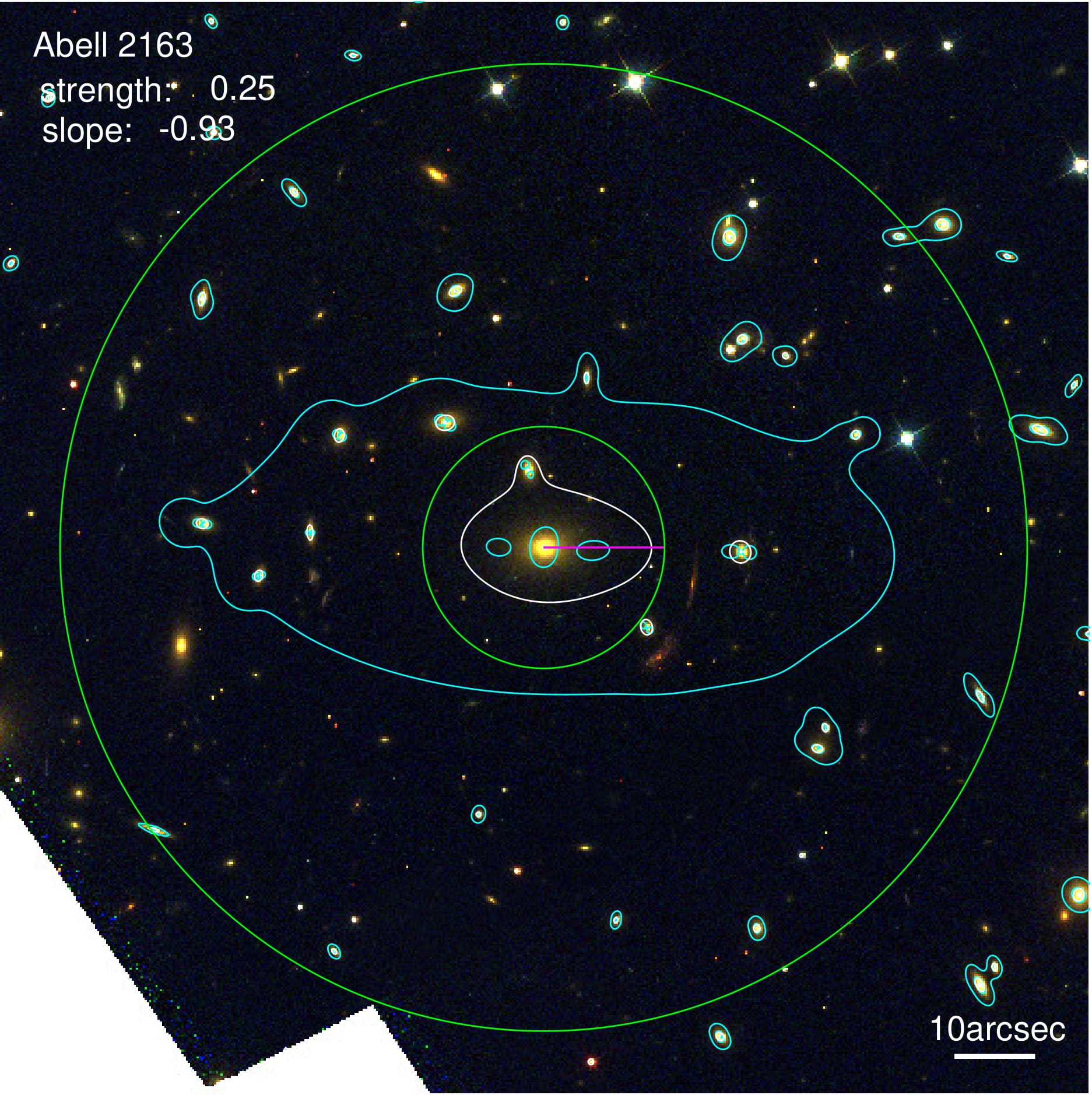}
    \includegraphics[width=0.23\textwidth]{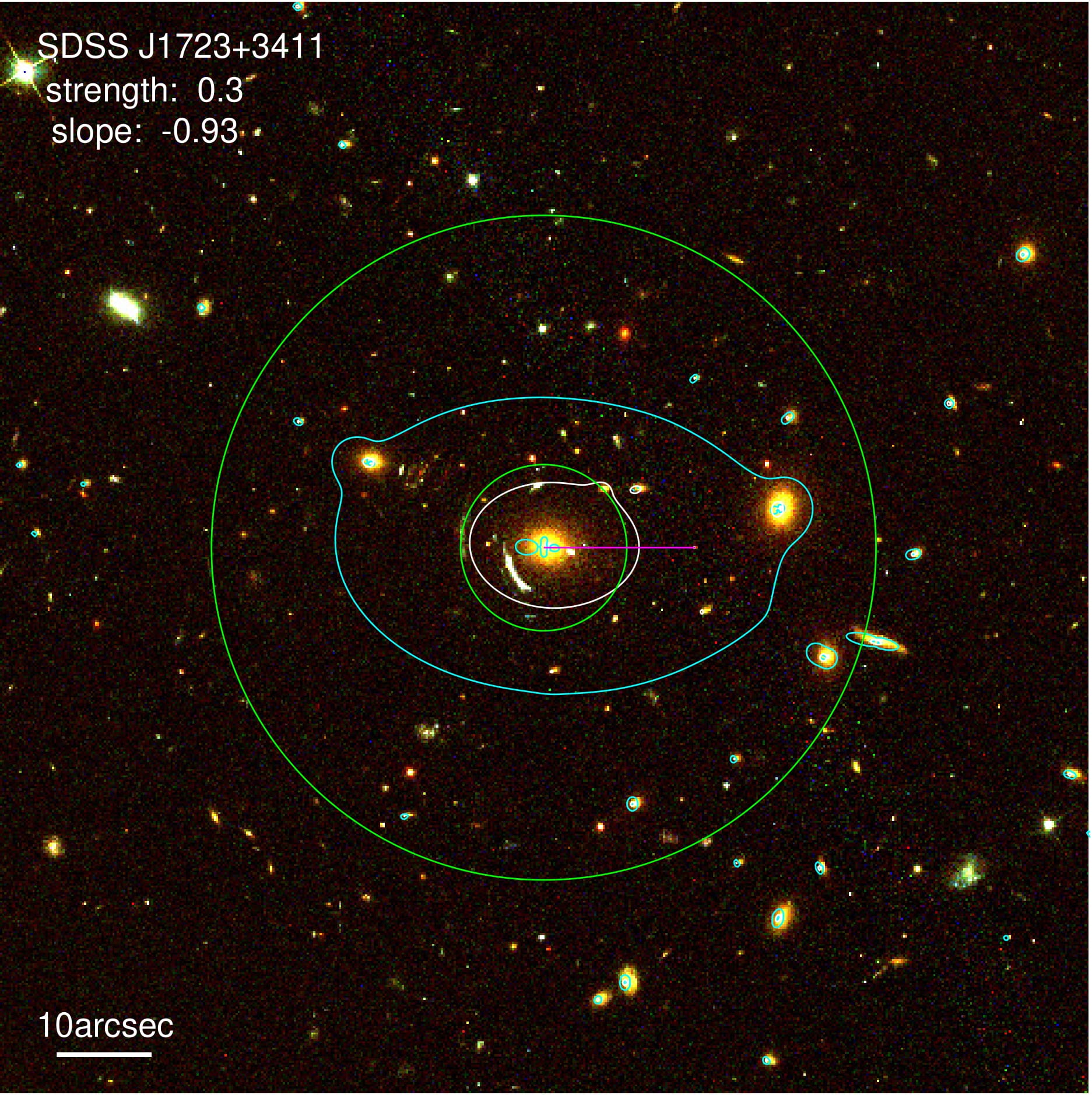}
    \includegraphics[width=0.23\textwidth]{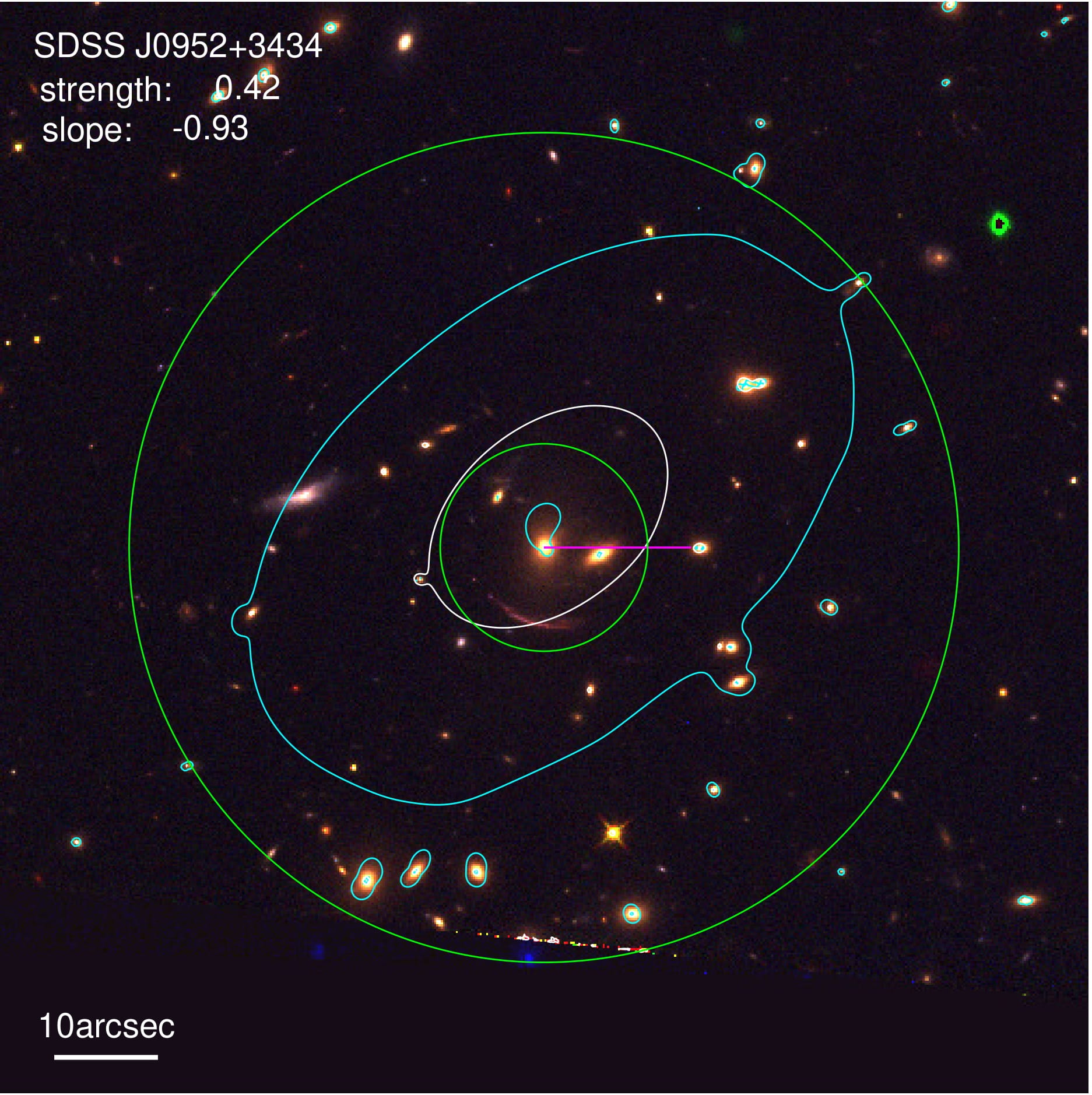}\\
    \includegraphics[width=0.23\textwidth]{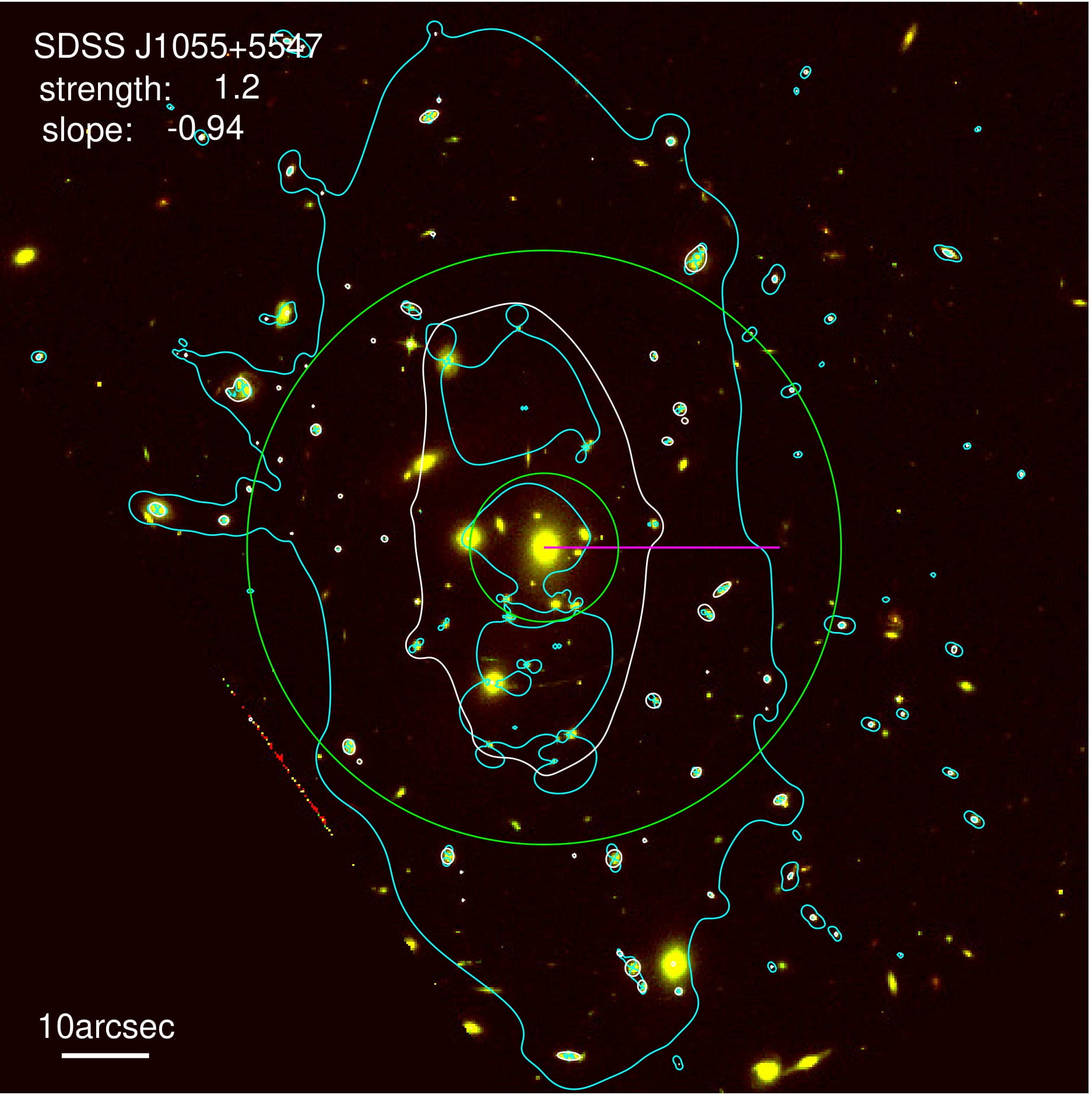}
    \includegraphics[width=0.23\textwidth]{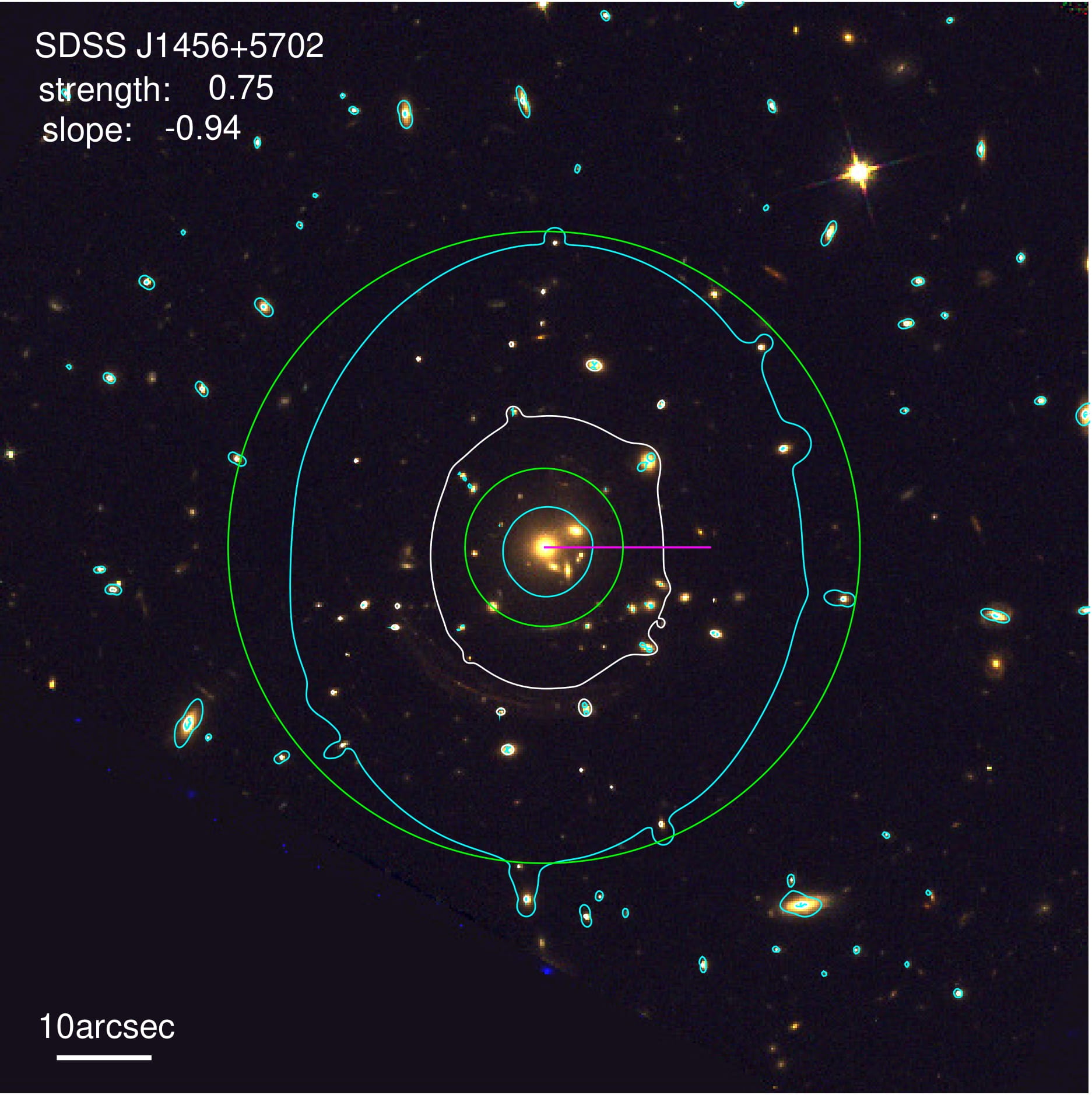}
    \includegraphics[width=0.23\textwidth]{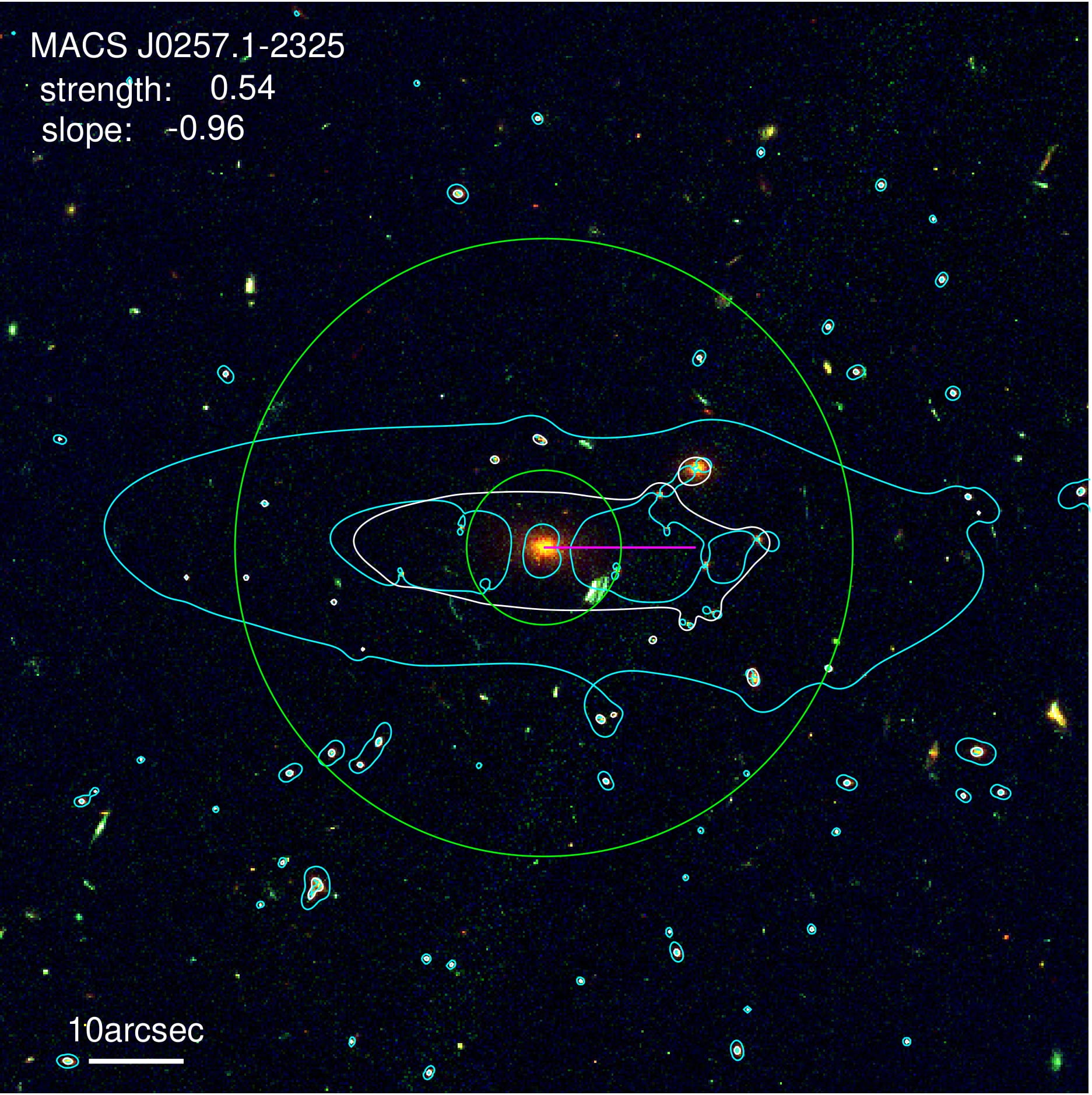}
    \includegraphics[width=0.23\textwidth]{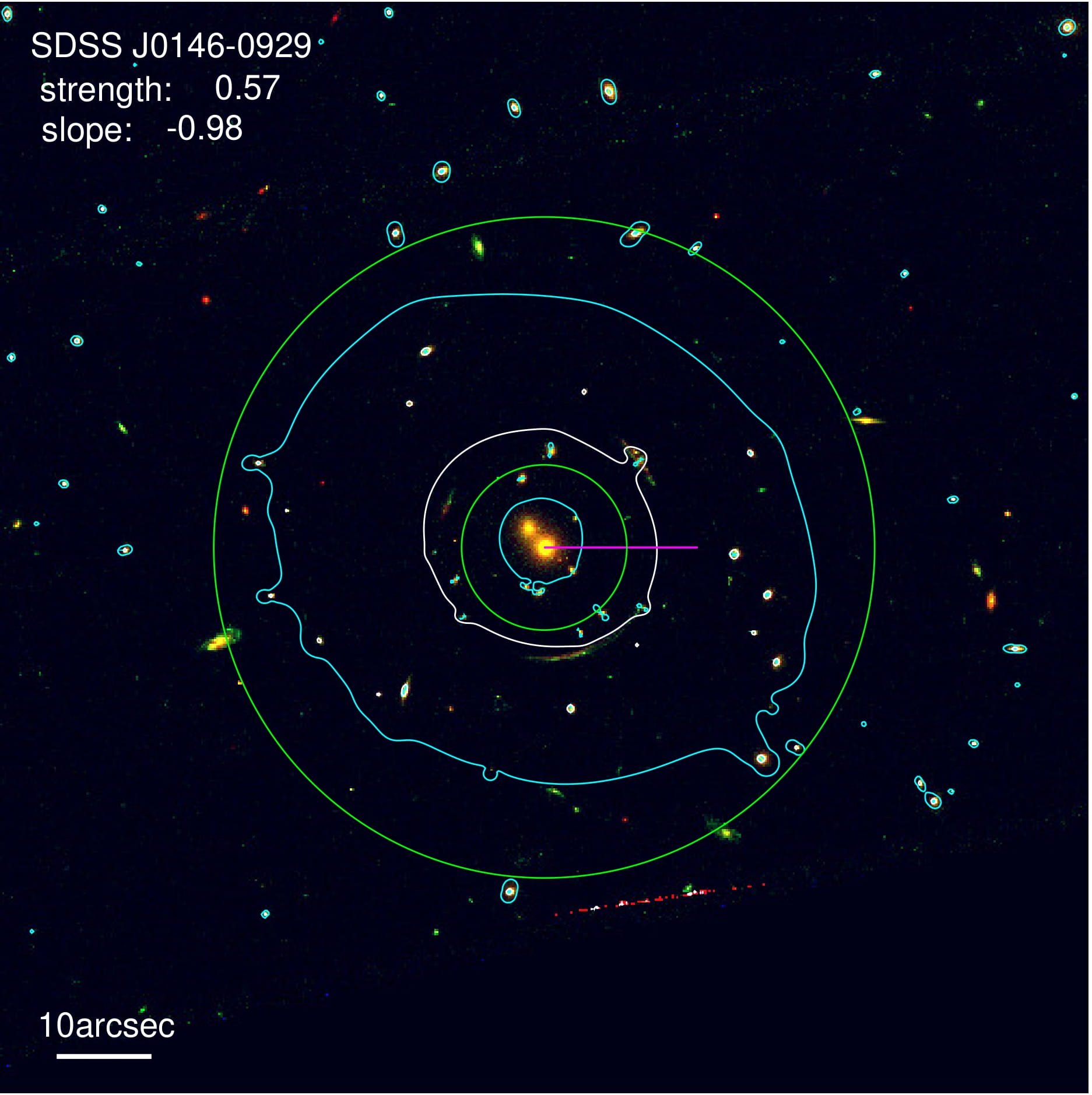}\\
    \includegraphics[width=0.23\textwidth]{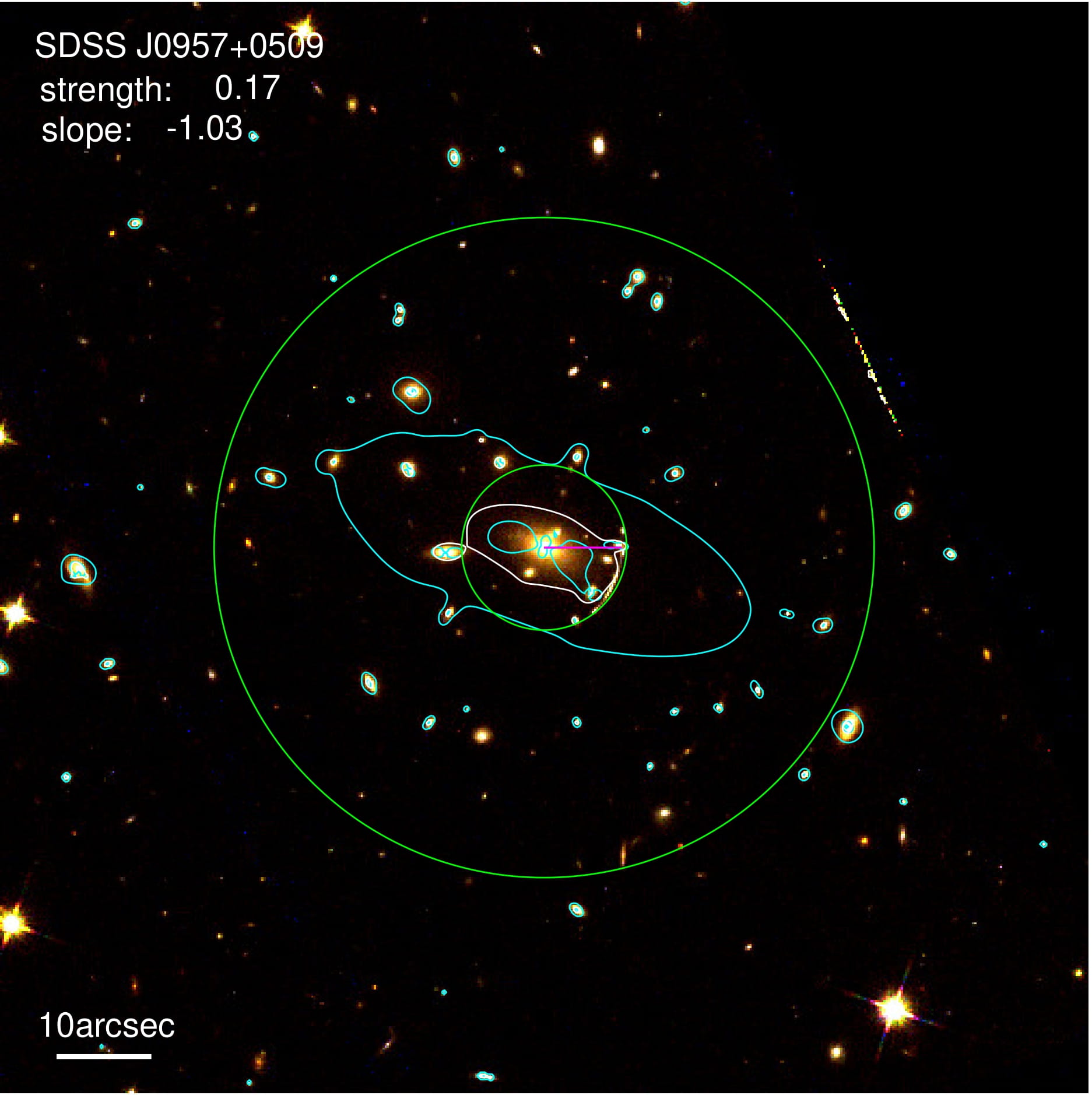}
    \includegraphics[width=0.23\textwidth]{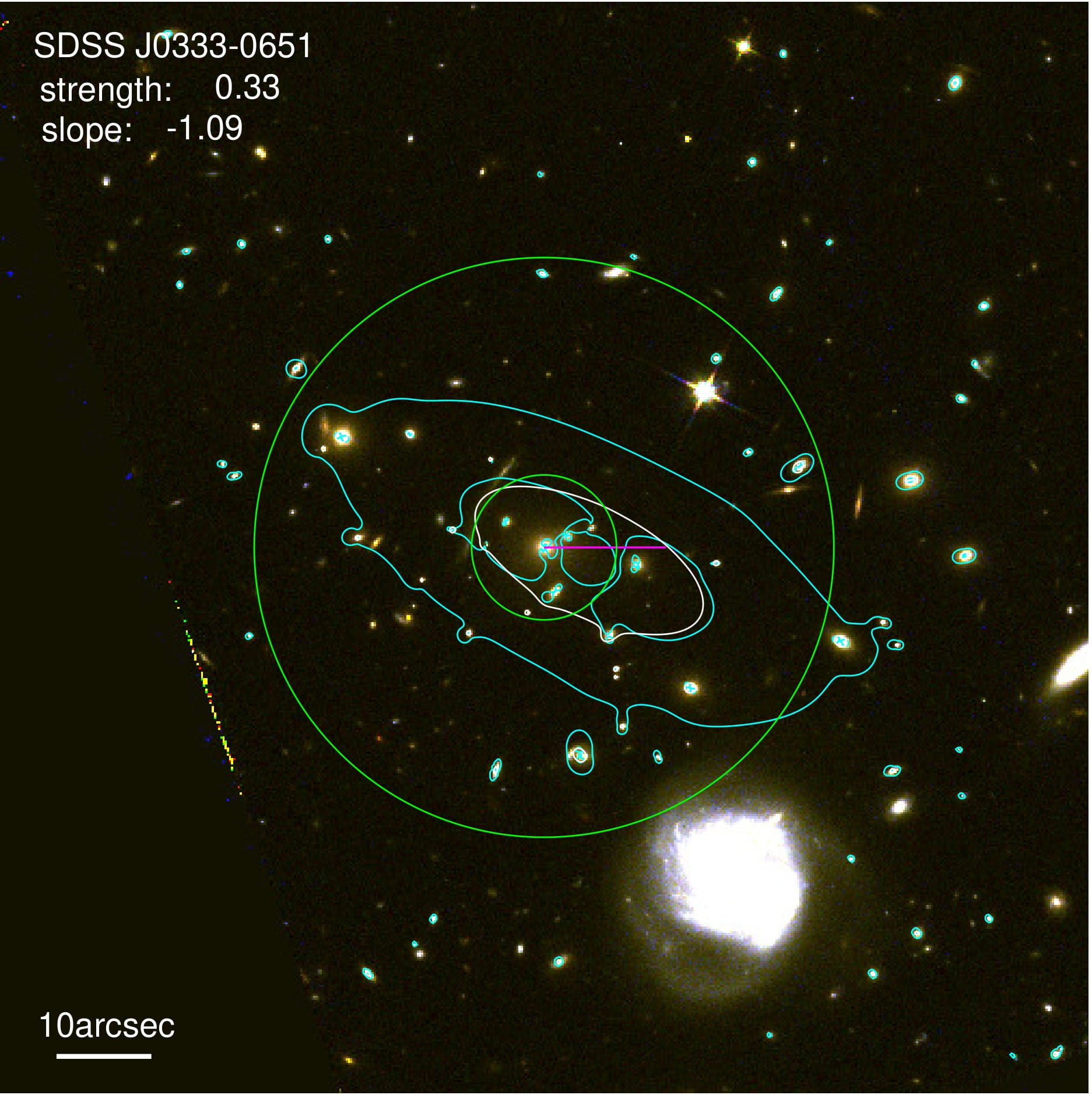}
    \includegraphics[width=0.23\textwidth]{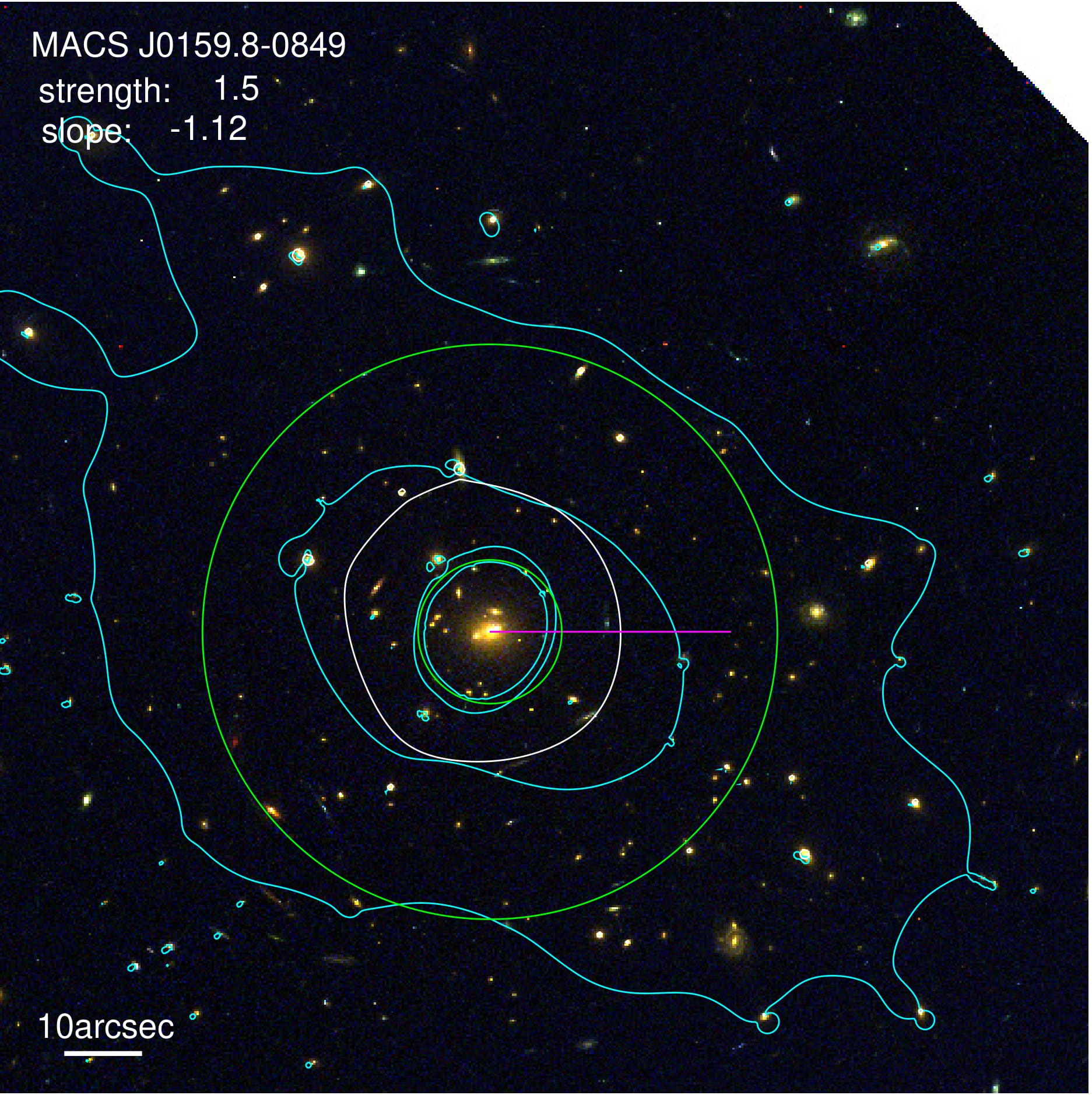}
    \includegraphics[width=0.23\textwidth]{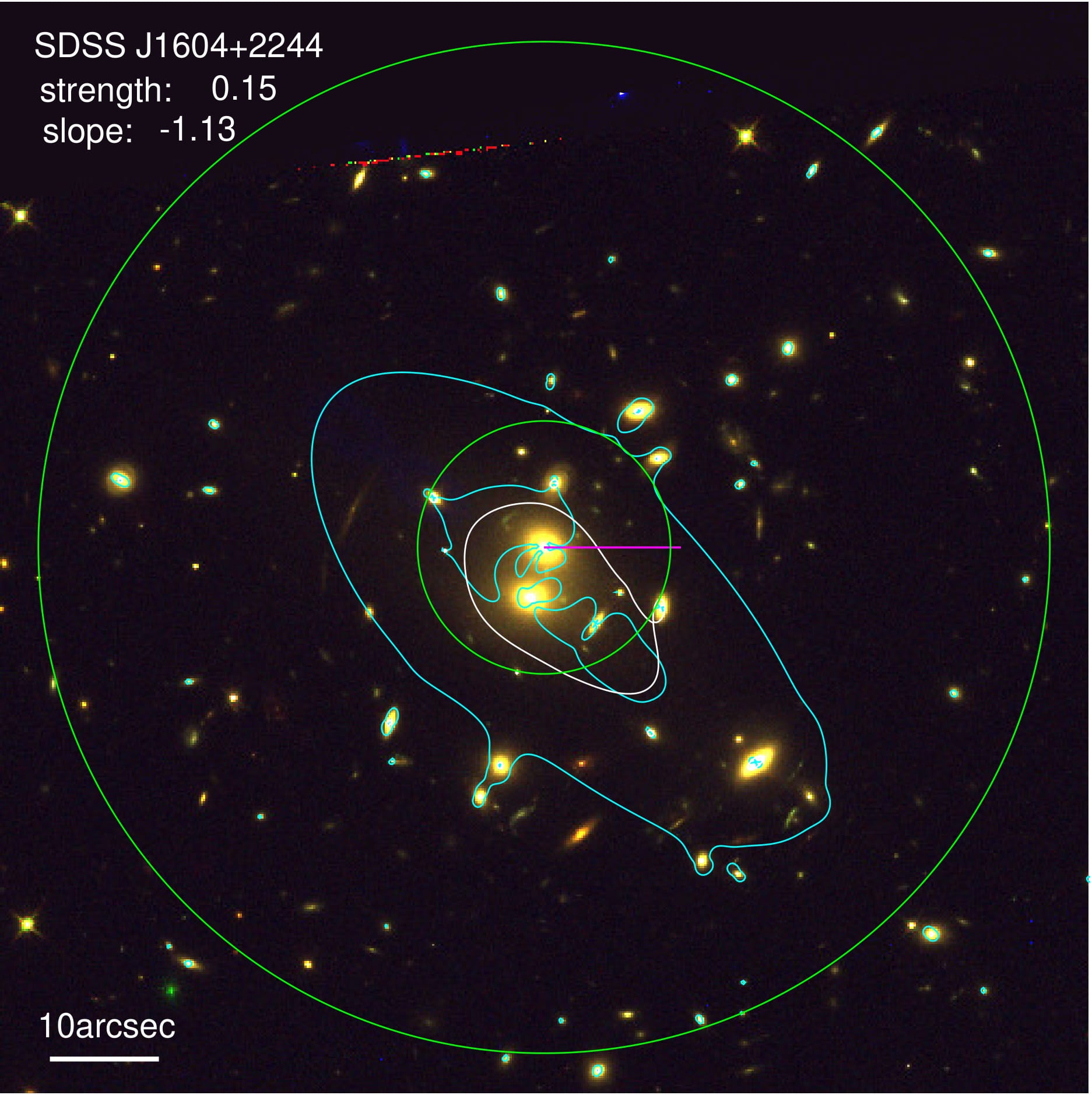}\\
    \includegraphics[width=0.23\textwidth]{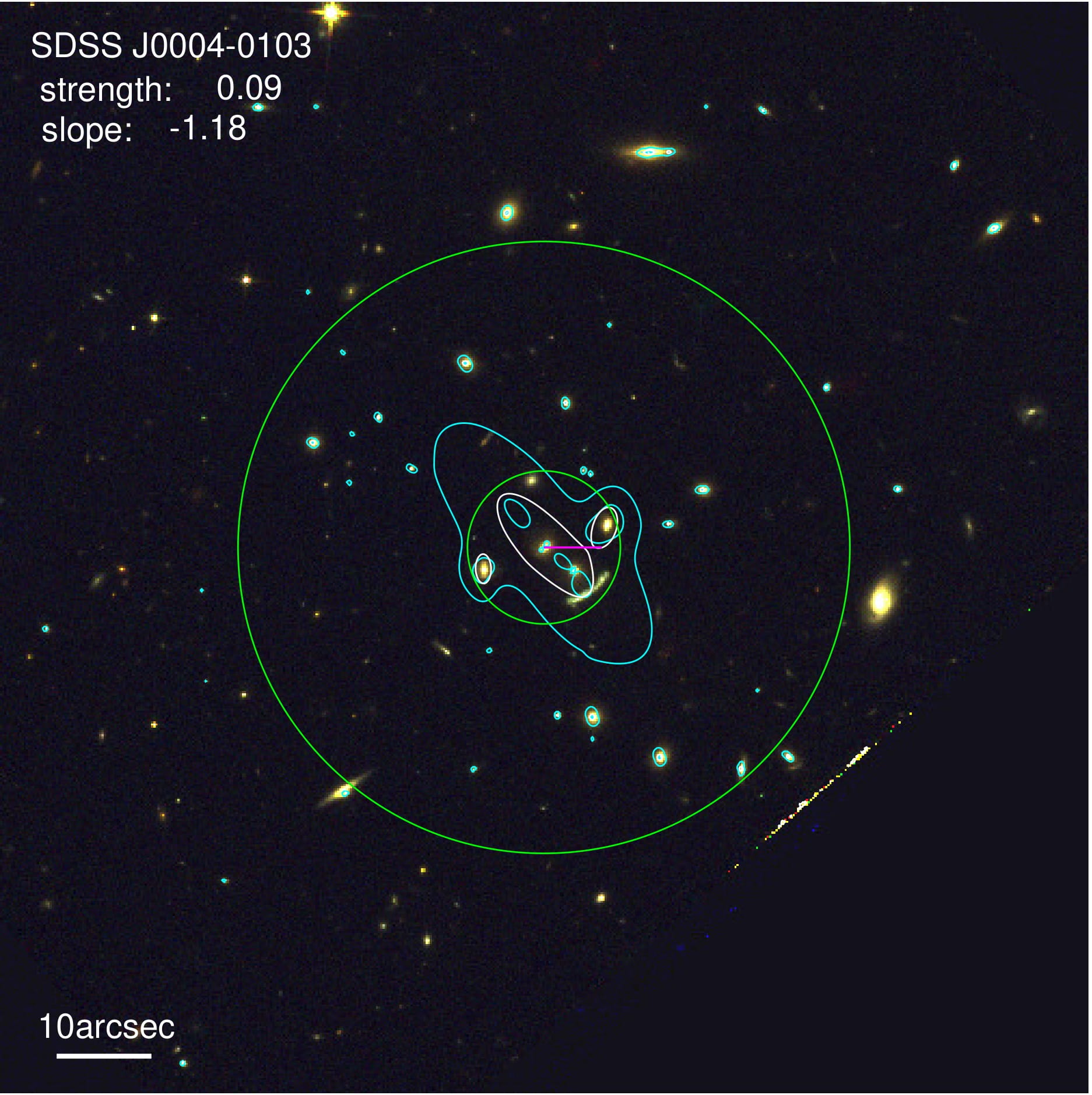}
    \includegraphics[width=0.23\textwidth]{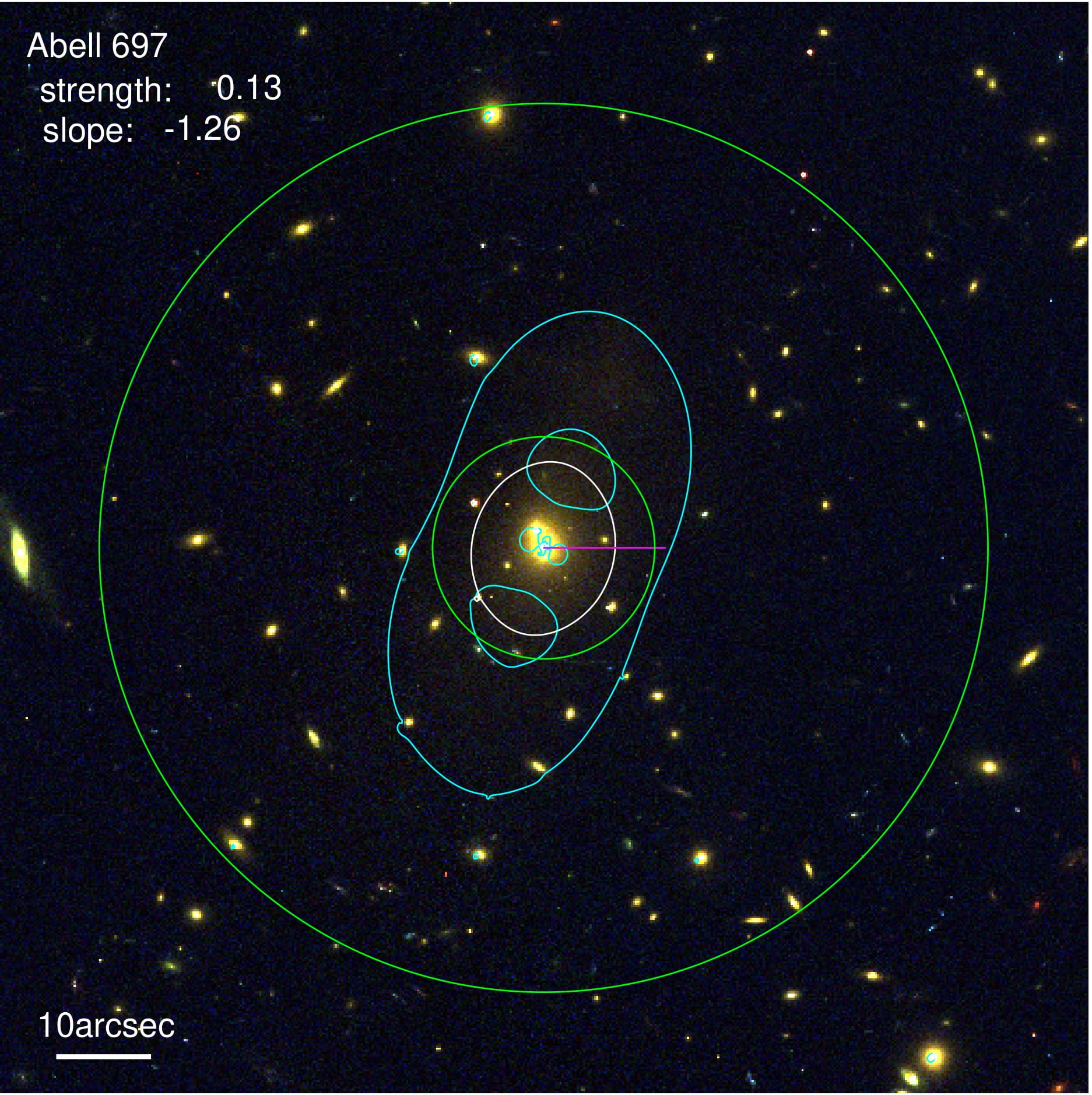}
    \includegraphics[width=0.23\textwidth]{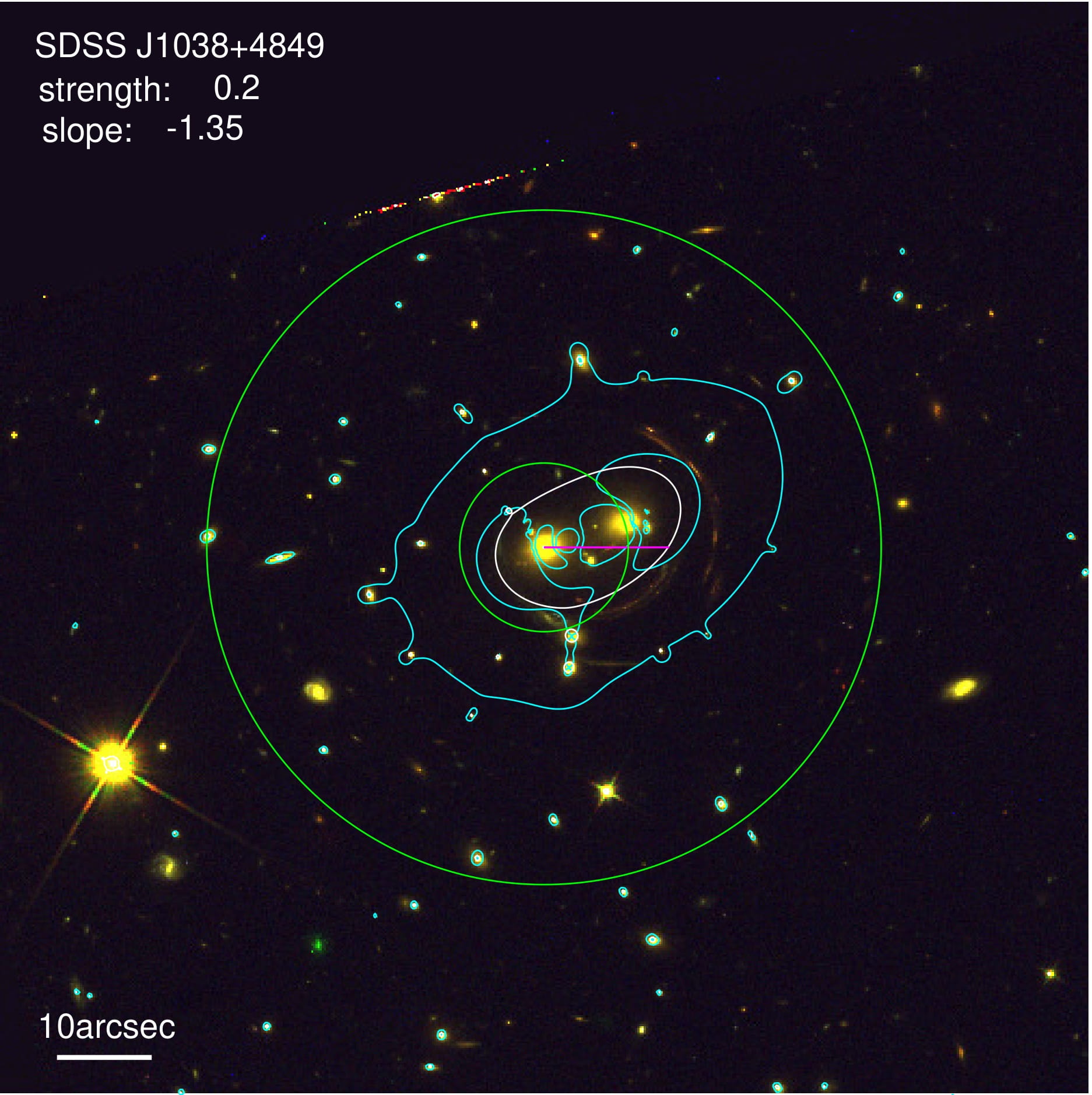}\\
\caption{Same as Figure \ref{fig:gallery1}, for the clusters PLCK~G171.9$+$40.7, Abell~2163, SDSS~J1723$+$3411, SDSS~J0952$+$3434, SDSS~J1055$+$5547, SDSS~J1456$+$5702, MACS~J0257.1$-$2325, SDSS~J0146$−$0929, SDSS~J0957$+$0509, SDSS~J0333$−$0651, MACS~J0159.8$-$0849, SDSS~J1604$+$2244, SDSS~J0004$−$0103, Abell~697, and SDSS~J1038$+$4849.
    }
    \label{fig:gallery4}
\end{figure*}

In Figures \ref{fig:gallery1}-\ref{fig:gallery4} we provide a gallery of color images for each cluster. Green circles mark $50$ kpc and $200$ kpc from the BCG assumed. We include the slope and normalized lensing strength. White contours mark the projected mass density at $\kappa = 1$, scaled to $D_{LS}/D_S = 1$. Cyan contours mark the magnification at $|\mu|\geq3$. The effective Einstein radius is shown in magenta. \Lenstool\ models are used where available (using the Sharon models for the HFF clusters), LTM if there is no \Lenstool\ model, and GLAFIC if there is no \Lenstool\ or LTM model. Clusters are ordered by slope, from flattest to steepest. 

\section{Comparisons with Non-normalized Lensing Strength}
\begin{figure*}
    \centering
    \includegraphics[width=\textwidth]{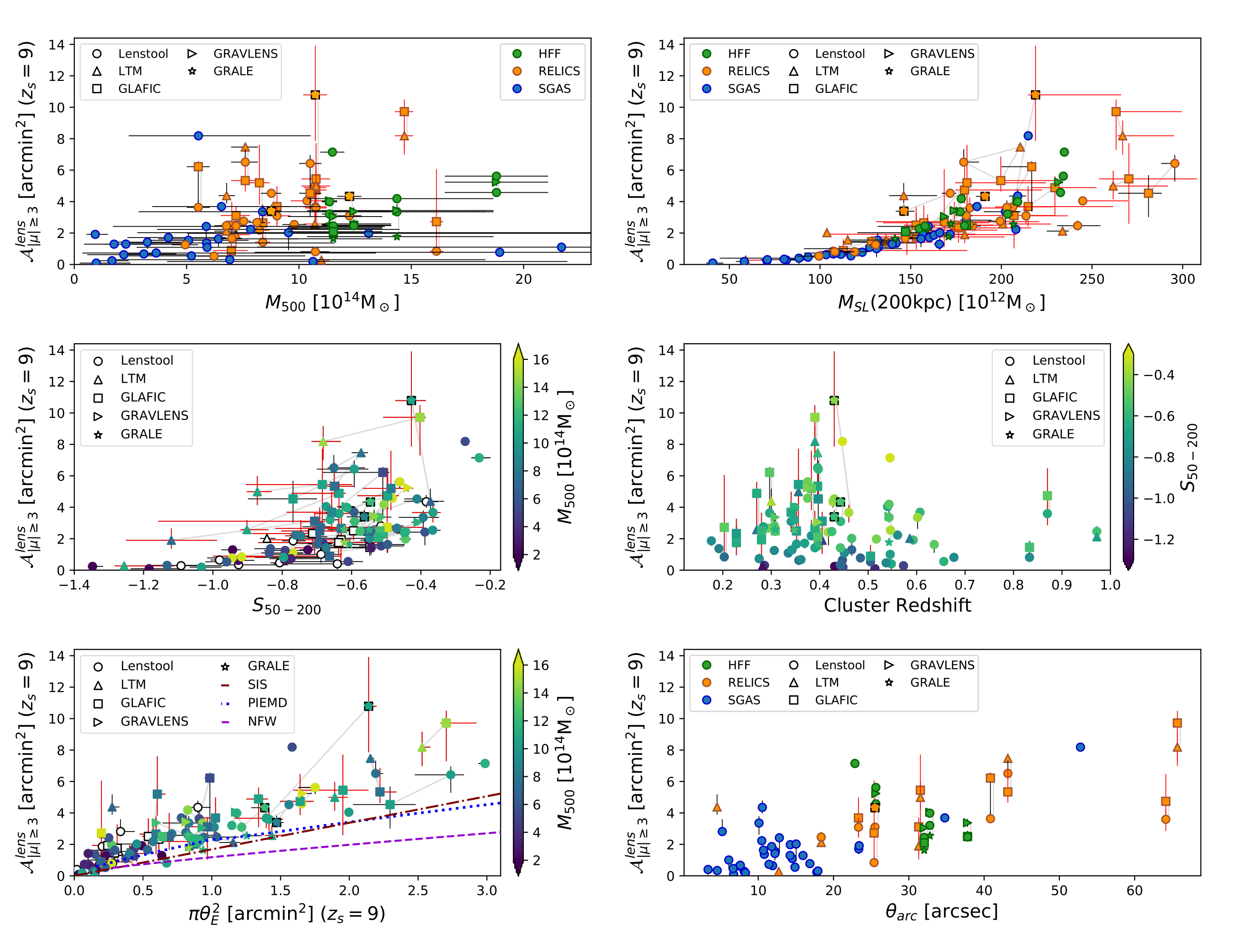}
    \caption{Non-normalized lensing strength (\Lstrength) compared with cluster properties. \textit{Top left:} Large scale mass estimate ($M_{500}$), which is measured independently of the lensing analyses.  \textit{Top Right:} Projected mass within $200$ kpc of the BCG (\MSL). \textit{Middle Left:} Inner slope of the projected mass density profile ($S_{50-200}$), color coded by $M_{500}$. Clusters without an $M_{500}$ estimate are plotted in white. \textit{Middle Right:} Cluster redshift, color coded by $S_{50-200}$. 
    \textit{Bottom Left:} Effective Einstein area (\Earea), color coded by $M_{500}$. Clusters without an $M_{500}$ estimate are plotted in white. The curves show the expected relationship for different potentials at a redshift of $z_{lens}=0.5$, and no ellipticity. Maroon shows an SIS distribution; Blue shows a PIEMD with a core radius of 40~kpc, and a cut radius of 1500kpc; Purple shows an NFW profile with a scale radius of 100~kpc. \textit{Bottom Right:} Distance between the BCG and farthest bright arc ($\theta_{arc}$). The shape of the data points indicates different lens modeling algorithms and light grey lines connect models of the same cluster modeled by different algorithms. The three GLAFIC models where 5\% was added to the upper error bar due to the limited field of view are given a dashed black border. In all panels, red error bars indicate models without spectroscopic constraints or if it is not known whether a spectroscopic constraint is used.}
    \label{fig:merged_notnormed_lens_strength_vs_stuff_plots}
\end{figure*}

In order to compare cluster properties to lensing strength, we corrected for the redshift dependence of area projection onto the sky, converting all lensing strengths (\Lstrength) to their corresponding values if the lensing strength area was projected onto the sky from a redshift of $0.5$ (\LstrengthN). In Figure \ref{fig:merged_notnormed_lens_strength_vs_stuff_plots} we provide comparisons to lensing strength with the non-normalized value.

\section{Comparisons with $M_{200}$}
\begin{figure*}
    \centering
    \includegraphics[width=\textwidth]{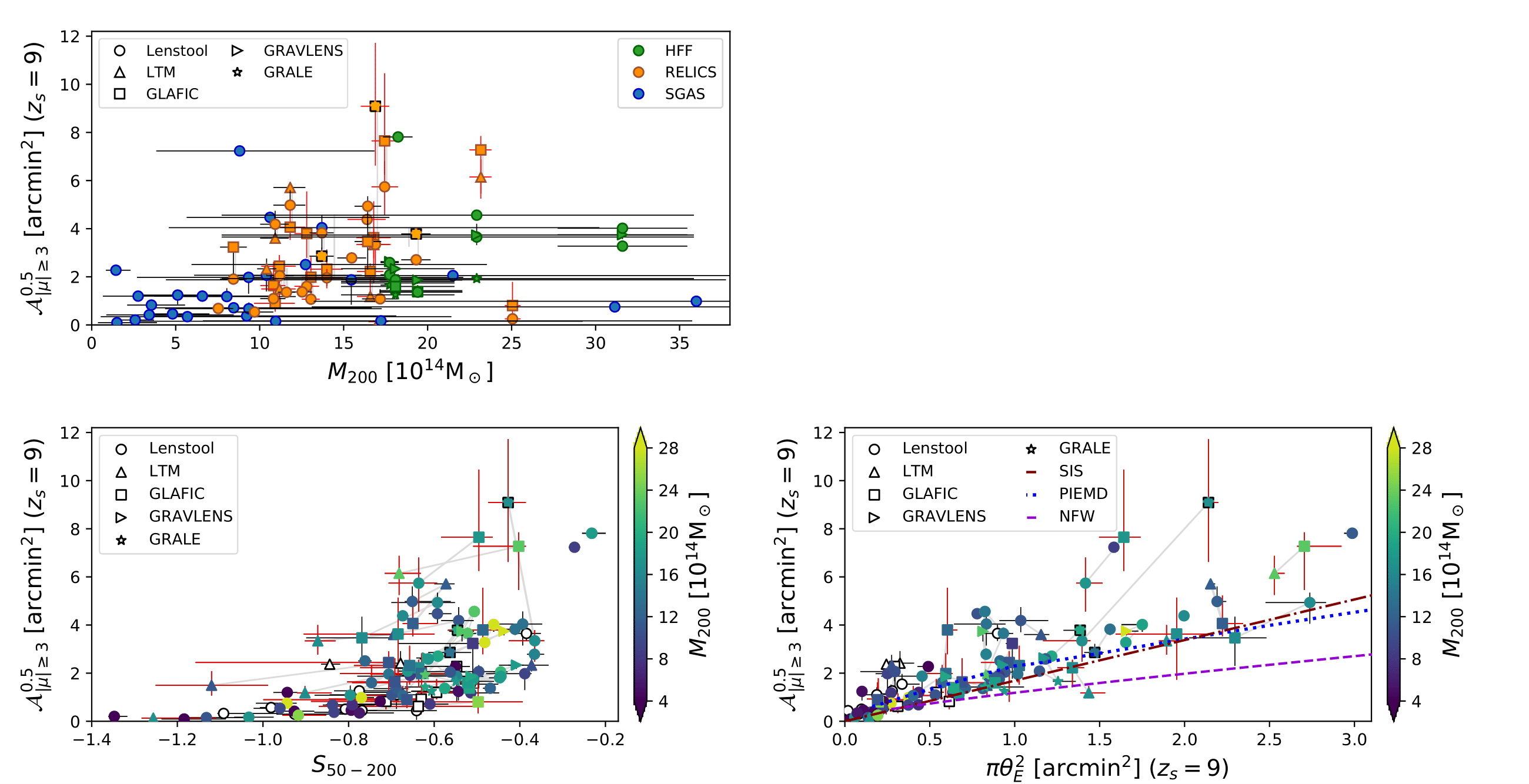}
    \caption{Normalized lensing strength (\LstrengthN) compared to cluster properties, with $M_{200}$ as the large scale mass estimate. \textit{Top left:} Large scale mass estimate ($M_{200}$), which is measured independently of the lensing analyses. \textit{Bottom Left:} Inner slope of the projected mass density profile ($S_{50-200}$), color coded by $M_{200}$. Clusters without an $M_{200}$ estimate are plotted in white. 
    \textit{Bottom Right:} Effective Einstein area (\Earea), color coded by $M_{200}$. Clusters without an $M_{200}$ estimate are plotted in white. The curves show the expected relationship for different potentials at a redshift of $z_{lens}=0.5$, and no ellipticity. Maroon shows an SIS distribution; Blue shows a PIEMD with a core radius of 40~kpc, and a cut radius of 1500~kpc; Purple shows an NFW profile with a scale radius of 100~kpc. The shape of the data points indicates different lens modeling algorithms and light grey lines connect models of the same cluster modeled by different algorithms. The three GLAFIC models where 5\% was added to the upper error bar due to the limited field of view are given a dashed black border. In all panels, red error bars indicate models without spectroscopic constraints or if it is not known whether a spectroscopic constraint is used.}
    \label{fig:m200merged_normed_lens_strength_vs_stuff_plots}
\end{figure*}

Throughout this paper we focus our analysis on $M_{500}$. In Figure \ref{fig:m200merged_normed_lens_strength_vs_stuff_plots}, we provide plots for comparison with $M_{200}$ for reference. To obtain an $M_{200}$ estimate for clusters included in the RELICS program we solve for the $M_{200}$ value corresponding to the \textit{Planck} SZ $M_{500}$ mass by means of the process outlined in Section \ref{sec:totalmass}. The same is done for Abell~S1063 and MACS~J0717.5$+$3745. For MACS~J1149.5$+$2223, MACS~J0416.1$-$2403, and Abell~2744, we use the $M_{500}$ computed from the X-ray temperature to solve for the corresponding $M_{200}$. For Abell~370, we use the $M_{200}$ estimate reported in \cite{2011ApJ...729..127U}. The median $M_{200}$ value is used for the SGAS clusters. 


\end{document}